\documentclass[a4paper]{article}
\pdfoutput=1
\usepackage[utf8]{inputenc}
\usepackage{geometry}
\usepackage{color}
\usepackage{mathtools}
\usepackage{amsmath}
\usepackage{amssymb}
\usepackage{setspace}
\usepackage{float}
\usepackage{jheppub}
\usepackage{subfig}
\usepackage{xcolor}
\usepackage{braket}
\usepackage{mathrsfs}
\allowdisplaybreaks

\def\beq{\begin{eqnarray}}
\def\eeq{\end{eqnarray}}

\newcommand{\reduced}{Fused }
\newcommand{\reducedWebs}{Fused-Webs }
\newcommand{\reducedWeb}{Fused-Web }

\newcommand{\txb}[1]{\textcolor{blue}{#1}}

\newcommand{\e}{\epsilon}

\newcommand{\as}{\alpha_s}

\title{Deciphering Colour Building Blocks of Massive Multiparton Amplitudes 
at 4-loops and beyond}
\author[]{Neelima Agarwal,}
\author[a]{Sourav Pal,}
\author[b]{Aditya Srivastav,}
\author[b]{Anurag Tripathi}
\affiliation[a]{Physical Research Laboratory, Ahmedabad - 380009, Gujarat, India}
\affiliation[b]{Department of Physics, Indian Institute of Technology Hyderabad, \\ Kandi, Sangareddy, Telangana State 502284, India}
\emailAdd{dragarwalphysics@gmail.com}
\emailAdd{sourav@prl.res.in}
\emailAdd{shrivastavadi333@gmail.com}
\emailAdd{tripathi.phy@iith.ac.in}

\abstract{
The soft function in non-abelian gauge theories exponentiate, and their logarithms can be organised in terms of the collections of Feynman diagrams called Cwebs. The colour factors that appear in the logarithm are controlled by the web mixing matrices. Direct construction of the diagonal blocks of Cwebs using the new concepts of Normal ordering, basis Cweb and Fused-Web was recently carried out  in~\cite{Agarwal:2022wyk}.  In this article we establish correspondence between the boomerang webs introduced in~\cite{Gardi:2021gzz} and non-boomerang Cwebs. We use this correspondence  together with Uniqueness theorem and Fused web formalism introduced in~\cite{Agarwal:2022wyk} to obtain the diagonal blocks of four general classes of Cwebs to all orders in perturbation theory which also cover all the four loop Boomerang Cwebs connecting four Wilson lines. We also fully construct the mixing matrix of a special Cweb to all orders in perturbation theory.} 
\usepackage{slashed}
\usepackage{lineno}
\begin{document}
	\maketitle
\section{Introduction}


The infrared (IR) structure of scattering amplitude  has a long history of almost a century~\cite{Bloch:1937pw,Sudakov:1954sw,Yennie:1961ad,Kinoshita:1962ur,Lee:1964is,Grammer:1973db,Mueller:1979ih,Collins:1980ih,Sen:1981sd,
	Sen:1982bt,Korchemsky:1987wg,Korchemsky:1988hd,Magnea:1990zb,
	Dixon:2008gr,Gardi:2009qi,Becher:2009qa,Feige:2014wja} and still remains an interesting object of study. These structures are universal, that is they are independent of the relevant hard scattering process. The universality of these structures gives us remarkable all order insights in their organisation at all orders in the perturbation theory. Although, these studies are completely theoretical, it has practical applications in the study of high energy scattering cross-sections in different colliders. These  IR singularities get cancelled in an infrared safe observable, such as total cross-section; however, they leave their signatures in large logarithms of kinematic invariants, which disturbs the perturbative expansions. It is the universality of the IR structure which helps in summing up these large logarithms at all orders in the perturbation theory to make a precise prediction. Along with these all order predictions, the knowledge about the structure of these singularities also helps in the fixed order calculations. The cancellation of these IR singularities for complicated observables in colliders is non-trivial task, and using the universality of the IR singularities, several efficient subtraction procedures are developed~\cite{GehrmannDeRidder:2005cm,Somogyi:2005xz,Catani:2007vq,
	Czakon:2010td,Boughezal:2015dva,Sborlini:2016hat,Caola:2017dug,
	Herzog:2018ily,Magnea:2018hab,Magnea:2018ebr,Capatti:2020xjc}.        

The factorization property of QCD in the IR limit enables us in studying these singular parts efficiently, without calculating the complicated hard parts for a process. The soft function, that controls the IR singular parts in a scattering process, can be expressed in terms of matrix elements of the Wilson lines. These matrix elements are also present in QCD based effective theories ~\cite{Manohar:2000dt,
	Brambilla:2004jw,Becher:2014oda}. The usual Wilson-line operators $\Phi ( \zeta )$  evaluated on smooth space-time contours $\zeta$ are defined as,
\begin{align}
\Phi \left(  \zeta \right) \, \equiv \, \mathcal{P} \exp \left[ {\rm i} g \!
\int_\zeta d x \cdot {\bf A} (x) \right] \, .
\label{genWL}
\end{align}
where ${\bf A}^\mu (x) = A^\mu_a (x) \, {\bf T}^a$ is a non-abelian gauge field, 
and ${\bf T}^a$ is a generator of the gauge algebra, which can be taken to belong
to any desired representation, and $ \mathcal{P} $ denotes path ordering of the gauge fields. The soft function describing the interaction of $ n $ hard particles in general can be expressed as, 
\begin{align}
{\cal S}_n \left( \zeta_i \right) \, \equiv \, \bra{0} \prod_{k = 1}^n
\Phi \left(  \zeta_k \right) \ket{0} \, ,
\label{genWLC}
\end{align}
These Wilson lines are semi-infinite Wilson lines along the direction of the hard particle, that is, the smooth contours run along $ \beta_k $, the velocities of the particles involved in this scattering, and has limit from origin to $ \infty $. Thus, one can write the soft function as, 
\begin{align}
{\cal S}_n \Big( \beta_i \cdot \beta_j, \as (\mu^2), \e \Big) \, \equiv \, 
\bra{0} \prod_{k = 1}^n \Phi_{\beta_k} \left( \infty, 0 \right) \ket{0} , \quad
\Phi_\beta \left( \infty, 0 \right) \, \equiv \, \mathcal{P} \exp \left[ {\rm i} g \!
\int_0^\infty d \lambda \, \beta \cdot {\bf A} (\lambda \beta) \right] ,
\label{softWLC}
\end{align}
The object $ \mathcal{S} $ suffers from both UV and IR (soft) singularities, and needs renormalization. In dimensional regularization $ d=4-2\e $, $ \mathcal{S} $  vanishes as it is made out of Wilson line correlators, that involve only scaleless integrals. Thus, in a renormalized theory, $ \mathcal{S} $ contains pure UV counterterms. 

The renormalized soft function obeys a renormalization group equation, and solving this equation results in an exponentiation of the form, 
\begin{align}
\mathcal{S}_n \Big( \beta_i \cdot \beta_j, \as (\mu^2), \e \Big) \, = \, 
\mathcal{P} \exp \left[ - \frac{1}{2} \int_{0}^{\mu^2} \frac{d \lambda^2}  
{\lambda^2} \, {\bf \Gamma}_n \Big( \beta_i \cdot \beta_j, \alpha_s (\lambda^2), 
\e \Big) \right]  \, ,
\label{eq:softmatr}
\end{align}    
where $ {\bf \Gamma}_n $ is known as the soft anomalous dimension. In case of multi-parton scatterings, the soft anomalous dimension is a matrix, which is an interesting theoretical object to  study, and has our main focus in this article. The perturbative calculation of soft-anomalous dimension using renormalization group approach has a history of more than twenty years. ${\bf \Gamma}_n$ was computed at one loop in~\cite{Kidonakis-1998} 
(see also~\cite{Korchemskaya:1994qp}); at two loops in the massless case 
in~\cite{Aybat:2006wq,Aybat:2006mz}, and in the massive case in~\cite{Mitov:2009sv,
	Ferroglia:2009ep,Ferroglia:2009ii,Kidonakis:2009ev,Chien:2011wz}; finally, at three 
loops in the massless case in~\cite{Almelid:2015jia,Almelid:2017qju}. The calculation of soft anomalous dimension at four loops is an ongoing effort. Several studies in this direction are available in the literature in~\cite{Becher:2019avh,Falcioni:2020lvv,Falcioni:2021buo,Falcioni:2021ymu,Catani:2019nqv,Moch:2017uml,Ahrens:2012qz,Moch:2018wjh,Chetyrkin:2017bjc,Das:2019btv,Das:2020adl,vonManteuffel:2020vjv,Henn:2019swt,Duhr:2022cob}.  The kinematic dependence of the soft anomalous dimension for scatterings, which involve only massless lines is restricted due to constraints discussed in  \cite{Gardi:2009qi,Becher:2009cu,Becher:2009qa}. However, these constraints do not hold true for scatterings involving massive particles. The state-of-the-art knowledge for soft anomalous dimension is known upto two loops for scatterings involving massive particle.

An alternative way of determining the soft anomalous dimension is using diagrammatic exponentiation. In terms of Feynman diagrams, the soft function has the form, 
\begin{align}
{\cal S}_n \left( \gamma_i \right) \, = \, \exp \Big[ {\cal W}_n \left( \gamma_i \right) 
\Big]  \, ,
\label{diaxp}
\end{align}
where $ {\cal W}_n \left( \gamma_i \right) $ is known as \textit{webs}, and can be directly computed using Feynman diagrams. Webs are defined as the connected photon sub-diagrams in the abelian gauge theory, while in non-abelian gauge theory, for two Wilson line processes, webs are defined as two-line irreducible diagrams, that is, diagrams that remains connected upon cutting the Wilson lines~\cite{Sterman-1981,
	Gatheral,Frenkel-1984}.  In case of multi-parton
scattering process, webs in non-abelian gauge theory are defined as sets of diagrams that differ from each other by the order of gluon attachments on each Wilson line~\cite{Mitov:2010rp,Gardi:2010rn}. The kinematics and the colour factors of diagrams in a web mix among themselves, through a web mixing matrix, which can be determined using a replica trick algorithm~\cite{Gardi:2010rn,Laenen:2008gt}. 

Cwebs or correlator webs were defined in \cite{Agarwal:2020nyc},  as the webs which are made out of gluon correlators instead of individual Feynman diagrams. 
Complete results for  the required colour building blocks for the calculation of the soft anomalous dimension matrix at four-loop order were presented in \cite{Agarwal:2020nyc,Agarwal:2021ais} (see also~\cite{Pal:2022fbn}). 
It is of interest to understand the underlying structure of the web mixing matrices and also to find ways of obtaining these mixing matrices directly without using
the replica trick algorithm. A few attempts towards this goal have been made in 
~\cite{Dukes:2013gea,Dukes:2013wa,Dukes:2016ger} using combinatorial objects \textit{posets}. 
Recently, we  proposed a new method of {\it Fused-Web} formalism  in~\cite{Agarwal:2022wyk}, which  provides insights into the diagonal blocks of several classes of mixing matrices for Cwebs for massless Wilson lines  -- it was  shown in~\cite{Agarwal:2022wyk} that these diagonal blocks are mixing matrices of the basis Cwebs -- and also readily gives the form of these blocks without any additional computation.

For scattering processes involving massive particles, such as top quark, whose masses cannot be ignored one needs to also 
consider webs whose diagrams contain self energy corrections on the Wilson lines. Those webs  which contain at least one self energy correction to one of the Wilson lines are called boomerang webs~\cite{Gardi:2021gzz}.
The exponentiated colour factors and kinematic contributions for Boomerang webs at three loops were computed in~\cite{Gardi:2021gzz}.

Following the definition of Cwebs \cite{Agarwal:2020nyc}, we define Boomerang Cwebs as those Cwebs that contain at least one two-point gluon correlator whose both ends are attached to the same Wilson line. The Boomerang Cwebs are all order objects having their own perturbative expansion.

In this article,  we establish the correspondence between the basis Cwebs  introduced in ~\cite{Agarwal:2022wyk} and boomerang webs and using this correspondence along with the the concepts of Normal ordering and Fused-Webs~\cite{Agarwal:2022wyk} we determine the diagonal blocks, and ranks of all the mixing matrices of general four classes of Cwebs to all orders in the perturbation theory that contains all  boomerang Cwebs that connect four Wilson lines at four loops. We also present the complete mixing matrix for one special class of Boomerang Cwebs at all orders in the perturbation theory. We also demonstrate the correspondence of the basis Cwebs introduced in~\cite{Agarwal:2022wyk} with the combinatorial objects called as nodes and edges in graph theory.

The paper is structured as follows. In section \ref{sec:review}, we review the concepts of Cwebs and the known properties of mixing matrices. In section \ref{sec:Fused-Webs}, we review the concepts of Normal ordering and Fused Cwebs. In section \ref{sec:Nodes}, we connect the concept of basis Cwebs to combinatorial objects nodes and edges.  In section \ref{sec:corres}, we establish the correspondence between Boomerang Cwebs and non-boomerang Cwebs. Further, in section \ref{sec:FormNnotation}, using the correspondence along with Fused web formalism and Uniqueness theorem, we determine the diagonal blocks and ranks of four classes of Cwebs at all orders in the perturbation theory which contains all the Boomerang Cwebs at four loops connecting four lines. 
Also in section \ref{sec:FormNnotation}, we directly construct the full mixing matrix of one special class of Boomerang Cweb at four loops connecting four lines by bypassing the replica trick.

 In the section \ref{sec:Boom-4loop-4line}, we have calculated the diagonal blocks and rank of mixing matrices for all the Boomerang Cwebs present at four loops that connect four Wilson lines.

\section{Cwebs and properties of mixing matrices}
\label{sec:review}

Cweb~\cite{Agarwal:2020nyc,Agarwal:2021him} is defined as a set of skeleton diagrams, built out of 
connected gluon correlators attached to Wilson lines, and closed under shuffle
of the gluon attachments to each Wilson line. 
These are not fixed-order quantities, but 
have their own perturbative expansion in powers of the gauge coupling $g$. 
We define Boomerang Cwebs as the Cwebs that contain at least one two-point gluon correlator whose both the ends are connected to a single non-lightlike Wilson line.

Throughout this article we adapt the notation~\cite{Agarwal:2020nyc,Agarwal:2021him,Agarwal:2022wyk} $W_n^{(c_2, \ldots , c_p)} (k_1, \ldots  , k_n)$ 
for a Cweb having $c_m$ $m$-point gluon correlators 
($m = 2, \ldots, p$).
Further, we choose  $k_1 \leq k_2 
\leq \ldots \leq k_n$, where $k_{i}$ denotes attachments on different Wilson lines.

\begin{figure}
	\centering
	\includegraphics[scale=0.4]{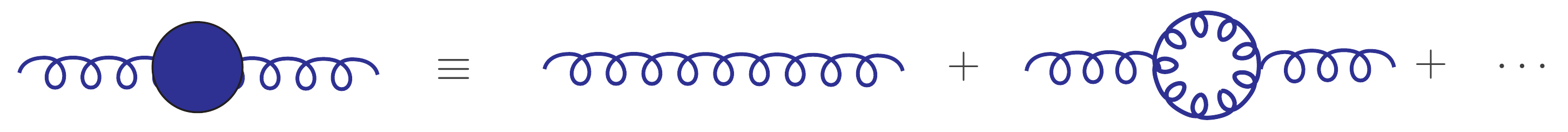}
	\caption{Perturbative expansion of a two point gluon correlator}
	\label{fig:correlator}
\end{figure}

Considering the fact that the 
perturbative expansion for an $m$-point connected gluon correlator starts 
at ${\cal O} (g^{m - 2})$, while each attachment to a Wilson line further carries a 
power of $g$, the perturbative expansion for a Cweb can be written as
\beq
W_n^{(c_2, \ldots , c_p)} (k_1, \ldots  , k_n)  \, = \, 
g^{\, \sum_{i = 1}^n k_i \, + \,  \sum_{r = 2}^p c_r (r - 2)} \, \sum_{j = 0}^\infty \,
W_{n, \, j}^{(c_2, \ldots , c_p)} (k_1, \ldots  , k_n) \, g^{2 j} \, ,
\label{pertCweb}
\eeq
which defines the perturbative coefficients $W_{n, \, j}^{(c_2, \ldots , c_p)}  
(k_1, \ldots  , k_n)$. 
The perturbative order of Cwebs is defined as the order at which they receive their lowest order contributions, which is given by the prefactor $ g^{\, \sum_{i = 1}^n k_i \, + \,  \sum_{r = 2}^p c_r (r - 2)} $. The remaining powers of $ g $ in eq.~\eqref{pertCweb} arise from the attachments within the \textit{blobs} (see fig.~\eqref{fig:correlator}) and they do not enter into the counting of order of Cwebs; for example, a two point correlator attached to Wilson lines contributes at order $ g^2 $. 

Cwebs are the proper building blocks of logarithm of the soft function; and are proved to be useful~\cite{Agarwal:2020nyc,Agarwal:2021him} in their organisation at higher perturbative orders. 
The logarithm of the Soft function is a sum over all the Cwebs at each perturbative order, 
\begin{align}
{\cal S} \, = \, \exp \left[ \sum_w
\sum_{d,d' \in  w} {\cal K} (d) \, R_w (d, d') \, C (d')
\right] \, .
\label{Snwebs}
\end{align}
Here $ {\cal K} (d) $ and $C(d)$ denote the kinematic and colour factors a diagram $ d $ in a  Cweb $w$. 
The quantity obtained after the application of web mixing matrix $ R $ on the colour factors of diagrams, is the exponentiated colour factor $\widetilde{C}$,   
\begin{align}
\widetilde{C} (d)  \, = \, \sum_{d'\in w} R_w (d, d') \, C(d') \, .
\label{eq:ecf}
\end{align}

In addition to the kinematic and colour factors, a  weight factor $ s(d) $ for a given diagram $ d $ is defined as  the numbers of ways in which the gluon correlators can be independently shrunk to the hard interaction vertex (origin). 
We can construct a column weight vector out of these $ s $-factors for a Cweb with $ n $ diagrams as 
\begin{align} 
S=\{s(d_1),s(d_2), \ldots,s(d_n)\}.
\end{align}
The diagrams of a Cweb were classified~\cite{Agarwal:2022wyk} based on their $s$-factors. A diagram $d$ is  reducible if $s(d) \neq 0$, whereas it is irreducible if $s(d) = 0$.  Further irreducible diagrams were classified in two categories. 

\noindent {Completely entangled diagram}: An irreducible diagram in which all the gluon correlators are entangled and thus none of them can be independently shrunk to the origin.  \vspace{0.1cm}\\ 
\noindent {Partially entangled diagram}: An irreducible diagram which has at least one gluon correlator which is not entangled with the other correlators. 

We adapt the notation of~\cite{Agarwal:2022wyk} to denote $ S $ and $ R $ of a Cweb compactly. Consider a Cweb with the column weight vector of the form, $S = \{s_1, \ldots, s_1, \ldots, s_i, \ldots, s_i, \ldots, s_l, \ldots s_l\},$
where, $ s_{1} < s_{2} < \cdots < s_{l}$. Here we have ordered the diagrams according to their $ s $-factors. $ S $ can be written as
\begin{align}\label{eq:S-definition}
S& = \left\{ (s_{1})_{{k_{1}}}, (s_{2})_{{k_{2}}}  , \ldots,  (s_{l})_{{k_{l}}}  \right\}\,.
\end{align}
Here first $ k_1 $ diagrams in the Cweb have weight $ s_1 $, followed by the  $k_{2}$ diagrams with weight $s_{2}$ and so forth. 
We denote the corresponding mixing matrix for the Cweb as,
\begin{align}
R\Big(  (s_{1})_{{k_{1}}}, (s_{2})_{{k_{2}}}  , \ldots,  (s_{l})_{{k_{l}}}  \Big).
\end{align}

\noindent The web mixing matrices are essential quantities for the determination of Wilson line correlators, and thus, the soft anomalous dimension matrix, and we now turn our focus on them.
\subsection*{Properties of mixing matrices}
General all order properties of the mixing matrices were first observed in \cite{Gardi:2010rn}, and were proven in \cite{Gardi:2011wa}. Further, a  conjecture regarding the columns of the mixing matrices was proposed in  \cite{Gardi:2011yz}. Based on the column weight vector of a Cweb, a uniqueness theorem was introduced in~\cite{Agarwal:2022wyk}. Below we list down these properties.

\begin{enumerate}
	\item {\it Idempotence:} These matrices are idempotent and act as projection operators:
	\begin{align}
	R^2 \, = \,  R \, .
	\label{eq:idempo} 
	\end{align}
	Thus their eigenvalues can only be either 0 or 1, which further implies that their trace is  equal to their rank.
	\item {\it Non-abelian exponentiation:} The general non-abelian exponentiation theorem~\cite{Gardi:2013ita} states that the colour factors that survive the above projection by $ R$ are the ones that are associated with a diagram that has only one gluon correlator. 
	\item {\it Row sum rule:} The elements of web mixing matrices obey the row sum rule 
	\begin{align}
	\sum_{d'} R (d, d') \, = \, 0 \,.
	\label{eq:rowsum}
	\end{align}
	\item {\it Column sum rule:}  
	The mixing matrices obey the following column sum conjecture:
	\begin{align}
	\sum_d s(d) R(d, d')\,=0\,.
	\label{eq:column-sum} 
	\end{align}
	\item {\it Uniqueness}:  For a given column weight vector $S=\{s(d_1),s(d_2), \ldots,s(d_n)\}$
	with all $s(d_{i}) \neq 0$, the mixing matrix is unique.
\end{enumerate}
The idempotence of mixing matrices implies that $R$ projects onto only those combinations of kinematic factors that do not contain ultraviolet sub-divergences\footnote{We refer here to UV divergences arising from sub-diagrams involving the Wilson lines: interactions away from the Wilson lines will still involve the usual gauge-theory UV divergences, which are dealt with by means of ordinary
	renormalization techniques.}. For the case of two Wilson lines, the absence of sub-divergences was proved in \cite{Gatheral,Frenkel-1984,Sterman-1981}. However, for more than two Wilson lines, the all order proof for column sum is not available, although this property has been verified upto four loops~\cite{Gardi:2011yz,Gardi:2013ita,Agarwal:2020nyc,Agarwal:2021him} for massless Wilson lines and upto three loops~\cite{Gardi:2021gzz} with massive lines.
The connection of the column-sum rule to UV sub-divergences is explained in a coordinate-space
picture in ~\cite{Erdogan:2014gha}. In coordinate-space, UV divergences 
arise from short distances between interaction vertices, and, thus the `shrinkable' 
correlators are naturally associated to UV sub-divergences,  eq.~(\ref{eq:column-sum}) 
guarantees that these correlators are projected out of the webs.

\section {Normal ordering and Fused-Webs}
\label{sec:Fused-Webs}
The Normal ordering of diagrams and Fused-webs were the two main ideas developed in~\cite{Agarwal:2022wyk} that shed light on the structure of the mixing matrices and were instrumental in the determination of diagonal blocks of mixing matrices for Cwebs. In this section, we briefly review these two concepts.   
\subsection* {Normal ordering}

The idea of normal
ordering of diagrams in a Cweb, that was introduced in~\cite{Agarwal:2022wyk},  makes the structure of the
web mixing matrix particularly transparent by reorganising its elements.
Consider a Cweb with $ n $ diagrams out of which $k$, $(l-k)$, and $ m $ are {\it completely entangled}, {\it partially entangled} and {\it reducible diagrams}, respectively.  The normal ordering is then defined as :

\begin{itemize} 
	\item [] {\it Normal order} :  The order in which first $k$ diagrams are completely entangled, followed by $(l-k)$
	partially entangled diagrams, then followed by reducible diagrams in ascending order of their $s$-factors.
\end{itemize}
Taking into account the fact (as can be readily inferred from the replica trick) that the exponentiated colour factors of reducible diagrams do not contain the colours of the irreducible diagrams, and those of completely entangled diagrams do not contain the colours of the partially entangled diagrams, the normal ordered mixing matrix takes the following form~\cite{Agarwal:2022wyk}:  
\begin{align}
R=\left(\begin{array}{cc}
A_{l\times l}  & B_{l\times m} \\
O_{m \times l} & D_{m\times m}
\end{array}\right)\,=\left(\begin{array}{c|c}
\begin{array}{cc}
\hspace*{-0.5cm}{I}_{k \times k} & (A_U)_{k\times (l-k)} \\
{O}_{(l-k)\times k} & \quad \, \, (A_L)_{(l-k) \times (l-k)}
\end{array} & B_{l\times m} \\ 
\hline
{O}_{m\times l } & D_{m\times m}
\end{array}\right)\,. 
\label{eq:R-gen-big-form}
\end{align}

The properties of block $ D $ were studied in~\cite{Agarwal:2022wyk} and it was shown that
$ D $ satisfies the known properties of a web mixing matrix with $S_D=\{ s_{l+1} \ldots s_{l+m}\} $. 
The diagrams present in $ D $ form a closed set under the action of replica ordering operator \textbf{R} (see appendix \ref{sec:repl}) in the same way as that of a Cweb with  $S_D=\{ s_{l+1} \ldots s_{l+m}\} $. Then mixing between the diagrams of $ D $ is same as that of the Cweb, and they have the same mixing matrix $ R(S_D) $. 
\subsection* {Fused Webs}
In this section, we review the Fused-Webs formalism~\cite{Agarwal:2022wyk} which proved to be useful in the  determination of  the diagonal blocks of $ A $ of the mixing matrix.

\begin{figure}
	\centering
	\subfloat[][]{\includegraphics[height=4cm,width=4cm]{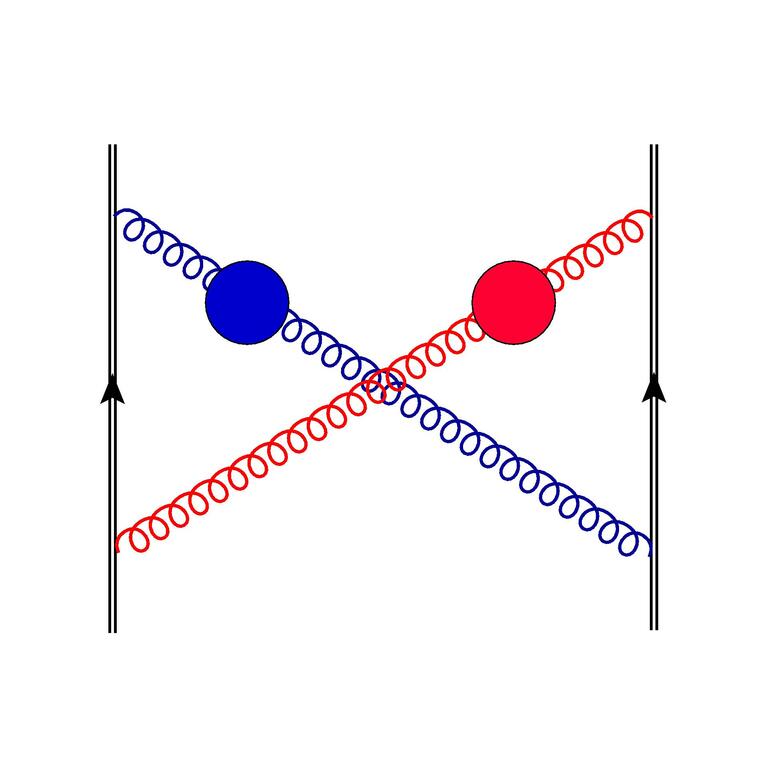} }
	\qquad 
	\subfloat[][]{\includegraphics[height=4cm,width=4cm]{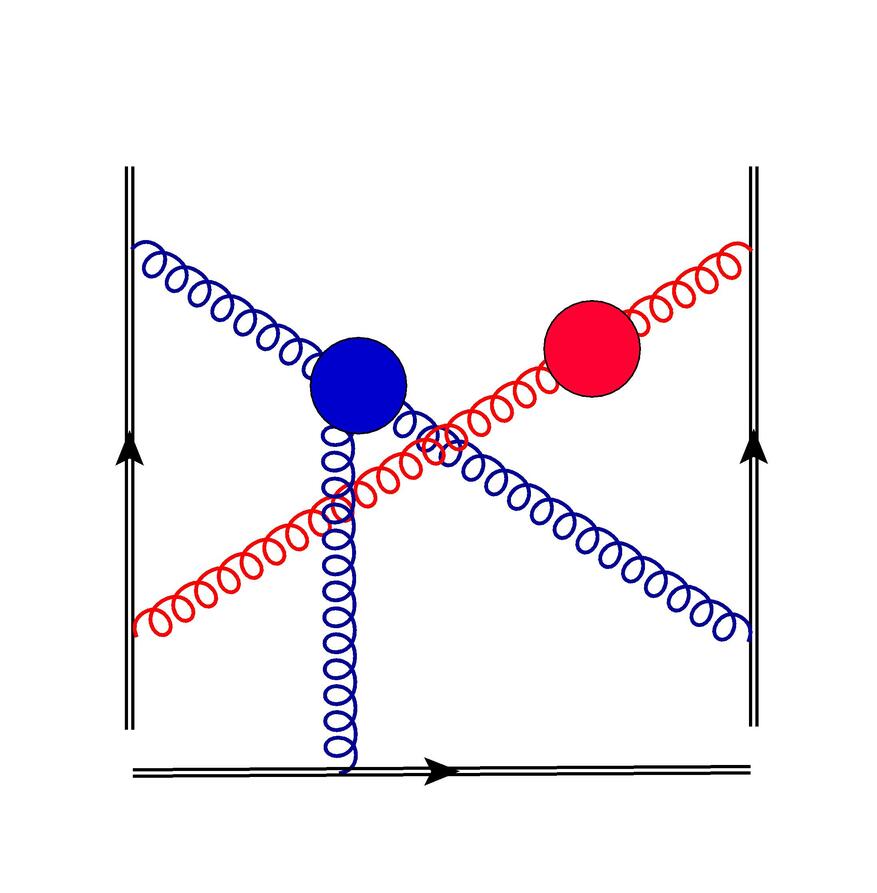} }
	\qquad 
	\subfloat[][]{\includegraphics[height=4cm,width=4cm]{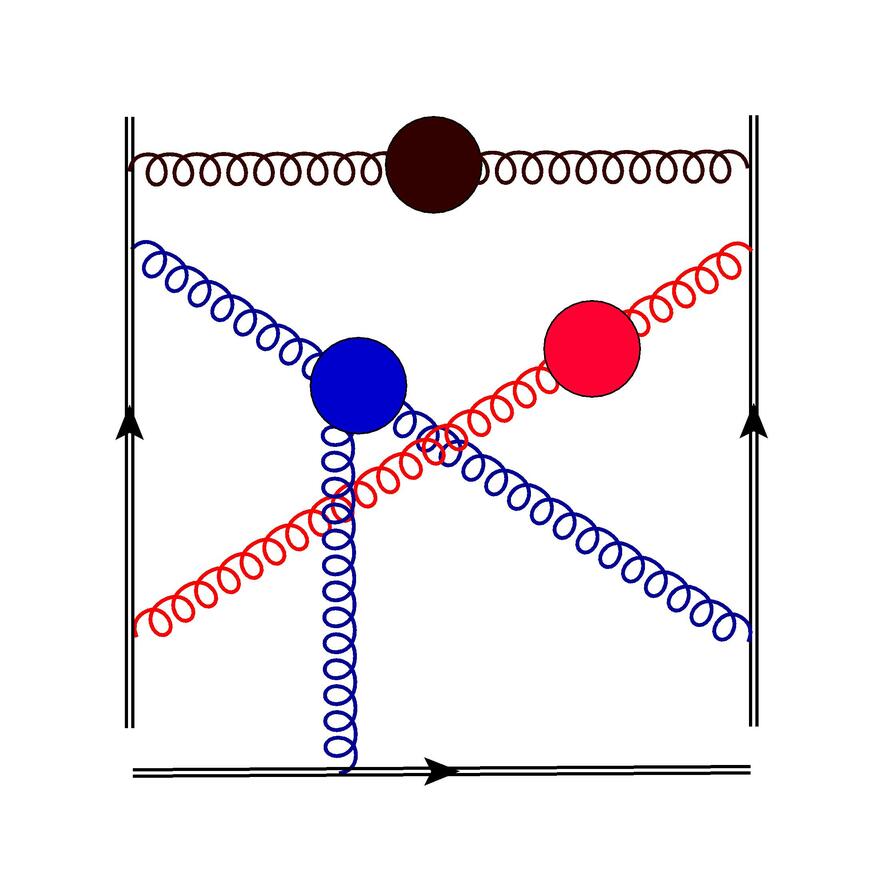} }	
	\quad 
	\subfloat[][]{\includegraphics[height=4cm,width=4cm]{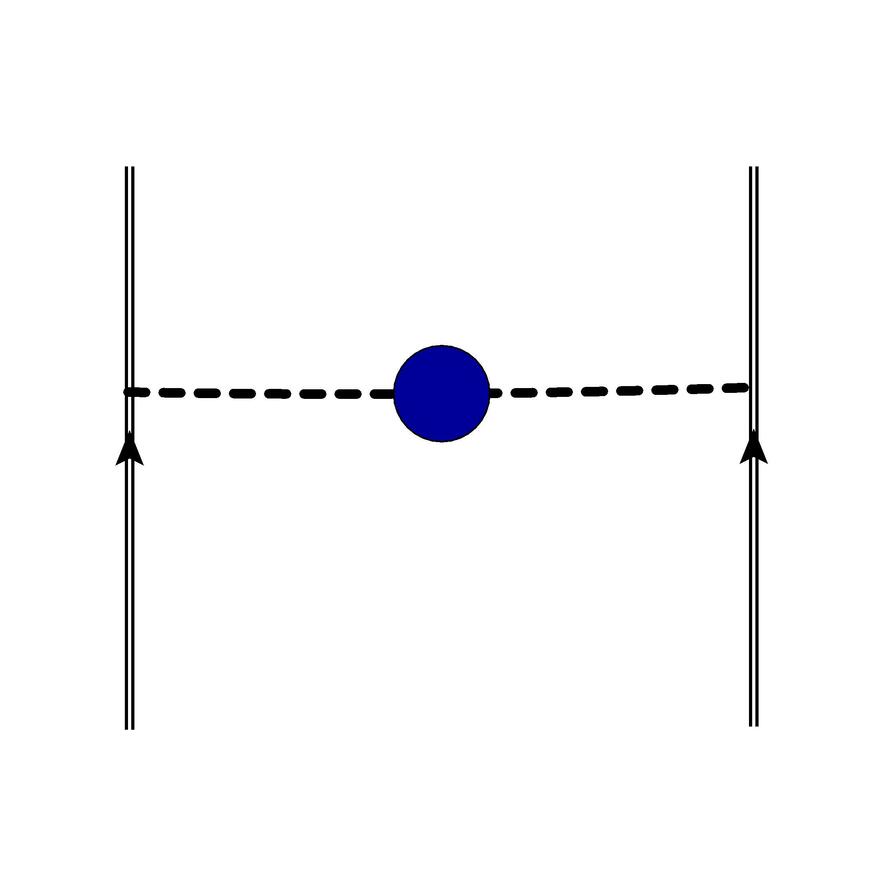} }
	\qquad 
	\subfloat[][]{\includegraphics[height=4cm,width=4cm]{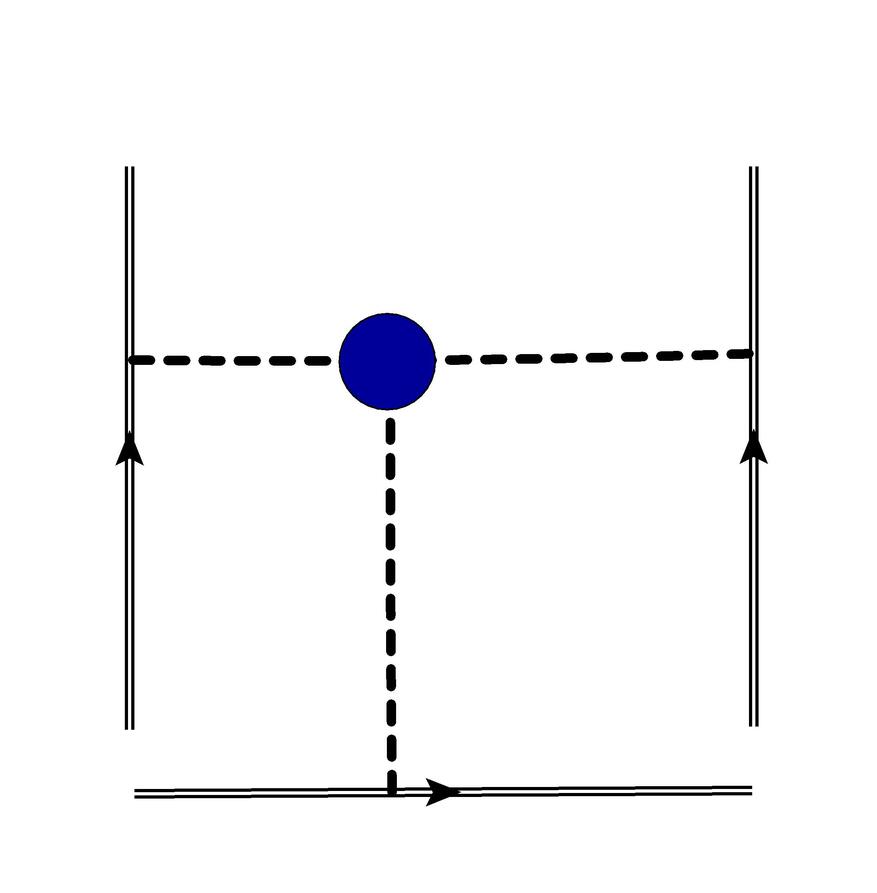} }
	\qquad  
	\subfloat[][]{\includegraphics[height=4cm,width=4cm]{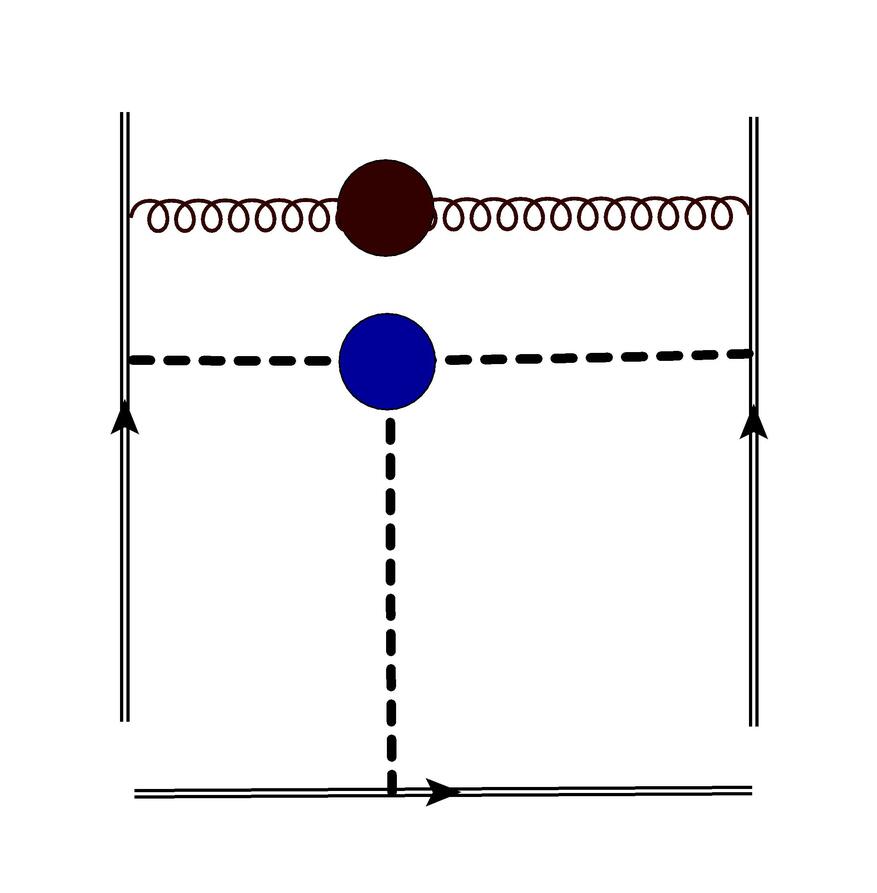} }
	\caption{Replacing an entangled piece by $ n $-point correlators.     
		Here, the diagrams in~(\textcolor{blue}{a}), (\textcolor{blue}{b}) and (\textcolor{blue}{c}) have an entangled piece made up of the blue and the red gluon correlators involving two and three Wilson lines. Therefore, replacing the entangled piece of diagram (\textcolor{blue}{a}) with a two point \reduced correlator; and that of diagram (\textcolor{blue}{b}), and (\textcolor{blue}{c}) with a three point \reduced correlator, provides the  \reduced diagrams (\textcolor{blue}{d}), (\textcolor{blue}{e}), and (\textcolor{blue}{f}) respectively. With this replacement, the diagrams (\textcolor{blue}{a}), (\textcolor{blue}{b}), and (\textcolor{blue}{c}), having $ s=0 $ correspond to  \reduced diagrams (\textcolor{blue}{d}), (\textcolor{blue}{e}), and (\textcolor{blue}{f}), with $ s=1 $, respectively. The colour factors of (\textcolor{blue}{d}), (\textcolor{blue}{e}) and (\textcolor{blue}{f}) are same as those of (\textcolor{blue}{a}), (\textcolor{blue}{b}) and (\textcolor{blue}{c}) respectively. 
	}
	\label{fig:irreducible-h1}
\end{figure}

Let us start with the definition of Fused diagrams. The diagram generated by replacing each entangled piece of an irreducible diagram involving $ n $-Wilson lines, with an $ n $-point correlator connecting the same $ n $-lines, is defined as \textit{\reduced diagram} associated with the irreducible diagram. This new $ n $-point correlator is called 
\textit{\reduced correlator}, which is denoted by the dotted lines\footnote{Denoting these new correlators by gluon lines introduces spurious problem of counting in the number of diagrams for a Cweb.}. These \reduced correlators are essentially gluon correlators. {However,} note that an $ n $-point \reduced correlator does not have the colour structure of an $ n $-point gluon correlator. The illustration of obtaining \reduced diagrams is shown in fig.~(\ref{fig:irreducible-h1}). 

The number of ways in which all the correlators of a \reduced diagram, including the \reduced correlators can be sequentially shrunk to the origin was defined~\cite{Agarwal:2022wyk} as the $ s $-factor of the \reduced diagram. 

Starting from a \reduced diagram we can generate a {\it fictitious} Cweb in the usual way by shuffling all the attachments on each Wilson line. This Cweb has its own mixing matrix $R_{\text{fict}}$ which can be put in the form of  eq.~(\ref{eq:R-gen-big-form}) by reordering the diagrams. The mixing between the \reduced diagrams of the {\it fictitious} Cweb that have $s \neq 0$, is given by the corresponding $D_{\text{fict}}$, and this corresponds to the mixing between the respective irreducible diagrams of the original Cweb. Thus, Fused-Webs were defined as the set of all reducible ($ s\neq0 $) \reduced diagrams that appear in the fictitious Cweb.

\noindent The mixing matrix of a \reducedWeb is given by $D_{\text{fict}}$. However, note that, even though the mixing matrix $D_{\text{fict}}$ satisfies all the properties of a web mixing matrix that are known to date, \reducedWeb is not really a Cweb as not all the diagrams of the fictitious Cweb are part of it.

The above concepts were used to provide an algorithm to obtain the diagonal blocks of $ A $. We outline the steps of the algorithm below. 
\begin{enumerate}
	\item  Identify the completely entangled diagrams of the Cweb. \reduced diagram of each of these forms a \reducedWeb which has only one diagram. The number of these completely entangled diagrams in a Cweb is the order of identity matrix appearing in the Normal ordered mixing matrix of the Cweb.
	\item  {Identify the \textit{distinct} entangled pieces appearing in partially entangled diagrams, and obtain \reduced diagram for each of these. }
	\item  Shuffle the attachments on each Wilson line of a \reduced diagram to generate the corresponding fictitious Cweb, the reducible diagrams of which will form the associated Fused-Web. 
	\item  Obtain the mixing matrix of the Fused-Web. For this we first find the weight vectors and then use the Uniqueness theorem. This 
	will correspond to the mixing between the associated partially entangled diagrams. 
	\item Order the diagrams of the Cweb associated with a Fused-Web, such that they appear next to each other. 
	\item Repeat steps 3, 4 and 5 for all distinct entangled pieces in a Cweb in order to compute the diagonal blocks of the matrix $ A $.    
\end{enumerate}

\section{Counting basis Cwebs: Connection with the Trees with Nodes}
\label{sec:Nodes}
\begin{figure}[t]
	\vspace{0.5cm}
	\captionsetup[subfloat]{labelformat=empty}
	\centering	
	\subfloat[][]{\includegraphics[scale=0.18]{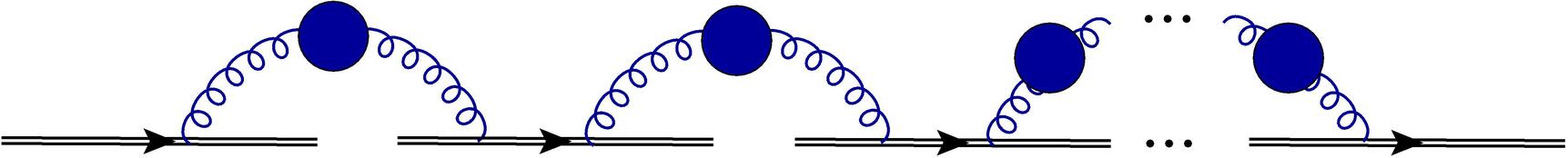} }
	\caption{A Cweb constructed out of $n$ two-point correlators connecting $n+1$ Wilson lines. This is one of the basis Cwebs at order $\alpha_{s}^{n}$.}
	\label{fig:MaximalCon}
\end{figure}

The study of basis Cwebs  from the combinatorial point of view  is important as their mixing matrices form the diagonals blocks of the mixing matrices of a general Cweb.
In this section we discuss a direct correspondence between the basis Cwebs and the combinatorial objects known as \textit{nodes}. A {\it basis Cweb} is defined~\cite{Agarwal:2022wyk} as follows:
\begin{itemize}
	\item [] {\it Basis Cweb}: Cweb formed by connecting $n$ two-point gluon correlators to $n+1$ Wilson lines.
\end{itemize}
For example, fig.~(\ref{fig:MaximalCon}) displays a basis Cweb \footnote{In this article we draw only one diagram and still call it a Cweb and it is understood that the other diagrams of the Cweb are obtained by shuffling all the attachments on each of the Wilson lines}.

In graph theory, a vertex or a node is defined as the fundamental unit of which the graphs are formed which are generally represented by points. There are two different types of nodes namely labelled and unlabelled. As the names suggest, the labelled nodes are distinguishable among themselves, whereas unlabelled are not. Further, an edge is defined as the connection between two nodes and is represented by a line. An object formed out of edges and nodes is called a graph in graph theory. There are different types of graphs, however for our analysis in this section we only discuss tree (open undirected graphs). Tree graphs are defined as the graphs in which there is only one edge between two nodes. Thus, a tree with $ n+1$ nodes has $ n $ edges. A simple example of a tree graph connecting three nodes with two edges is shown in fig.~(\ref{fig:nodes-example}).

Now, we relate these objects of graph theory to the basis Cwebs. We start by noting that the Cwebs which differ only by a permutation of their Wilson lines are structurally identical~\cite{Agarwal:2020nyc,Agarwal:2021him}. Hence, in the language of graph theory, Wilson lines behave as unlabelled nodes. Further, for basis Cwebs the Wilson lines can only be connected by two-point gluon correlators which is equivalent of connecting two nodes by a single edge. Thus obtaining the trees for a given number of nodes is equivalent of obtaining the basis Cwebs for a given number of Wilson lines.

\begin{figure}[t]
	
	\centering \subfloat[][]{\includegraphics[scale=0.07]{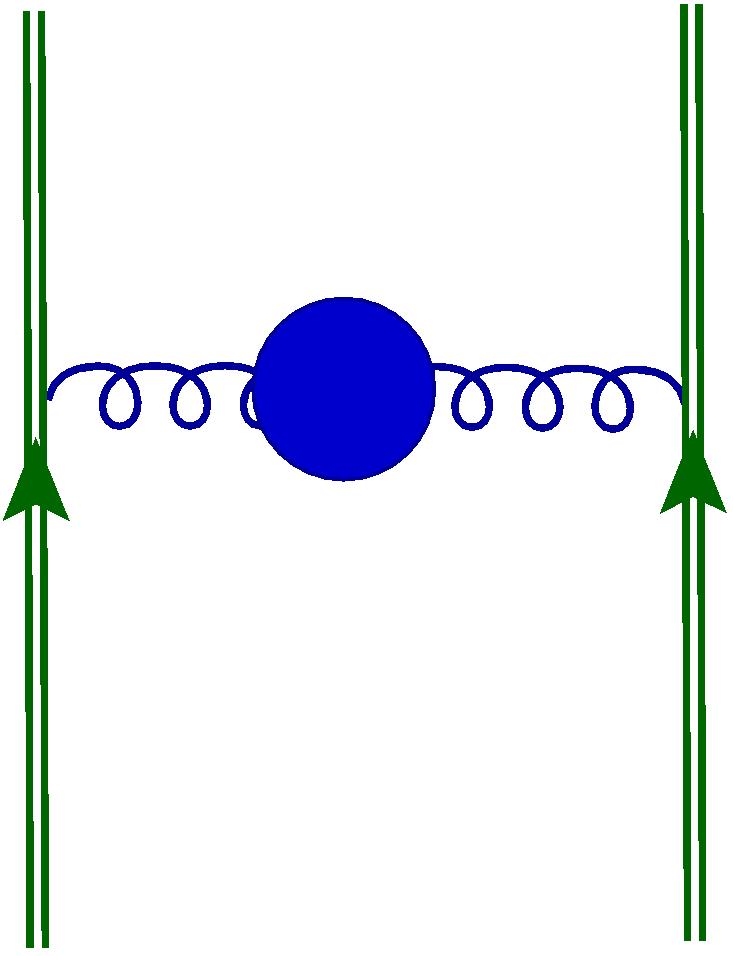} }
	\qquad \qquad \qquad\qquad
	\subfloat[][]{\includegraphics[scale=0.07]{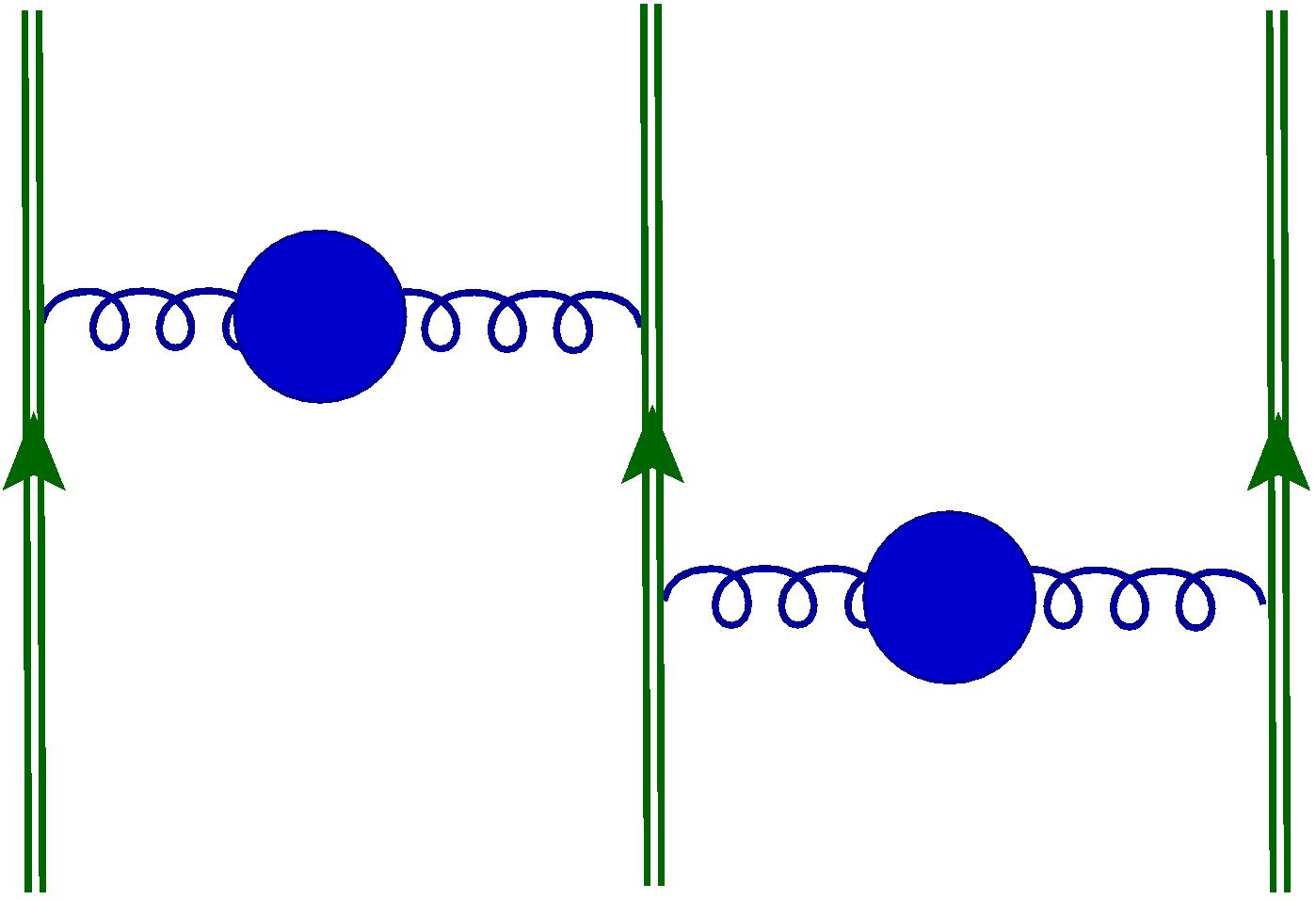} }
	\qquad\\
	\subfloat[][]{\includegraphics[scale=0.07]{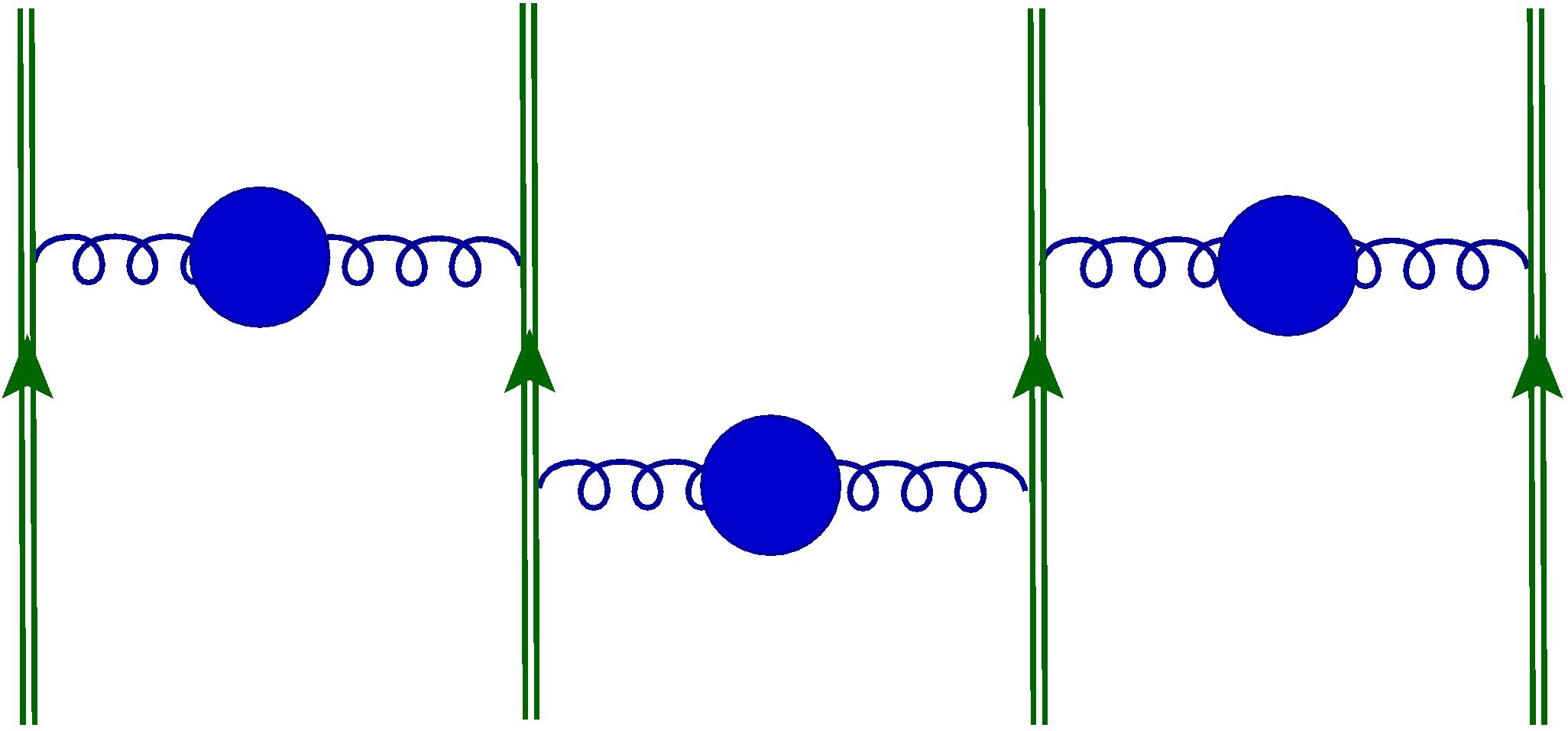} }
	\qquad \qquad\qquad
	\subfloat[][]{\includegraphics[scale=0.04,angle=90]{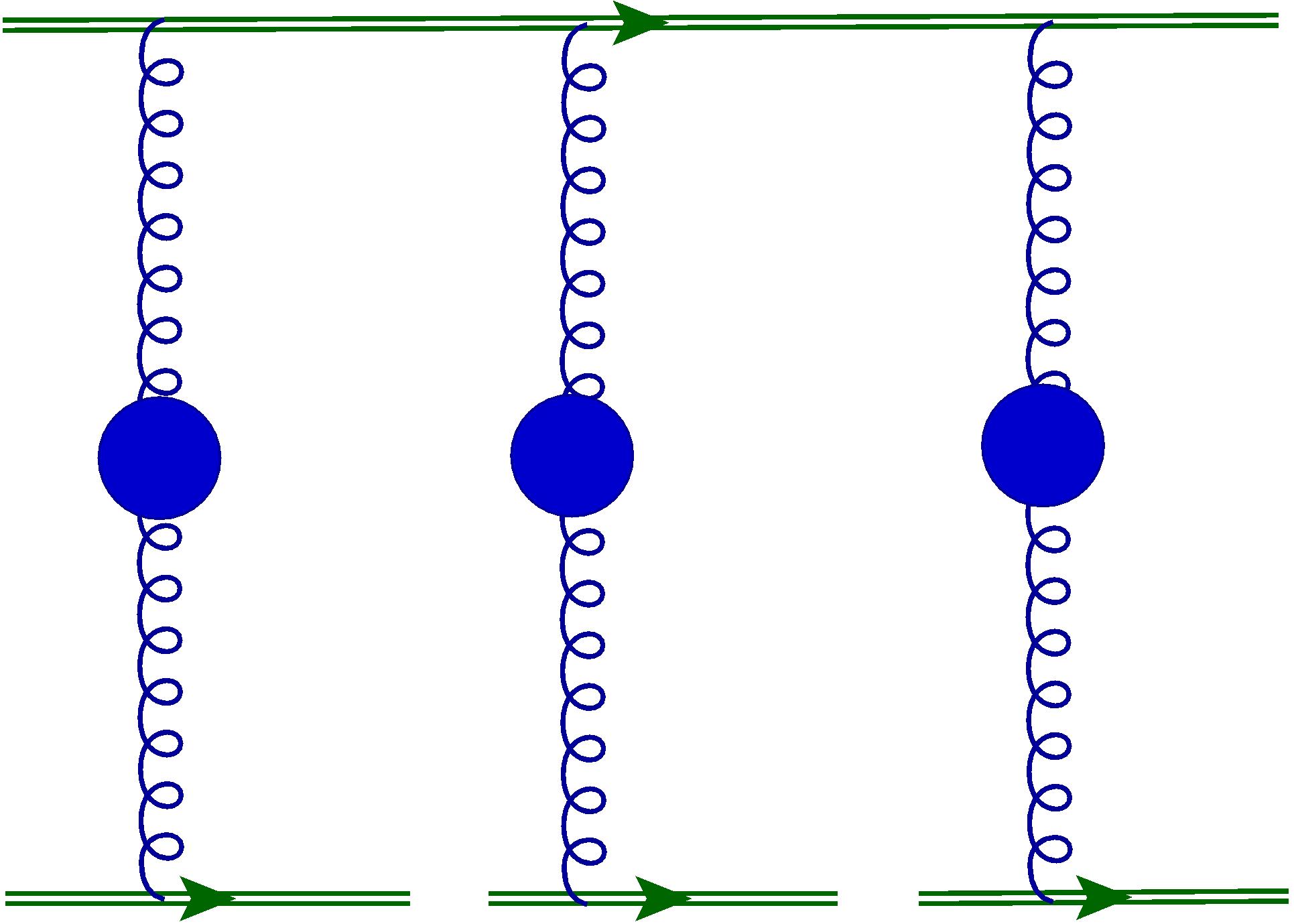} }
	\caption{Basis Cwebs upto three loops}
	\label{fig:Basis-Upto-4-nodes}
\end{figure}

\begin{figure}[b]
	
	\centering
	\subfloat[][]{\includegraphics[scale=0.07]{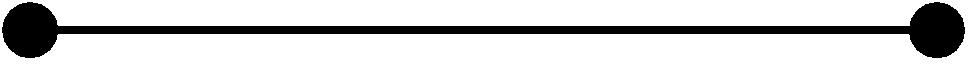} }
	\qquad \qquad 
	\subfloat[][]{\includegraphics[scale=0.06]{Node2} } \\
	\hspace{-1.0cm}
	\subfloat[][]{\includegraphics[scale=0.055]{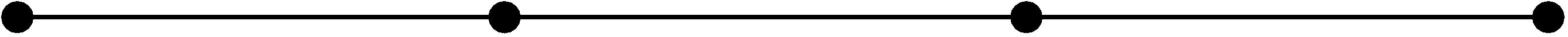} }
	\qquad \qquad 
	\subfloat[][]{\includegraphics[scale=0.06]{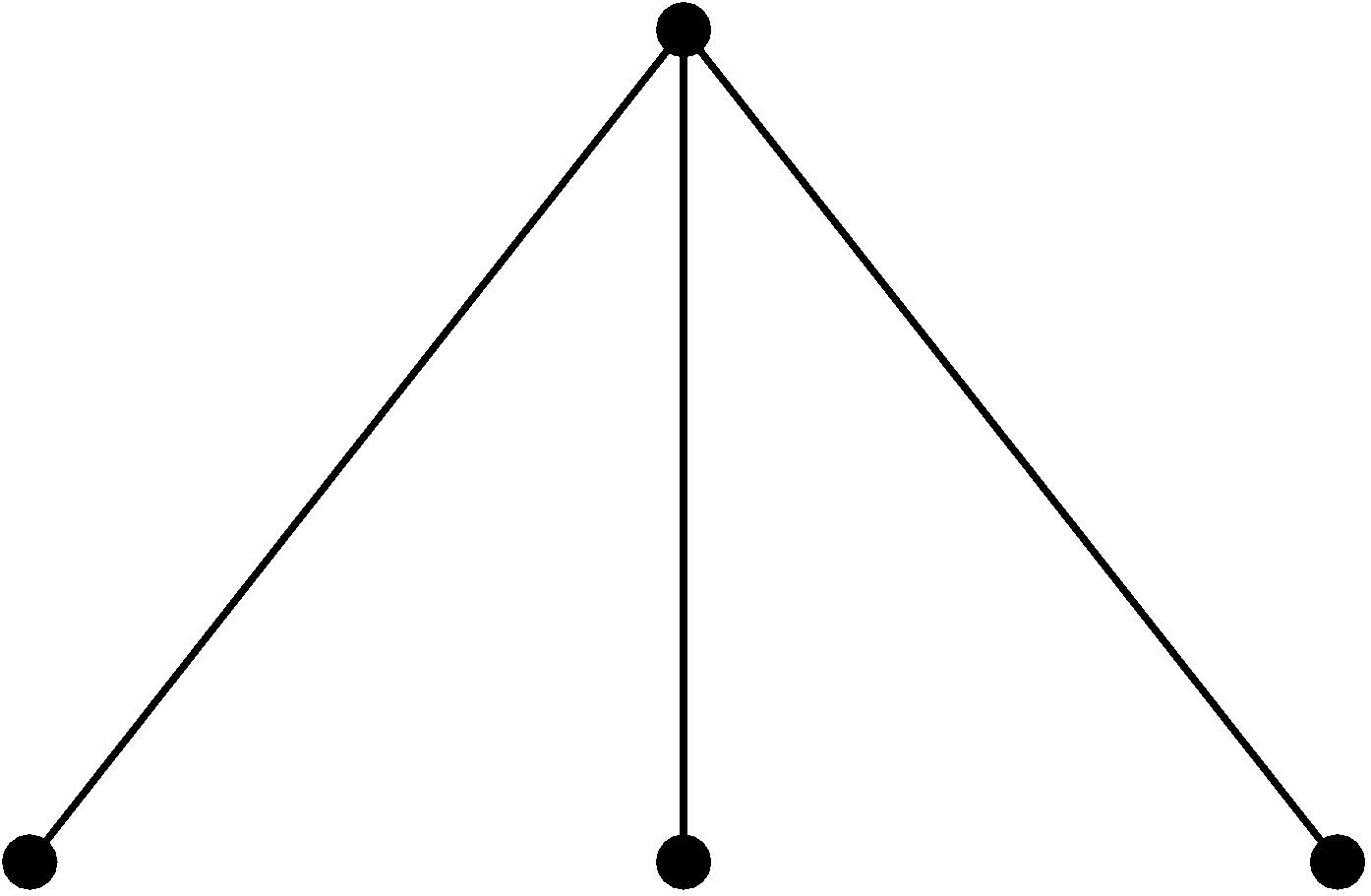} }
	\caption{Tree with unlabelled nodes.}
	\label{fig:Tree-Upto-4-nodes}
\end{figure}

Now, let us count the number of basis Cwebs:
\begin{enumerate}
	\item Let us start with one loop basis Cweb, which contains two Wilson lines connected by a single two-point gluon correlator, shown in fig.~(\ref{fig:Basis-Upto-4-nodes}\txb{a}). In graph theory, this is a tree with two unlabelled nodes and one edge, shown in fig.~(\ref{fig:Tree-Upto-4-nodes}\txb{a}).
	\item
	At two loops, we have only one basis Cweb shown in fig.~(\ref{fig:Basis-Upto-4-nodes}\txb{b}). It has three Wilson lines which are connected by two gluons. In graph theory, this is equivalent of a graph with two nodes and three edges, which is shown in fig.~(\ref{fig:Tree-Upto-4-nodes}\txb{b}).  
	
	\begin{figure}[H]
		\centering
		\includegraphics[width=0.3\linewidth]{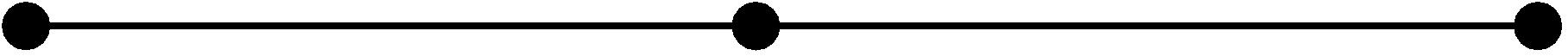} 
		\qquad 
		\caption{Example of a tree with three unlabelled nodes and two edges. }
		\label{fig:nodes-example}
	\end{figure}
	
	\item
	There are two basis Cwebs at three loops have four Wilson line connected by three gluons. These are shown in figs.~(\ref{fig:Basis-Upto-4-nodes}\txb{c}) and~(\ref{fig:Basis-Upto-4-nodes}\txb{d}).  
	This is equivalent of connecting four nodes with three edges which gives two trees shown in figs.~(\ref{fig:Tree-Upto-4-nodes}\txb{c}) and figs.~(\ref{fig:Tree-Upto-4-nodes}\txb{d}). It can be realized that these are the only possible trees with two, three and four nodes.  
	\item
	Beyond three loops, we find that there are three, six and eleven basis Cwebs at four, five and six loops respectively using the recursive algorithm of~\cite{Agarwal:2020nyc}.
\end{enumerate}

We see from the above that each of the trees connecting $ n $ unlabelled nodes has a one to one correspondence to one of the basis Cwebs connecting $ n $ lines.  

In combinatorial mathematics the trees for a given number of nodes $ n $ are known~\cite{Nodes}. The first few entries of this sequence for 
$ n=\{0,1,2,3,4,5,6,7,8,9,10, \cdots\} $ are $\{1,1,1, 1, 2, 3, 6, 11, 23, 47,\cdots\}$. The sequence for the number of basis Cwebs starts at $ n=2 $ and we have included the first two entries --  0 and 1,  for the completeness of the sequence. With this analysis, we conclude that the structures of basis Cwebs at any arbitrary loop order can be determined as the structures of trees with unlabelled nodes are known.


\section{Information exchange between Boomerang Cwebs and non-Boomerang Cwebs}
\label{sec:corres}

There is a flow of data from Boomerang Cwebs  to (non-Boomerang) basis Cwebs and vice versa.  

\subsection*{Form basis Cwebs to Boomerang Cwebs} 

Recall that the definition of basis Cweb allows us to draw a general structure with $ n $ two-point gluon correlators connecting $ (n+1) $ Wilson lines as shown in fig.~\eqref{fig:MaximalCon}. The symmetry factor $ s $ for this is non-zero as any rearrangement that lead to an irreducible diagram cannot connect all the $n+1$ Wilson lines. Also, any rearrangement of the attachments on and across the Wilson lines, keeping the number of lines fixed, gives another basis Cweb.

\begin{figure}[t]
	\vspace{0.5cm}
	\captionsetup[subfloat]{labelformat=empty}
	\centering	
	\subfloat[][]{\includegraphics[scale=0.08]{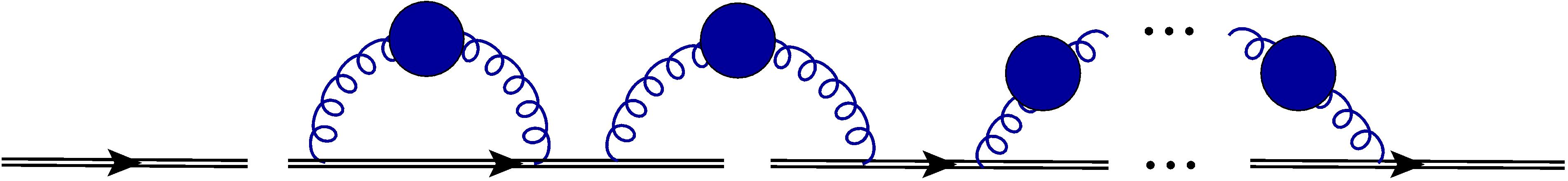} }
	\caption{A boomerang Cweb constructed out of $ n $ two point correlators connecting $ n $ Wilson lines, derived from  one of the basis Cwebs at $ \mathcal{O}(\alpha_{s}^n) $.}
	\label{fig:BMaximalCon}
\end{figure}

\noindent As noted earlier, when the Wilson lines are massive, the diagrams that have boomerang correlators are also present. At $ \mathcal{O}(\alpha_{s}^n)  $, one can at most connect $ n $ Wilson lines to form a boomerang Cweb. 
Such a boomerang Cweb is shown in fig.~(\ref{fig:BMaximalCon}). 
The boomerang Cwebs of fig.~(\ref{fig:BMaximalCon}) can be obtained from that in fig.~(\ref{fig:MaximalCon}) by shifting one of the attachments of a two point correlator such that both of its ends attach on the same Wilson line. 
Note that all the diagrams of a basis Cweb are reducible, that is each of the diagram $d_{i}$ in a basis Cweb satisfies $s(d_{i}) \neq 0$. Contrary to this 
the boomerang Cweb shown in fig.~(\ref{fig:BMaximalCon}) has both reducible and irreducible diagrams;
the reducible diagrams are the ones for which none of the gluon attachments fall between the two ends of a boomerang correlator. Thus, for the purpose of 
counting the number of shuffles, or equivalently, the number of reducible diagrams, the two ends of the boomerang can be viewed as one single attachment on the concerned Wilson line.

Once the diagrams are normal ordered, the mixing matrix of the above boomerang Cweb  takes  the general form eq.~\eqref{eq:R-gen-big-form}. Recall that the block $ D $ generates  the mixing among the reducible diagrams. From uniqueness theorem and properties of $ D $ found in~\cite{Agarwal:2022wyk} it follows that $ D $ is a  mixing matrix of a  basis Cweb. This basis Cweb has the same number of diagrams as the number of reducible diagrams in the boomerang Cweb that connects the maximum number of Wilson lines at a given order. Thus, we have shown that the block $D$ of the mixing matrix of any  boomerang Cweb that connect highest possible  number of Wilson lines at a given order (these contain only two-point gluon correlators)  is the same as the mixing matrix of the basis Cwebs at that order.

The above correspondence also holds for  Cwebs with a self-energy correlator (for which all the gluons of the correlator are attached to the same Wilson line) because, the mixing of reducible diagrams for this and the boomerang  Cweb --- obtained by replacing the self-energy correlator with a boomerang --- is the same.

\begin{figure}[H]

	\centering
	\subfloat[][]{\includegraphics[scale=0.06]{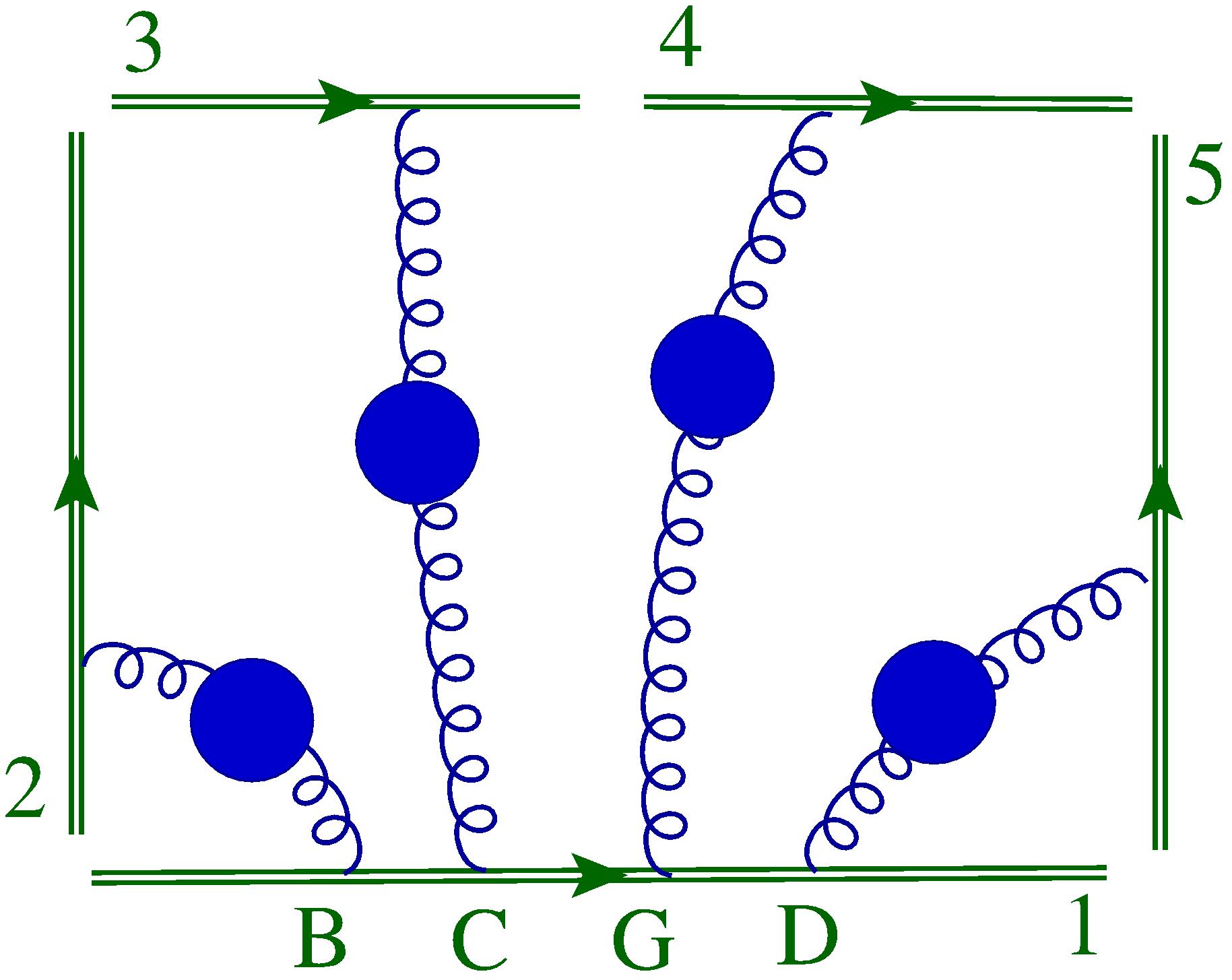} }
	\qquad
	\subfloat[][]{\includegraphics[scale=0.06]{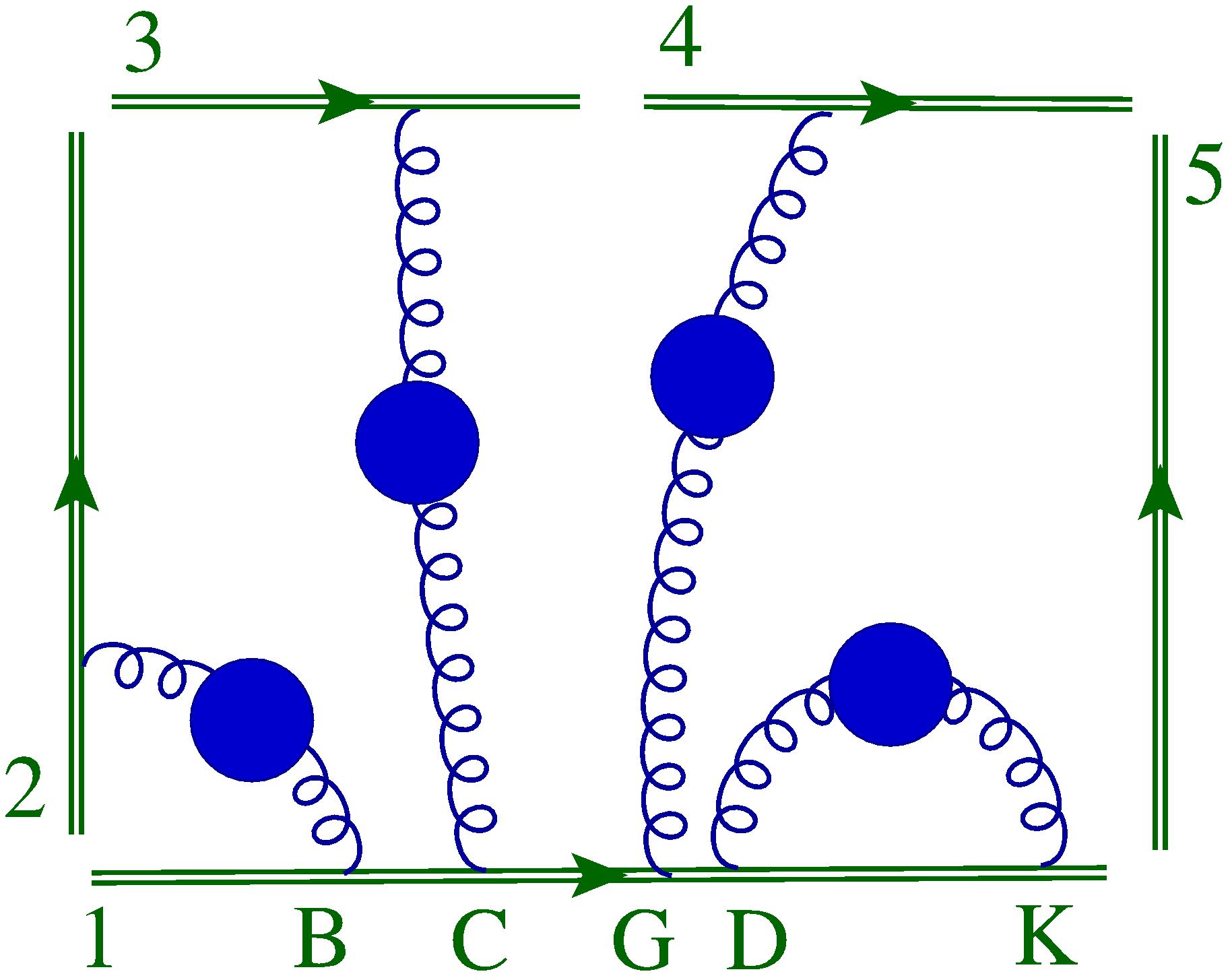} }
	\qquad
	\subfloat[][]{\includegraphics[scale=0.06]{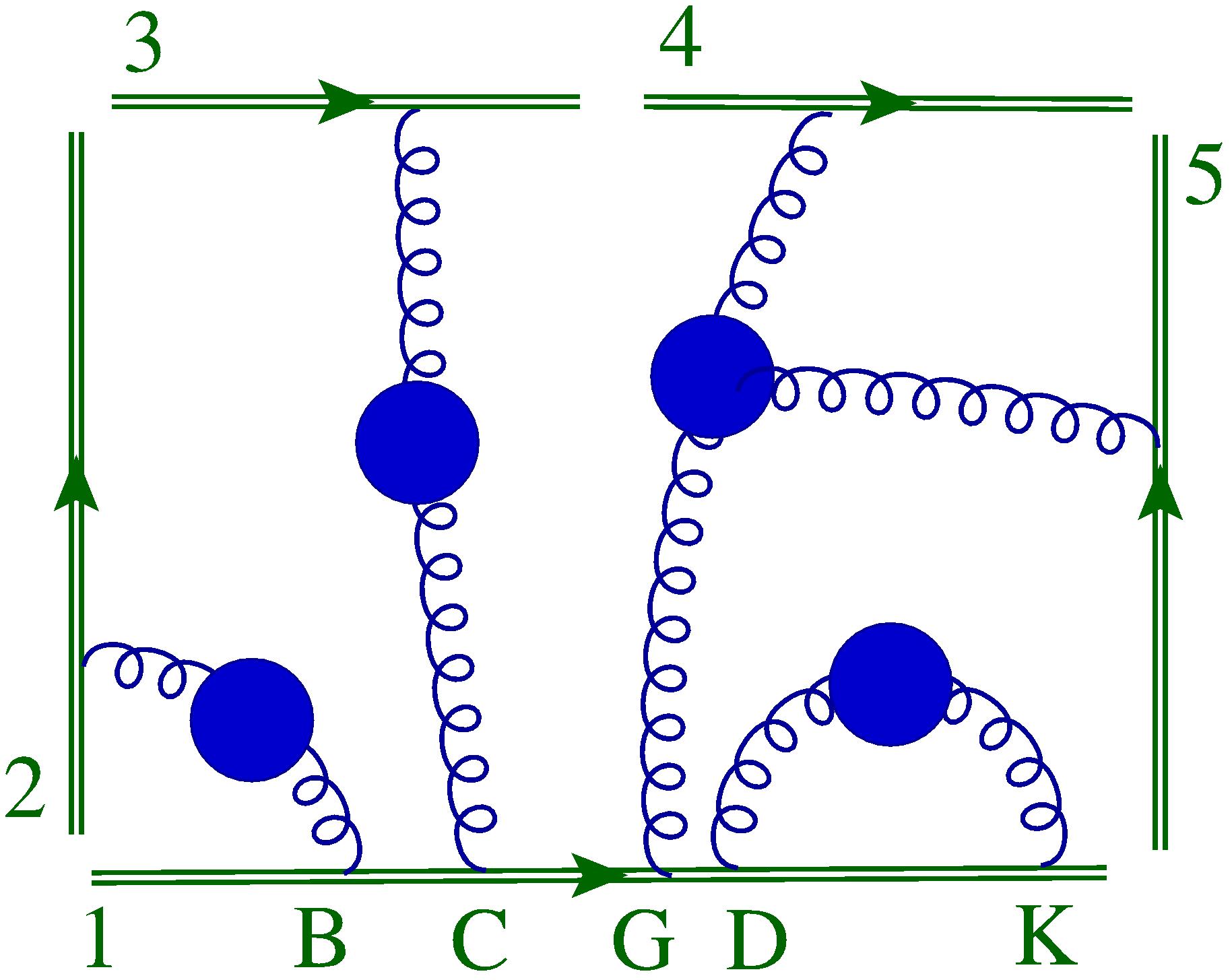} }
	\caption{Cwebs with same reducible diagrams}
	\label{fig:Corres1}
\end{figure}

We illustrate this correspondence by using a simple example of a basis Cweb shown in fig.~(\ref{fig:Corres1}\txb{a}). It  has shuffle only on line 1 and other lines are symmetric with respect to the gluon attachments. One can turn this into a boomerang Cweb by connecting both ends of any of the four two-point gluon correlators to line 1,  one of which results in fig.~(\ref{fig:Corres1}\txb{b}). Further a higher order member of this boomerang Cweb is shown in fig.~(\ref{fig:Corres1}\txb{c}). The basis Cweb has $4!$ diagrams each with $ s=1 $. The boomerang Cwebs of figs.~(\ref{fig:Corres1}\txb{b}) and~(\ref{fig:Corres1}\txb{c}) also have 4! reducible diagrams each with $ s=1 $. These reducible diagrams can be generated by shuffling the attachments on line 1, and considering both the attachments of boomerang gluon as one attachment. From the uniqueness theorem and the fact that $ D $ is itself a mixing matrix~\cite{Agarwal:2022wyk} of basis Cwebs, we get the $ D $ block for the boomerang Cwebs,
\begin{align}
D  = R(1_{24})\,.
\end{align} 

\subsection*{From Boomerang Cwebs to basis Cwebs} 
\begin{figure}[b]
	\centering
	\subfloat[][]{\includegraphics[scale=0.06]{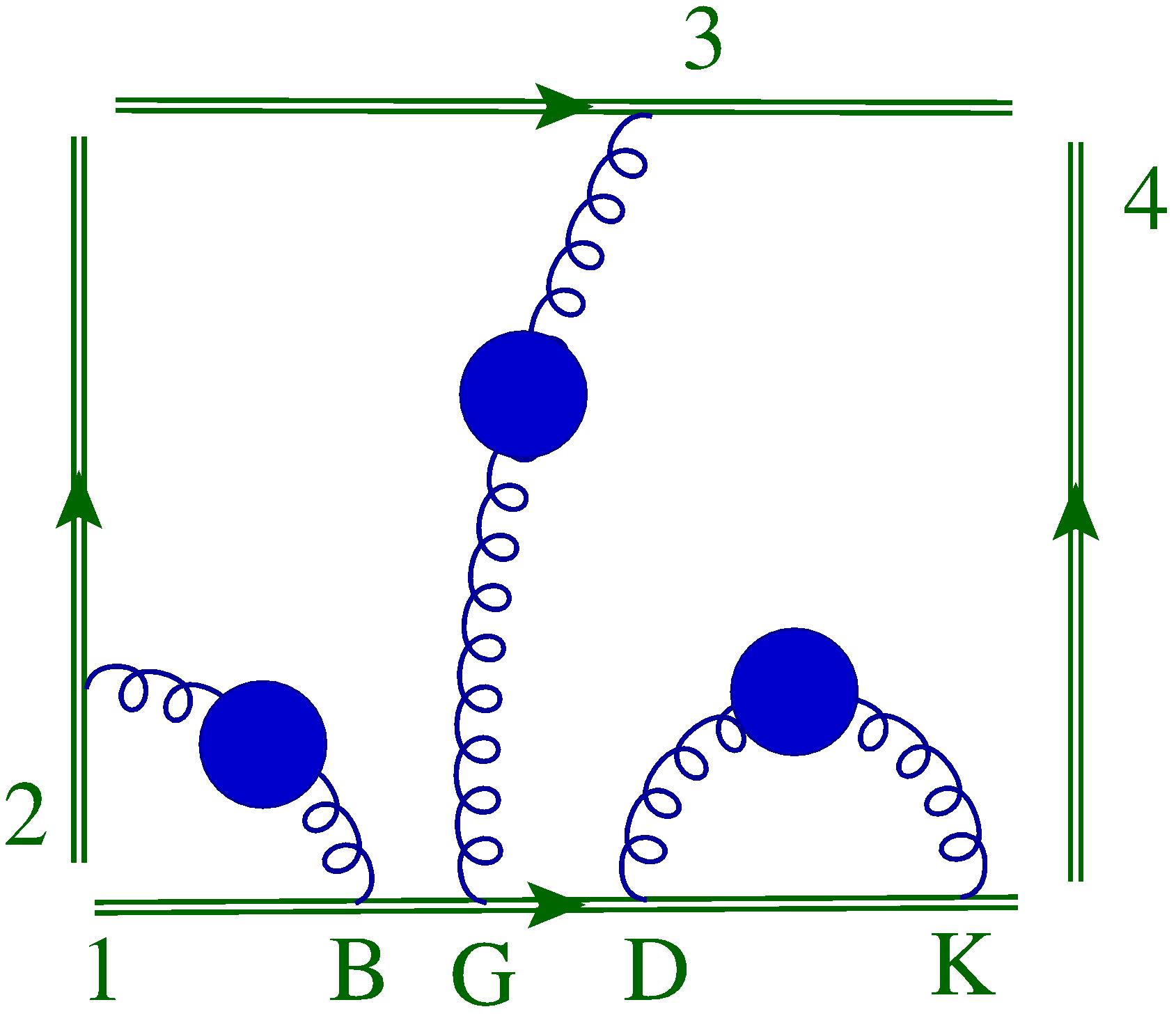} }
	\qquad  \qquad
	\subfloat[][]{\includegraphics[scale=0.06]{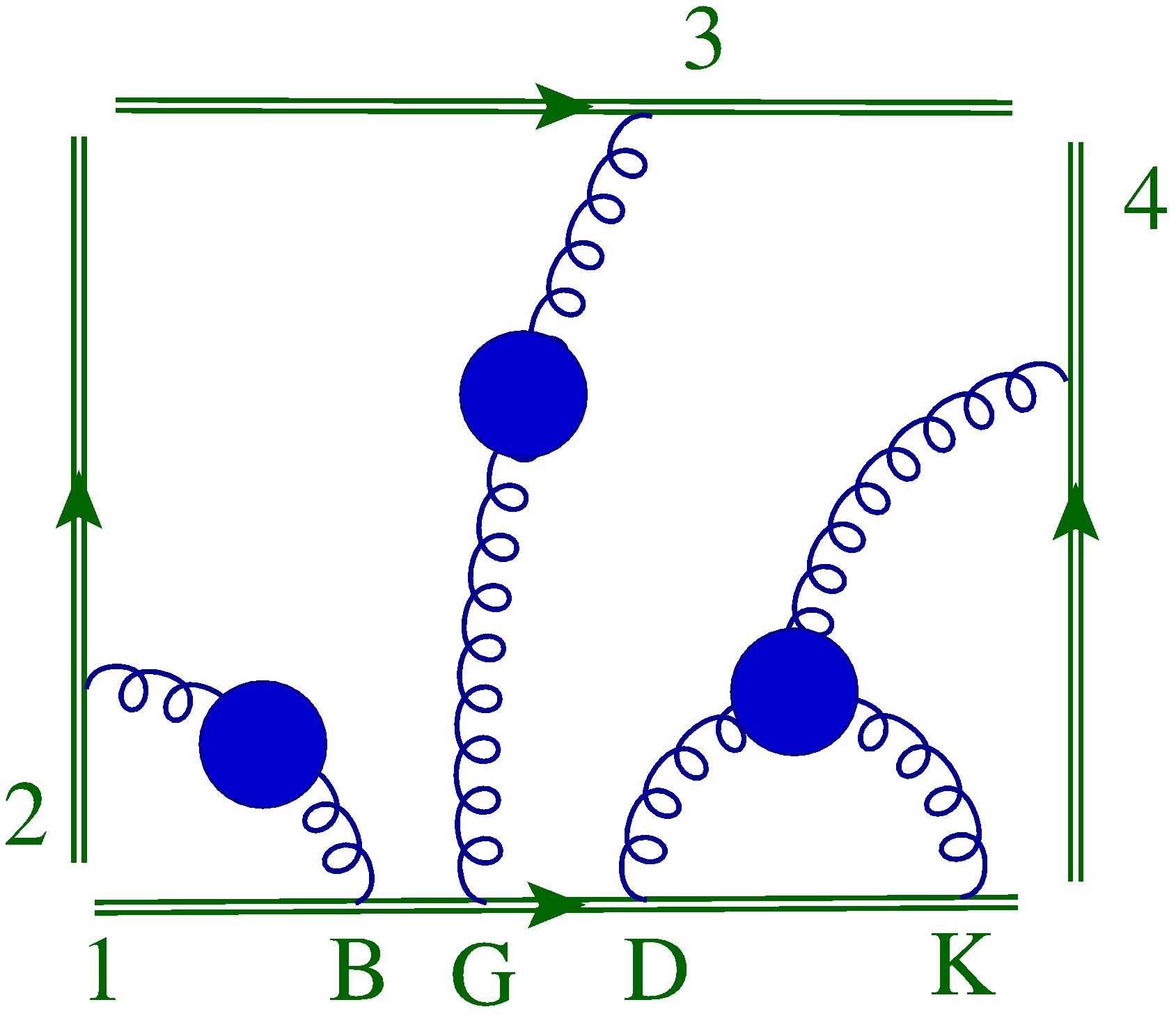} }
	\caption{Family of a boomerang Cweb}
	\label{fig:Corres2}
\end{figure}
Now, we discuss another correspondence between boomerang and non-boomerang Cwebs. We note that, one can obtain non-Boomerang Cwebs from Boomerang Cwebs by attaching a gluon from the boomerang correlator to a new Wilson line as shown in fig.~(\ref{fig:Corres2}).  Now, using the fact that the family of Cwebs  has same mixing matrix due to same shuffle, both Boomerang Cwebs and non-Boomerang Cwebs obtained this way have same mixing matrices as they belong to the same family. Thus, the two Cwebs shown in fig.~\eqref{fig:Corres2} have the same mixing matrix.


\section{Direct construction of Boomerang Cwebs}
\label{sec:FormNnotation}

The computation of mixing matrices for boomerang Cwebs, its rank  which gives the number of exponentiated colour factors can be found  using replica trick formalism. 
However,  a direct construction is desirable using the known properties of the mixing matrices as it sheds light on the structure of these matrices. A few attempts towards this goal  have been made in the literature in 
~\cite{Dukes:2013gea,Dukes:2013wa,Dukes:2016ger,Agarwal:2021him,Agarwal:2022wyk}. Recently several new concepts such as  basis Cwebs, Normal ordering and Fused-Webs were introduced  in~\cite{Agarwal:2022wyk}, which proved to be very useful  in determining 
the diagonal blocks of several classes of mixing matrices for Cwebs for the case of massless Wilson lines,  which readily yield the number of exponentiated colour factors for a given web if certain data is already available. These concepts were discussed briefly in section \ref{sec:review}.

The Normal ordered form of eq.~\eqref{eq:R-gen-big-form} and the properties of the matrix $ D $ allow us to obtain the diagonals blocks of mixing matrices for boomerang Cwebs at order $\alpha_s^{n}$ and beyond using the basis Cwebs upto $ \alpha_s^n $. Towards this end, we introduce few objects and notations 
that would be helpful.  In the next subsection, we will also apply these concepts to completely determine the mixing matrix of a special class of boomerang Cwebs at all orders.

Suppose there are several gluon attachments on a Wilson line. If $A_{1}, A_{2},  \ldots A_{n}$ denote attachments that originate from the same gluon correlator and there no attachments from any other correlator between $A_{1}$ and $A_{n}$, then we denote these attachments as $\overline{A_{1}, \ldots, A_{n}}$.

\begin{figure}[t]

	\centering
	\subfloat[][]{\includegraphics[scale=0.055]{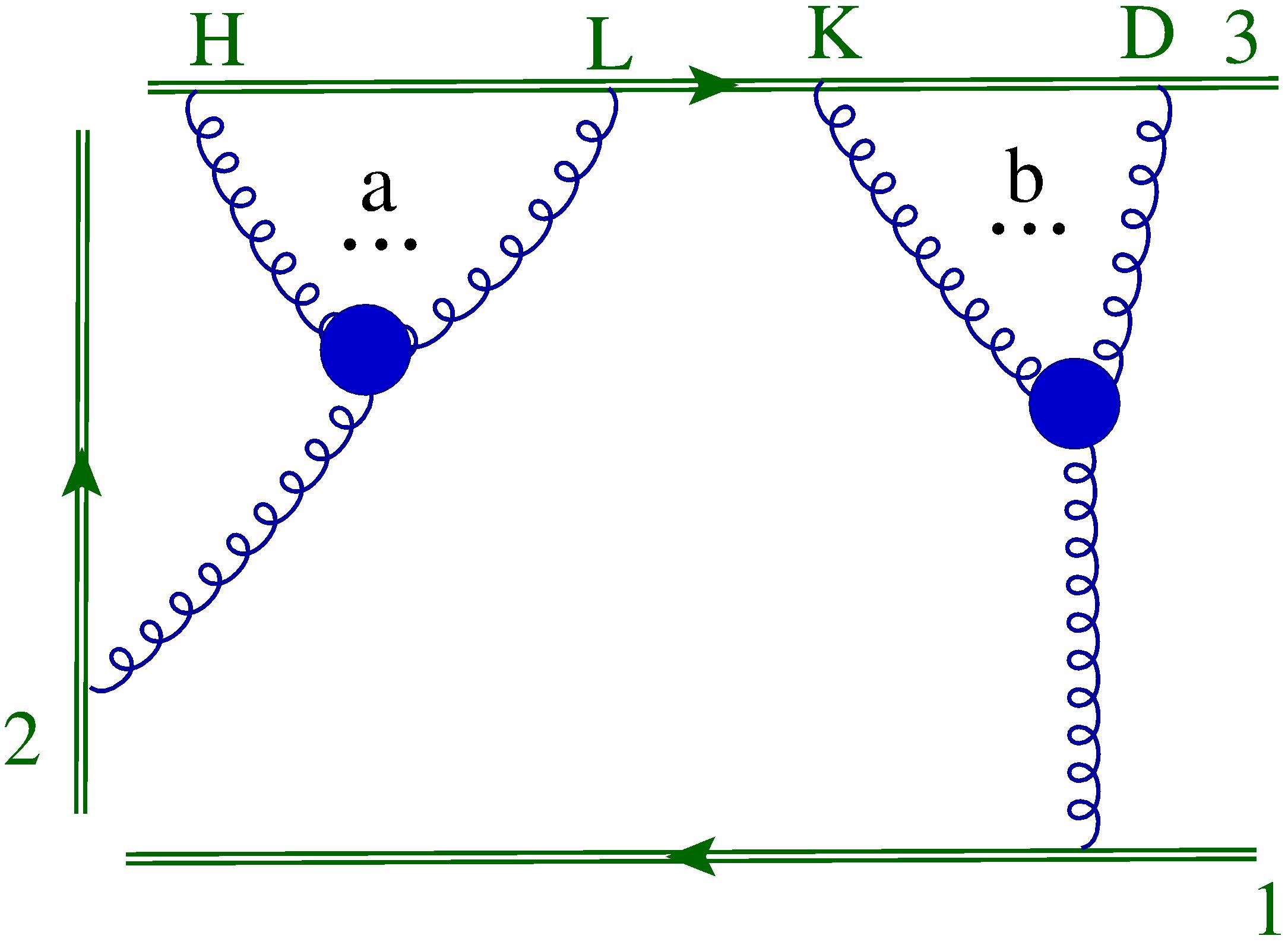} }
	\qquad \qquad
	\subfloat[][]{\includegraphics[scale=0.055]{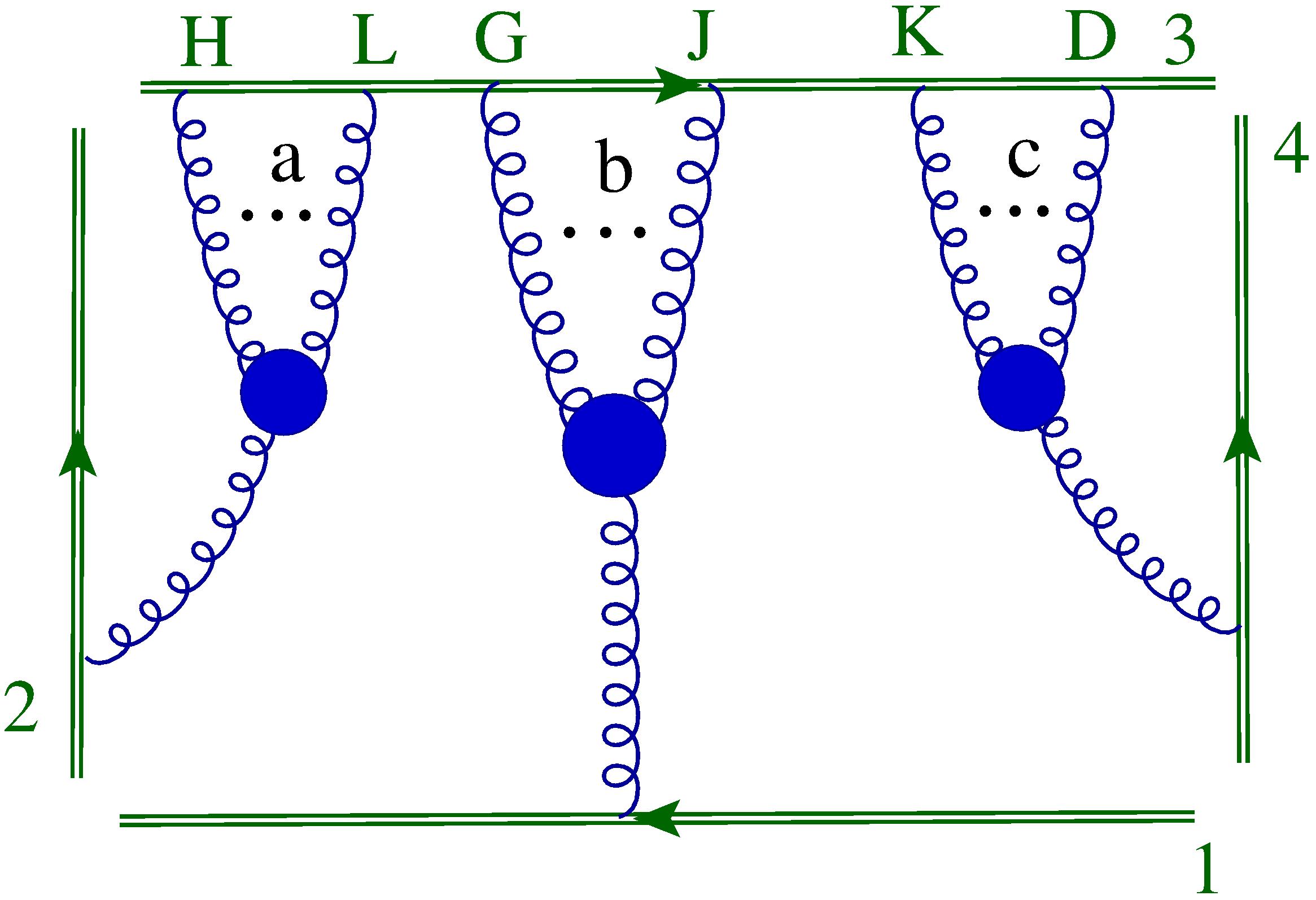} }
	\caption{Cwebs having shuffle of $ a $, $ b $ and $ c $ attachments at same lines}
	\label{fig:count-shuffle}
\end{figure}

A useful quantity is the number of shuffles for which the correlators attached to a Wilson line remain entangled, which we will denote by $E(a_{1}, \ldots a_{n})$ where $a_{j}$ denotes the number of attachments from a gluon correlator.
This equals the total number of shuffles on the Wilson line minus the number of shuffles for which  the correlators get disentangled.  If $ a $ gluons from one correlator and $ b $ gluons from a different correlator are attached to the same Wilson line, as shown in fig.~(\ref{fig:count-shuffle}\txb{a}), then the number of ways in which these correlators remain entangled is, 
\begin{align}
\label{eq:Num-two-entangled0}
E(a,b)=	(a\,\Pi\,b) - 2&=\; \dfrac{(a+b)!}{a! \,b!} -2\,,
\end{align}
where the first term indicates the total number of shuffles possible on the given Wilson line, and the second term \textit{i.e.}, $ 2 $ represents the total number of reducible diagrams.

If there are $a$, $b $ and $ c $ attachments from three different correlators to a Wilson line, as shown in fig.~(\ref{fig:count-shuffle}\txb{b}), then the number of ways in which these correlators are entangled is given by
\begin{align}
\label{eq:Num-three-entangled}
E(a,b,c) &=\;  (a\,\Pi\,b\,\Pi\,c ) - 2[(a\,\Pi\,b)-2]- 2[(b\,\Pi\,c)-2]- 2[(c\,\Pi\,a)-2] -6 \\
&=\;\dfrac{(a+b+c)!}{a!\,b!\,c!}+6 \nonumber \\
& -2 \left[ \dfrac{(a+b)!}{a!\,b!} + \dfrac{(b+c)!}{b!\,c!} + \dfrac{(c+a)!}{c!\,a!} \right] \,.
\end{align}
where the first term represents total number of shuffles among the three correlators whereas second, third and fourth terms represent the number of ways in which attachments from two of the three correlators remain entangled; fifth term ie $ 6 $ or $ 3! $ represents the number of ways in which all the three correlators are completely disentangled. If there are attachments from several correlators on more than one Wilson line then  the number of shuffles for which the correlators are entangled on each line are multiplied together.

In the next subsection, we describe the calculation of diagonal blocks for four special classes of Cwebs. These four special classes are considered because if one sits down and draws the Boomerang Webs that appear at four loops connecting four Wilson lines, one finds that these classes are sufficient to determine their diagonal blocks. In subsection \ref{explicit} we fully construct the mixing matrix of a class of Cwebs to all orders.

\subsection{Diagonal blocks of four special classes}
\label{sec:general-class}

\subsubsection{Cweb $\text{W}\,_{n+2}^{(n,1)}(1,1,\dots,1,n+2)$}
\begin{figure}[H]
	\captionsetup[subfloat]{labelformat=empty}
	\centering
	\subfloat[][]{\includegraphics[scale=0.064]{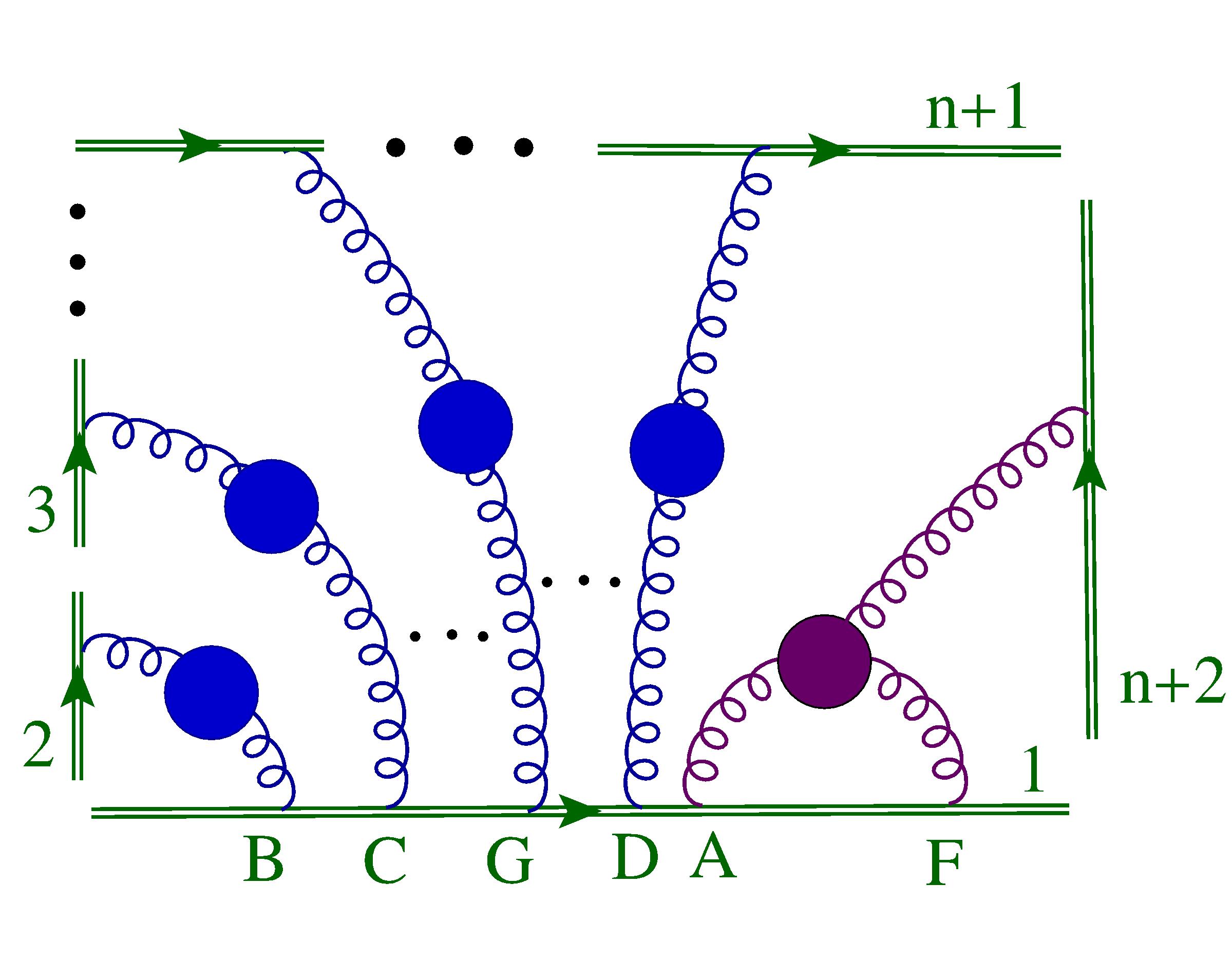} }
	\caption{Diagram of the general Cweb $\text{W}\,_{n+1}^{(n+1)}(1,1,1,\dots,n+2)$}
	\label{fig:Sclass1}
\end{figure}

\begin{figure}[t]
	\centering
	\subfloat[][]{\includegraphics[scale=0.055]{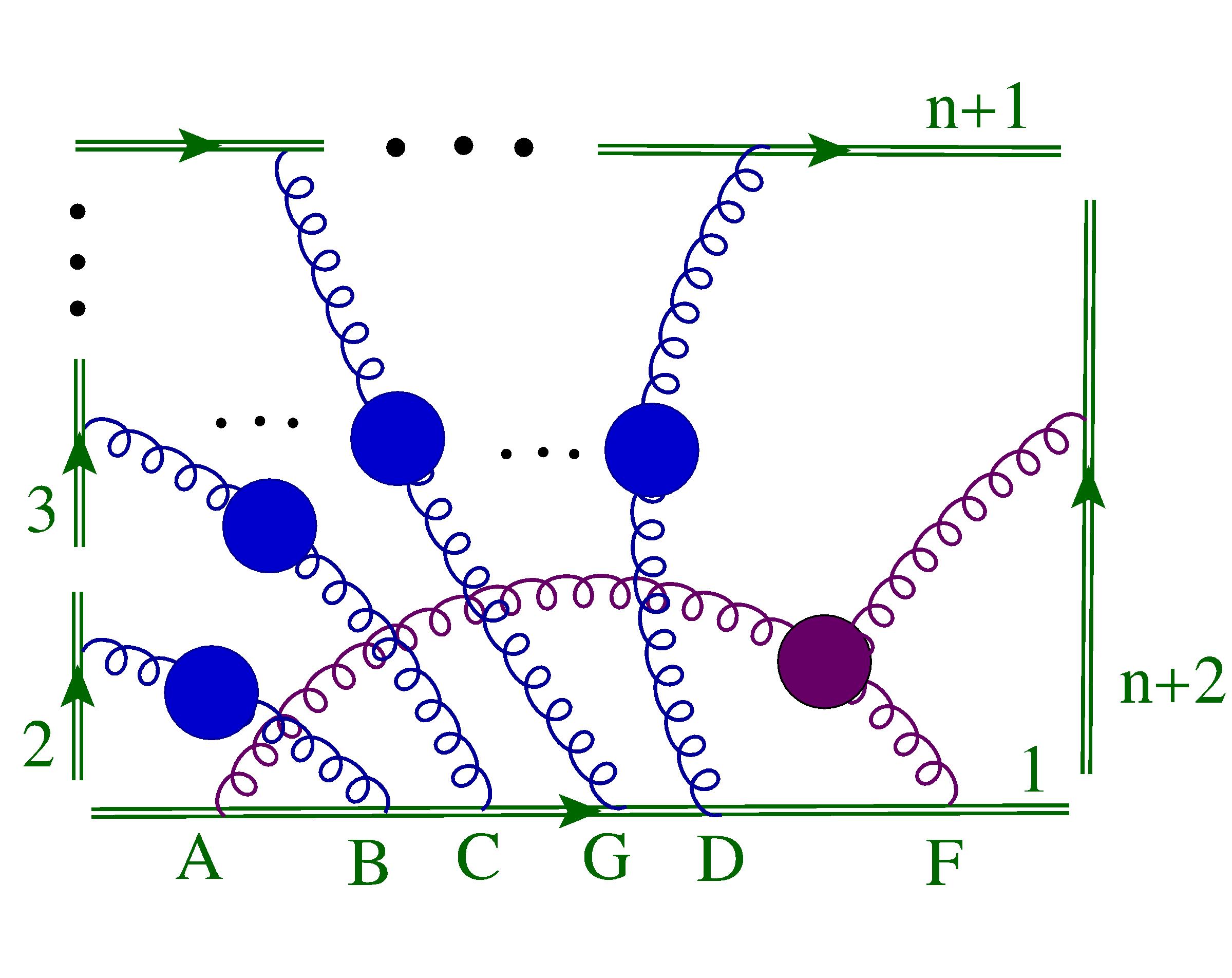} }
	\hspace{0.01cm}
	\subfloat[][]{\includegraphics[scale=0.055]{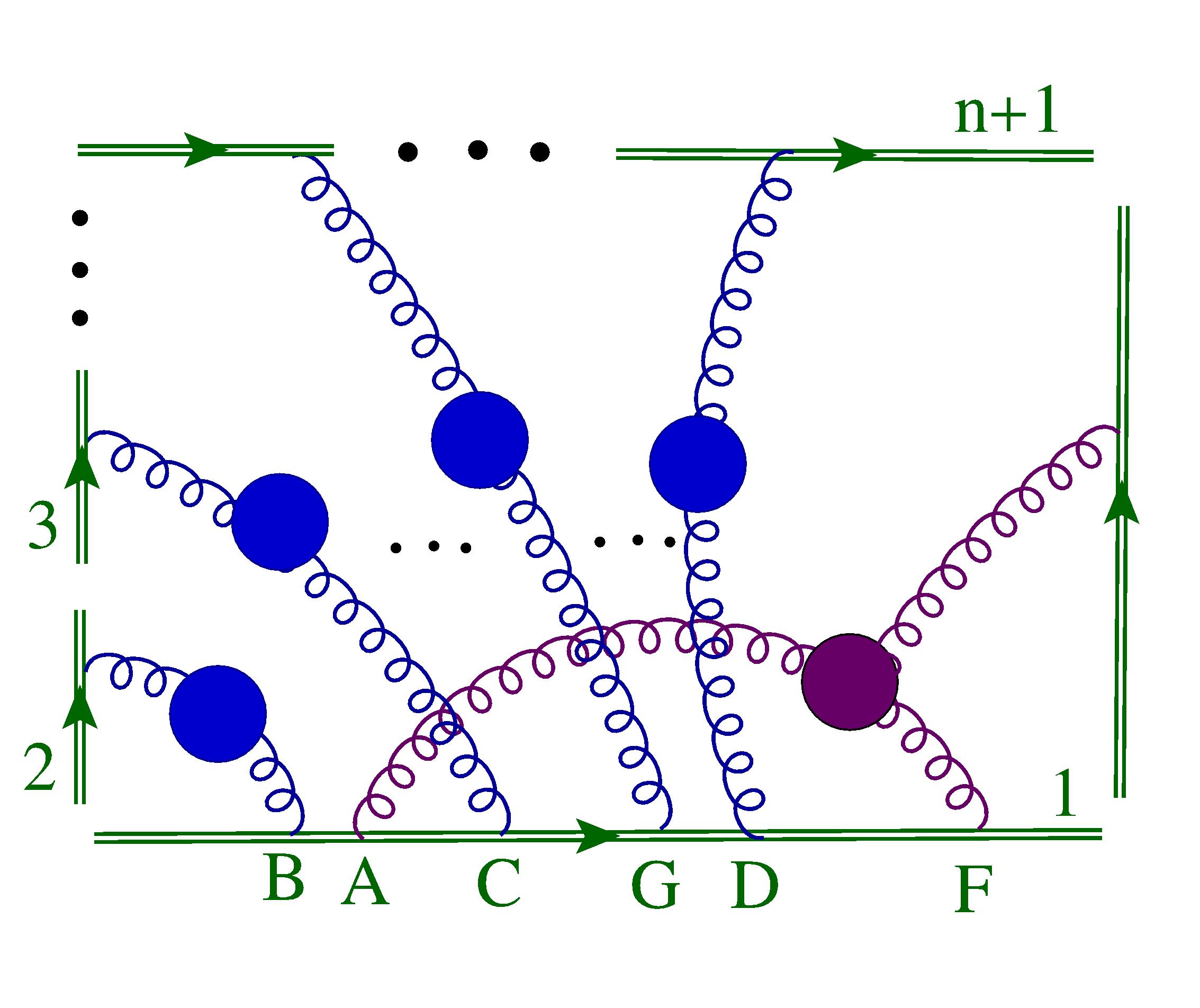} }
	\hspace{0.01cm}
	\subfloat[][]{\includegraphics[scale=0.055]{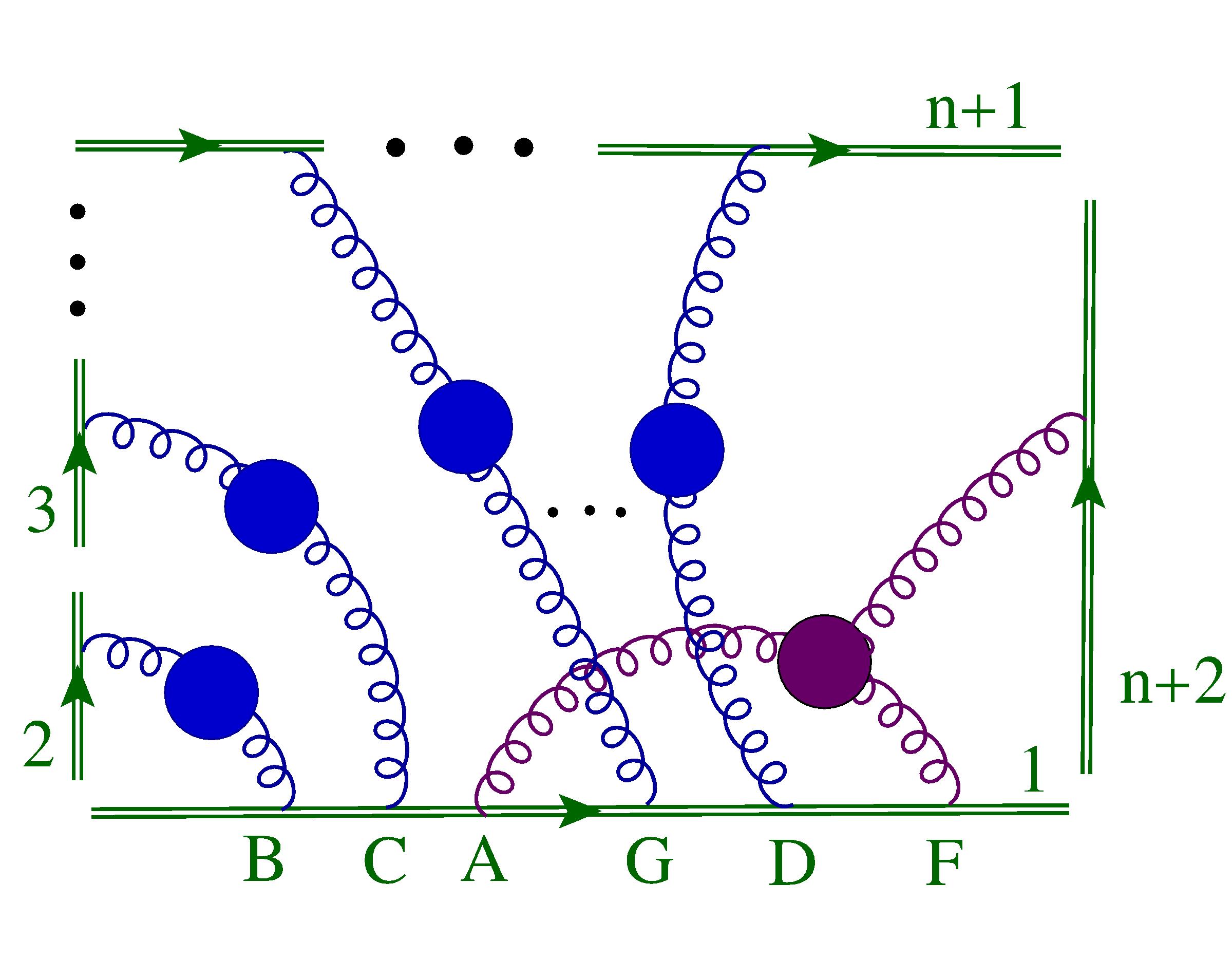} }
	\hspace{0.01cm}
	\subfloat[][]{\includegraphics[scale=0.055]{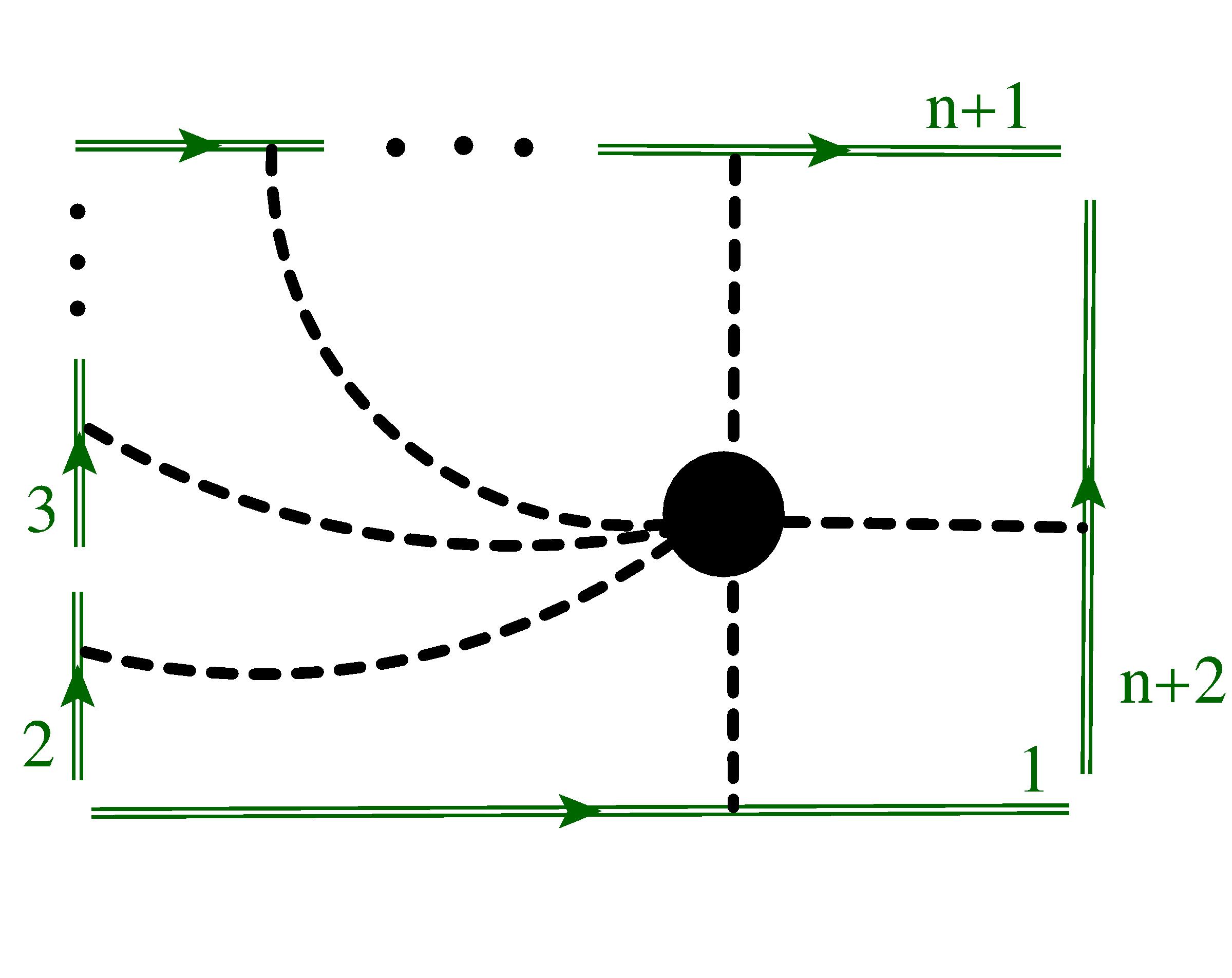} }
	\hspace{0.01cm}
	\subfloat[][]{\includegraphics[scale=0.055]{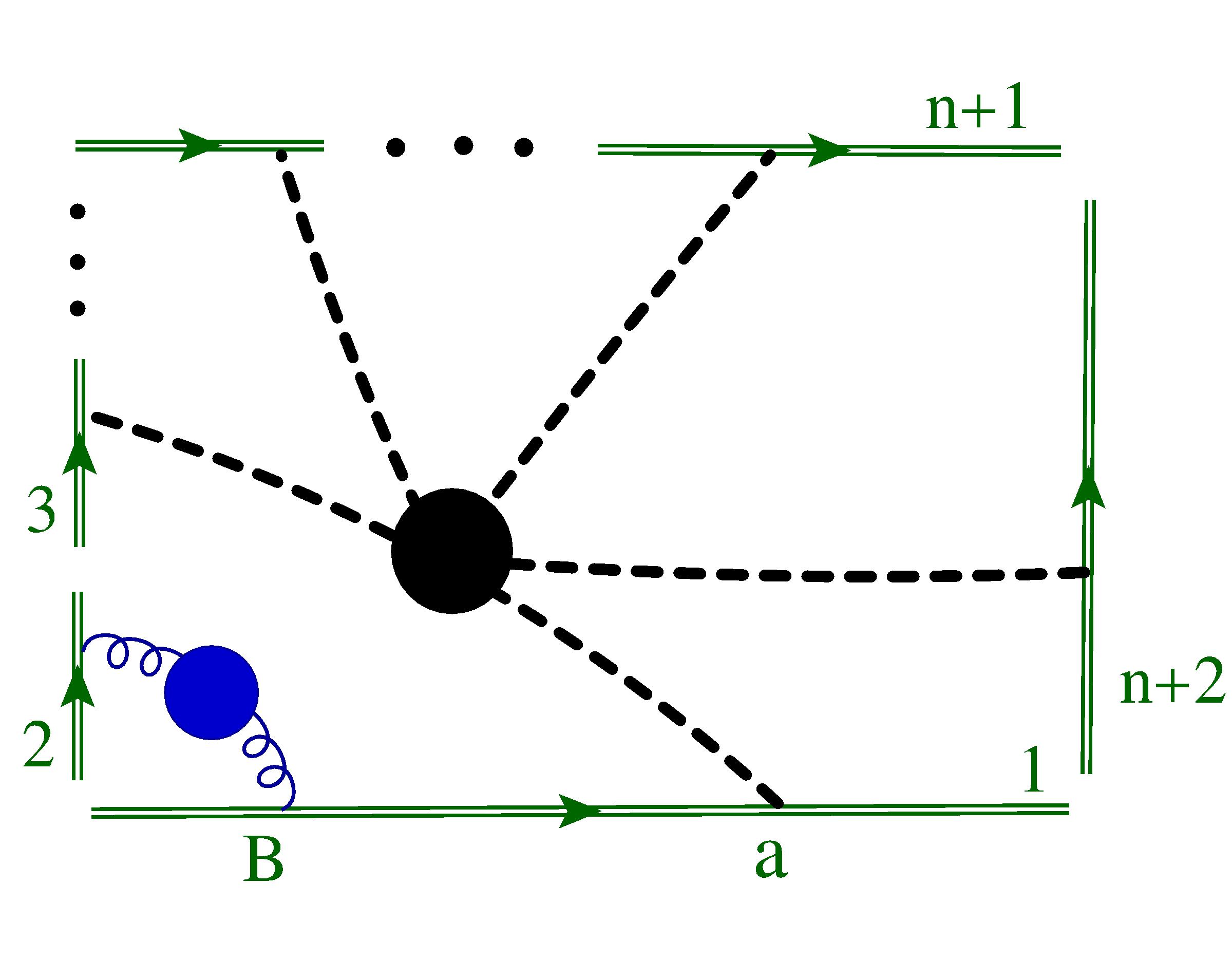} }
	\hspace{0.01cm}
	\subfloat[][]{\includegraphics[scale=0.055]{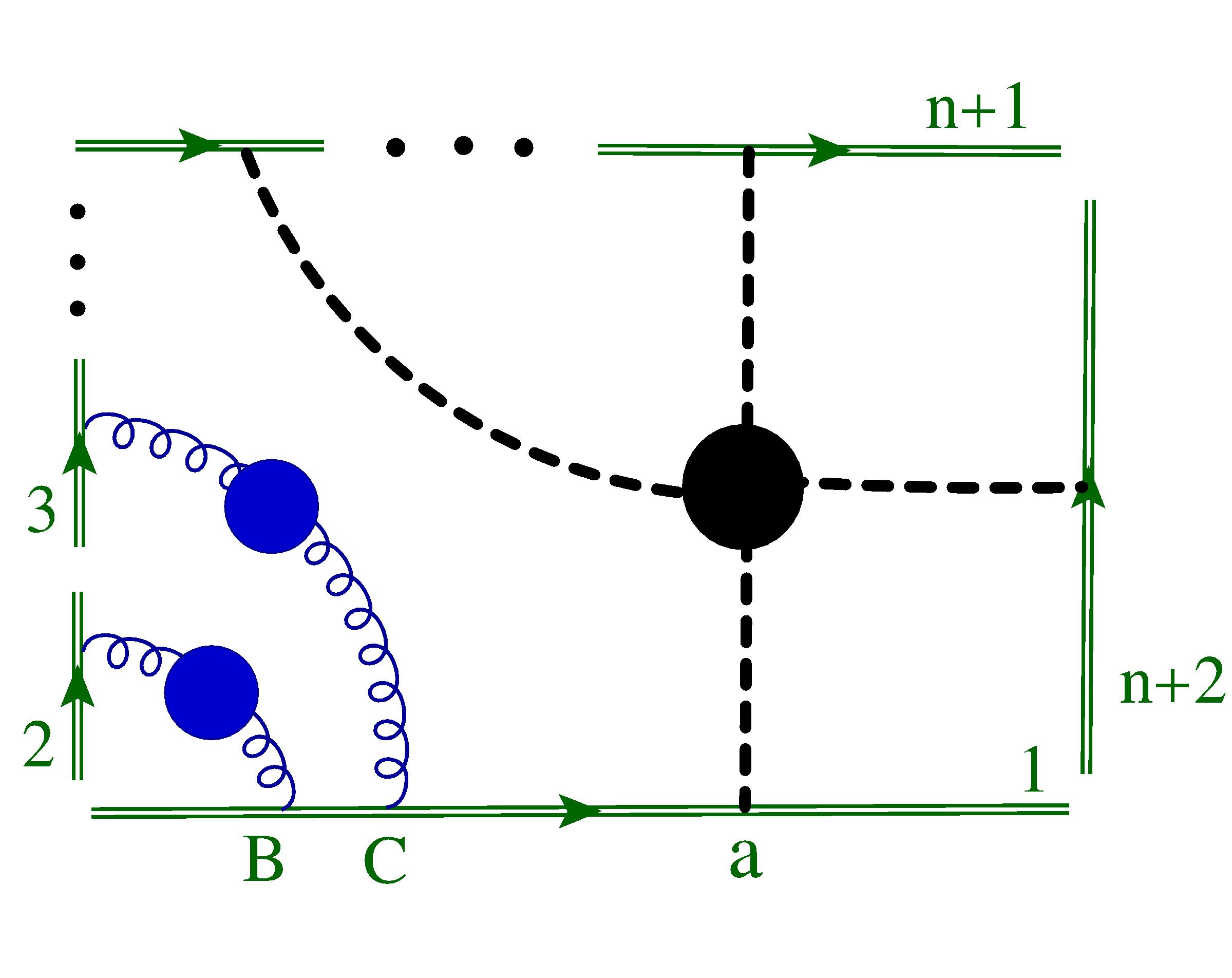} }
	\caption{Fused-Web for completely and partially entangled diagrams of Cweb $\text{W}\,_{n+1}^{(n+1)}(1,1,1,\dots,n+2)$}
	\label{fig:Sclass1-R}
\end{figure}

Cwebs of this class contain one three-point and $ n $ two-point gluon correlators connecting $ (n+2) $ Wilson lines. In this Cweb, the shuffle is possible only on one Wilson line, and the other lines have single attachments. A general Cweb of this class is shown in fig.~(\ref{fig:Sclass1}). The shuffle of attachments on line $ 1 $ generates all the diagrams for this Cweb. The total number of diagrams is  
\begin{align}
\frac{(n+2)!}{2!} \,.\nonumber
\end{align}
Note that the reducible diagrams can only be generated by considering the three point correlator as a single entity. If we do so then the reducible diagrams of the Cweb become same as that of $ \text{W}_{n+2}^{(n+1)}(1,1,\cdots,n+1) $, with $ S=\{1,1,1,\cdots,1\} $~\cite{Dukes:2013gea}. The total number of reducible diagrams for this Cweb is $ (n+1)! $.
Thus the column weight vector $ S_D=\{1_{(n+1)!}\} $, is associated with matrix $ D $. Further, from the Uniqueness theorem we can write 
\begin{align}
D = R(1_{(n+1)!})\,\qquad;\qquad r(D)\;=\; r(R(1_{(n+1)!}))\;=\;n!\,.
\end{align}
The explicit form and rank of $ R(1_{(n+1)!}) $ for any arbitrary $ n $ has been computed~\cite{Dukes:2013gea} without applying the replica trick algorithm. For $n=1$, $2$ and $3$, we get  $  R(1_{2!})  $, $  R(1_{3!})  $ and $  R(1_{4!}) $  which appear as mixing matrices of basis Cwebs at two, three and four loops connecting three, four and five Wilson lines respectively, see appendix~\textcolor{blue}{C} of~\cite{Agarwal:2022wyk}.
\\
After computing the block $ D $, we can determine the diagonal blocks of $ A $ using Fused-Webs. The form of $ A $ has a similar structure as that of $ R $,
\begin{align}\label{eq:A-GENform}
A=\left(\begin{array}{cc}
I&A_U\\
O&A_L
\end{array}\right)\,.
\end{align} 
The order of identity matrix is determined by the number of completely entangled diagrams of a Cweb. For completely entangled diagrams in this Cweb, the attachments of all the two point gluon correlators on line 1 are in between the two attachments of the three-point gluon correlator, as shown in fig.~(\ref{fig:Sclass1-R}\txb{a}).  

These attachments have $ n! $ shuffle, within the two attachments of three point correlator and each of these shuffle corresponds to an ordering of attachments of a completely entangled diagram whose Fused-Web is shown in fig~(\ref{fig:Sclass1-R}\txb{d}). Thus the order of identity matrix is $ n! $\,. 
\\

Further, to determine the diagonal blocks of $ A_L $, we have to consider partially entangled diagrams of this Cweb. Any diagram of this Cweb is partially entangled when at least one of the two-point gluon correlator lies outside the attachments of three-point gluon correlator. 

Consider a general case, where $ k $ two-point gluon correlators $ (k<n) $ are outside the two attachments of the three point correlator. There are $ \,^nC_k $  possibilities in which $ k $ two-point gluon correlators are chosen and they can be placed outside the three point correlator on line $ 1 $ in $ k! $ ways. Further, the remaining $ (n-k) $ gluon correlators can be attached in $ (n-k)! $ ways within the attachments of three point correlator. For example if we take $ k=2 $ as shown in fig.~(\ref{fig:Sclass1-R}\txb{c}) then the  two two-point gluon correlators can be placed outside the three point correlator in $ 2! $ ways, and the remaining $ (n-2) $ attachments inside the boomerang will shuffle in $ (n-2)! $ ways. Thus for general $k$, number of partially entangled diagrams is,
\begin{align}
\,^nC_k\,(n-k)! \,.\nonumber
\end{align}
There is a Fused diagram corresponding to all completely and partially entangled diagrams. The shuffle of attachments of each Fused diagram generates the corresponding Fused-Web with $ S=\{1_{(k+1)!} \}  $. This Fused-Web is the member of family $ f(1_{(k+1)!}) $ with basis Cweb $\text{W}\,_{k+1}^{(k+2)}(1,1,\dots,k+1)$. From the Uniqueness theorem, we can directly write the mixing matrix for this Fused-Cweb as $ R(1_{(k+1)!}) $.
\\
\\
The above discussion is illustrated explicitly for $ k=1,2 $ in fig.~(\ref{fig:Sclass1-R}). The details of the Fused-Webs and associated mixing matrices  for any general $ k $ are given in table \ref{tab:Sclass1}. 

The mixing matrices of Fused-Webs appear as the diagonal blocks of $ A $. The rank of $ A $ is given by the sum of product of rank of these matrices and the number of times they appear on the diagonals. Using table~\ref{tab:Sclass1}, the rank of $ A $ for this class of Cwebs is given by, 
\begin{align}
r(A) &=\;\, ^nC_{0}\,n!\times 0!\,+\, \,^nC_{1}\,(n-1)!\times 1!\,+\,\,^nC_{2}\,(n-2)!\times 2!\,+\,\dots\,+\,^nC_{n-1}\,1!\times(n-1)!\nonumber\\
&=\; \, ^nC_{0}\,n!\times 0! + \sum_{1}^{n-1} \, ^nC_{k}\,(n-k)!\times k!\nonumber\\
&=\;\, n! + n!(n-1)\;=\;n\times n! 
\end{align}
The rank of $ R $ is given by the sum of ranks of $ A $ and $ D $~\cite{Agarwal:2022wyk}, thus,
\begin{align}
r(R)\;&=\, n\times n!\;+\; n! \;=\; (n+1)!\,.
\end{align}
\begin{table}[H]
	\begin{center}
		\begin{tabular}{|c|c|c|c|c|c|}
			\hline
			two point correlators & Irreducible & Fused-Web &Mixing matrix & Rank &Number of ways  \\ 
			outside three point correlator&  diagram & &of Fused-Web & & for configuration   \\
			\hline
			$ 0 $ &\ref{fig:Sclass1-R}\txb{a}&\ref{fig:Sclass1-R}\txb{d}&	$ \textbf{I} $ & 0! & $ \,^nC_{0}\,n! $  \\  \hline
			$ 1 $&\ref{fig:Sclass1-R}\txb{b} &\ref{fig:Sclass1-R}\txb{e}& $ R(1_{2!}) $& $ 1! $ &  $ \,^nC_{1}\,(n-1)! $ \\ 
			\hline
			$ 2 $&\ref{fig:Sclass1-R}\txb{c}&\ref{fig:Sclass1-R}\txb{f}&$ R(1_{3!}) $ & $ 2! $&  $ \,^nC_{2}\,(n-2)! $ \\ 
			\hline
			$ k $ &&&$ R(1_{(k+1)!}) $ & $ k! $&   $ \,^nC_{k}\,(n-k)! $ \\ 
			\hline
			\vdots&\vdots&\vdots& \vdots & \vdots & \vdots \\
			\hline
			$ n-1 $&&&$ R(1_{(n)!}) $ & $ (n-1)! $&  $ \,^nC_{n-1}\,1! $\\ 
			&  &&& &\\\hline
		\end{tabular}	
	\end{center}
	\caption{Fused-Webs and their mixing matrices for Cweb $\text{W}\,_{n+1}^{(n+1)}(1,1,1,\dots,n+2)$}
	\label{tab:Sclass1}
\end{table}
\noindent 
By removing the gluon attachment on line $ n+2 $ in the diagram of Cweb, the three point correlator can be turned into a boomerang. Based on the correspondence established in the previous section we can make the predictions for boomerang Cwebs of this family as well.

The following table shows the ranks of mixing matrices for boomerang Cwebs that appear upto four loops and belong to this category. 
\begin{table}[H]
	\begin{center}
		\begin{tabular}{|c|l|c|c|c|c|c|}
			\hline
			Value of $ n $ &  Boomerang Cweb & Attachment on leg 1 & Loop order  & Matrix $ D $ & $ r(A) $ & $ r(R) $  \\ \hline
			1  & $\;\text{W}\,_{2}^{(2)}(1,3)$& 3 & $ \mathcal{O}(g^4) $ & $ R(1_{2!}) $ & 1 & 2  \\
			\hline
			2  & $\;\text{W}\,_{3}^{(3)}(1,1,4)$& 4 & $ \mathcal{O}(g^6) $ & $ R(1_{3!}) $ & 4 & 6  \\
			\hline
			3  & $\;\text{W}\,_{4}^{(4)}(1,1,1,5)$& 5 & $ \mathcal{O}(g^8) $ & $ R(1_{4!}) $ & 18 & 24  \\
			\hline
		\end{tabular}	
	\end{center}
	\caption{Fused-Webs and their mixing matrices for Cweb $\text{W}\,_{n+1}^{(n+1)}(1,1,1,\dots,n+2)$}
	\label{tab:Rank-prediction-111n}
\end{table}
The families of above Cwebs also have same ranks. Thus, boomerang Cweb $ \text{W}\,_{4}^{(1,0,1)}(1,1,1,3) $, shown in fig.~(\ref{fig:4legsWeb6}) being a higher order member in family of $  \text{W}\,_{2}^{(2)}(1,3) $ has rank 2. Similarly, boomerang Cweb $ \text{W}\,_{4,\text{I}}^{(2,1)}(1,1,1,4) $, shown in fig.~(\ref{fig:six-one-web4-5-av}) is the member of family of $ \text{W}\,_{3}^{(3)}(1,1,4) $. Finally, the third boomerang Cweb $\text{W}\,_{4}^{(4)}(1,1,1,5)$ of the above table appears at four loops and connects four lines, as shown in fig.~(\ref{fig:six-one-web4-7-av}). The Fused-Web formalism for each of these Cweb  is shown in section~\ref{sec:Boom-4loop-4line}. 


\subsubsection{Cweb $\text{W}\,_{5}^{(1,1,1,1)}(1,1,1,k+p,m+l+n)$}
\begin{figure}[H]
	\captionsetup[subfloat]{labelformat=empty}
	\centering
	\subfloat[][]{\includegraphics[scale=0.07]{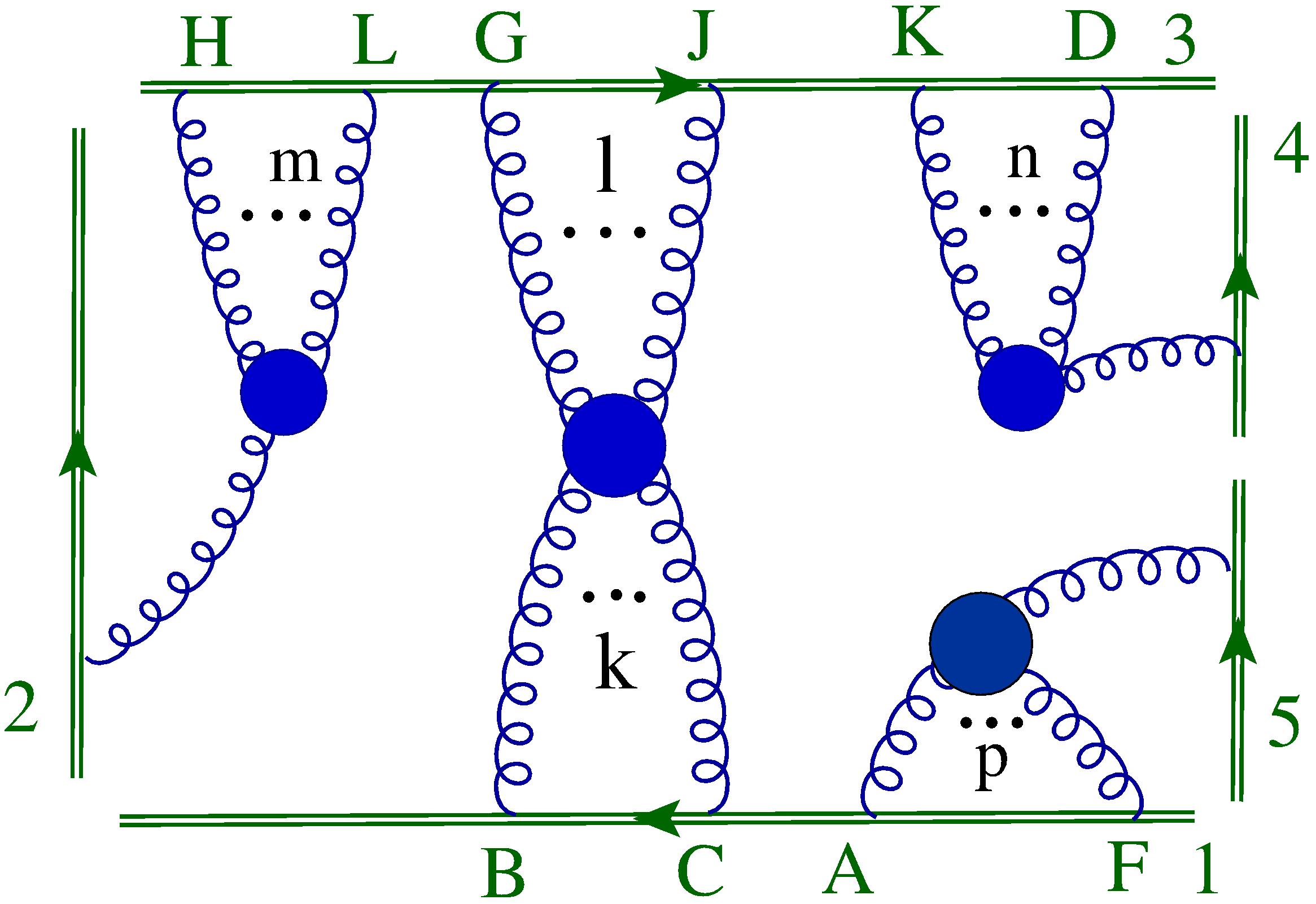} }
	\caption{Diagram of the general Cweb $\text{W}\,_{5}^{(1,1,1,1)}(1,1,1,k+p,m+l+n)$}
	\label{fig:2Sclass2}
\end{figure}

A general Cweb of this class is shown in fig.~(\ref{fig:2Sclass2}) which contains $ (m+1) $-point, $ (n+1) $-point, $ (p+1) $-point and a $ (l+k) $-point gluon correlators connecting five Wilson lines. The reducible diagrams for this Cweb can be obtained by considering the gluon attachments that are coming from the same correlator as a single piece and can be denoted by  $ \overline{C\cdots B} $, defined in section \ref{sec:FormNnotation}. The sequences of reducible diagrams for this Cweb and their corresponding $ s $-factors are given in the following table. 
\begin{table}[t]
	\begin{center}
		\begin{tabular}{|c|c|}
			\hline 
			\textbf{Sequences}  & \textbf{s-factors}  \\ 
			\hline 
			$\lbrace\lbrace \overline{F \cdots A},\,\overline{C\cdots B}\rbrace,\,\lbrace \overline{G\cdots J},\, \overline{H\cdots L},\,\overline{K\cdots D}\rbrace\rbrace$  &  1\\ \hline
			$\lbrace\lbrace \overline{F \cdots A},\,\overline{C\cdots B}\rbrace, \lbrace \overline{G\cdots J},\,\overline{K\cdots D},\, \overline{H\cdots L}\rbrace\rbrace$  &  1 \\ \hline
			$\lbrace\lbrace \overline{C\cdots B},\,\overline{F \cdots A}\rbrace, \lbrace  \overline{H\cdots L},\,\overline{K\cdots D},\,\overline{G\cdots J}\rbrace\rbrace$  &  1\\ \hline
			$\lbrace\lbrace \overline{C\cdots B},\,\overline{F \cdots A}\rbrace, \lbrace \overline{K\cdots D} ,\,\overline{H\cdots L},\,\overline{G\cdots J}\rbrace\rbrace$  & 1 \\ \hline
			$\lbrace\lbrace \overline{F \cdots A},\,\overline{C\cdots B}\rbrace, \lbrace  \overline{H\cdots L},\,\overline{G\cdots J},\,\overline{K\cdots D}\rbrace\rbrace$  & 2 \\ \hline
			$\lbrace\lbrace \overline{F \cdots A},\,\overline{C\cdots B}\rbrace, \lbrace  \overline{H\cdots L},\,\overline{K\cdots D},\,\overline{G\cdots J}\rbrace\rbrace$  & 2 \\ \hline
			$\lbrace\lbrace \overline{F \cdots A},\,\overline{C\cdots   B}\rbrace, \lbrace \overline{K\cdots   D} ,\,\overline{H\cdots L},\,\overline{G\cdots   J}\rbrace\rbrace$  & 2 \\ \hline 
			$\lbrace\lbrace \overline{F \cdots A},\,\overline{C\cdots   B}\rbrace, \lbrace \overline{K\cdots   D},\,\overline{G\cdots   J} ,\,\overline{H\cdots L}\rbrace\rbrace$  & 2 \\ \hline 
			$\lbrace\lbrace \overline{C\cdots   B},\,\overline{F \cdots A}\rbrace, \lbrace  \overline{H\cdots L},\,\overline{G\cdots   J},\,\overline{K\cdots   D}\rbrace\rbrace$  & 2 \\ \hline 
			$\lbrace\lbrace \overline{C\cdots   B},\,\overline{F \cdots A}\rbrace, \lbrace \overline{G\cdots   J},\, \overline{H\cdots L},\,\overline{K\cdots   D}\rbrace\rbrace$  & 2 \\ \hline 
			$\lbrace\lbrace \overline{C\cdots   B},\,\overline{F \cdots A}\rbrace, \lbrace \overline{G\cdots   J},\,\overline{K\cdots   D} ,\,\overline{H\cdots L}\rbrace\rbrace$  & 2 \\ \hline 
			$\lbrace\lbrace \overline{C\cdots   B},\,\overline{F \cdots A}\rbrace, \lbrace \overline{K\cdots   D},\,\overline{G\cdots   J},\, \overline{H\cdots L}\rbrace\rbrace$  & 2 \\ \hline  		
		\end{tabular}
	\end{center}
	\caption{Reducible diagrams and their $ s $-factors}
	\label{tab:2Sclass2-Reducible-D}
\end{table}

\noindent The $ s $-factors of the reducible diagrams, the Uniqueness theorem and the property of $ D $ fix, 
\begin{align}\label{eq:2Sclass2-rankD}
D &=\;R(1_4,2_8)\,, \qquad\qquad r(D)\;=\;r(R(1_4,2_8))\;=\;2\,.
\end{align}    
Note that the Cweb corresponding to the basis matrix $ R(1_4,2_8) $ appears for $ \text{W}_5^{(4)}(1,1,1,2,3)$ at four loops connecting five massless Wilson lines. 
\begin{figure}[t]
	\centering
	\subfloat[][]{\includegraphics[scale=0.055]{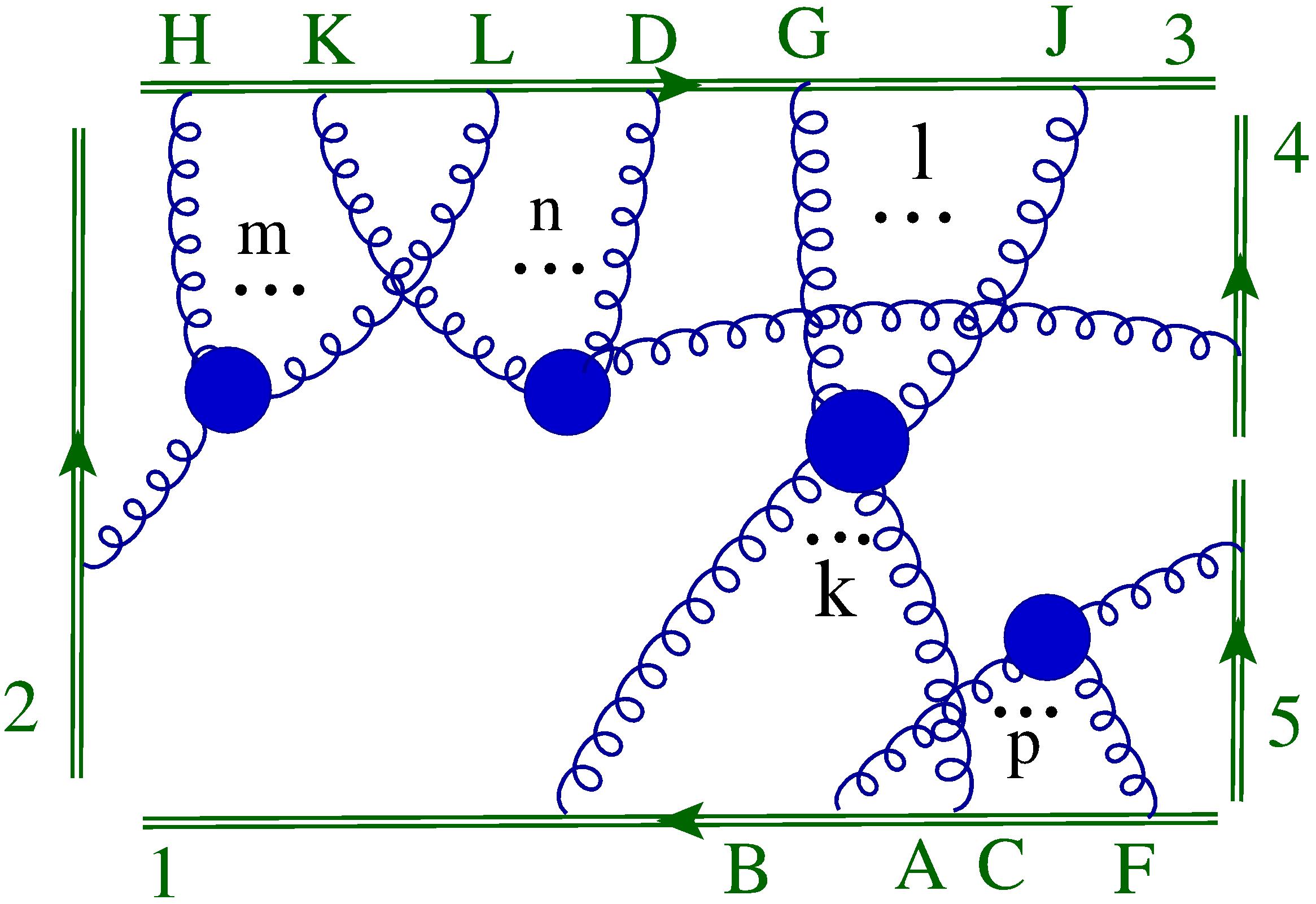} }
	\quad
	\subfloat[][]{\includegraphics[scale=0.055]{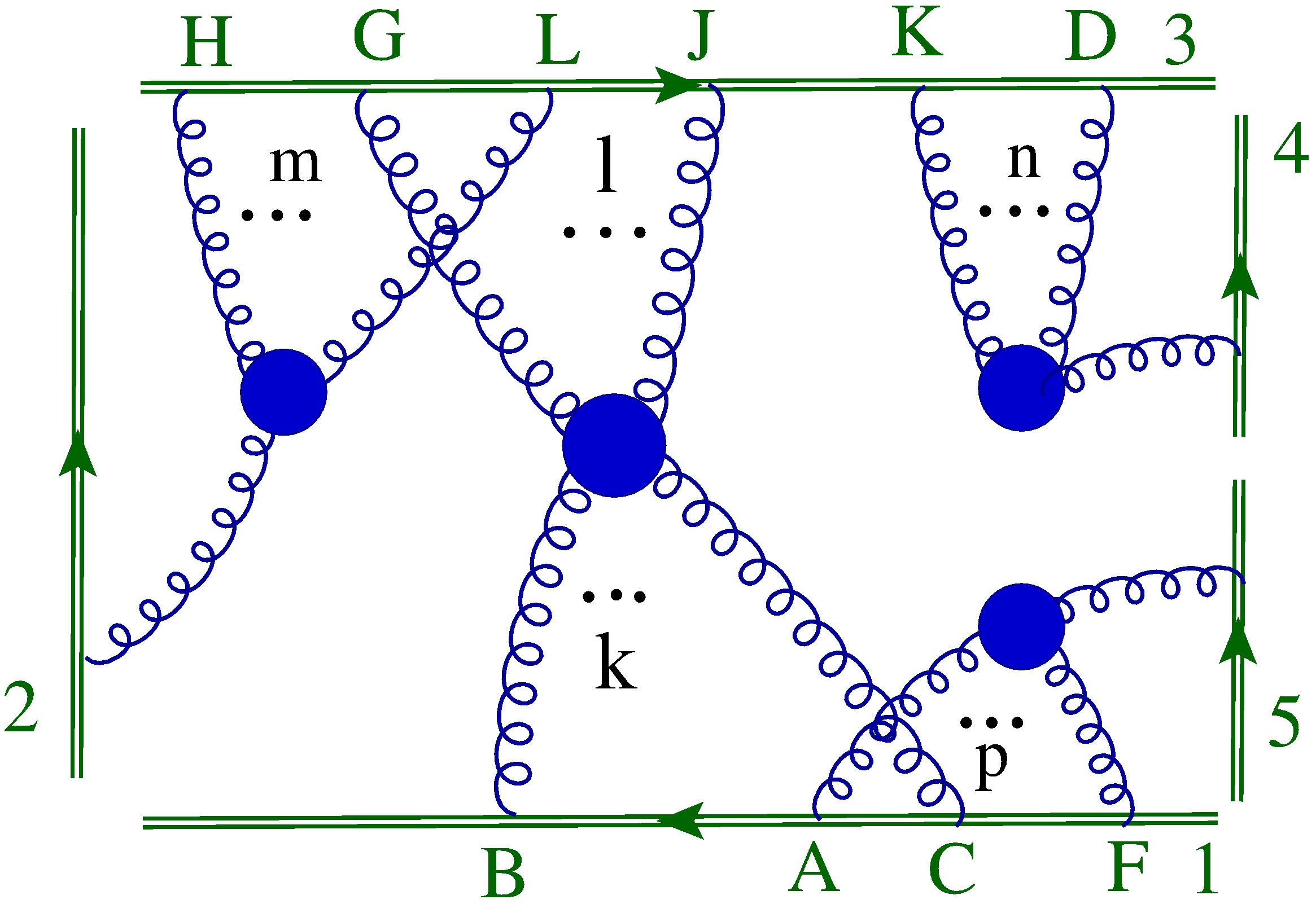} }
	\quad
	\subfloat[][]{\includegraphics[scale=0.055]{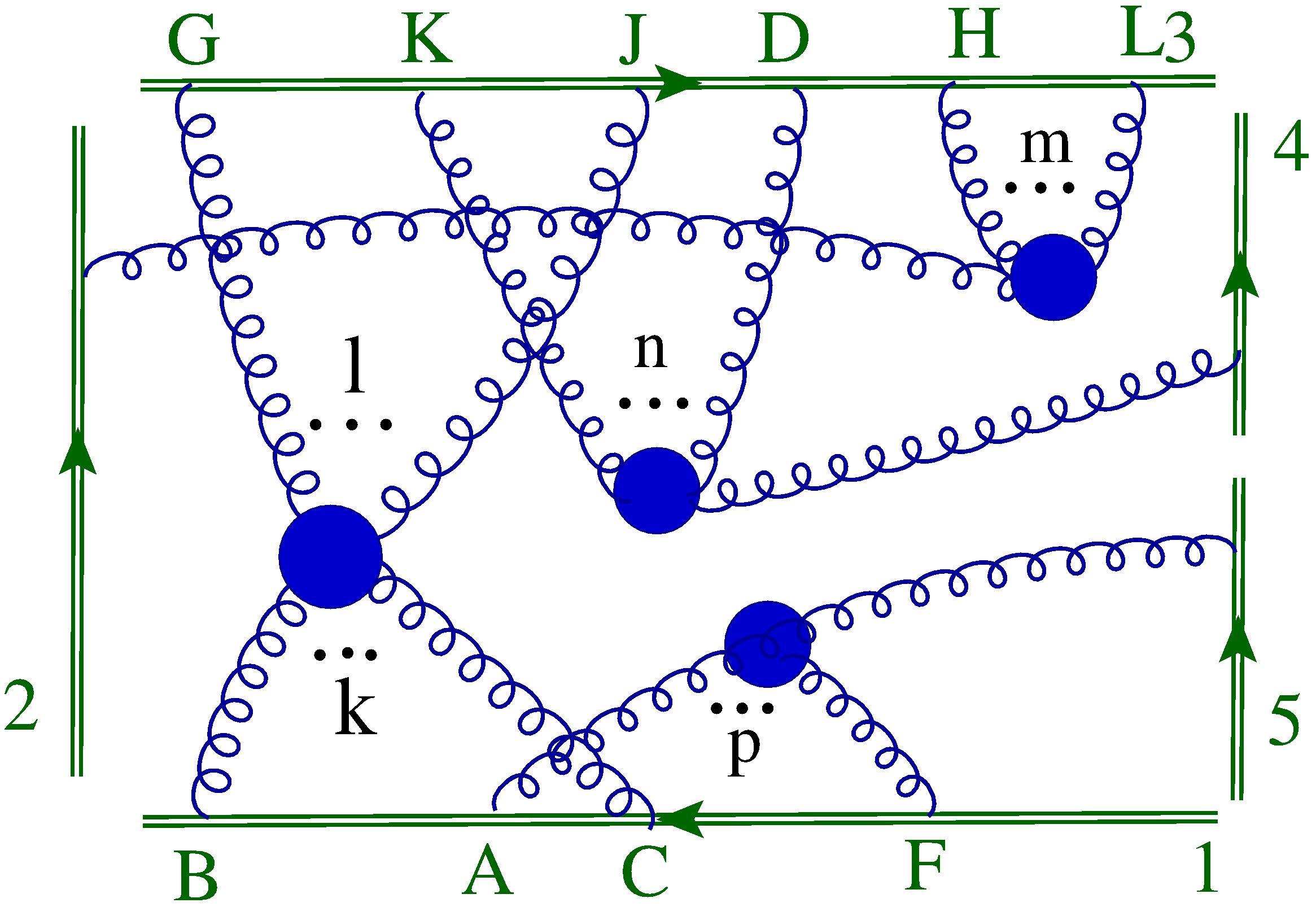} }
	\quad
	\subfloat[][]{\includegraphics[scale=0.055]{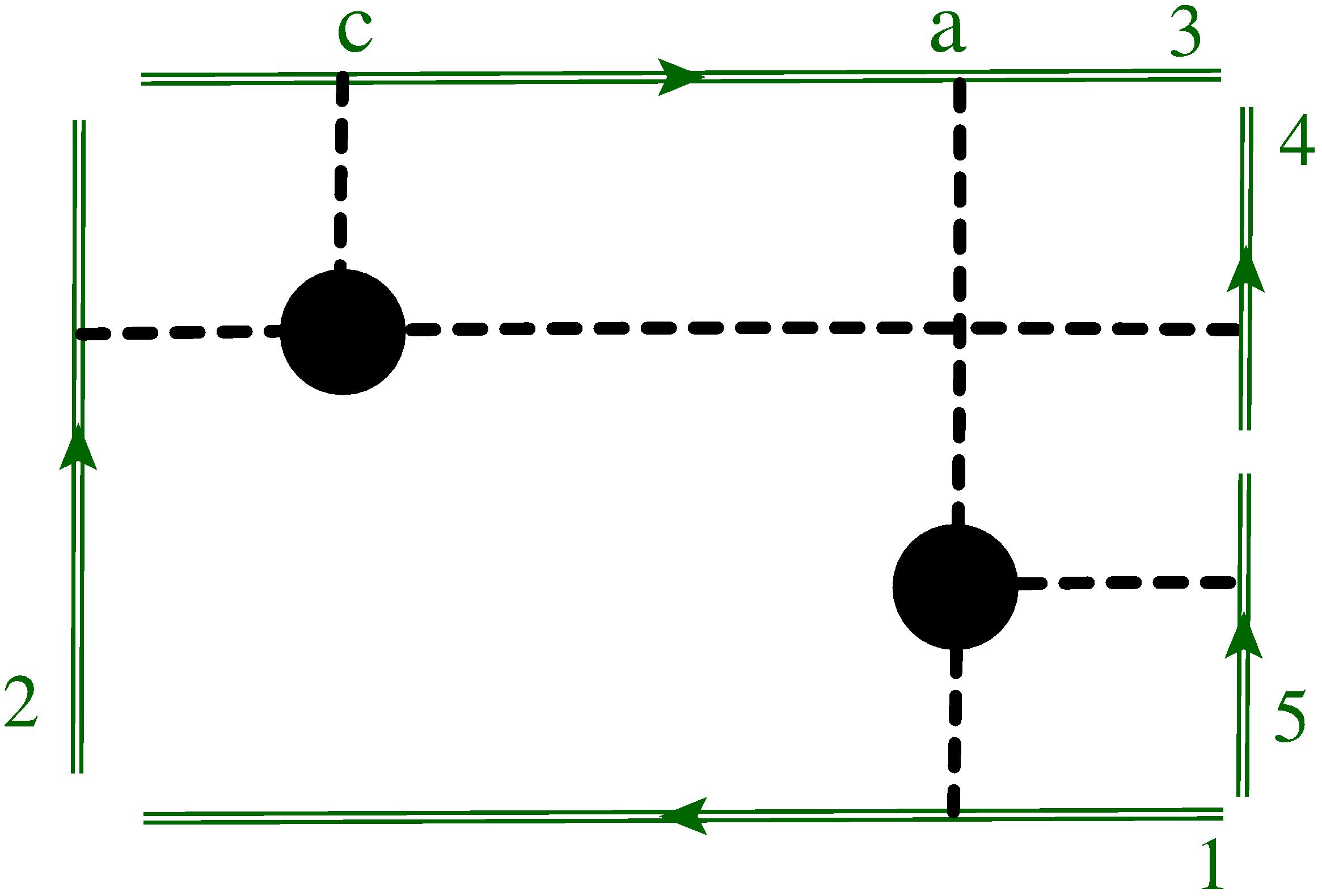} }
	\quad
	\subfloat[][]{\includegraphics[scale=0.055]{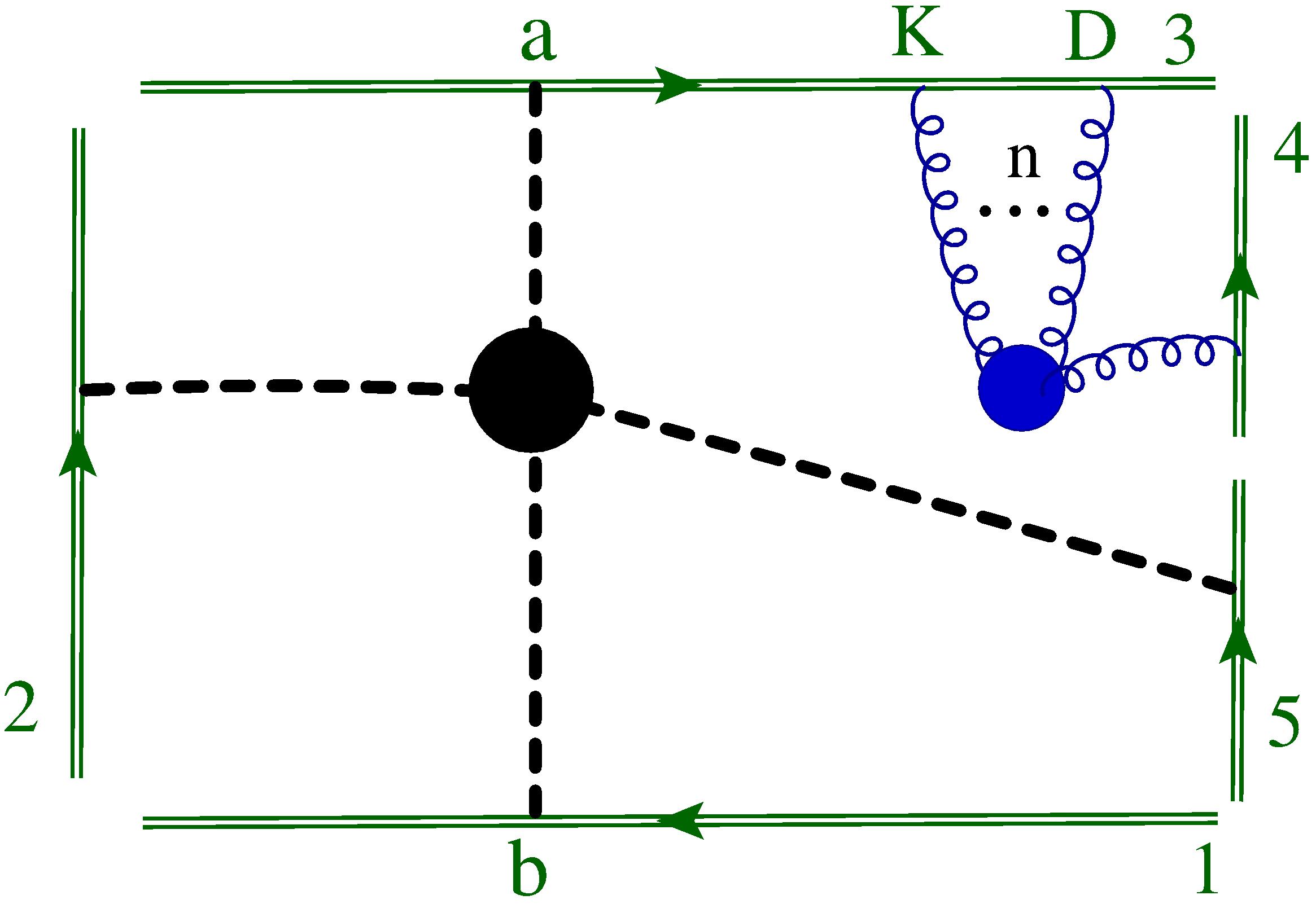} }
	\quad
	\subfloat[][]{\includegraphics[scale=0.055]{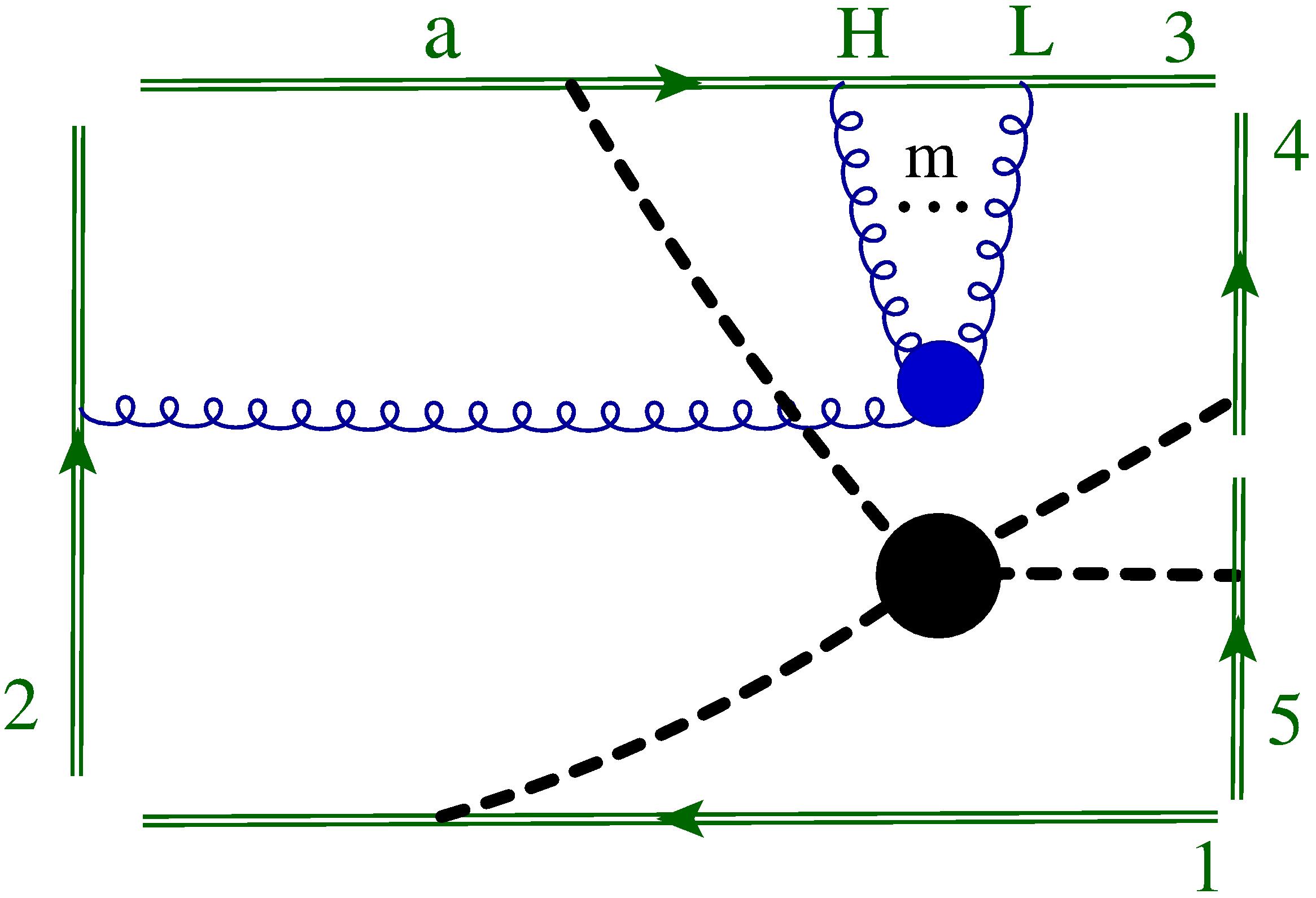} }
	\caption{Fused-Webs for Cweb $\text{W}\,_{5}^{(1,1,1,1)}(1,1,1,k+p,m+l+n)$}
	\label{fig:2Sclass2-A}
\end{figure}
To determine the diagonal blocks of the mixing matrix for this Cweb, we now need to calculate $ A $. 

\begin{figure}
	\centering
	\subfloat[][]{\includegraphics[scale=0.055]{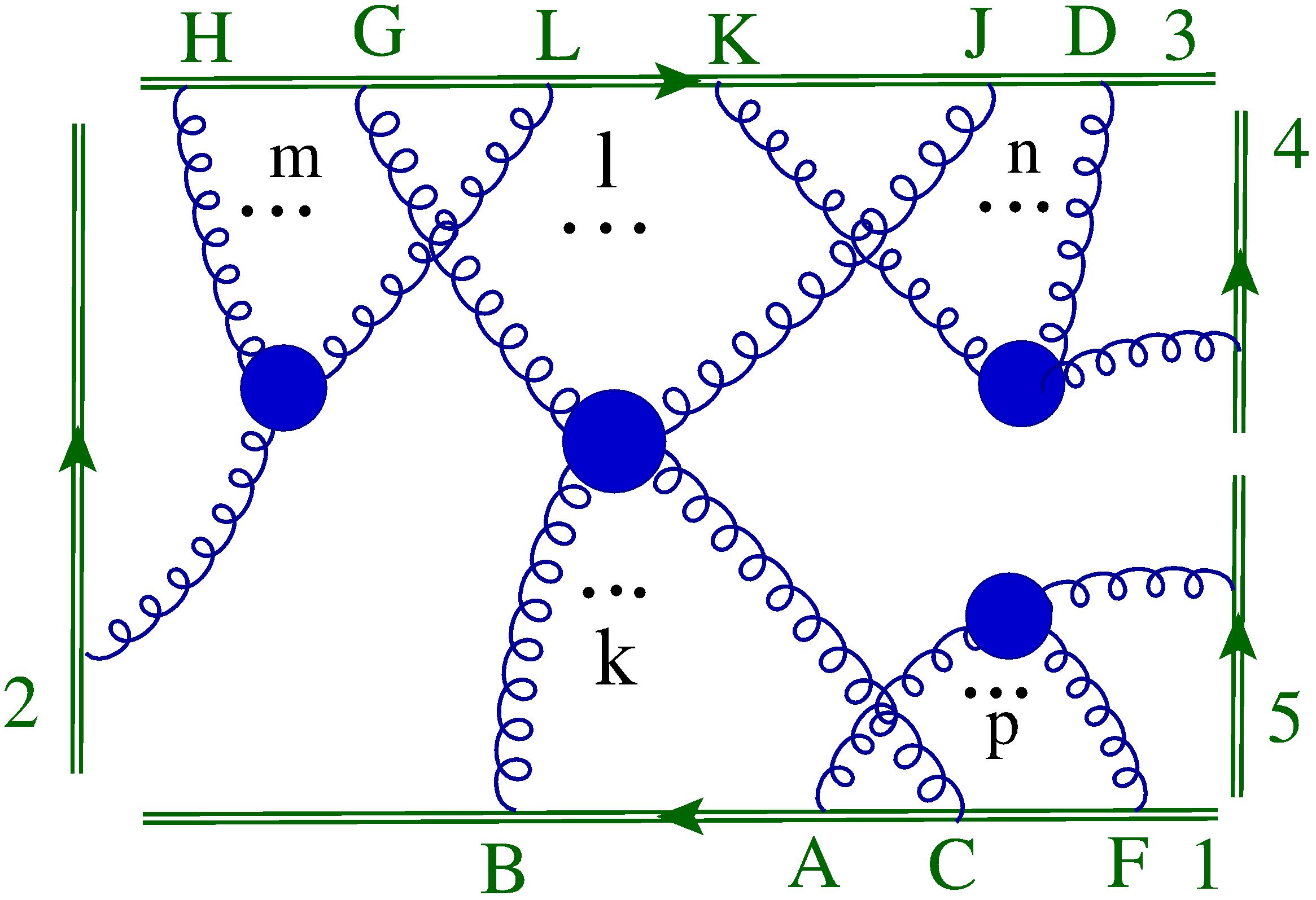} }
	\quad
	\subfloat[][]{\includegraphics[scale=0.055]{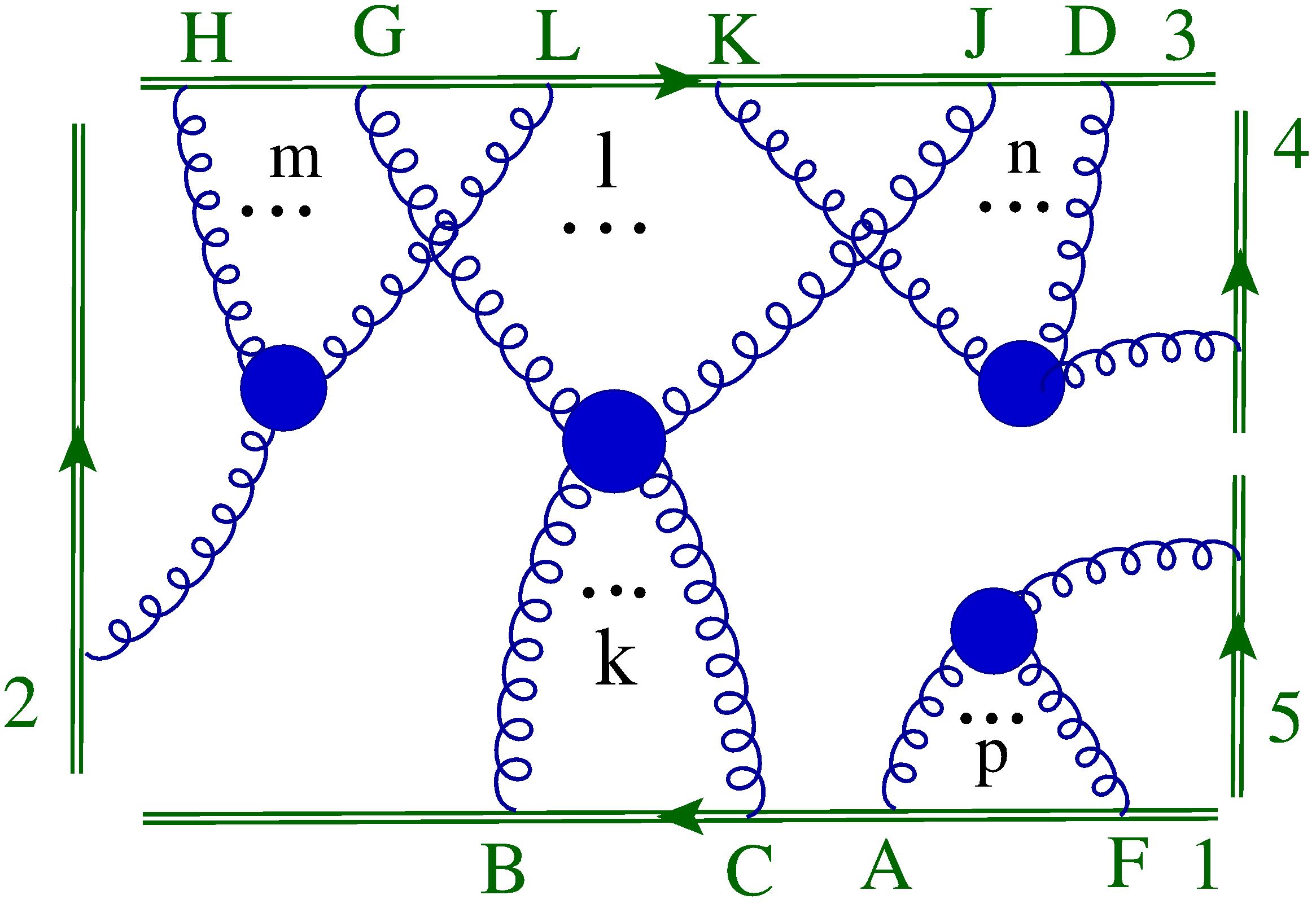} }
	\quad
	\subfloat[][]{\includegraphics[scale=0.055]{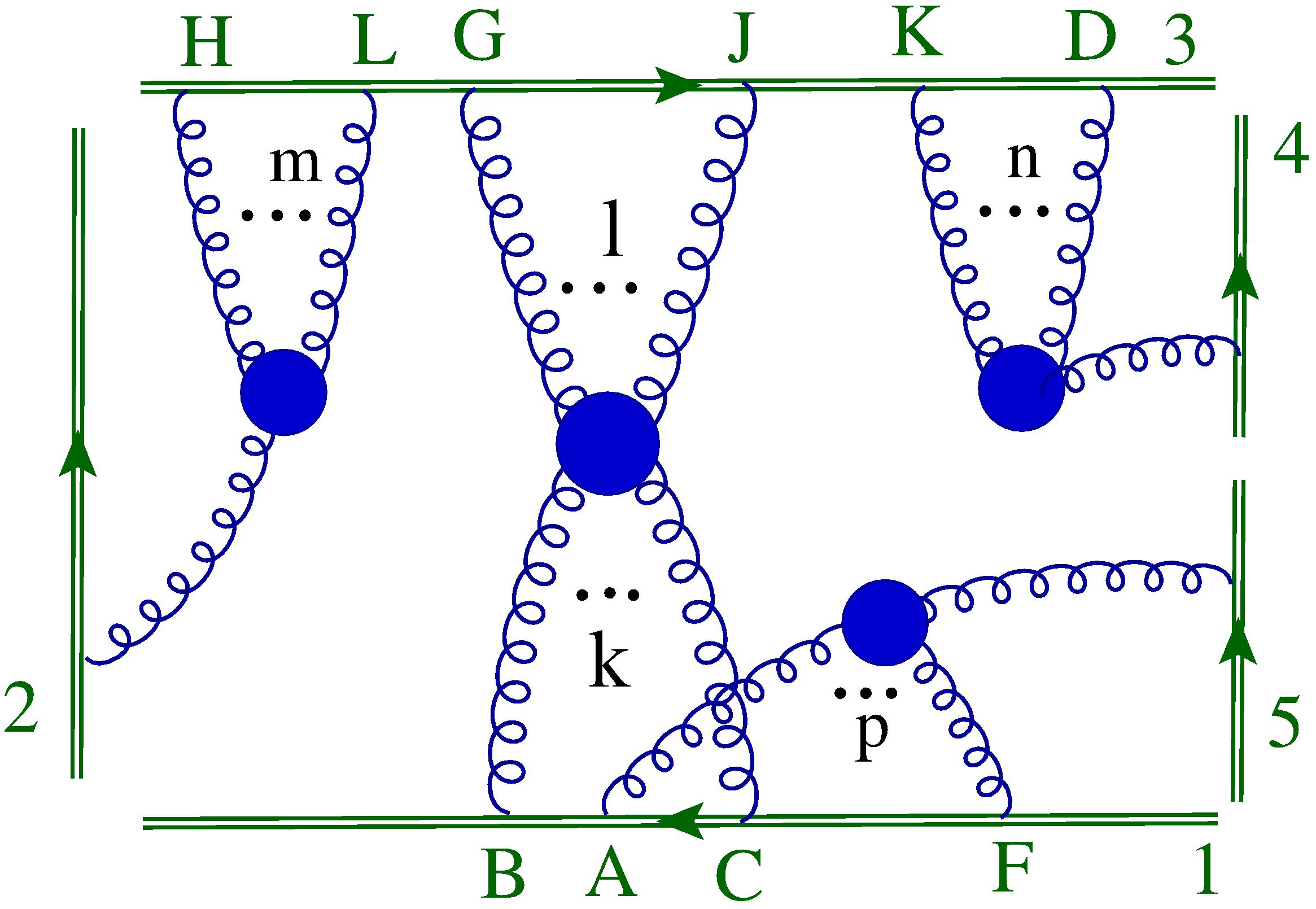} }
	\quad
	\subfloat[][]{\includegraphics[scale=0.055]{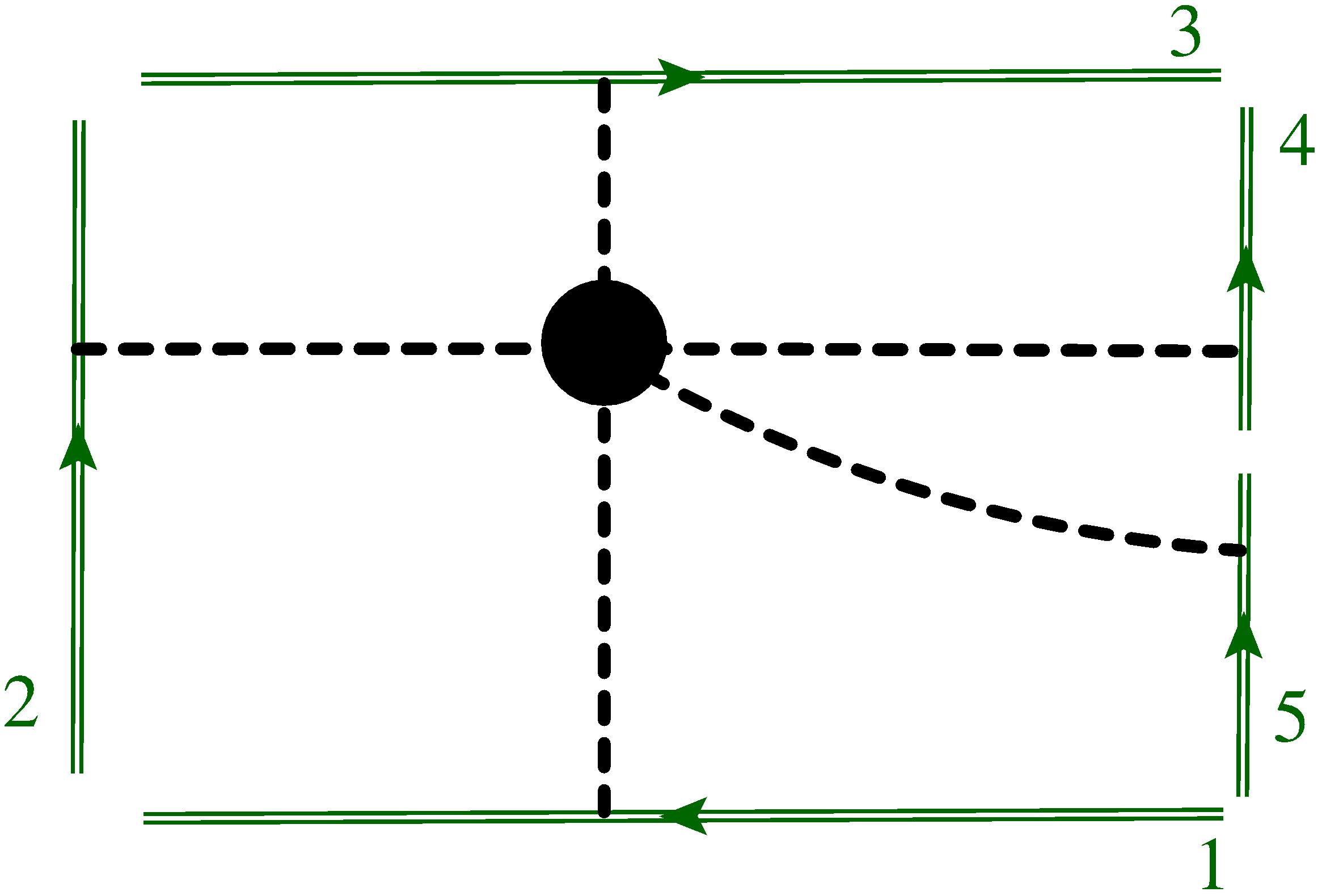} }
	\quad
	\subfloat[][]{\includegraphics[scale=0.055]{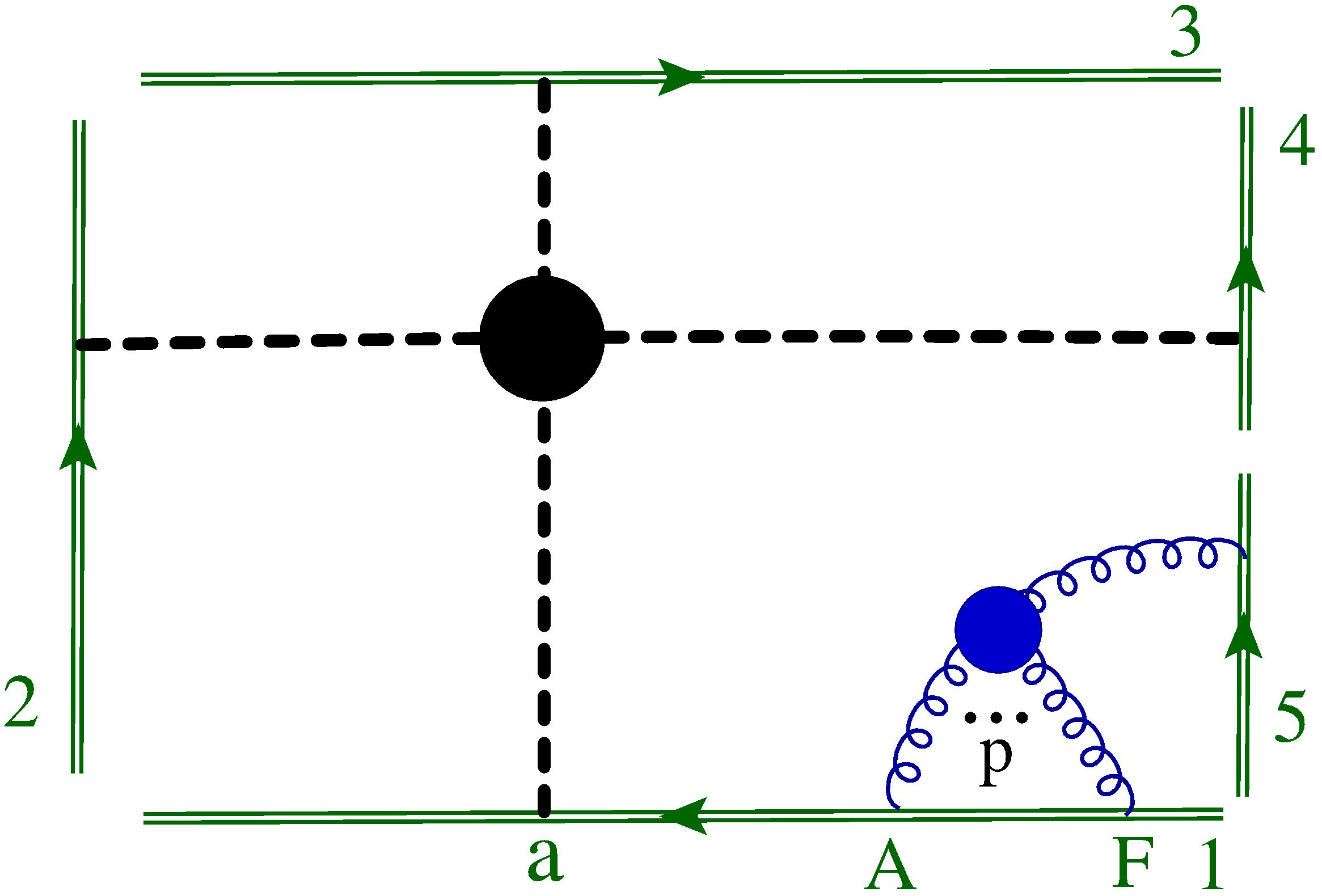} }
	\quad
	\subfloat[][]{\includegraphics[scale=0.055]{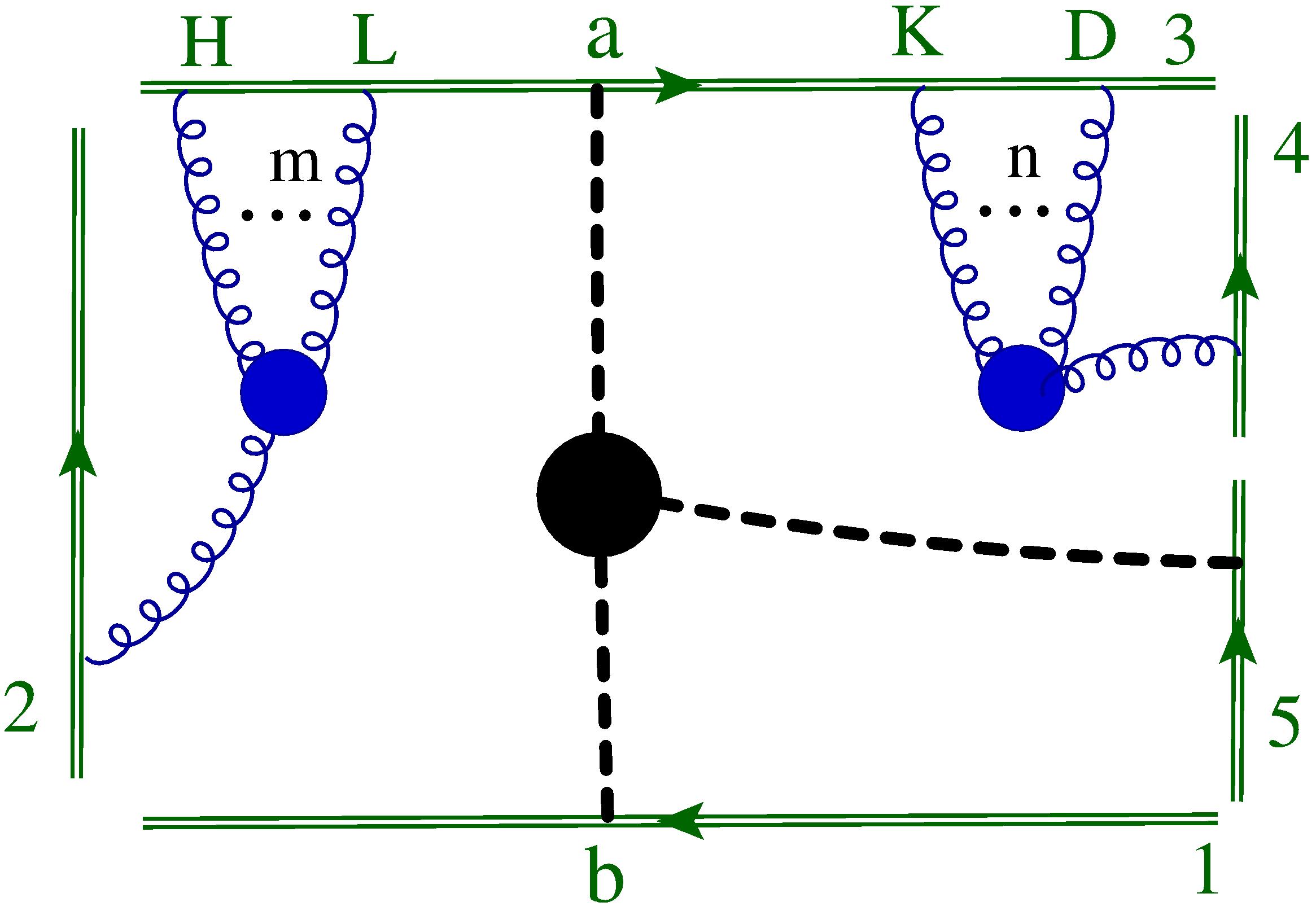} }
	\caption{Fused-Webs for Cweb $\text{W}\,_{5}^{(1,1,1,1)}(1,1,1,k+p,m+l+n)$}
	\label{fig:2Sclass2-C}
\end{figure}

We start with the completely entangled diagrams which appear when all the four gluon correlators present in this Cweb are entangled with each other as shown in fig.~(\ref{fig:2Sclass2-C}\txb{a}). Using eqs.~\eqref{eq:Num-two-entangled0} and \eqref{eq:Num-three-entangled}, the number of completely entangled diagrams is given by,
\begin{align}
((k \,\Pi \,p) -2!)\bigg((l \,\Pi \,m\,\Pi \,n)  
-2\{(l \,\Pi \,m) 
+(m \,\Pi \,n)+(l \,\Pi \,n)\}+6\bigg)\,. \nonumber
\end{align}  
Each of these form a Fused diagram, shown in fig.~(\ref{fig:2Sclass2-C}\txb{d}), and corresponds to identities of the mixing matrix.

The number of partially entangled diagrams for different entanglements can appear in four different ways.

\begin{figure}
	\centering
	\subfloat[][]{\includegraphics[scale=0.055]{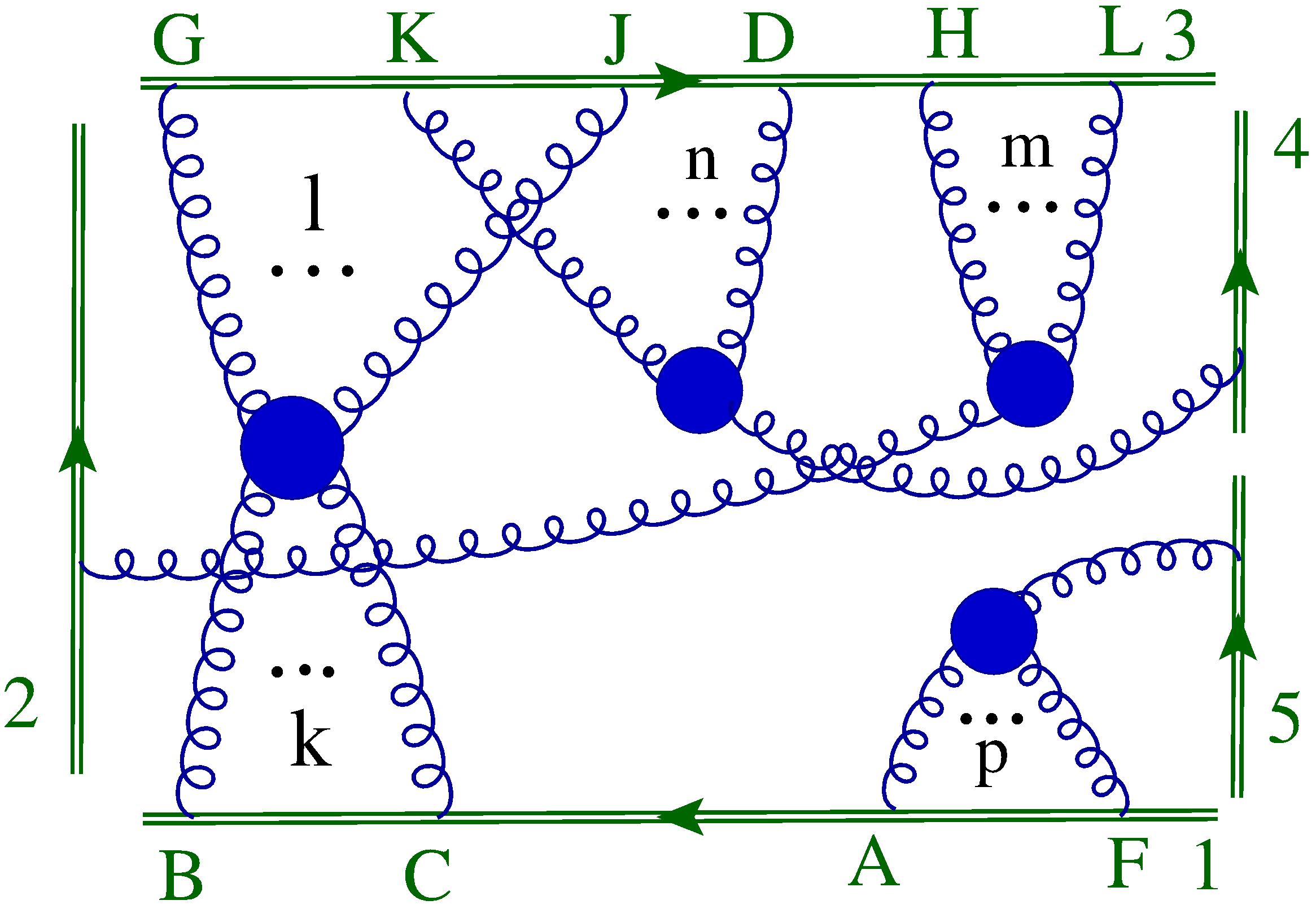} }
	\quad	
	\subfloat[][]{\includegraphics[scale=0.055]{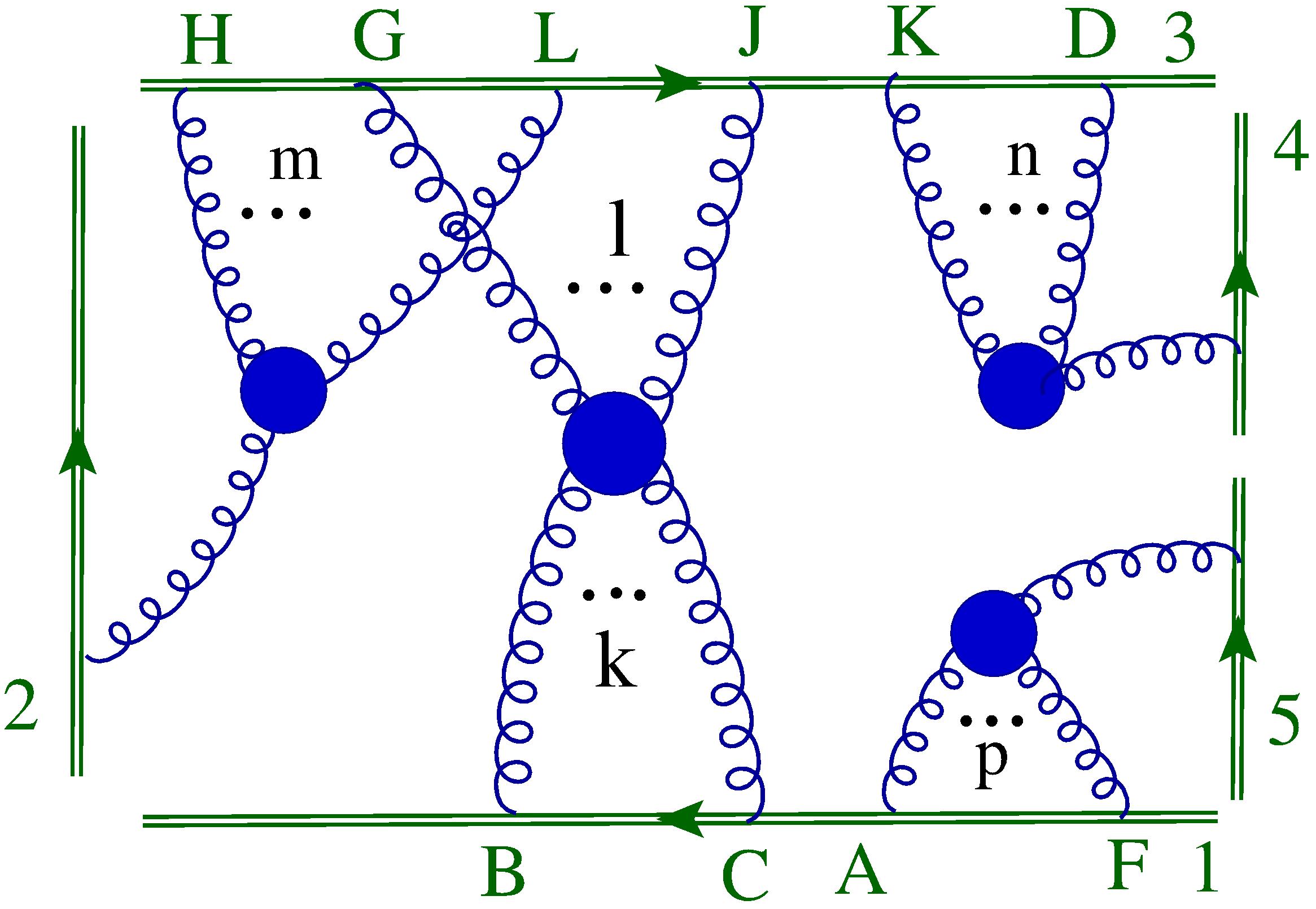} }
	\quad
	\subfloat[][]{\includegraphics[scale=0.055]{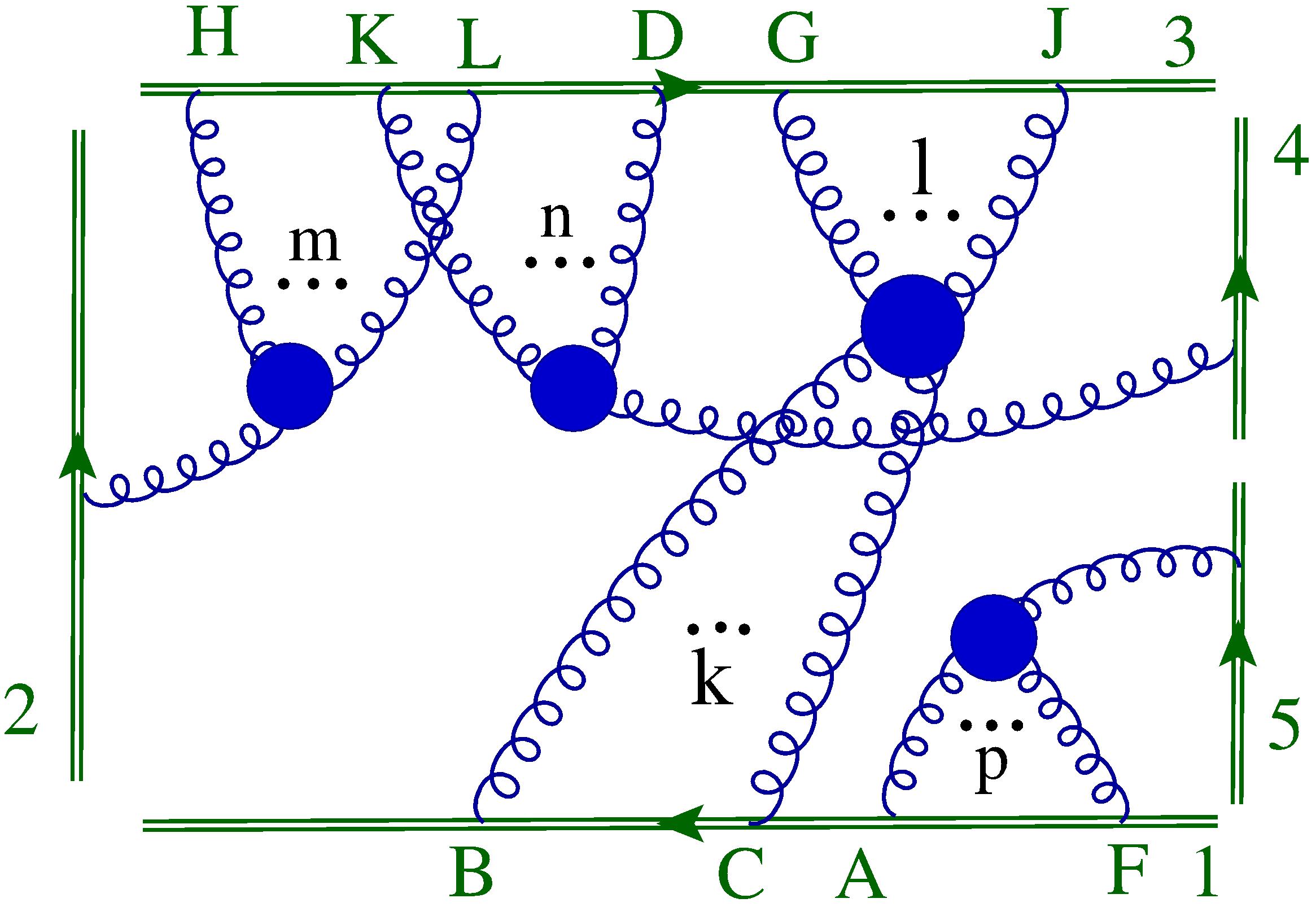} }
	\hspace{0.01cm}
	\subfloat[][]{\includegraphics[scale=0.055]{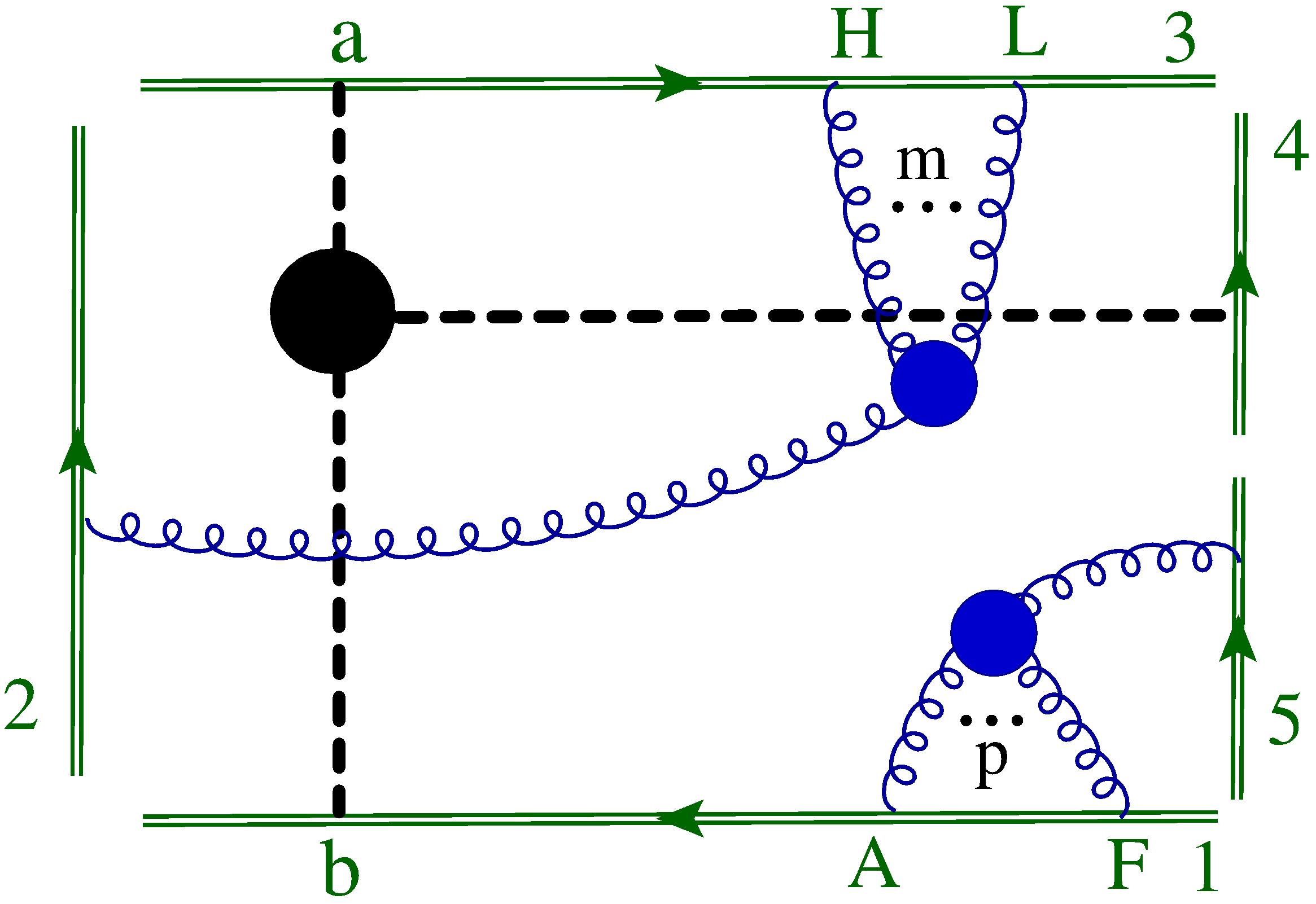} }
	\hspace{0.01cm}
	\subfloat[][]{\includegraphics[scale=0.055]{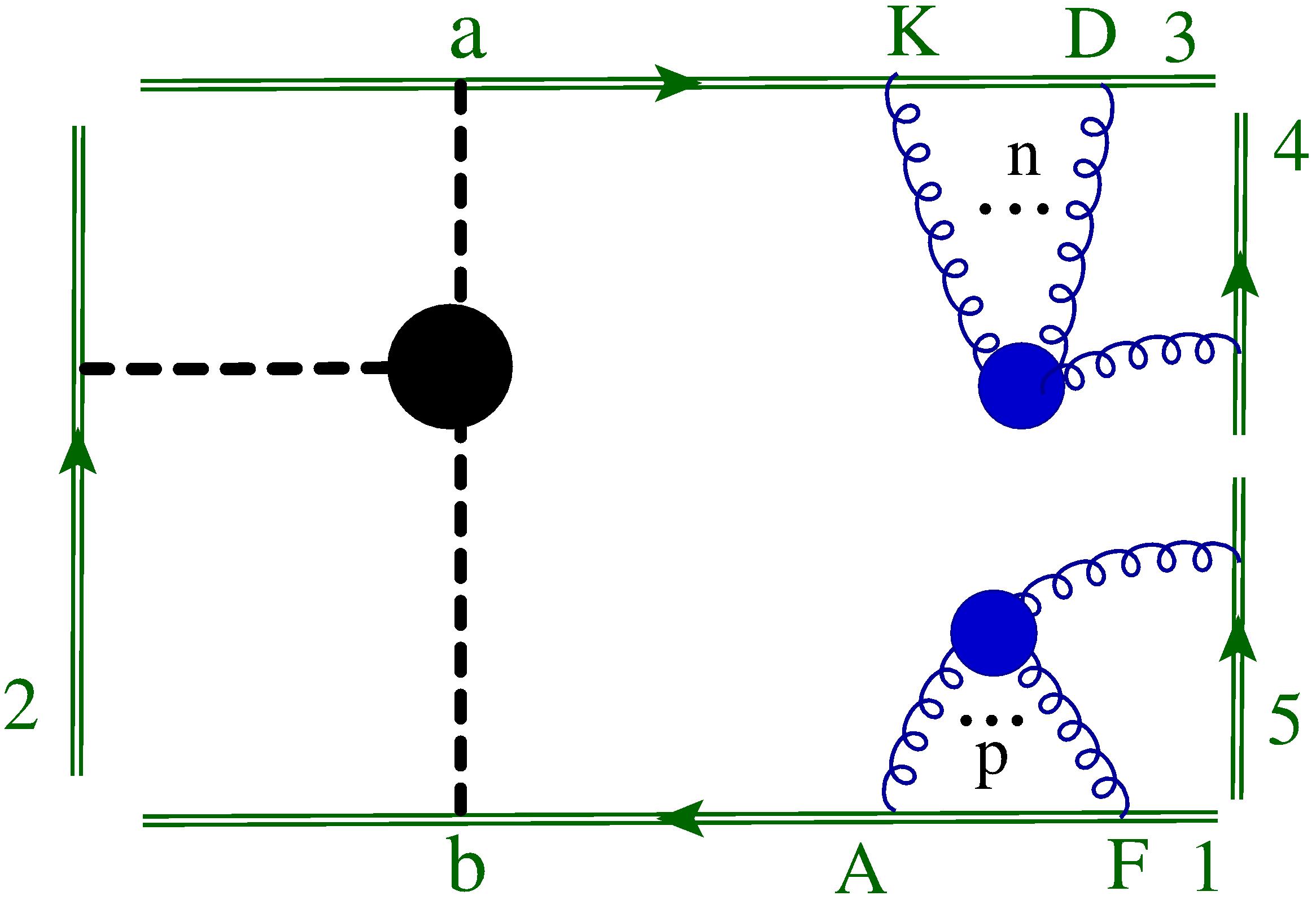} }
	\hspace{0.01cm}
	\subfloat[][]{\includegraphics[scale=0.055]{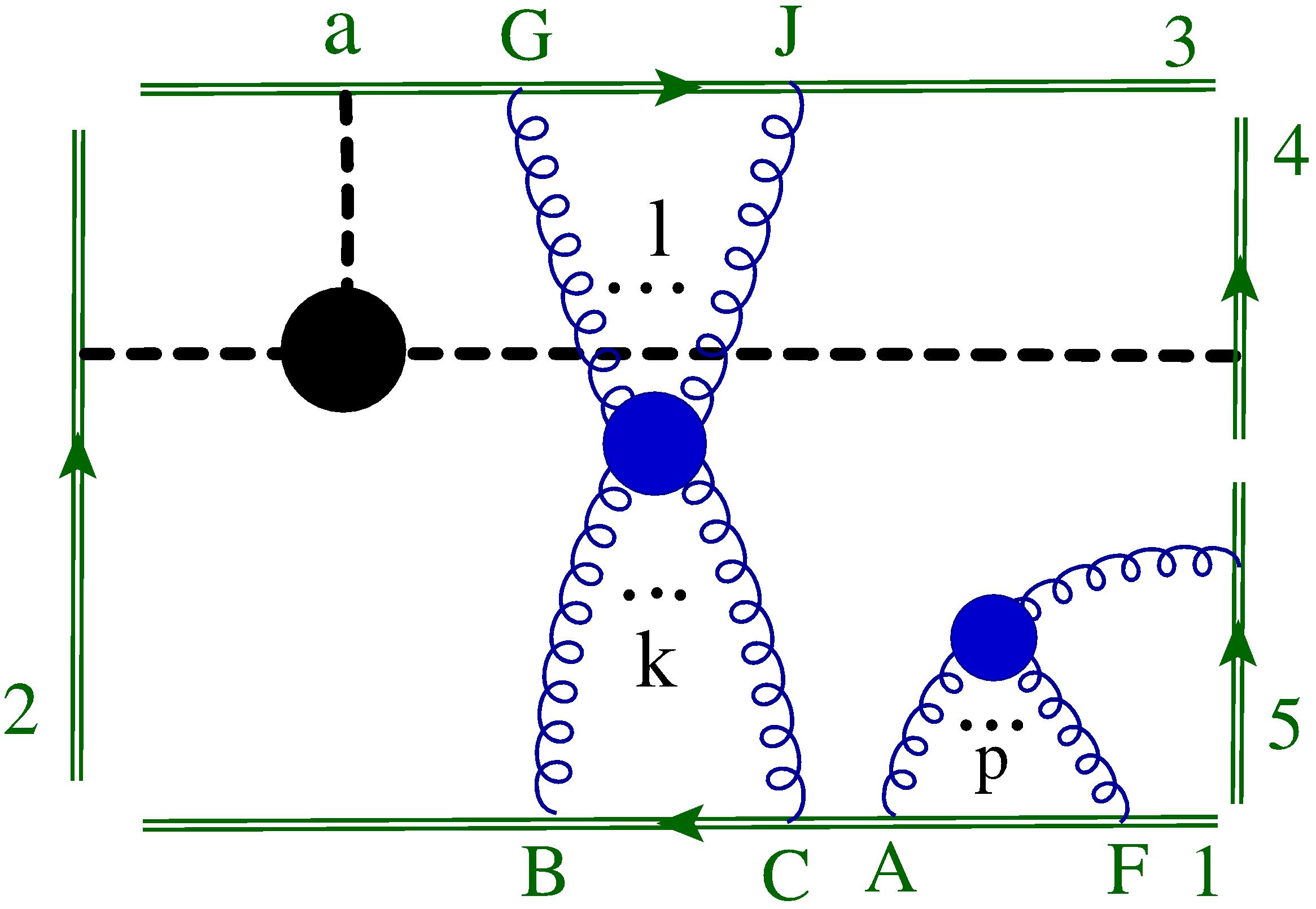} }
	\caption{Fused-Webs for Cweb $\text{W}\,_{5}^{(1,1,1,1)}(1,1,1,k+p,m+l+n)$}
	\label{fig:2Sclass2-B}
\end{figure}

\begin{itemize}
	
	\item First type of partially entangled diagrams appear when two out of three correlators are entangled on Wilson line $3$ and and both the correlators are entangled on line $1$. One such diagram, shown in fig.~(\ref{fig:2Sclass2-A}\txb{b}) involves $ (l+k) $ and $ (m+1) $-point correlators entangled on line 3, and $ (p+1) $-point correlator entangled with the $ (l+k) $-point gluon correlator on line 1. There are three such possibilities in which this type of entanglement can appear.
	The correlator which is not entangled on line $3$ is placed to the right of the entangled correlators.
	Thus, the number of diagrams with this type of entanglement is given by,
	\begin{align}
	(k\,\Pi\,p-2)\bigg((l\,\Pi\, m-2)+(l\,\Pi\, n-2)+(m\,\Pi\, n-2)\bigg)\,,\nonumber
	\end{align}
	The Fused-diagrams for this type of entanglement are shown in fig.~\eqref{fig:2Sclass2-A}.
	The shuffle of the correlators for each of these diagrams generates a Fused-Web with $ S=\{1_2\} $ and mixing matrix $ R(1_2) $.
	
	\item In the second type of entanglement, two of the three correlators are entangled on line $ 3 $ and there is no entanglement on line 1. A diagram of this configuration is shown in fig.~(\ref{fig:2Sclass2-B}\txb{a}), where $ (l+k) $ and $ (n+1) $-point gluon correlators are entangled with each other. There are again three such possibilities in which two of the three correlators on line $ 3 $ are entangled. 
	Also, the correlator which is not entangled on line $3$ and the $ (p+1) $-point correlator on line $1$ are placed to the right of entangled piece and $ (k+l) $-point gluon correlator respectively.  
	Using eq.~\eqref{eq:Num-two-entangled0}, the number of diagrams which involve this type of entanglement is given by, 
	\begin{align}
	(l\,\Pi\, m-2)+(l\,\Pi\, n-2)+(m\,\Pi\, n-2)\,.\nonumber
	\end{align}
	The Fused diagrams for this type of entanglement are shown in fig.~\eqref{fig:2Sclass2-B}.
	The shuffle of the correlators for each of these diagrams generates a Fused-Web with $ S=\{1_2,2_2\} $ and mixing matrix $ R(1_2,2_2) $. 
	
	\item In the third type of entanglement, all three correlators on line $ 3 $ are entangled and there is no entanglement on line $ 1 $. An example of this kind of entanglement is shown in fig.~(\ref{fig:2Sclass2-C}\txb{b}). The number of diagrams in which this entanglement appears is given by,
	\begin{align}
	(l\,\Pi\,m\,\Pi\,n ) - 2[(l\,\Pi\,m) + (l\,\Pi\,n) + (m\,\Pi\,n)] + 6 \nonumber
	\end{align}.
	where we have considered that the $ (p+1) $-point correlator on line $ 1 $ is placed to the right of $ (l+k) $-point gluon correlator. 
	The Fused diagram for this entanglement is shown in fig.~(\ref{fig:2Sclass2-C}\txb{e}). The shuffle of attachments generates a Fused-Web with $ S=\{1_2\} $ and mixing matrix $ R(1_2) $.	 
	
	\item Finally, in the fourth type of entanglement, the $ (p+1) $-point and $ (k+l) $-point correlators on line $ 1 $ are entangled and there is no entanglement on line $3$, shown in fig.~(\ref{fig:2Sclass2-C}\txb{c}). The number of diagrams in which this entanglement appears is given by, 
	\begin{align}
	(k\,\Pi\,p)-2,
	\end{align} 
	Here we have considered that  $ (m+1) $ and $ (n+1) $-point correlators are placed to the left and the right of entangled correlator, respectively. 
	Each of these diagrams generates Fused diagram of fig.~(\ref{fig:2Sclass2-C}\txb{f}). The shuffle of correlators for each of these Fused diagrams generates a Fused-Web with $ S=\{1_6\} $ and mixing matrix $ R(1_6) $.   
	
\end{itemize}

The details of each type of entanglement, their associated Fused-Webs and the mixing matrices are listed in table~\ref{tab:2Sclass2-Fused}.

\begin{table}
	\begin{center}
		\resizebox{0.97\textwidth}{!}{
			\begin{tabular}{|c|c|c|c|c|c|c|}
				\hline
				Entangled & Diagram & Fused&  Diagrams in  &{ $ s $-factors}   & $ R $ & Number of ways \\ 
				correlators&  &Web & Fused-Web &    & & for configuration \\
				\hline
				& &   &  & &  & $ [(k \,\Pi \,p) -2!]\times[(l \,\Pi \,m\,\Pi \,n)  $\\ 
				All& \ref{fig:2Sclass2-C}\txb{a}&\ref{fig:2Sclass2-C}\txb{d}  &   - & $ 1 $ & $  I $   & $    -2\{(l \,\Pi \,m)$\\ 
				&&&&&& $ +(m \,\Pi \,n)+(l \,\Pi \,n)\}+6]   $\\\hline
				
				$ (m+1),(k+l), $ &\ref{fig:2Sclass2-A}\txb{b} & \ref{fig:2Sclass2-A}\txb{e}   & $ \{a,\,KD\} $ & 1 & $ R(1_2) $ &$ \,\{(k \,\Pi \,p) -2\}   $\\ 
				$ (p+1) $& &                                                                 & $ \{KD,\,a\} $& 1 &  &         $ \times\{(l \,\Pi \,m) -2\} $\\ \hline
				
				$ (n+1),(k+l), $ &\ref{fig:2Sclass2-A}\txb{c} & \ref{fig:2Sclass2-A}\txb{f}   & $ \{a,\,HL\} $ & 1 & $ R(1_2) $ &$ \,\{(k \,\Pi \,p) -2\}   $\\ 
				$ (p+1) $&  &			                                                    &$ \{HL,\,a\} $ & 1 &  & $ \times\{(l \,\Pi \,n) -2\} $\\ \hline
				
				$ (m+1),(n+1), $ &\ref{fig:2Sclass2-A}\txb{a} & \ref{fig:2Sclass2-A}\txb{d}   & $ \{a,\,c\} $ & 1 & $ R(1_2) $ &$ \,\{(k \,\Pi \,p) -2\}   $\\ 
				$ (p+1) $,$(k+l)$&  &			                                                    &$ \{c,\,a\} $ & 1 &  & $ \times\{(m \,\Pi \,n) -2\} $\\ \hline
				
				$ (n+1),(k+l), $ & \ref{fig:2Sclass2-B}\txb{a}& \ref{fig:2Sclass2-B}\txb{d}   & $ \{a,\,HL\},\,\{b,\,AF\} $ & 2 & $ R(1_2,2_2) $ &$ \{(l \,\Pi \,n) -2\} $\\ 
				&&                                                         & $ \{a,\,HL\},\,\{AF,\,b\} $ & 1 &  & \\ 
				&&& $ \{HL,\,a\},\,\{b,\,AF\} $ & 1 &  & \\ 
				&&& $ \{HL,\,a\},\,\{AF,\,b\} $ & 2 &  & \\ \hline
				$ (m+1),(k+l), $ &\ref{fig:2Sclass2-B}\txb{b}& \ref{fig:2Sclass2-B}\txb{e}    & $ \{a,\,KD\},\,\{b,\,AF\} $ & 2 & $ R(1_2,2_2) $ &$ \{(l \,\Pi \,m) -2\} $\\ 
				&&                                                         & $ \{a,\,KD\},\,\{AF,\,b\} $ & 1 &  & \\
				&&& $ \{KD,\,a\},\,\{b,\,AF\} $ & 1 &  & \\ 
				&&& $ \{KD,\,a\},\,\{AF,\,b\} $ & 2 &  & \\ \hline
				$ (n+1),(m+1), $ &\ref{fig:2Sclass2-B}\txb{c}& \ref{fig:2Sclass2-B}\txb{f}    & $ \{a,\,GJ\},\,\{BC,\,AF\} $ & 1 & $ R(1_2,2_2) $ &$ \{(m \,\Pi \,n) -2\} $\\ 
				&&                                                         & $ \{a,\,GJ\},\,\{AF,\,BC\} $ & 2 &  & \\ 
				&& & $ \{GJ,\,a\},\,\{BC,\,AF\} $ & 2 &  & \\   
				&& & $ \{GJ,\,a\},\,\{AF,\,BC\} $ & 1 &  & \\\hline
				$ (n+1),(k+l), $ &\ref{fig:2Sclass2-C}\txb{b}& \ref{fig:2Sclass2-C}\txb{e}    & $ \{a,\,AF\} $ & 1 & $ R(1_2) $ &$ (l \,\Pi \,m\,\Pi \,n)  -2\{(l \,\Pi \,m)  $\\ 
				$ (m+1) $         & &                                                       & $ \{AF,\,a\} $ & 1 &  &$ +(m \,\Pi \,n)+(l \,\Pi \,n)\} + 6$ \\ \hline
				
				$ (k+l), $  $ (p+1) $      &\ref{fig:2Sclass2-C}\txb{c} & \ref{fig:2Sclass2-C}\txb{f}   & $ \{a,\,HL,\,KD\} $ & 1 & $ R(1_6) $ &$ \{(k \,\Pi \,p) -2\}  $\\ 
				&  &                                                               & $ \{a,\,KD,\,HL\} $ & 1 &  & \\ 
				&& &$ \{HL,\,a,\,KD\} $ & 1 &  &\\
				&& &$ \{HL,\,KD,\,a\} $ & 1 &  &\\
				&& &$ \{KD,\,a,\,HL\} $ & 1 &  &\\
				&& &$ \{KD,\,HL,\,a\} $ & 1 &  &\\\hline
		\end{tabular}}	
	\end{center}
	\caption{\reducedWebs and their mixing matrices for Cweb}
	\label{tab:2Sclass2-Fused}
\end{table}
The diagonal blocks of the mixing matrix for this class of Cweb is then given by,
\begin{align}
R\,=\,\left(\begin{array}{c|cc|c}
\textbf{I}& & \cdots& B\\ 
\textbf{O}& &\begin{array}{ccccccc}
R\,(1_2) & & &  & & \\
& \ddots &  & & & \\
\vdots &  &  & R\,(1_2,2_2) & & \\
&  &  & &  \ddots & & \\
&  &  & &  & R(1_6) & \\  
\end{array}\\
\hline
& & \cdots& R(1_4,2_8)\\ 
\end{array}\right)\,,
\end{align}

\noindent 
Using the data given in table~\ref{tab:2Sclass2-Fused} the rank of $ A $  is given by,

\begin{align}
r(A) &=\; r(I)\,\bigg( (k \,\Pi \,p) -2! \bigg)\, \bigg(  (l \,\Pi \,m\,\Pi \,n)  -2\{(l \,\Pi \,m) +(m \,\Pi \,n)+(l \,\Pi \,n)\} + 6 \bigg) \nonumber\\
&\;+\; r(R(1_2))\,\bigg((k \,\Pi \,p) -2!\bigg) \, \bigg(  (l \,\Pi \,m) -2! \;+\; (m \,\Pi \,n) -2! +  (l \,\Pi \,n) -2! \bigg) \nonumber\\
&\;+\; r(R(1_2,2_2))\,\bigg( (l \,\Pi \,n) -2!
\;+\;(l \,\Pi \,m) -2!  
\;+\;((m \,\Pi \,n) -2!\bigg)  \nonumber\\
&\;+\; r(R(1_2))\,\bigg( (l \,\Pi \,m\,\Pi \,n)  -2\{(l \,\Pi \,m) +(m \,\Pi \,n)+(l \,\Pi \,n)\} + 6 \bigg)\;+\; r(R(1_6))\bigg((k \,\Pi \,p) -2!\bigg)\nonumber
\end{align}  

Using the ranks of the mixing matrices of basis Cwebs, we get,
\begin{align}
r(A)&\;=\;\bigg( (k\,\Pi\,p)  - 1\bigg)\bigg( (l\,\Pi\,m\Pi\,n) - (l \,\Pi \,m) - (m \,\Pi \,n) - (l \,\Pi \,n)  +2   \bigg)  \;-\;2\,.\nonumber
\end{align}
Now using the rank of $ D $, given in eq.~\eqref{eq:2Sclass2-rankD}, the rank of $ R $ becomes
\begin{align}
r(R) &=\; r(A)\;+\;r(R(1_4,2_8))\nonumber\\
&=\; \bigg( (k\,\Pi\,p)  - 1\bigg)\bigg( (l\,\Pi\,m\Pi\,n) - (l \,\Pi \,m) - (m \,\Pi \,n) - (l \,\Pi \,n)  +2   \bigg)\,.
\end{align}
The correspondence described in the beginning of this section allow us to predict the rank of boomerang webs as well.
Thus, few Cwebs, their attachment content, perturbative order and the rank of the mixing matrices of this class is given in table~\ref{tab:Sclass-3-Rank-prediction}. Note that this class of Cwebs appears at four loops and beyond.

\begin{table}[t]
	\begin{center}
		\begin{tabular}{|c|c|c|c|c|l|c|c|}
			\hline
			Value of  & Value of  &  Value of  &	Value of  &	Value of  &  Boomerang   & Loop  & $ r(R) $  \\ 
			$ k $ &    $ l $     &   $ m $     & $ n $         & $ p $    & Cweb        & order & \\
			\hline
			1&1&1&1&2  & $\;\text{W}\,^{(4)}_{4}(1,1,3,3)$& $ \mathcal{O}(g^8) $ & 4  \\
			\hline
			1&1&1&2&1  & $\;\text{W}\,^{(4)}_{4}(1,1,2,4)$& $ \mathcal{O}(g^8) $ & 6  \\
			\hline
			1&1&1&2&2  & $\;\text{W}\,^{(3,1)}_{4}(1,1,3,4)$& $ \mathcal{O}(g^{10}) $& 12  \\
			\hline
			1&1&2&2&2  & $\;\text{W}\,^{(2,2)}_{4}(1,1,3,5)$& $ \mathcal{O}(g^{12}) $  & 40 \\
			\hline
			1&2&2&2&2  & $\;\text{W}\,^{(1,3)}_{4}(1,1,3,6)$& $ \mathcal{O}(g^{12}) $  & 148 \\
			\hline	
		\end{tabular}	
	\end{center}
	\caption{Boomerang Cwebs present at four loops and beyond}
	\label{tab:Sclass-3-Rank-prediction}
\end{table}
The first two boomerang Cwebs listed above are shown in fig.~(\ref{fig:4legsWeb1}\txb{a}) and~(\ref{fig:4legsWeb2}\txb{a}) appear at four loops connecting Wilson four lines. The Fused-Web formalism for each of these web  is shown in section~\ref{sec:Boom-4loop-4line}. 

\subsubsection{Cweb $\text{W}\,_{4}^{(1,1,1)}(1,1,l+p,k+q)$}

\begin{figure}[b]
	\captionsetup[subfloat]{labelformat=empty}
	\centering
	\subfloat[][]{\includegraphics[scale=0.064]{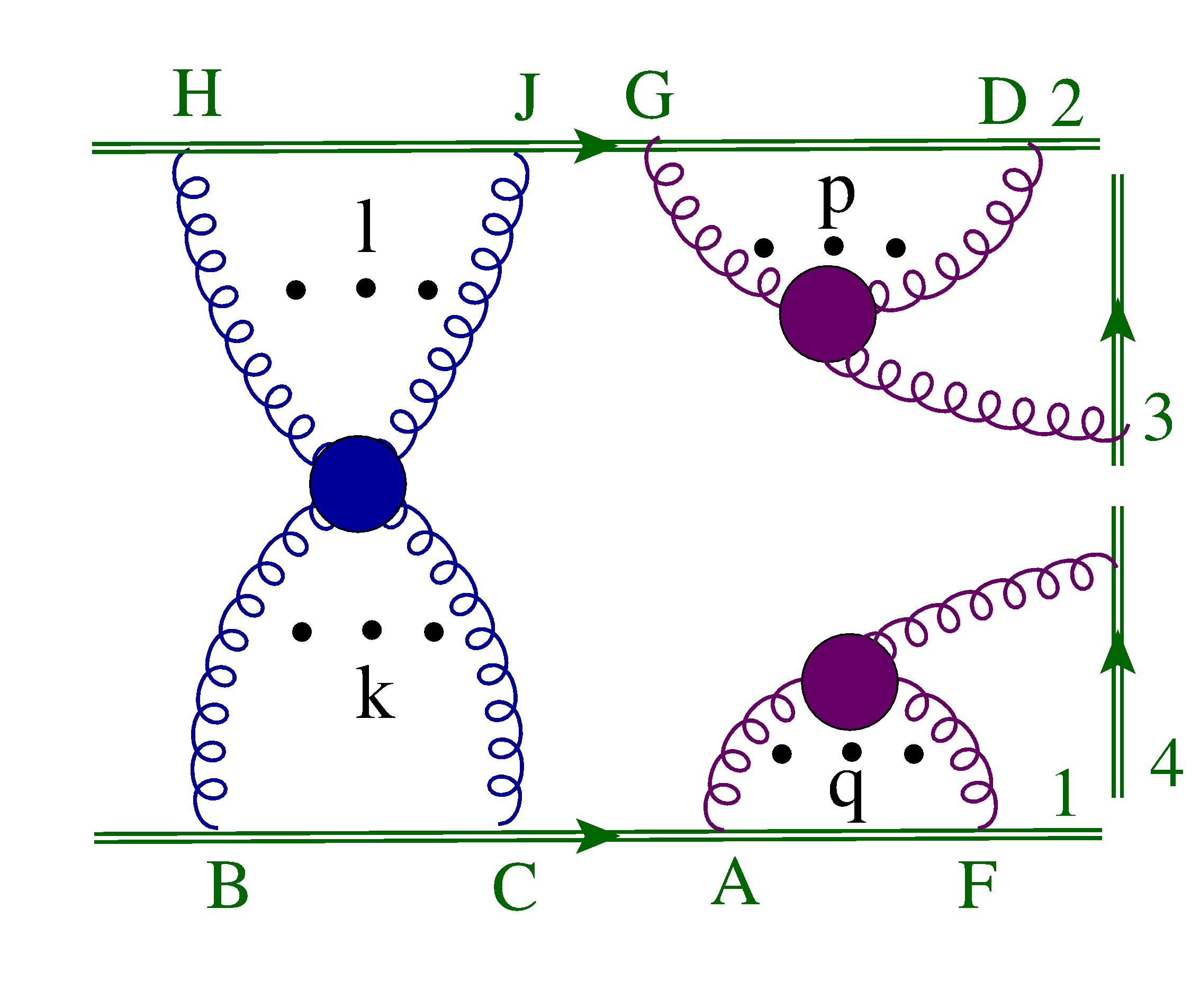} }
	\caption{Diagram of the general Cweb $\text{W}\,_{4}^{(1,1,1)}(1,1,l+p,k+q)$}
	\label{fig:2Sclass3}
\end{figure}

The general form of this class of Cwebs is  shown in fig.~(\ref{fig:2Sclass3}). It contains one $(k+l) $-point, $ (p+1) $-point and a $ (q+1) $-point gluon correlator with $ k $ and $ l $ attachments on Wilson line $1$ and $2$ respectively. The shuffle of attachments on both the Wilson lines generates all diagrams of the Cweb, out of which only four are reducible ($ s \neq 0 $). Shuffle for the reducible diagrams, represented with attachments $ \{\overline{H\cdots J},\overline{G\cdots D}\} $ and $ \{\overline{B\cdots C},\overline{A\cdots F}\}$ on line 1 and 2  respectively, are given in the following table.

\begin{table}[H]
	\begin{center}
		\begin{tabular}{|c|c|}
			\hline 
			\textbf{Sequences}  & \textbf{s-factors}  \\ 
			\hline
			$ \{\overline{H\cdots J},\overline{G\cdots D}\}, \{\overline{A\cdots F},\overline{B\cdots C}\} $  & 1 \\ \hline
			$ \{\overline{G\cdots D},\overline{H\cdots J}\}, \{\overline{B\cdots C},\overline{A\cdots F}\}$  & 1 \\ \hline
			$ \{\overline{H\cdots J},\overline{G\cdots D}\}, \{\overline{B\cdots C},\overline{A\cdots F}\}$  & 2 \\ \hline
			$ \{\overline{G\cdots D},\overline{H\cdots J}\}, \{\overline{A\cdots F},\overline{B\cdots C}\}$  & 2 \\ \hline
		\end{tabular}
	\end{center}
	\caption{Reducible diagrams of the Cweb and their $ s $-factors.}
	\label{tab:2Sclass3-redu-S}
\end{table}
\noindent From Uniqueness theorem the diagonal block $ D $ is given by,
\begin{align}
D &=\;R(1_2,2_2)\qquad ;\qquad r(D)=r(R(1_2,2_2))\;=\;1.
\end{align}
Note that $R(1_2,2_2)$  is a basis Cweb at $3$ loops connecting $4$ Wilson lines.

\begin{figure}

	\centering
	\subfloat[][]{\includegraphics[scale=0.06]{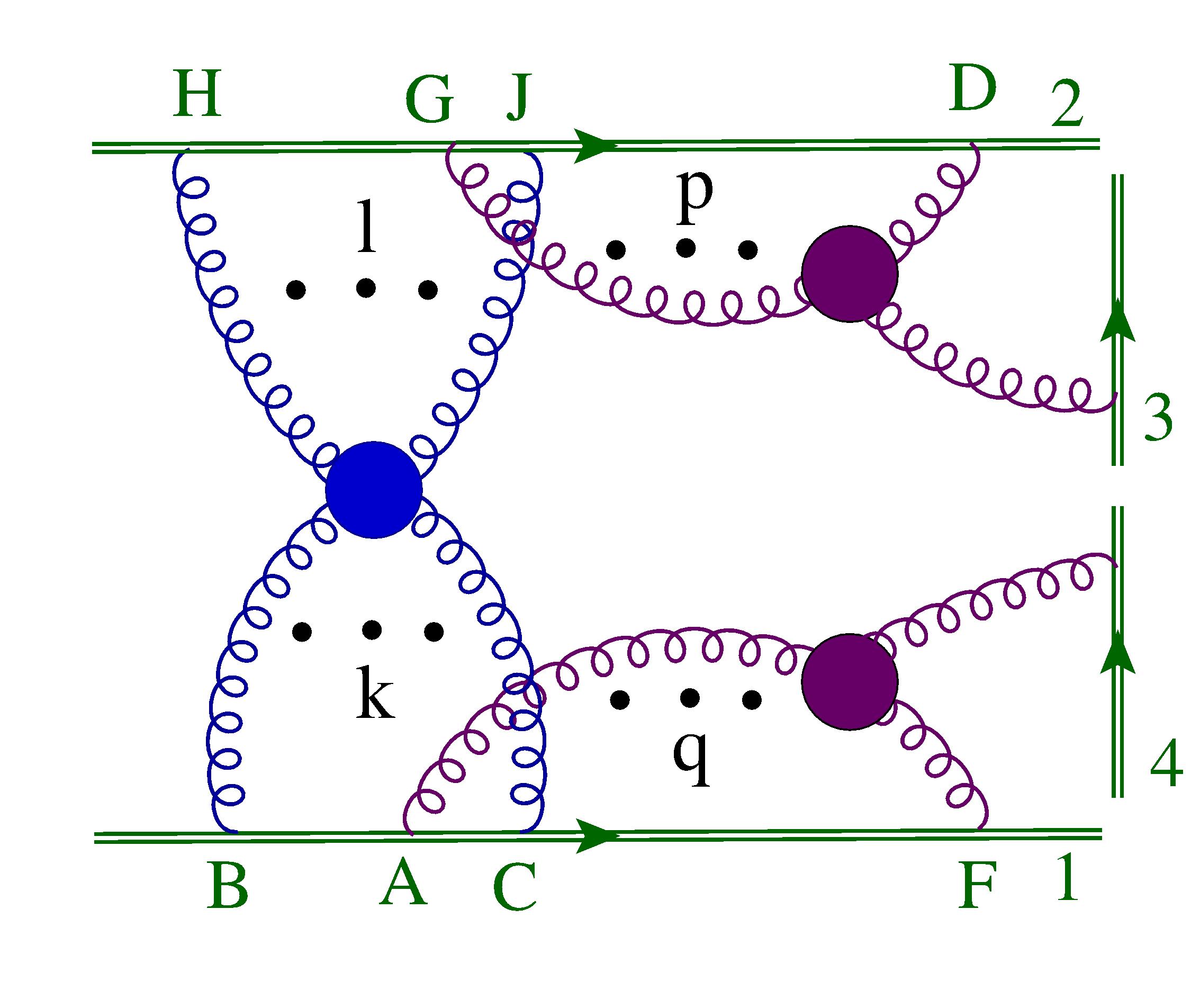} }
	\hspace{0.01cm}
	\subfloat[][]{\includegraphics[scale=0.06]{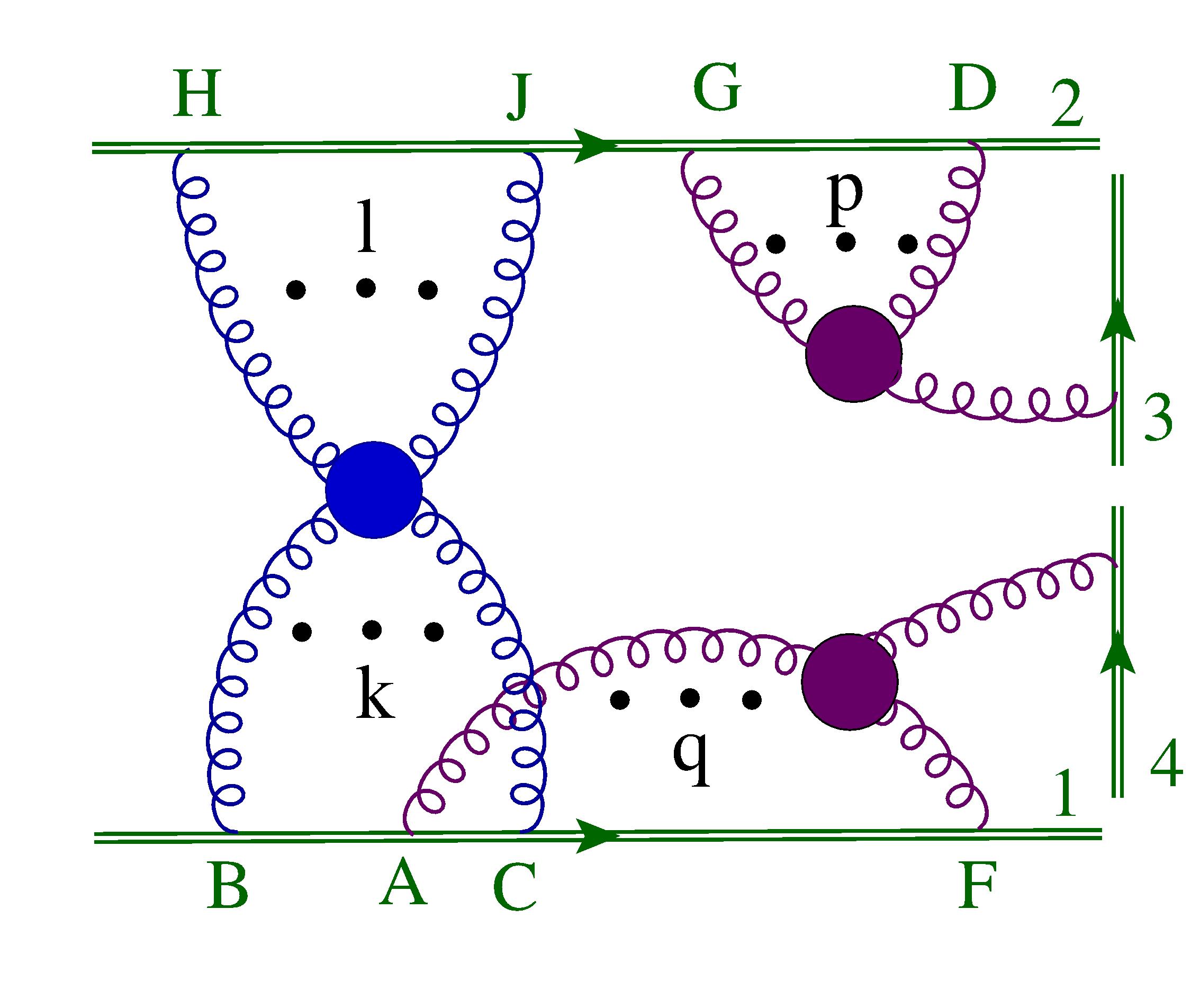} }
	\hspace{0.01cm}
	\subfloat[][]{\includegraphics[scale=0.06]{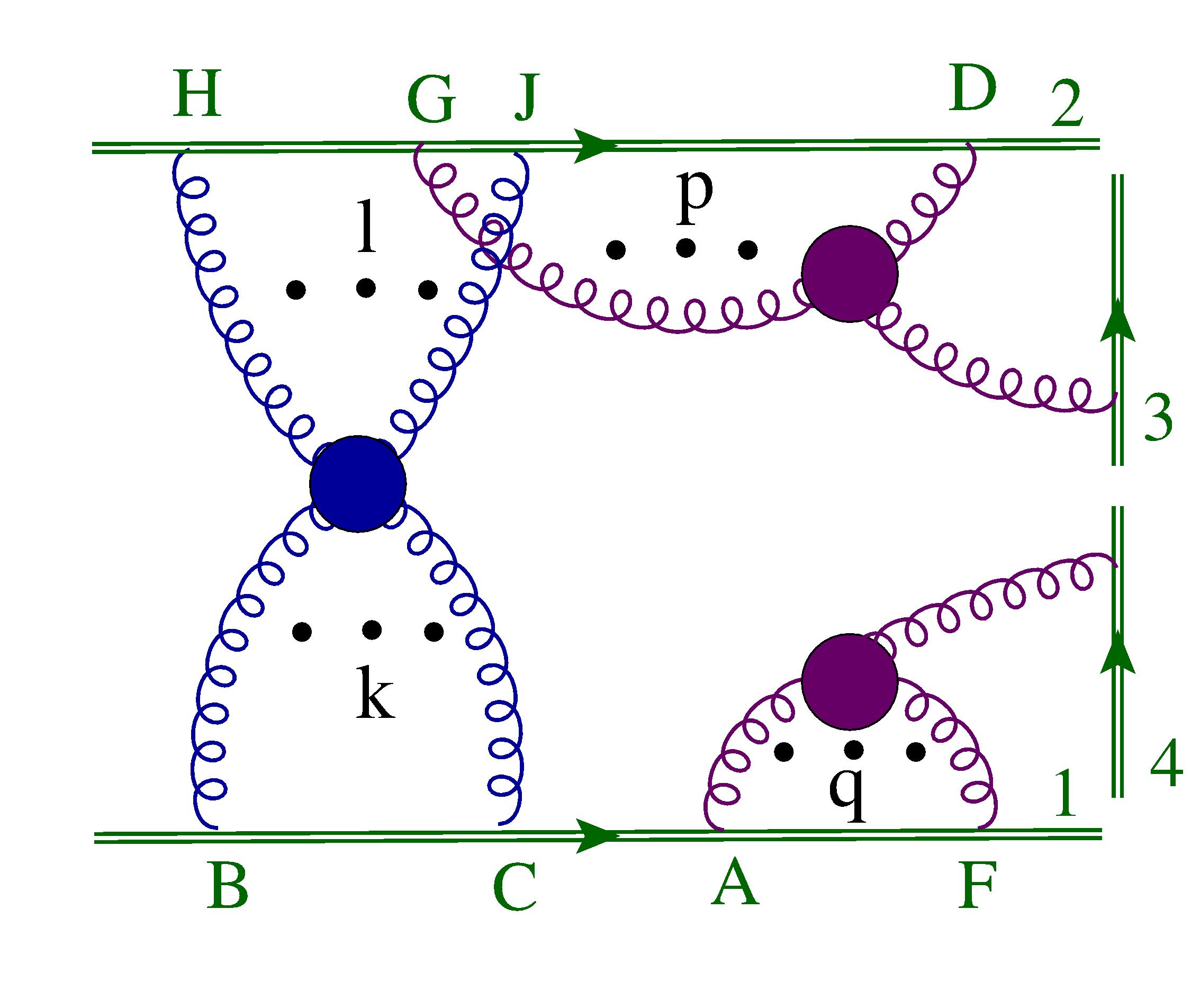} }
	\hspace{0.01cm}
	\subfloat[][]{\includegraphics[scale=0.06]{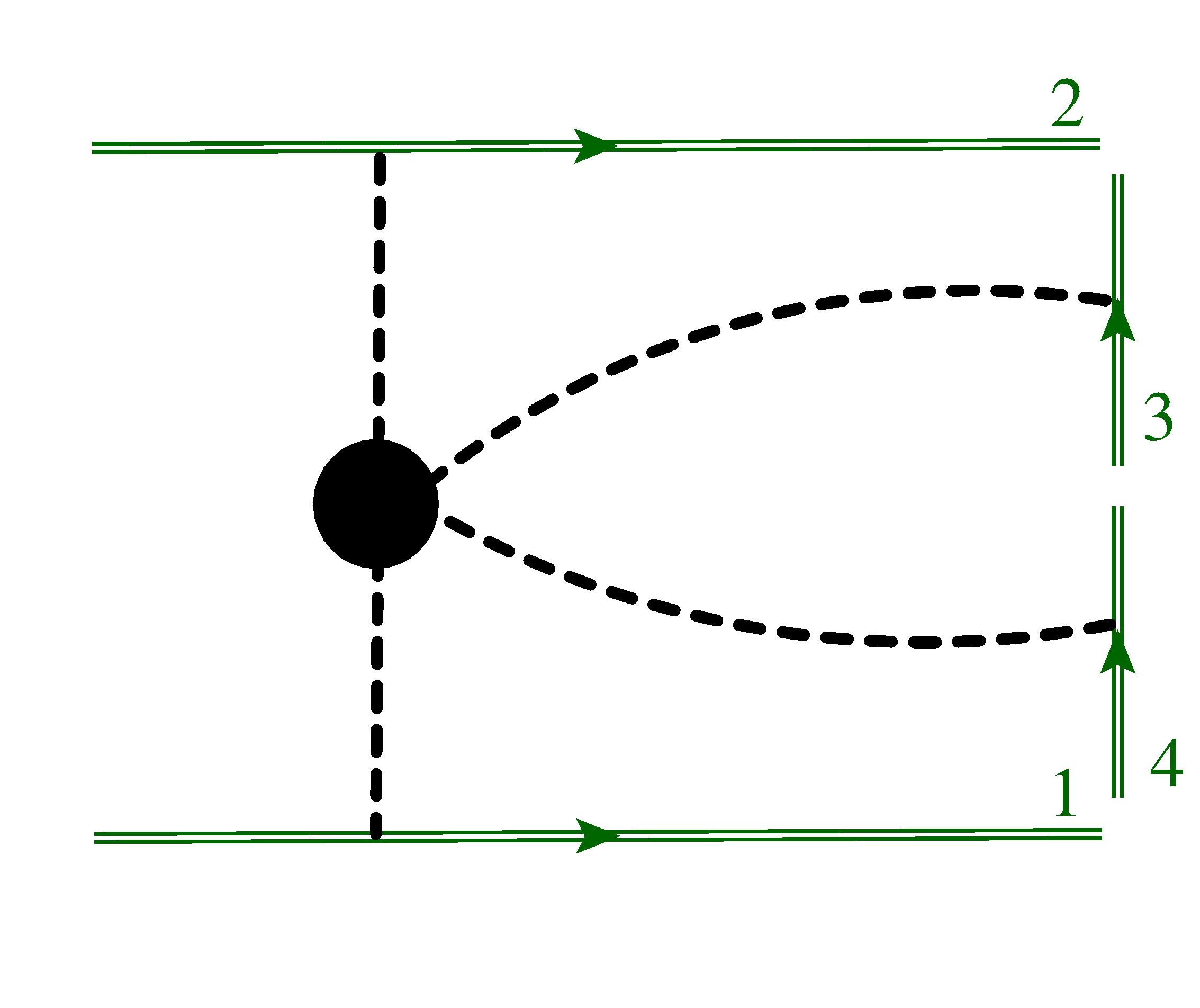} }
	\hspace{0.01cm}
	\subfloat[][]{\includegraphics[scale=0.06]{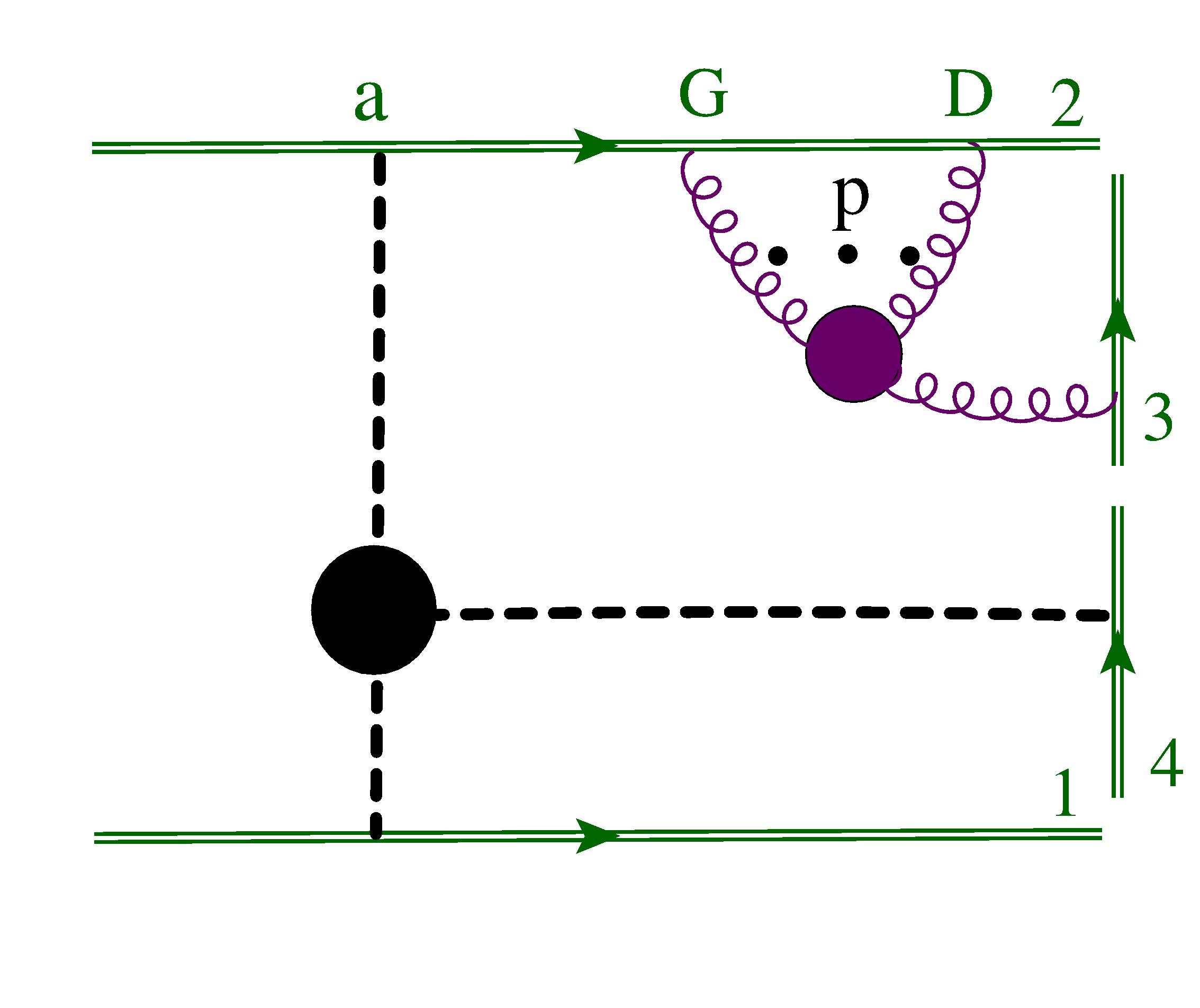} }
	\hspace{0.01cm}
	\subfloat[][]{\includegraphics[scale=0.06]{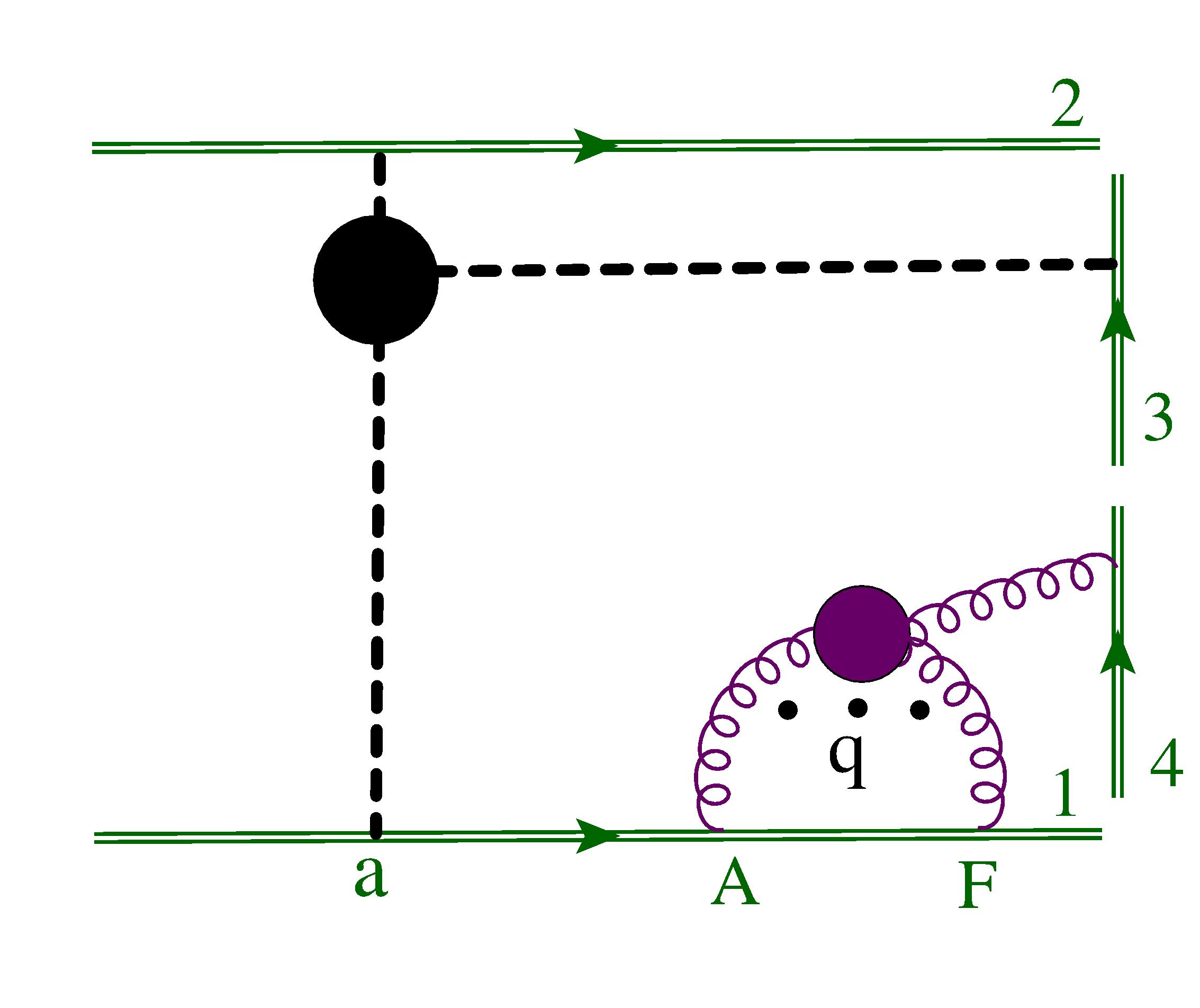} }
	\caption{Fused-Webs for Cweb $\text{W}\,_{4}^{(1,1,1)}(1,1,l+p,k+q)$}
	\label{fig:2Sclass3-R(I-2)}
\end{figure}

We now proceed to determine the diagonal blocks of $ A $ using Fused-Webs. 
The completely entangled diagrams for this Cwebs have all the three correlators entangled as shown in fig.~(\ref{fig:2Sclass3-R(I-2)}\txb{a}). 
From eq.~\eqref{eq:Num-two-entangled0}, the number of completely entangled diagrams is given by,
\begin{align}\label{eq:2Sclass3-Numb-Identity} 
\{(k \,\Pi \,q) -2\} \times \{  (l \,\Pi \,p) -2 \} &=\left(\dfrac{(k+q)!}{k! \,q!} -2\right) \times \left(\dfrac{(l+p)!}{l!\, p!} -2\right)\,,
\end{align}
where the first and second factor give the number of ways in which correlators remain entangled on line $ 1 $ and $ 2 $ respectively. Thus, the order of identity is given by eq.~\eqref{eq:2Sclass3-Numb-Identity}.

Next, we consider the partially entangled diagrams. There are two kind of entanglements that appear in partially entangled diagrams: 
\begin{itemize}
	\item The $ (l+k) $-point gluon correlator gets entangled with $ (q+1) $-point correlator on line 1 and $ (p+1) $-point correlator on line 2 is placed to the right of entangled correlators, as shown in fig.~(\ref{fig:2Sclass3-R(I-2)}\txb{b}). The number of such diagrams in which this entanglement appears, is given by, 
	\begin{align}
	\left(\dfrac{(k+q)!}{k! \,q!} -2\right) \,.\nonumber
	\end{align}
	Every such diagram has a corresponding Fused diagram, shown in fig.~(\ref{fig:2Sclass3-R(I-2)}\txb{e}) and the shuffle of its correlators generates the Fused-Web with $ S=\{1_2\} $ and the mixing matrix $ R(1_2) $.
	\item Entanglement of the $ (l+k) $-point gluon correlator with $ (p+1) $-point correlator on line 2 and again remaining correlator is placed to the right of entangled correlators on line 1, as shown in fig.~(\ref{fig:2Sclass3-R(I-2)}\txb{c}). The number of such diagrams with this entanglement is given by,
	\begin{align}
	\left(\dfrac{(l+p)!}{l! \,p!} -2\right) \,.\nonumber
	\end{align}
	The Fused diagrams for each of these diagrams are same, and is given in fig.~(\ref{fig:2Sclass3-R(I-2)}\txb{f}). It again generates the Fused-Web with $ S=\{1_2\} $ and the mixing matrix $R(1_2) $.
\end{itemize} 
The diagrams in Fused-Webs, their mixing matrices and number of times they appear, are listed  in table~\ref{tab:2Sclass3-Fused}.
\begin{table}
	\begin{center}
		\begin{tabular}{|c|c|c|c|c|c|}
			\hline
			Entanglements & Fused  &  Diagrams in  & $ s $-factors   & $ R $ & Number of ways \\ 
			&  Web& Fused-Web &    & & for configuration \\
			\hline
			Completely entangled  & \ref{fig:2Sclass3-R(I-2)}\txb{d}    & - & $ 1 $ & $  I $ & $ \{(k \,\Pi \,q) -2\} \times \{  (l \,\Pi \,p) -2 \} $\\ 
			&  & &  &  & \\ \hline
			First partial entanglement  & \ref{fig:2Sclass3-R(I-2)}\txb{e}    & $ \{a,\,GD\} $ & 1 & $ R(1_2) $ &$ \{(k \,\Pi \,q) -2\}   $\\ 
			&  & $ \{GD,\,a\} $& 1 &  & \\ \hline
			Second partial entanglement & \ref{fig:2Sclass3-R(I-2)}\txb{f}   & $ \{a,\,AF\} $ & 1 & $ R(1_2) $ &$ \{  (l \,\Pi \,p) -2 \} $\\ 
			&  & $ \{AF,\,a\} $ & 1 &  & \\ \hline
		\end{tabular}	
	\end{center}
	\caption{\reducedWebs and their mixing matrices for Cweb $\text{W}\,_{4}^{(1,1,1)}(1,1,l+p,k+q)$}
	\label{tab:2Sclass3-Fused}
\end{table}
Using this table, the diagonal blocks of the mixing matrix for this class of Cwebs are constructed and are given by, 
\begin{align}
R\,=\,\left(\begin{array}{c|cc|c}
\textbf{I}& & \cdots & B\\ 
\hline
\textbf{O}& & \begin{array}{ccccccc}
R\,(1_2) & & &  & & \\
& \ddots &  & & & \\
\vdots &  &  & R\,(1_2) & & \\
&  &  & &  \ddots & & \\  
\end{array} \\
\hline 	 
& & & R(1_2,2_2)
\end{array}\right)\,,
\end{align}
The rank of block $ A $ is then given by, 
\begin{align}
r(A) &=\; r(I)\times \{(k \,\Pi \,q) -2\} \times \{  (l \,\Pi \,p) -2 \} \;+\; r(R(1_2))\times  \{(k \,\Pi \,q) -2\}   \;+\; r(R(1_2))\times  \{(l \,\Pi \,p) -2\} \nonumber 
\end{align}	
Now, using eq.~\eqref{eq:Num-two-entangled0} and the ranks of mixing matrices of basis Cwebs, above equation takes the form
\begin{align}
r(A) \;=\;	\frac{(k+q)!\; (l+p)!}{k! \;l!\; p!\; q!}-\frac{(k+q)!}{k!\; q!}-\frac{(l+p)!}{l!\; p!}
\end{align}
The rank of the mixing matrices is the sum of that of $ A $ and $ D $ as follows
\begin{align}
r(R)&=\;r(A) + r(D)\;=\;	\frac{(k+q)!\; (l+p)!}{k! \;l!\; p!\; q!}-\frac{(k+q)!}{k!\; q!}-\frac{(l+p)!}{l!\; p!}\;+\;1
\end{align}
Based on the correspondence mentioned in the previous section, we list down the Cwebs that fall under this class in the following table.
\begin{table}[H]
	\begin{center}
		\begin{tabular}{|c|c|c|c|l|c|c|}
			\hline
			Value of  &	Value of &	Value of &	Value of &  Boomerang   & Loop  & $ r(R) $  \\ 
			$ k $     &$ l $     &   $ p $   &  $ q $    &  Cweb        & order & \\
			\hline
			1&1&1&2  & $\;\text{W}\,^{(3)}_{3}(1,2,3)$& $ \mathcal{O}(g^6) $ & 2  \\
			\hline
			1&1&2&2  & $\;\text{W}\,^{(3)}_{2}(3,3)$& $ \mathcal{O}(g^6) $ & 4  \\
			\hline
			1&2&2&2  & $\;\text{W}\,^{(2,1)}_{2}(3,4)$& $ \mathcal{O}(g^8) $  & 10  \\
			\hline
			2&2&2&2  & $\;\text{W}\,^{(2,0,1)}_{2}(4,4)$ & $ \mathcal{O}(g^{10}) $  & 25 \\
			\hline
		\end{tabular}	
	\end{center}
	\caption{Cwebs of this class present at different loops}
	\label{tab:Sclass-2-Rank-prediction}
\end{table}

The family of first boomerang Cweb in the table~\ref{tab:Sclass-2-Rank-prediction} has two boomerang Cwebs $ \text{W}^{(2,1)}_{4,\text{I}}(1,1,2,3) $ and $ \text{W}^{(2,1)}_{4,\text{II}}(1,1,2,3) $, shown in figs.~(\ref{fig:six-one-web4-8-av}\txb{a}) and~(\ref{fig:six-one-web4-4-av}\txb{a}) respectively, appearing at four loops connecting four lines.

\subsubsection{Cweb $\text{W}\,_{5}^{(1,1,1,1)}(1,1,k+p,m+l,q+n)$}

\begin{figure}[H]
	\captionsetup[subfloat]{labelformat=empty}
	\centering
	\subfloat[][]{\includegraphics[scale=0.06]{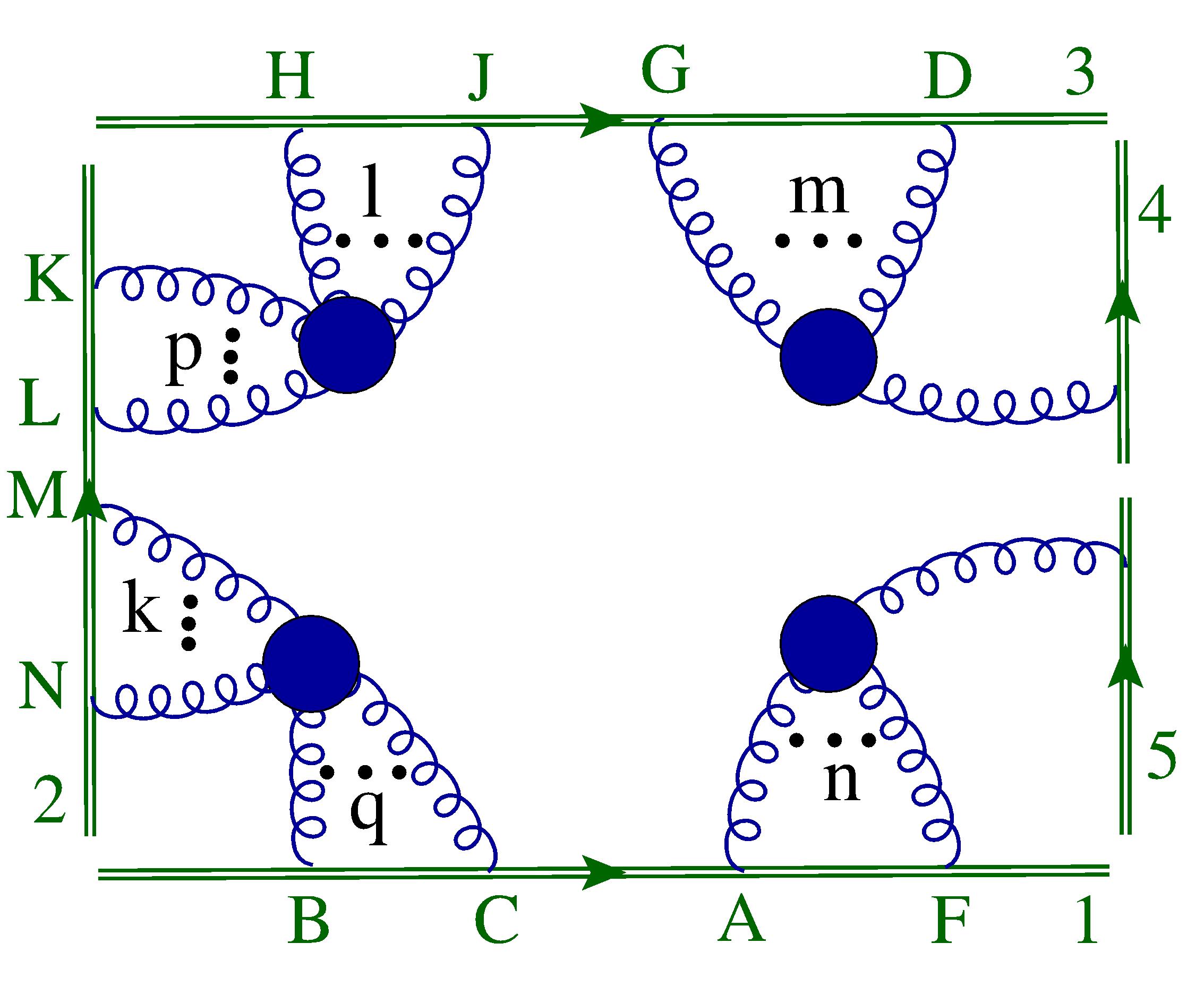} }
	\caption{Diagram of the general Cweb $\text{W}\,_{5}^{(1,1,1,1)}(1,1,k+p,m+l,q+n)$}
	\label{fig:2Sclass4}
\end{figure}

A general structure of this class is shown in fig.~(\ref{fig:2Sclass4}). This Cweb has $ (l+p) $-point, $ (k+q) $-point, $ (m+1) $-point and one $ (n+1)  $-point gluon correlator connecting five Wilson lines. The shuffle of attachments from different correlator on each Wilson line generates all diagrams of the Cweb. Order of attachments and $ s $-factors of the  reducible diagrams are given in table~\ref{tab:2Sclass4-shuffle}. The Fused-Web formalism for each of these web  is shown in section~\ref{sec:Boom-4loop-4line}. 

The uniqueness theorem tells that the $ D $ block of the associated mixing matrix is given by,
\begin{align}
D &=\; R(1_2,2_2,3_3,4_2);\qquad \qquad r(D)\,=\,r(R(1_2,2_2,3_3,4_2))\,=\,1\,.
\end{align}
 
\begin{figure}
	\centering
	\subfloat[][]{\includegraphics[scale=0.06]{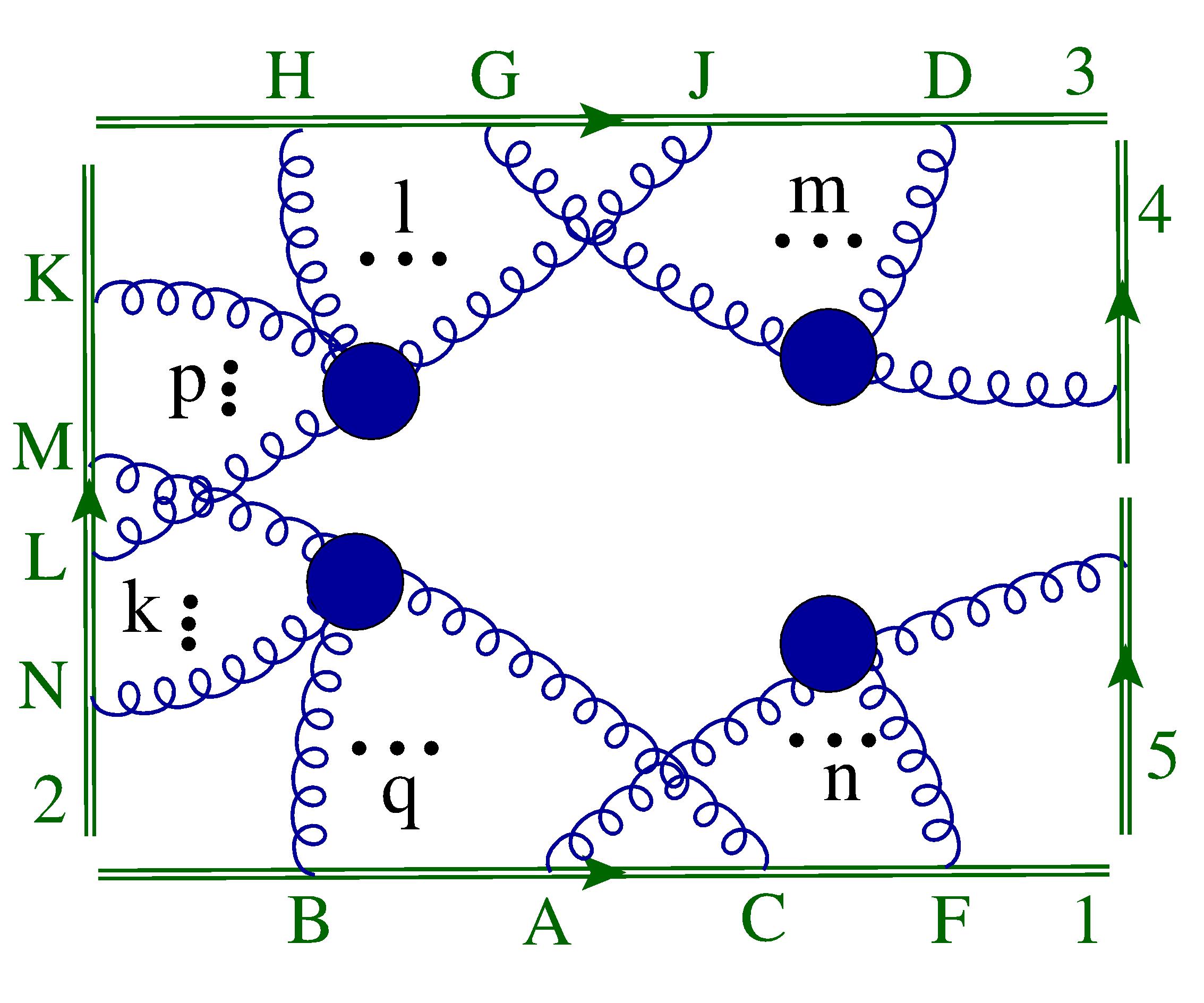} }
	\qquad
	\subfloat[][]{\includegraphics[scale=0.06]{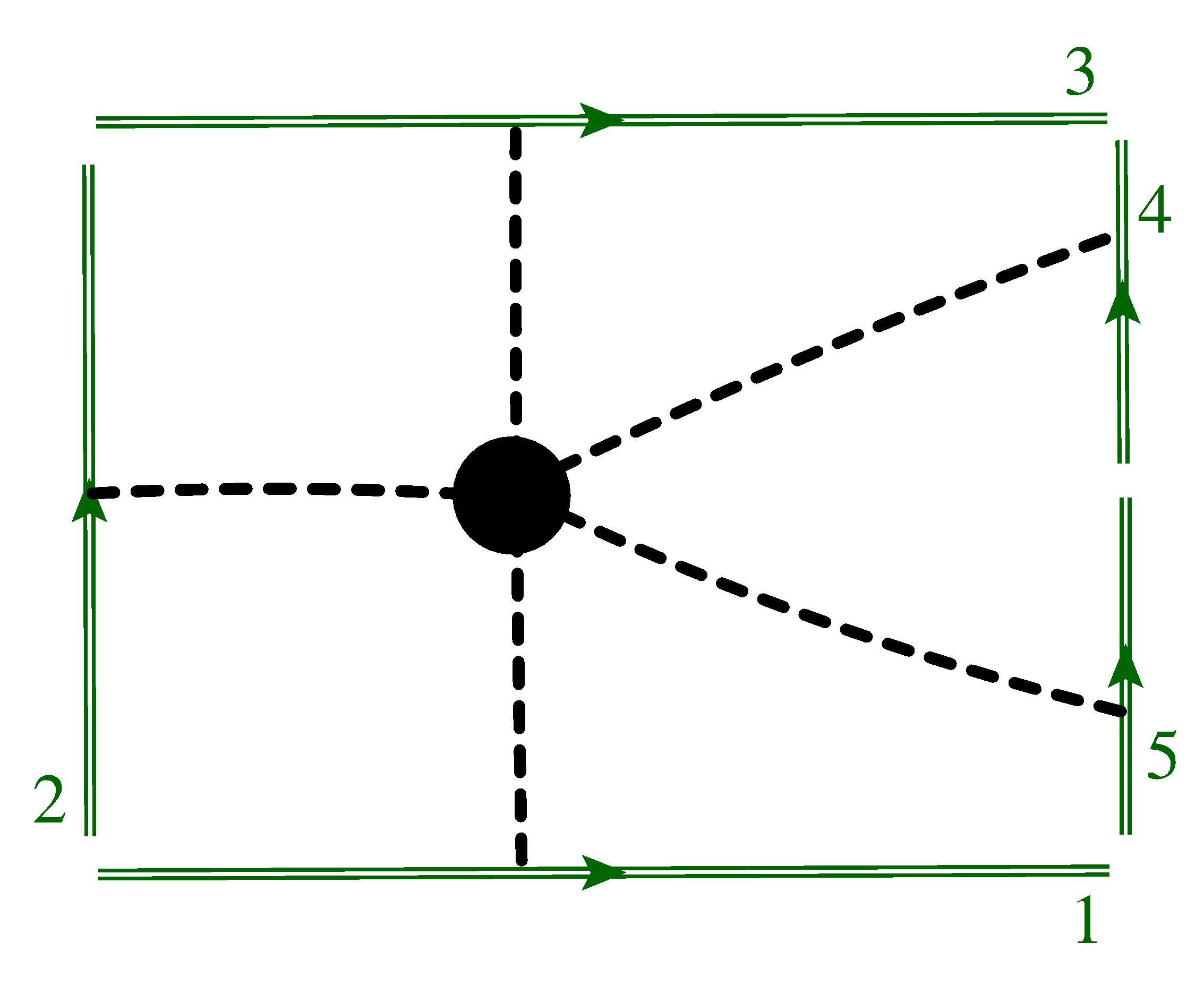} }
	\caption{Fused-Webs for Cweb $\text{W}\,_{5}^{(1,1,1,1)}(1,1,k+p,m+l,q+n)$}
	\label{fig:2Sclass4-a}
\end{figure}

\begin{figure}

	\centering
	\subfloat[][]{\includegraphics[scale=0.055]{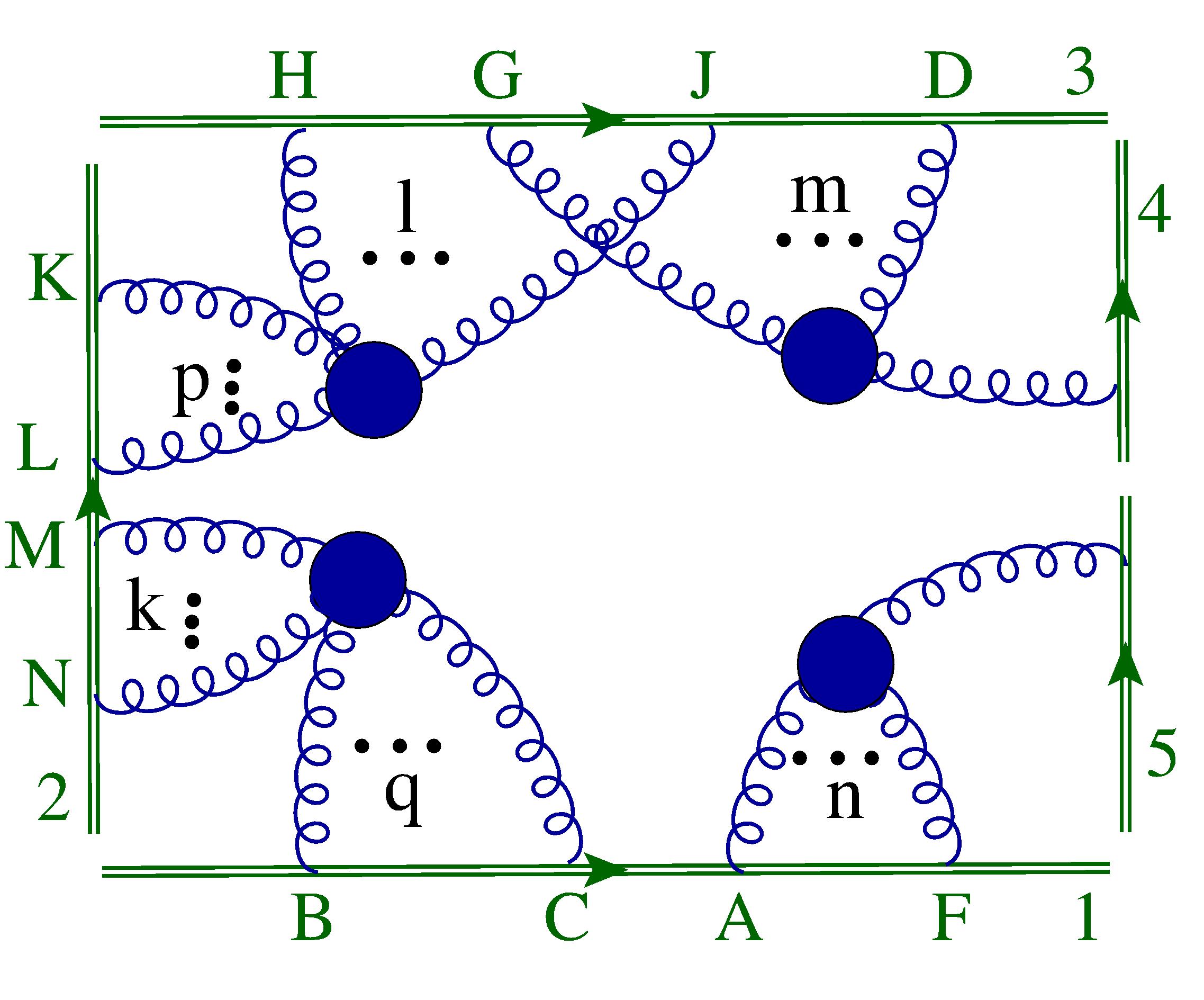} }
	\qquad
	\subfloat[][]{\includegraphics[scale=0.055]{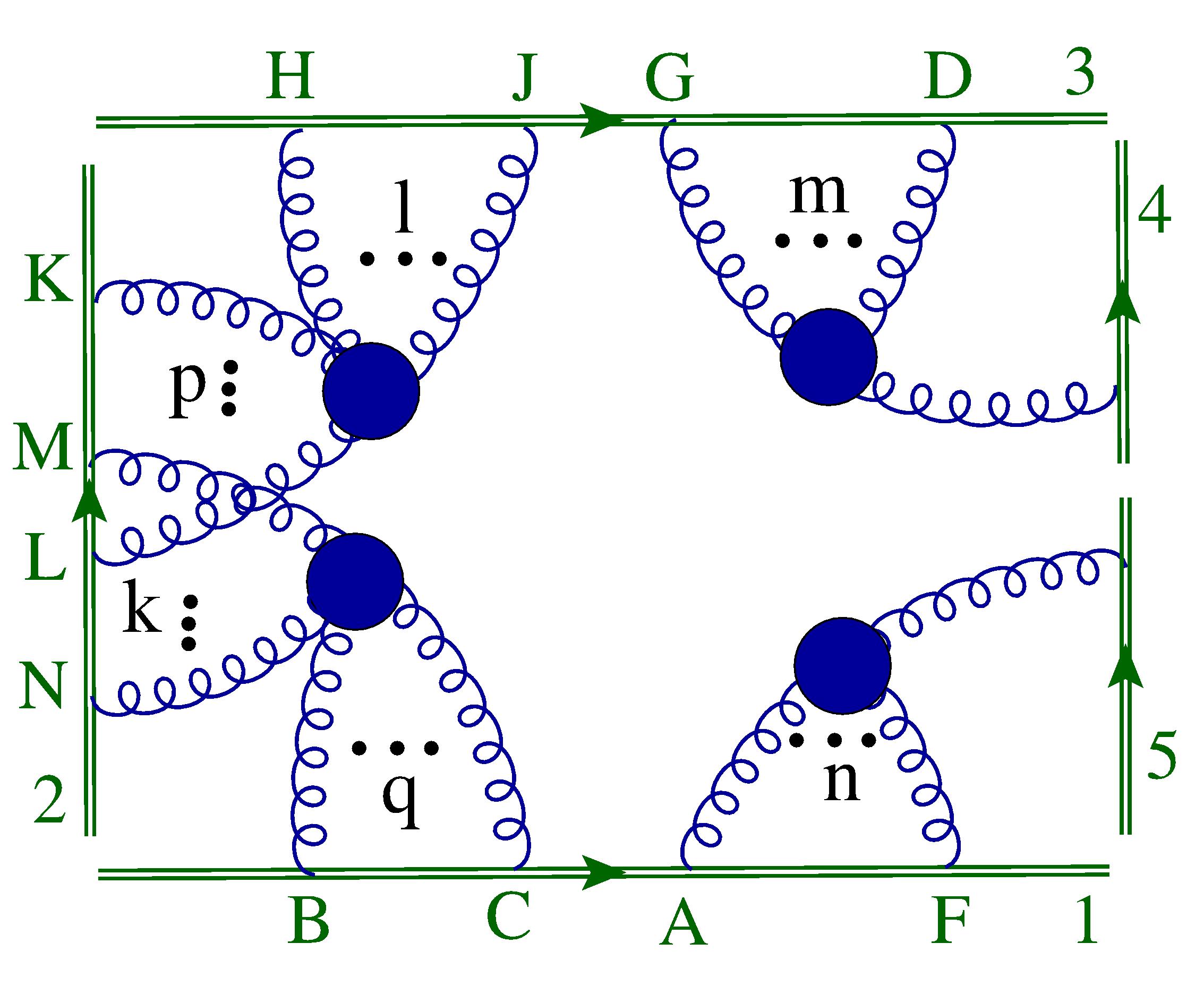} }
	\qquad
	\subfloat[][]{\includegraphics[scale=0.055]{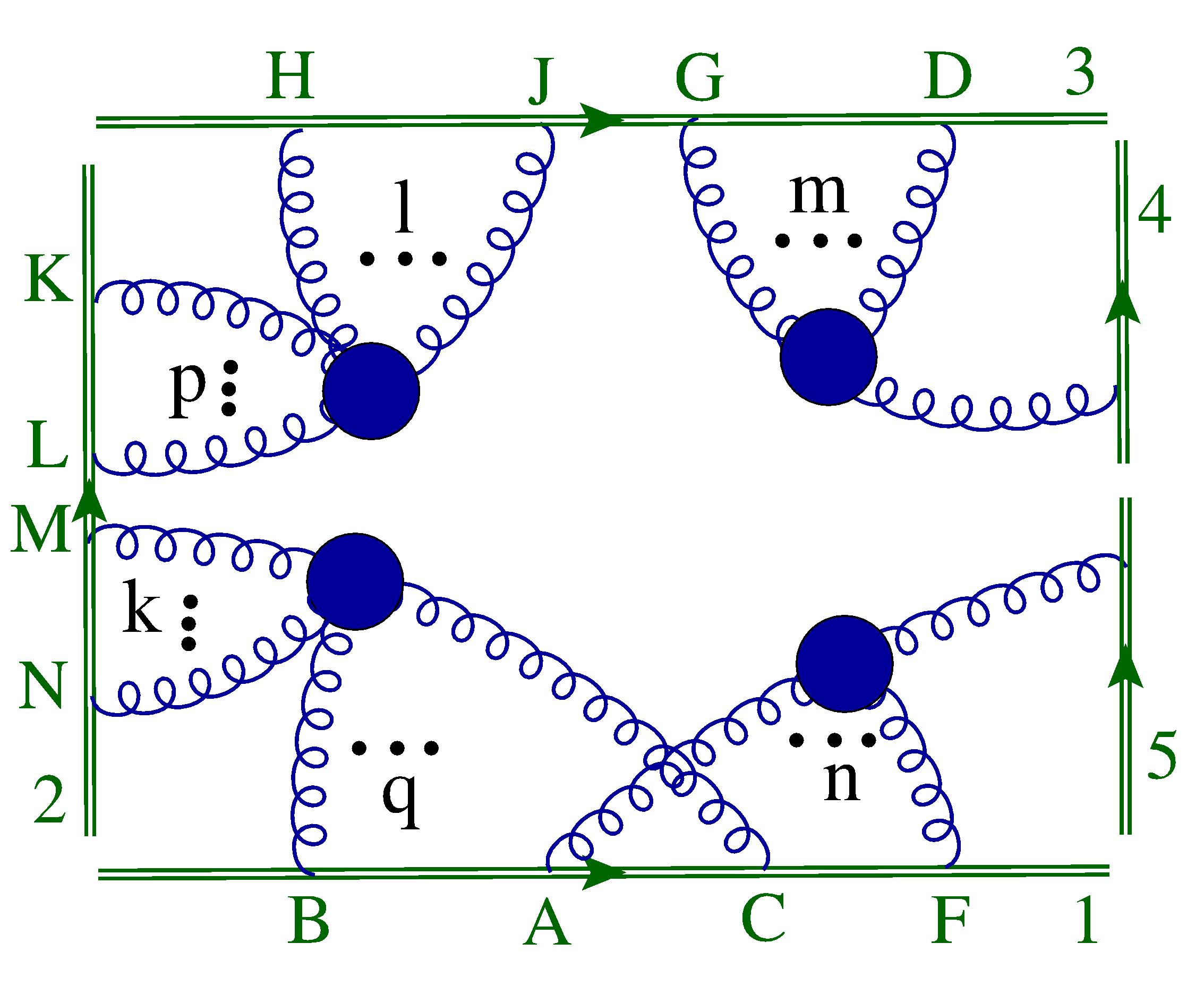} }
	\qquad
	\subfloat[][]{\includegraphics[scale=0.055]{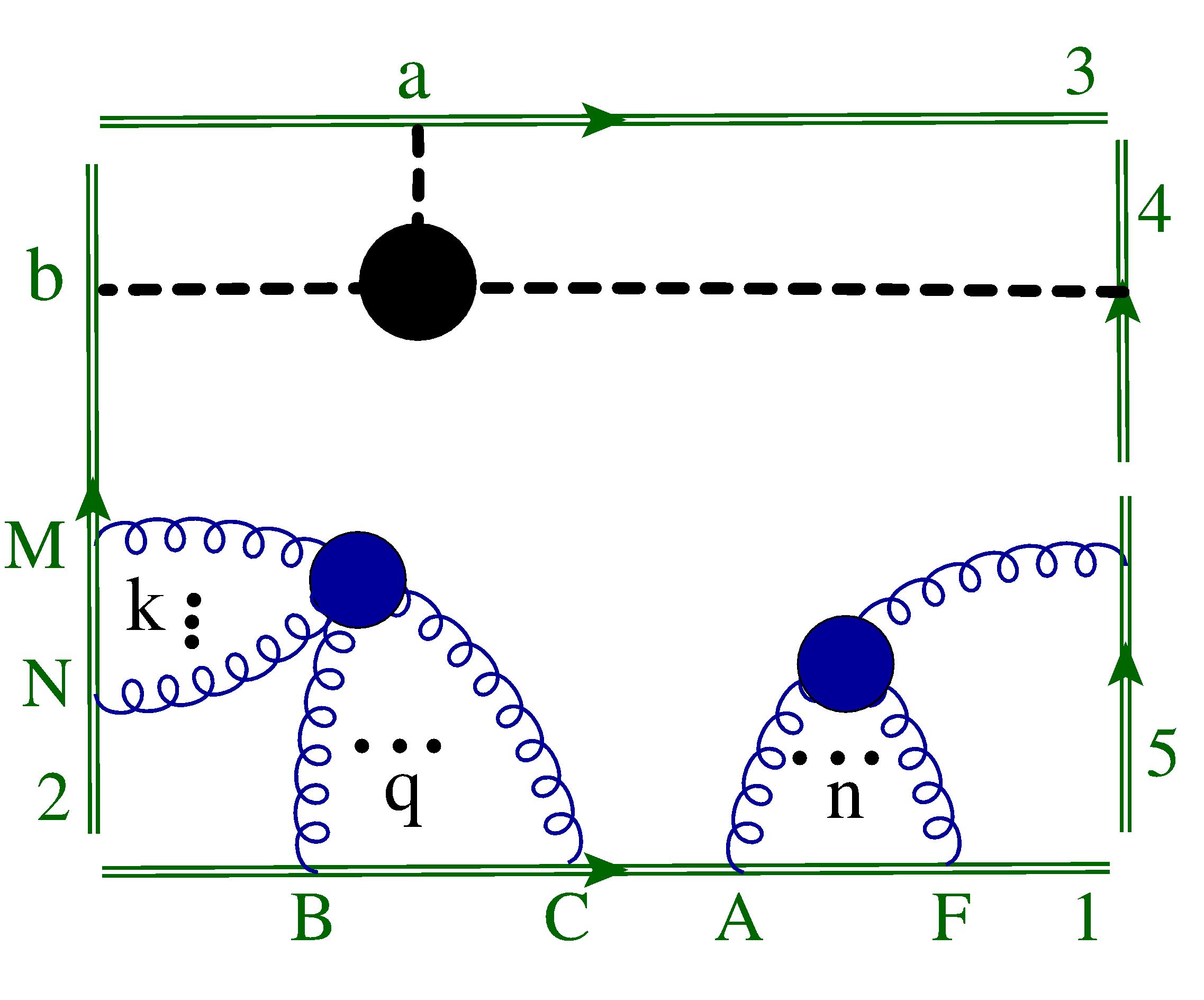} }
	\qquad
	\subfloat[][]{\includegraphics[scale=0.055]{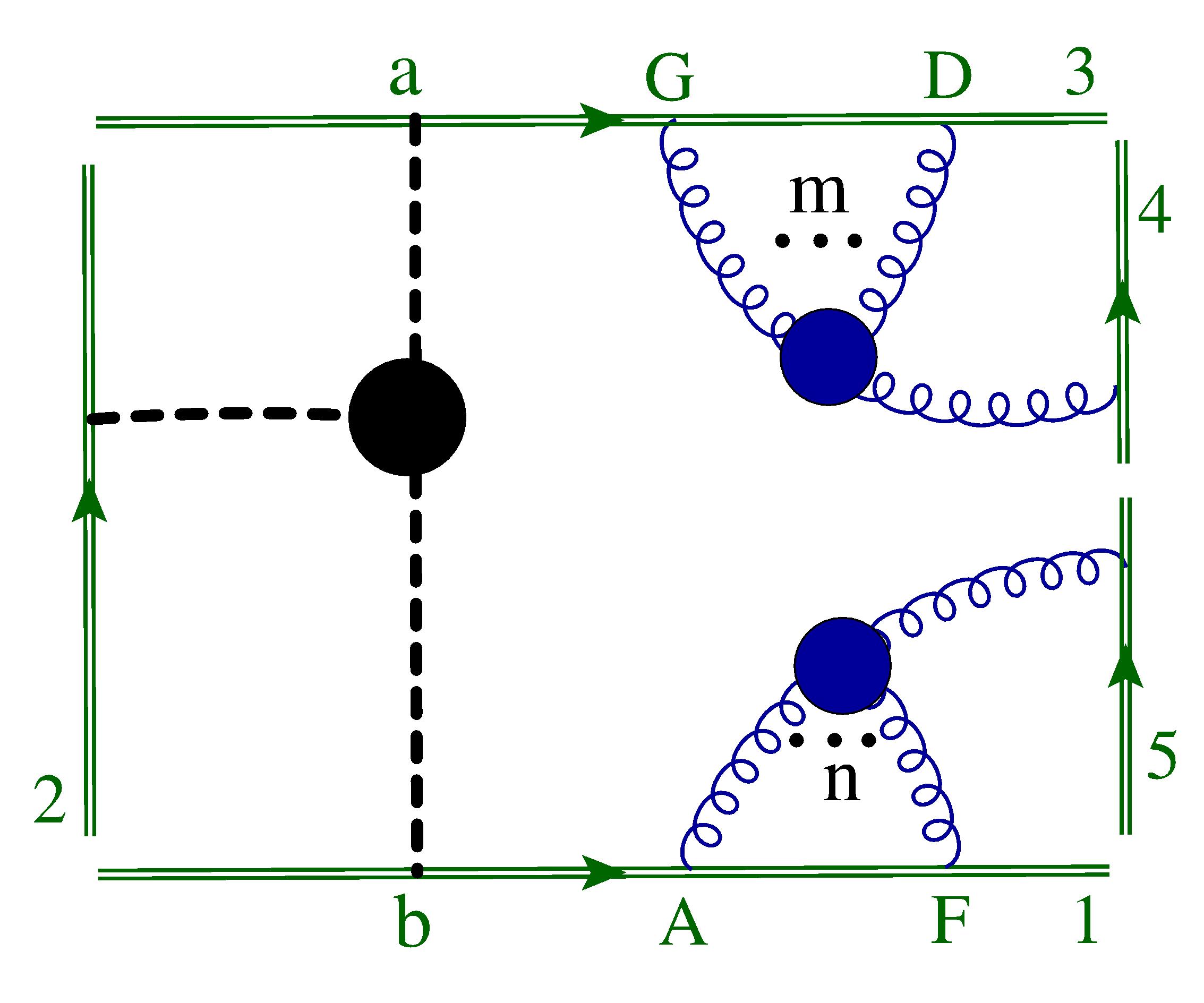} }
	\qquad
	\subfloat[][]{\includegraphics[scale=0.055]{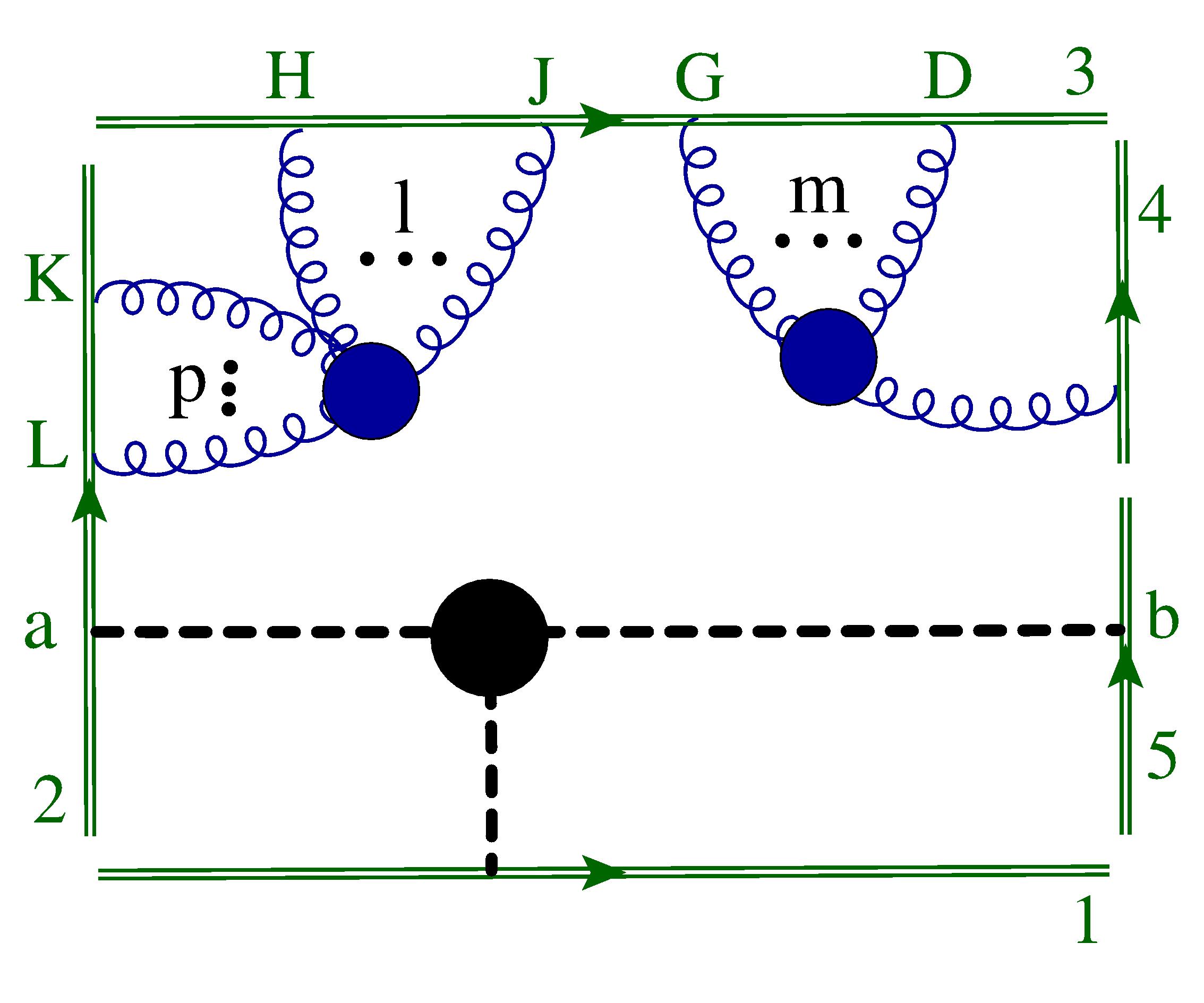} }
	\caption{Fused-Webs for Cweb $\text{W}\,_{5}^{(1,1,1,1)}(1,1,k+p,m+l,q+n)$}
	\label{fig:2Sclass4-b}
\end{figure}

\begin{figure}

	\centering
	\subfloat[][]{\includegraphics[scale=0.055]{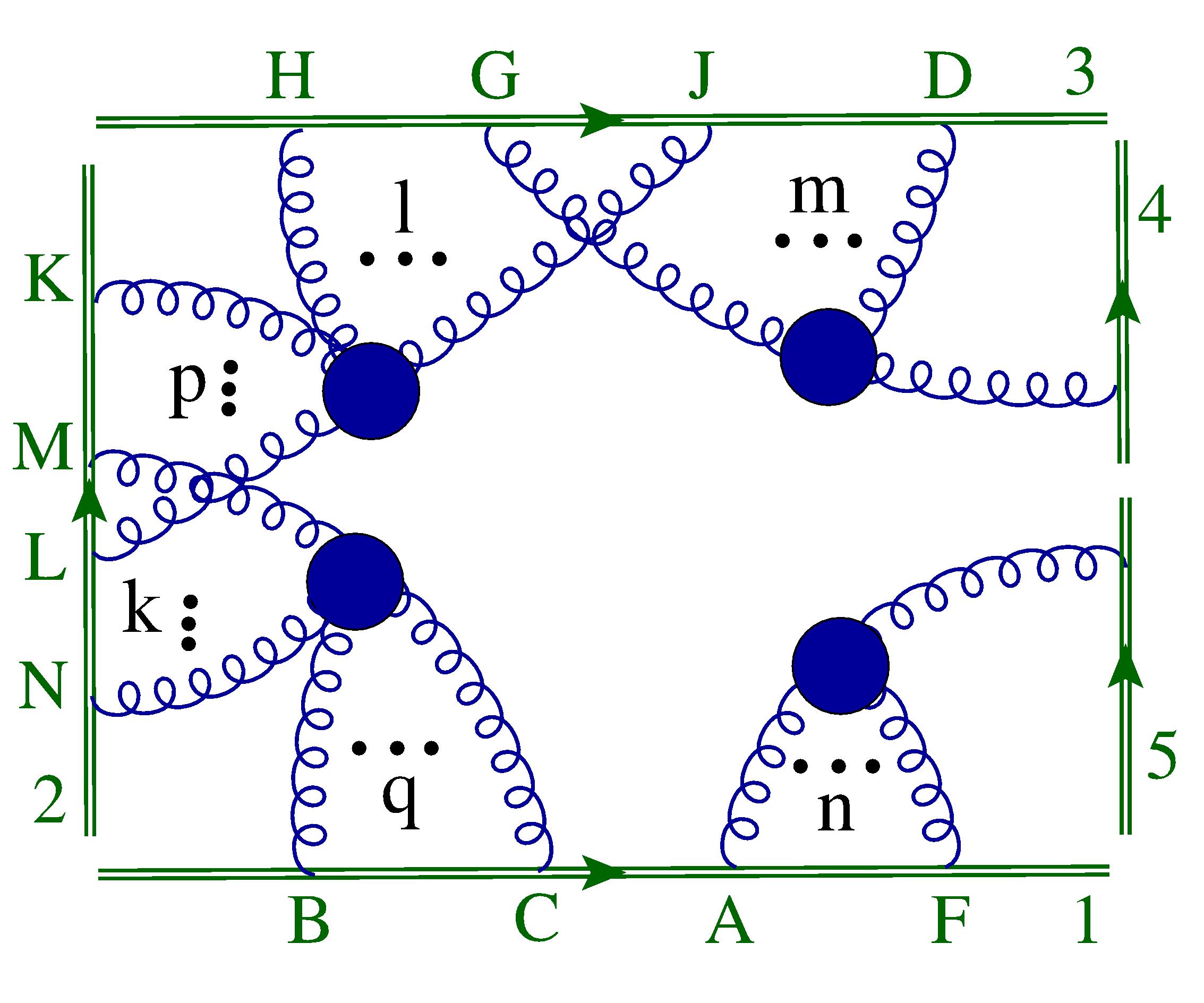} }
	\qquad
	\subfloat[][]{\includegraphics[scale=0.055]{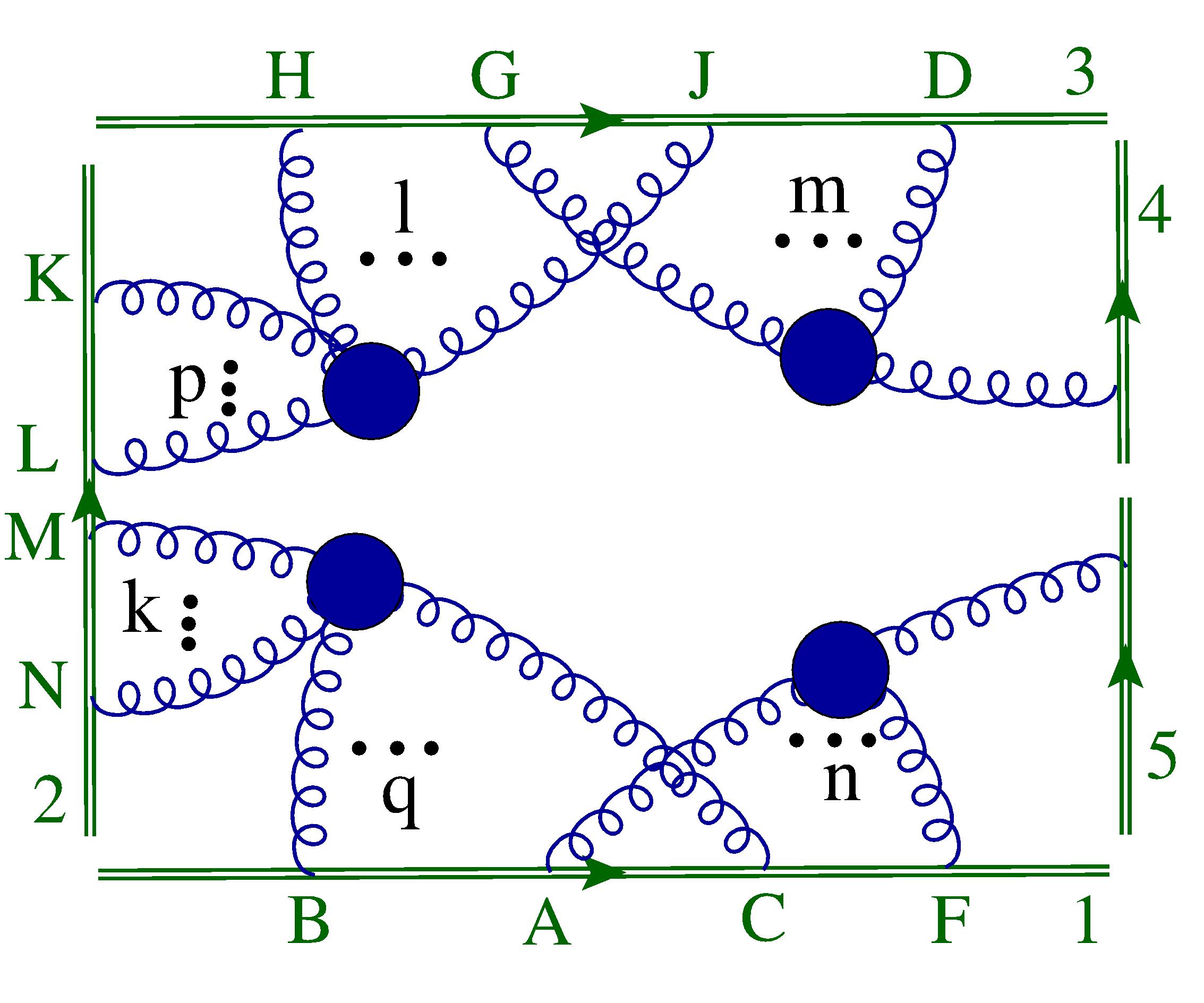} }
	\qquad
	\subfloat[][]{\includegraphics[scale=0.055]{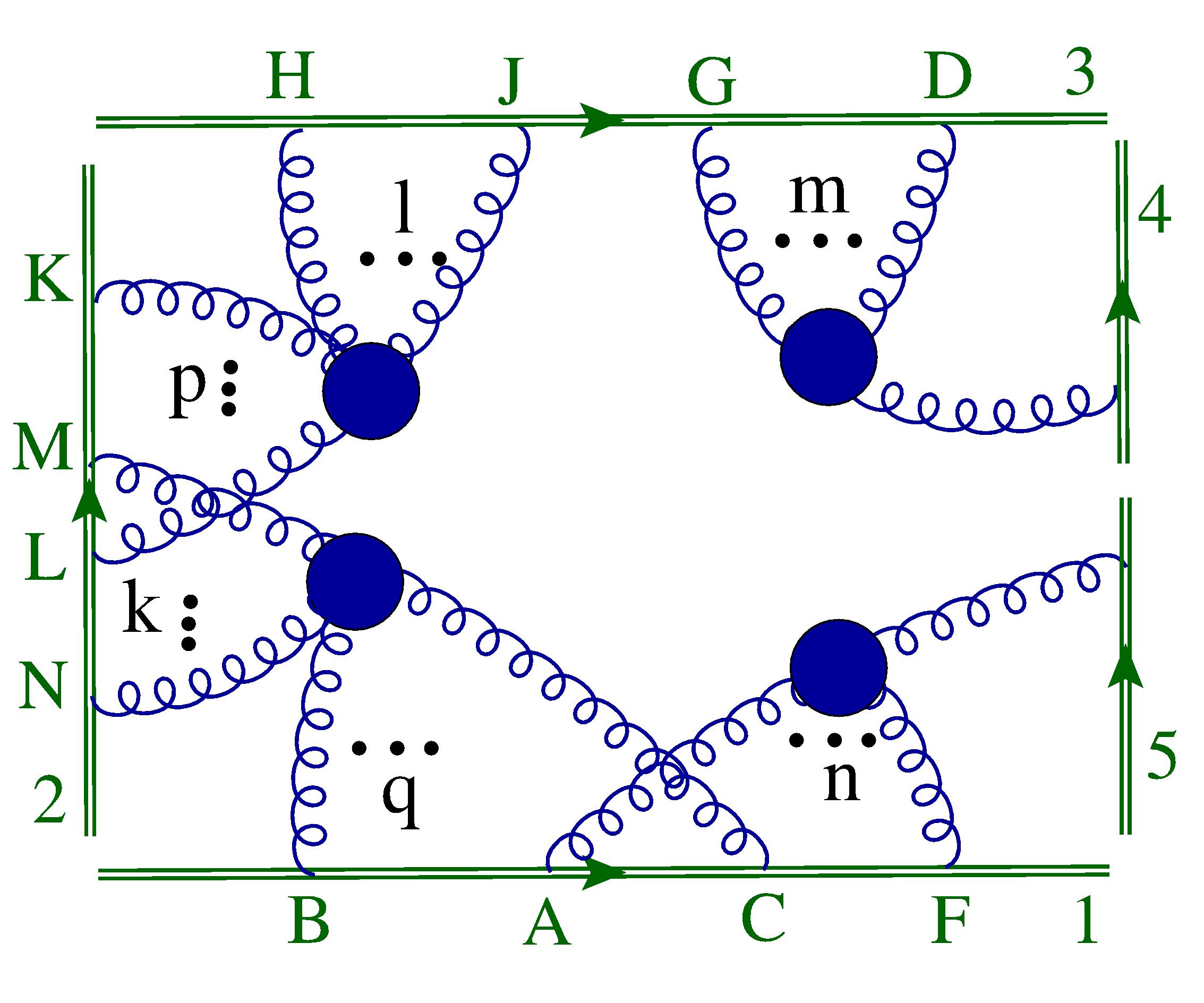} }
	\qquad
	\subfloat[][]{\includegraphics[scale=0.055]{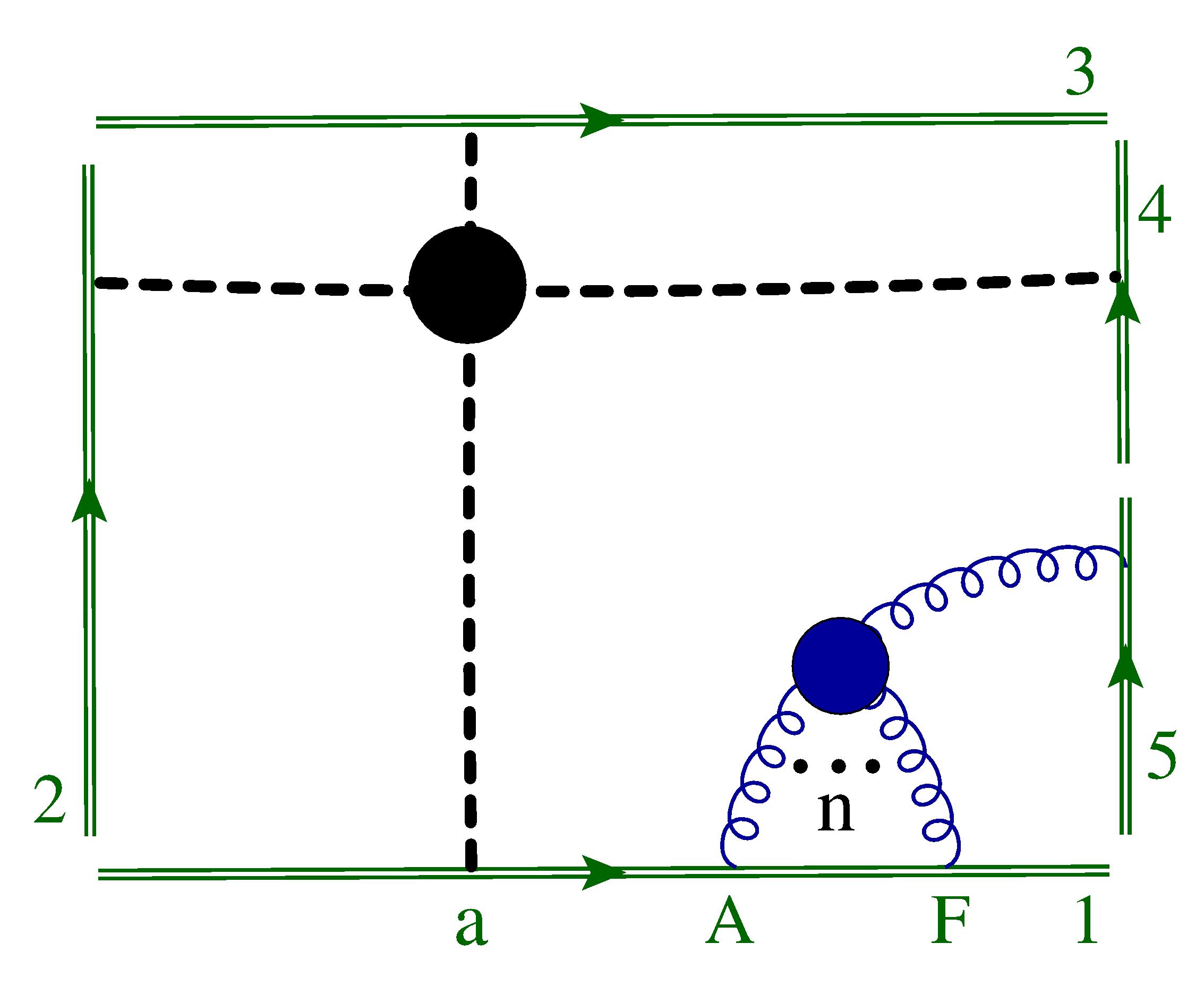} }
	\qquad
	\subfloat[][]{\includegraphics[scale=0.055]{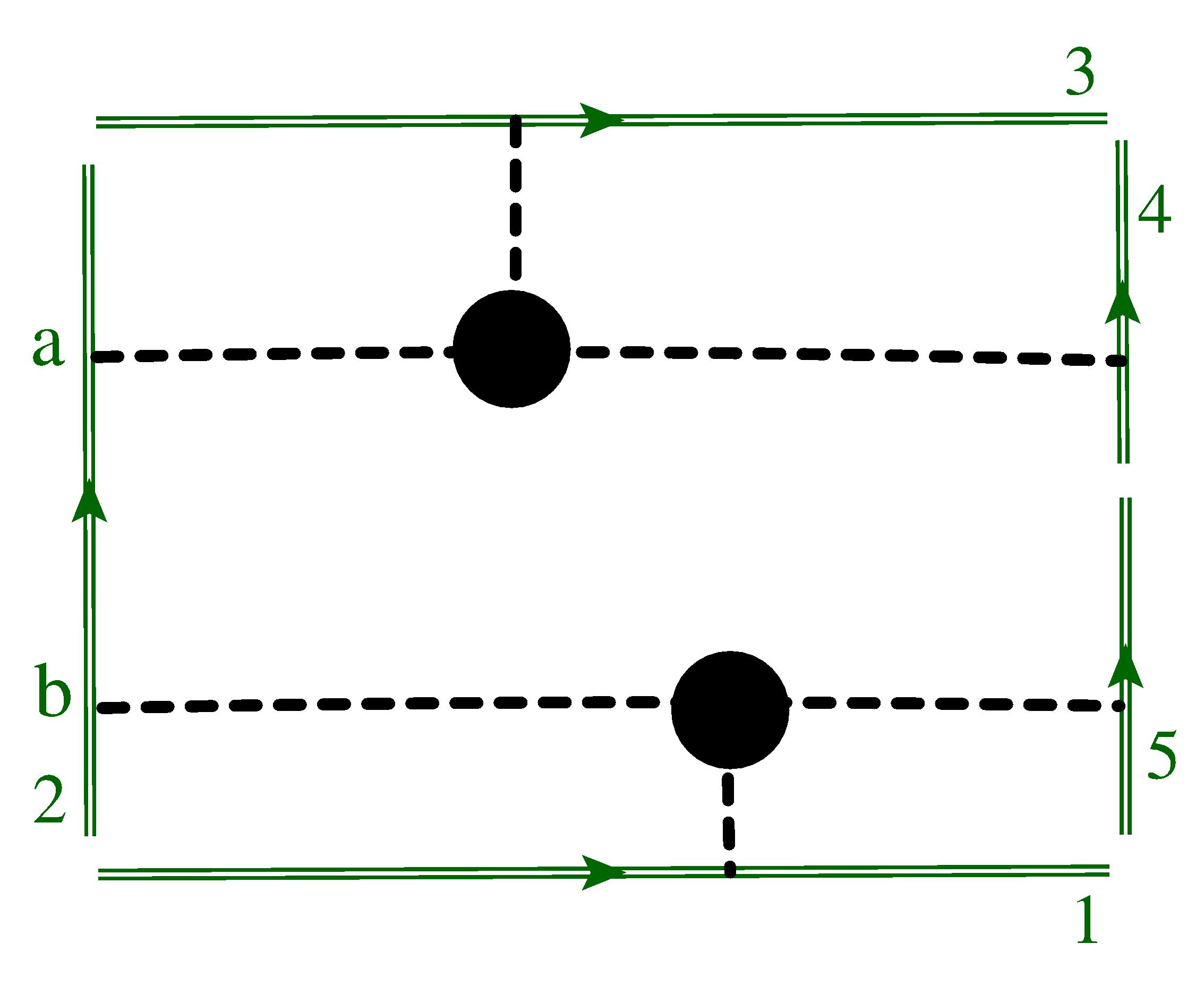} }
	\qquad
	\subfloat[][]{\includegraphics[scale=0.055]{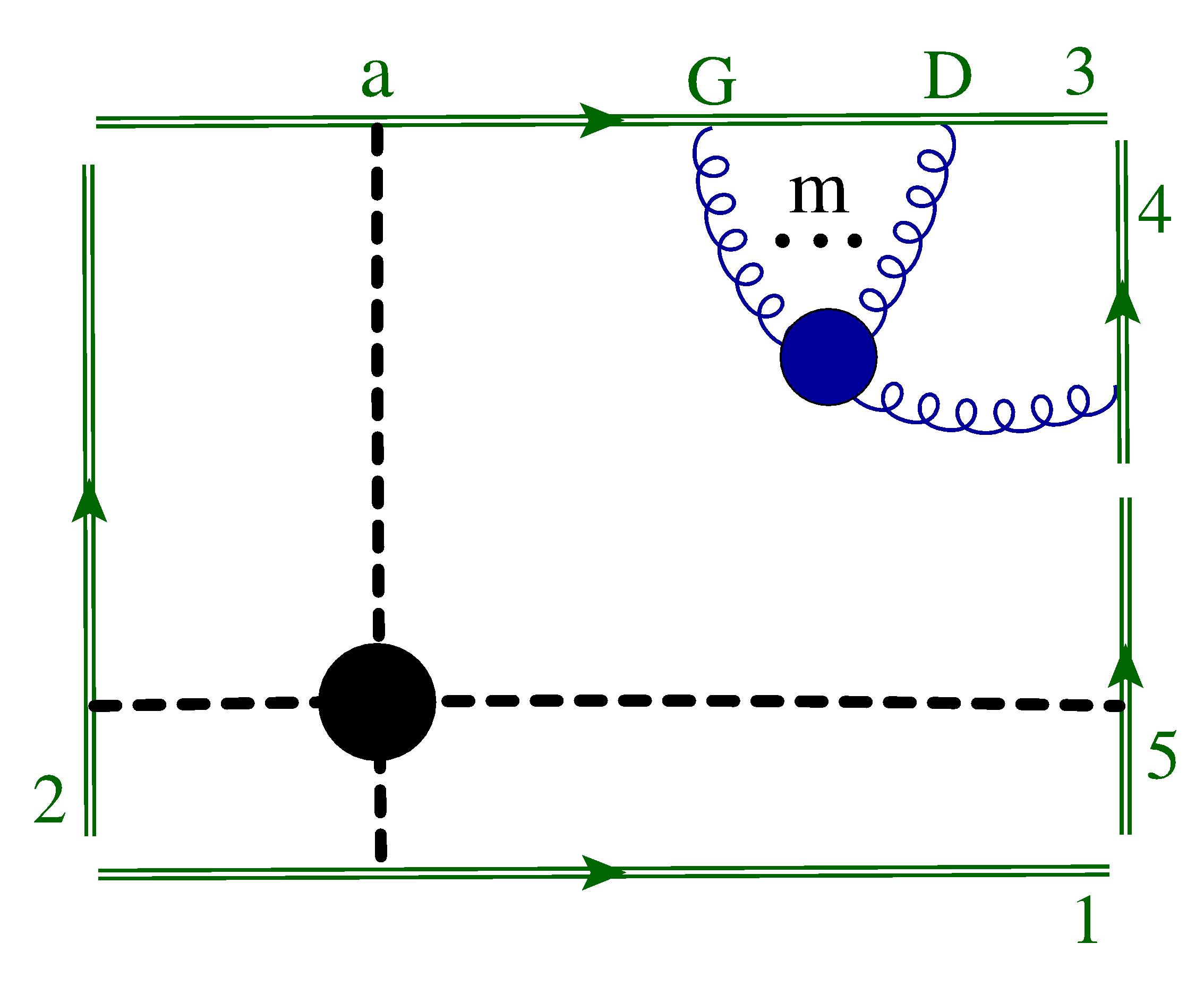} }
	\caption{Fused-Webs for Cweb $\text{W}\,_{5}^{(1,1,1,1)}(1,1,k+p,m+l,q+n)$}
	\label{fig:2Sclass4-c}
\end{figure}

\begin{table}
	\begin{center}
		\begin{tabular}{ | c | c | c |}
			\hline
			\textbf{Diagrams} & \textbf{Sequences} & \textbf{S-factors} \\ \hline
			$C_3$ & $\lbrace \lbrace \overline{N\cdots M},\;\overline{L\cdots K}\rbrace,  \lbrace \overline{H\cdots J},\;\overline{G \cdots D} \rbrace, \lbrace \overline{A\cdots F},\;\overline{B\cdots C}\rbrace \rbrace$ & 1 \\ 
			\hline
			$C_6$ & $\lbrace \lbrace \overline{L\cdots K},\;\overline{N\cdots M}\rbrace,  \lbrace \overline{G \cdots D},\;\overline{H\cdots J} \rbrace, \lbrace \overline{B\cdots C},\;\overline{A\cdots F}\rbrace \rbrace$ & 1 \\ 		
			\hline
			$C_4$ & $\lbrace \lbrace \overline{N\cdots M},\;\overline{L\cdots K}\rbrace,  \lbrace \overline{H\cdots J},\;\overline{G \cdots D} \rbrace, \lbrace \overline{B\cdots C},\;\overline{A\cdots F}\rbrace \rbrace$ & 2 \\ 
			\hline
			$C_5$ & $\lbrace \lbrace \overline{L\cdots K},\;\overline{N\cdots M}\rbrace,  \lbrace \overline{G \cdots D},\;\overline{H\cdots J} \rbrace, \lbrace \overline{A\cdots F},\;\overline{B\cdots C}\rbrace \rbrace$ & 2 \\ 
			\hline
			$C_1$ & $\lbrace \lbrace \overline{N\cdots M},\;\overline{L\cdots K}\rbrace,  \lbrace \overline{G \cdots D},\;\overline{H\cdots J} \rbrace, \lbrace \overline{A\cdots F},\;\overline{B\cdots C}\rbrace \rbrace$ & 3 \\ 
			\hline
			$C_8$ & $\lbrace \lbrace ,\;\overline{L\cdots K}\overline{N\cdots M}\rbrace,  \lbrace \overline{H\cdots J},\;\overline{G \cdots D} \rbrace, \lbrace \overline{B\cdots C},\;\overline{A\cdots F}\rbrace \rbrace$ & 3 \\ 
			\hline
			$C_7$ & $\lbrace \lbrace,\; \overline{L\cdots K}\overline{N\cdots M}\rbrace,  \lbrace \overline{H\cdots J},\;\overline{G \cdots D} \rbrace, \lbrace \overline{A\cdots F},\;\overline{B\cdots C}\rbrace \rbrace$ & 4 \\ 
			\hline
			$C_2$ & $\lbrace \lbrace,\; \overline{N\cdots M}\overline{L\cdots K}\rbrace,  \lbrace \overline{G \cdots D},\;\overline{H\cdots J} \rbrace, \lbrace \overline{B\cdots C},\;\overline{A\cdots F}\rbrace \rbrace$ & 4 \\
			\hline
		\end{tabular}
	\end{center}
	\caption{Web $ W_{5}^{(1,1,1,1)}(1,1,k+p,m+l,q+n) $}\label{tab:2Sclass4-shuffle}
\end{table}
In order to determine the rank of mixing matrix we need to calculate the rank of $ A $ block of $ R $. For this we require the order of identity matrix in $ A $, which is the number of completely entangled diagrams. One of these diagram and its Fused diagram are shown in fig.~(\ref{fig:2Sclass4-a}). The number of entangled diagrams, determined by using the eq.~(\ref{eq:Num-two-entangled0}) for each of the Wilson lines, is given as,
\begin{align}
\bigg(l\,\Pi\,m \;-2\bigg)\;\bigg(p\,\Pi\,k \;-2\bigg)\;\bigg(q\,\Pi\,n \;-2\bigg)\,.
\end{align}
Each of these diagrams has the same associated Fused diagram, shown in fig.~(\ref{fig:2Sclass4-a}\txb{b}). The shuffle of which forms the Fused-Web having one diagram. 
Next we have two types of partially entangled diagrams. 
\begin{itemize}
	\item The first case involves entanglement among the correlators at two Wilson lines simultaneously. There are three such possibilities, as shown in fig.~(\ref{fig:2Sclass4-c}\txb{a}), (\ref{fig:2Sclass4-c}\txb{b}) and (\ref{fig:2Sclass4-c}\txb{c}) with the remaining correlator placed towards the head of each Wilson line. The sum of number of such diagrams for all the three possibilities is given by,
	\begin{align}
	\bigg(l\,\Pi\,m \;-2\bigg)\;\bigg(p\,\Pi\,k \;-2\bigg)\;+\bigg(l\,\Pi\,m \;-2\bigg)\;\bigg(q\,\Pi\,n \;-2\bigg)\;+\bigg(p\,\Pi\,k \;-2\bigg)\;\bigg(q\,\Pi\,n \;-2\bigg)\;\nonumber
	\end{align} 
	The Fused diagrams for each of three cases are shown in fig.~(\ref{fig:2Sclass4-c}\txb{d}), (\ref{fig:2Sclass4-c}\txb{e}) and (\ref{fig:2Sclass4-c}\txb{f}) respectively. Each of the Fused diagrams form a Fused-Web by the shuffle of its attachment. Every such Fused-Web has two diagrams with $ s=1 $, that has a mixing matrix $ R(1_2) $, given by the Uniqueness theorem.
	
	\item In the second case, entanglement occurs between correlators only at one Wilson line. Again there are three possibilities. The first case shown in fig.~(\ref{fig:2Sclass4-b}\txb{a}) entangles $ (l+p) $ with $ (m+1) $-point correlator on line 3. The number of such diagrams with $ (p+l) $-point correlator is placed right to the $ (k+q) $-point correlator on line 2, and on line 1, $(n+1) $-point correlator is placed right to the $ (k+q) $-point correlator, is given by
	\begin{align}
	\bigg(l\,\Pi\,m \;-2\bigg)\,.
	\end{align} 
	The corresponding Fused diagram shown in fig.(\ref{fig:2Sclass4-b}\txb{d}) generate the Fused-Web with four diagrams having following order of attachments and $ s $-factors.
	\begin{table}[H]
		\begin{center}
			\begin{tabular}{|c|c|c|}
				\hline 
				\textbf{Diagrams}  & \textbf{Sequences}  & \textbf{s-factors}  \\ 
				\hline
				$C_{1}$  & $\{\overline{B\cdots C},\;\overline{A\cdots F}\},\; \{b,\;\overline{N\cdots M}\}$  & 1 \\ \hline
				$C_{2}$  & $\{\overline{A\cdots F},\;\overline{B\cdots C}\},\; \{\;\overline{N\cdots M},\;b\}$ & 1 \\ \hline
				$C_{3}$  & $\{\overline{B\cdots C},\;\overline{A\cdots F}\},\; \{\overline{N\cdots M},\;b\}$  & 2 \\ \hline
				$C_{4}$  & $\{\overline{A\cdots F},\;\overline{B\cdots C},\},\; \{b,\;\overline{N\cdots M}\}$  & 2 \\ \hline
			\end{tabular}
		\end{center}
	\end{table}
	The $ R $ for this Fused-Web is $ R(1_2,2_2) $.
	The second case is shown in fig.~(\ref{fig:2Sclass4-b}\txb{b}) where $ (l+p) $ and $ (k+q) $-point correlators are entangled on line 2. The number of such diagrams with both the remaining correlators placed after the entangled correlators on line 1 and 3, is given by
	\begin{align}
	\bigg(p\,\Pi\,k \;-2\bigg)\,.
	\end{align}
	The Fused diagram for each of these is same and  displayed in fig.(\ref{fig:2Sclass4-b}\txb{e}). The shuffle of its attachments generate Fused-Webs with four diagrams whose order of attachments and $ s $-factor are given in the following table
	\begin{table}[H]
		\begin{center}
			\begin{tabular}{|c|c|c|}
				\hline 
				\textbf{Diagrams}  & \textbf{Sequences}  & \textbf{s-factors}  \\ 
				\hline
				$C_{1}$  & $\{b,\;\overline{A\cdots F}\},\; \{\overline{G\cdots D},\;a\}$  & 1 \\ \hline
				$C_{2}$  & $\{\overline{A\cdots F},\;b,\},\; \{a,\;\overline{G\cdots D}\}$  & 1 \\ \hline
				$C_{3}$  & $\{b,\;\overline{A\cdots F}\},\; \{a,\;\overline{G\cdots D}\}$  & 2 \\ \hline
				$C_{4}$  & $\{\overline{A\cdots F},\;b\},\; \{\overline{G\cdots D},\;a\}$ & 2 \\ \hline
			\end{tabular}
		\end{center}
	\end{table}
	Following the uniqueness theorem, the mixing matrix for the Fused-Web is $ R(1_2,2_2) $. The third possibility involves the attachments from $ (k+q) $ and $ (n+1) $-point correlator on line 1 as shown in fig.~(\ref{fig:2Sclass4-b}\txb{c}). The number of such diagrams with attachments from $ (m+1) $ and $ (l+p) $-point correlators placed right to the $  (l+p)  $ and $ (q+k) $ on line 3 and 2 respectively, is given as 
	\begin{align}
	\bigg(q\,\Pi\,n \;-2\bigg)\,.	
	\end{align}
	Each of these has same Fused diagram shown in fig.~(\ref{fig:2Sclass4-b}\txb{f}). The Fused-Web generated by the shuffle of its attachments has four diagrams with following $ s $-factors. 
	\begin{table}[H]
		\begin{center}
			\begin{tabular}{|c|c|c|}
				\hline 
				\textbf{Diagrams}  & \textbf{Sequences}  & \textbf{s-factors}  \\ 
				\hline
				$C_{1}$  & $\{\overline{G\cdots D},\;\overline{H\cdots J}\},\; \{\overline{K\cdots L},\;a\}$  & 1 \\ \hline
				$C_{2}$  & $\{\overline{H\cdots J},\;\overline{G\cdots D}\},\; \{\overline{K\cdots L},\;a\}$ & 1 \\ \hline
				$C_{3}$  & $\{\overline{H\cdots J},\;\overline{G\cdots D}\},\; \{\overline{K\cdots L},\;a\}$  & 2 \\ \hline
				$C_{4}$  & $\{\overline{G\cdots D},\;\overline{H\cdots J}\},\; \{a,\;\overline{K\cdots L}\}$  & 2 \\ \hline
			\end{tabular}
		\end{center}
	\end{table}
	The uniqueness theorem tells that these $ s $-factors correspond to a unique mixing matrix $ R(1_2,2_2) $. 
\end{itemize}
Therefore the rank of block $ A $ for this class of Cwebs is,
\begin{align}
r(A)&=\; r(I) \big(l\,\Pi\,m \;-2\big)\;\big(p\,\Pi\,k \;-2\big)\;\big(q\,\Pi\,n \;-2\big)\,\nonumber\\
&\quad \;+\, r(R(1_2))\bigg[ \big(l\,\Pi\,m \;-2\big)\;\big(p\,\Pi\,k \;-2\big)\;+\big(l\,\Pi\,m \;-2\big)\;\big(q\,\Pi\,n \;-2\big)\;+\big(p\,\Pi\,k \;-2\big)\;\big(q\,\Pi\,n \;-2\big)\bigg]\nonumber\\
& \quad \;+\;r(R(1_2,2_2)) \bigg[\big(l\,\Pi\,m \;-2\big)\;+\;\big(p\,\Pi\,k \;-2\big)\;+\;\big(q\,\Pi\,n \;-2\big)\bigg]
\end{align}
Upon substituting the ranks of Identity and basis matrices we get,

\begin{align}
r(A) &=\; \frac{(k+p)!\, (l+m)!\, (n+q)!}{k! \,l!\, m!\, n!\, p!\, q!}-\frac{(k+p)!\, (l+m)!}{k!\, l!\, m!\, p!}-\frac{(k+p)! \,(n+q)!}{k! \,n!\, p!\, q!}+\frac{(k+p)!}{k!\, p!}\nonumber\\
&
\qquad  
-\frac{(l+m)! \,(n+q)!}{l!\, m!\, n!\, q!}+\frac{(l+m)!}{l!\, m!}+\frac{(n+q)!}{n!\, q!}-2
\end{align}
The rank of $ D $ for this Cweb is $ 1 $. Thus, we can write the rank of the mixing matrix of this class of Cweb as
\begin{align}
r(R) &=\; \frac{(k+p)!\, (l+m)!\, (n+q)!}{k! \,l!\, m!\, n!\, p!\, q!}-\frac{(k+p)!\, (l+m)!}{k!\, l!\, m!\, p!}-\frac{(k+p)! \,(n+q)!}{k! \,n!\, p!\, q!}+\frac{(k+p)!}{k!\, p!}\nonumber\\
&
\qquad  
-\frac{(l+m)! \,(n+q)!}{l!\, m!\, n!\, q!}+\frac{(l+m)!}{l!\, m!}+\frac{(n+q)!}{n!\, q!}-1	
\end{align}
The correspondence described in the previous section guides us to predict the rank of boomerang Cwebs along with the non Boomerang ones. A boomerang Cweb $ \text{W}_{4}^{(4)}(1,2,2,3) $, shown in fig.~(\ref{fig:4legsWeb3}\txb{a}) appearing at four loops connecting four lines belong to this class. The mixing matrix corresponding to this has rank 2 as predicted from above expression by choosing the values $  n=2 $ and $ l=m=k=q=p=1 $. The Fused-Web formalism for this web is shown in section~\ref{sec:Boom-4loop-4line}. 

We end this section by concluding that we have provided four all order general structures of Cwebs. The ranks of the mixing matrices for these are predicted using the formalism developed in~\cite{Agarwal:2022wyk}. These four classes suffice to determine the diagonal blocks and ranks of mixing matrices of the boomerang Cwebs appearing at four loops connecting four Wilson lines.

\subsection{Explicit form of mixing matrix for $\text{W}\,_{2}^{(1,1)}(1,k+2)$}
\label{explicit}
We start our discussion with a class of Cwebs $\text{W}\,_{2}^{(1,1)}(1,k+2)$ as the explicit form of the mixing matrix for this class can be obtained using the properties and the concepts discussed in section~\ref{sec:Fused-Webs}.
\begin{figure}[b]
	\captionsetup[subfloat]{labelformat=empty}
	\centering
	\subfloat[][]{\includegraphics[scale=0.064]{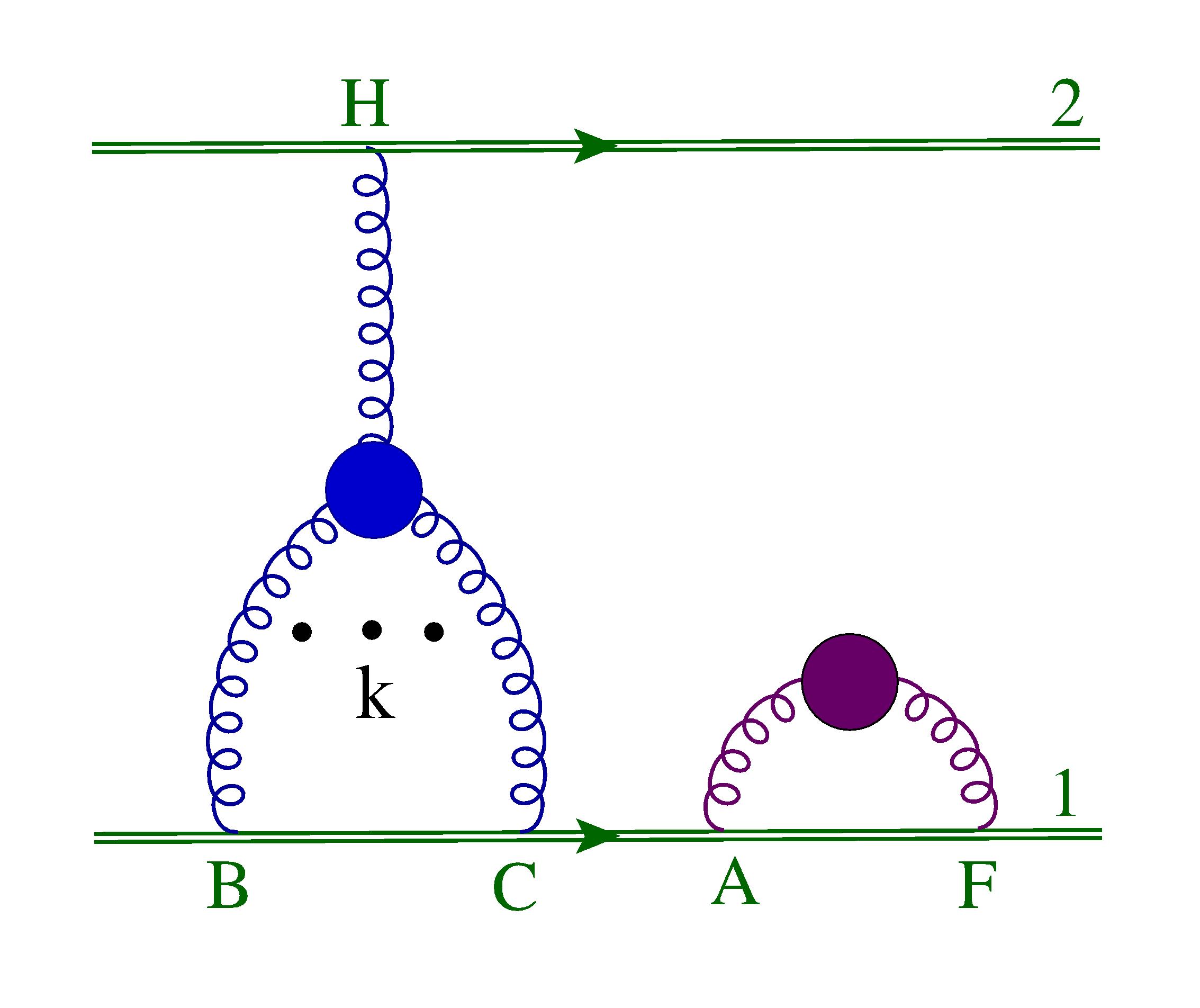} }
	\caption{Diagram of the general Cweb $\text{W}\,_{2}^{(1,1)}(1,k+2)$}
	\label{fig:Sclass1-Sub}
\end{figure}

In this class, we have one $ (k+1) $-point gluon correlator along with one Boomerang correlator. Its general structure is shown in fig.~(\ref{fig:Sclass1-Sub}). The shuffle of correlators on Wilson line 1 generates all the the diagrams of Cwebs, which can either be completely entangled or  reducible. 
There are only two possible reducible diagrams in which the boomerang is either placed left or right to the (k+1)-point correlator on line 1. 
Both of these diagrams have $ s=1 $, and from the Uniqueness theorem, we can write the explicit form of $ D $  as $ R(1_2) $. Thus
\begin{align}
D\;=\;R(1_2)\,,\qquad \text{and} \qquad r(D)\;=\;r(R(1_2))\;=\;1\,.
\label{eq:rD-Sclass-1}
\end{align}

In this case $ A $ contains only the identity matrix as there are no partially entangled diagrams. Henceforth it is sufficient to determine the number of completely entangled diagrams of the Cweb to determine the rank of the mixing matrix~\cite{Agarwal:2022wyk}.
As mentioned earlier, the number of ways in which two correlators of attachments $ a $ and $ b $ on the same line remain entangled is given in eq.~\eqref{eq:Num-two-entangled0}.
In this case $ a=k $ and $ b=2 $, so the number of completely entangled diagrams and the rank of $ A $ is given by
\begin{align}
r(A)=(k\,\Pi\,2) - 2&=\; 
\dfrac{(k+1)\,(k+2)}{2}-2  \,\,.
\end{align}  
Thus, using the above equation and eq.~\eqref{eq:rD-Sclass-1} rank of $ R $ is given by,
\begin{align}\label{eq:Sclass-Sub-rank}
r(R) 
&=\;\dfrac{(k+1)\,(k+2)}{2}-1\,.
\end{align}

Now, we can calculate the off-diagonal entries of $ R $ using the known properties of the mixing matrices described in section~\ref{sec:review}. Once diagonals are known, we can write down the explicit form of the mixing matrix $ R $ for this class as,
\begin{align}
R\,=\,\left(\begin{array}{cccc}
\textbf{I}_{l} && & \begin{array}{cc}
c_{1}&  b_{1} \\ 
c_{2} &   b_{2} \\
\vdots & \vdots \\
c_{l} &   b_{l}
\end{array}\\
\textbf{O}_{2\times l}& && \begin{array}{c}
R\,(1_2)
\end{array}	 \\
& & 
\end{array}\right)\,\qquad ;\qquad l\,\equiv\,\dfrac{(k+1)\,(k+2)}{2}-2 \,,
\end{align}
where $ b_i $ and $ c_i $ are unknowns. Also note that, here we have denoted the order of identity by $ l $, which is the number of completely entangled diagrams. 
Now, to determine these unknowns, we use the known properties of the mixing matrices. The row-sum rule fixes,
\begin{align}
b_j\,&=-1-c_j, \mkern-18mu & 1 \leq j \leq l, 
\label{eq:row-sum}	
\end{align}
and the idempotence property fixes,
\begin{align}
c_j=-\frac{1}{2}\quad 1\leq j \leq l \,.
\end{align}
With this we have uniquely fixed all the elements of the mixing matrix for this class of boomerang Cwebs, and it is then given by
\begin{align}
R\,=\,\left(\begin{array}{cccccccc}
1&0& 0 &\ldots & 0 & \, & -1/2 &  -1/2  \\
0&1& 0 &\ldots & 0 & \, &  -1/2 &   -1/2 \\
0&0& 1 &\ldots & 0 & \, &  -1/2 &   -1/2 \\
\vdots &\vdots & \vdots &\vdots & \vdots & \vdots &  \vdots &\vdots \\
0&0& 0 &\ldots & 1 & \, & -1/2 &   -1/2 \\
0&0& 0 &\ldots & 0 & \, & 1/2 &   -1/2 \\
0&0& 0 &\ldots & 0 & \, & -1/2 &   1/2 \\
\end{array}\right)\,.
\label{eq:R-gen-final}
\end{align}

We now conclude this section. In this section, we have discussed four classes of Cwebs for which the diagonals of the mixing matrices can be constructed using the Normal ordering, Fused-Webs, uniqueness theorem and properties of mixing matrices. We again emphasize that these four classes are sufficient to calculate diagonal blocks of all eight boomerang Cwebs occurring at  four loops that connect four Wilson lines. We have also obtained explicit form a special class of boomerang Cweb at all orders in the perturbation theory. We verify the results of this section for boomerang Cwebs at four loops in the next section.

\section{Diagonal blocks of boomerang Cwebs at four loops connecting 4 Wilson lines}\label{sec:Boom-4loop-4line}

In this section, we calculate the diagonal blocks of mixing matrices of boomerang Cwebs that appear at four loops and connect four lines. There are total eight such boomerang Cwebs at four loops which are generated using the algorithm presented in~\cite{Agarwal:2020nyc,Agarwal:2021him}. 
We show in this section that the mixing matrices for all eight boomerang Cwebs at four loops connecting four Wilson lines have general form of the four special classes shown in the previous section.

\vspace{0.5cm}
\textbf{1.}\, $ \textbf{W}^{(1,0,1)}_{4}(1,1,1,3) $\vspace{0.2cm}\\

This Cweb has three diagrams, out of which one is completely entangled and two are reducible. The diagrams after normal ordering and their corresponding $ s $-factors are presented in the following table.

\begin{figure}[H]

	\centering
	\subfloat[][]{\includegraphics[height=4cm,width=4cm]{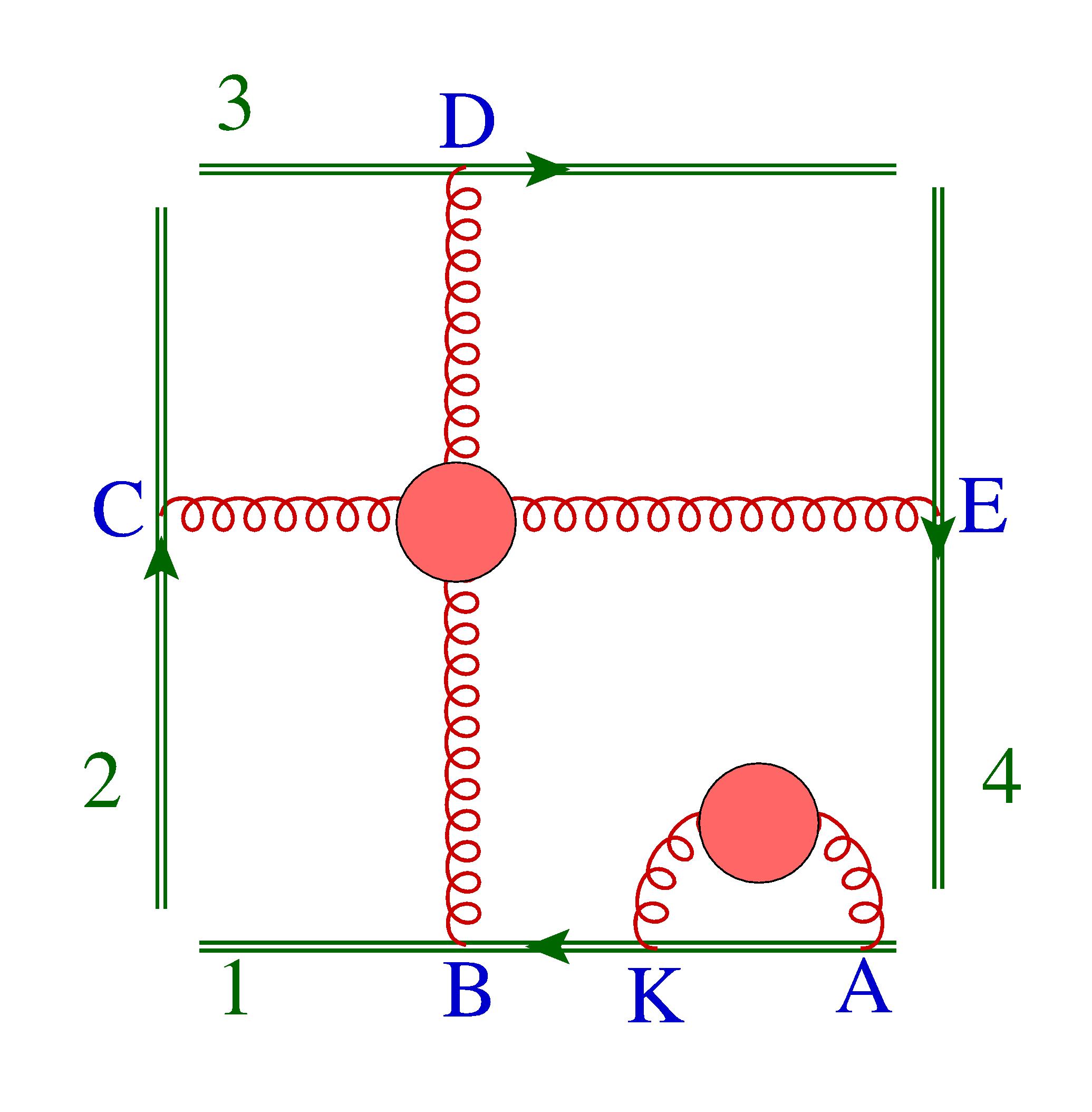} }
	\caption{A diagram of Cweb $ W^{(1,0,1)}_{4}(1,1,1,3) $}
	\label{fig:4legsWeb6}
\end{figure}

\begin{table}[H]
	\begin{center}
		\begin{tabular}{|c|c|c|}
			\hline 
			\textbf{Diagrams}  & \textbf{Sequences}  & \textbf{s-factors}  \\ 
			\hline
			$C_{1}$  & $\lbrace ABK\rbrace$  & 0 \\ \hline
			$C_{2}$  & $\lbrace AKB\rbrace$  & 1 \\ \hline
			$C_{3}$  & $\lbrace BAK\rbrace$  & 1 \\ \hline
		\end{tabular}
	\end{center}
	\caption{Normal ordered diagrams of Cweb $ W^{(1,0,1)}_{4}(1,1,1,3) $ }
	\label{tab:4legsWeb6}
\end{table}
We show only $ C_2 $ in fig.~\eqref{fig:4legsWeb6}. The number of diagrams and the dimension of the mixing matrix for this Cweb is three, thus the mixing matrix for this Cweb is given by~\cite{Agarwal:2022wyk,Agarwal:2021him},  
\begin{align}
R=\left(\begin{array}{ccc}
1 & -\frac{1}{2} & -\frac{1}{2} \\
0 & \frac{1}{2} & -\frac{1}{2} \\
0 & -\frac{1}{2} & \frac{1}{2} 
\end{array}\right )
\end{align}

\vspace{0.5cm}
\textbf{2.}\, $ \textbf{W}^{(2,1)}_{4,\text{II}}(1,1,2,3) $ \\

\begin{figure}[H]

	\centering
	\subfloat[][]{\includegraphics[height=4cm,width=4cm]{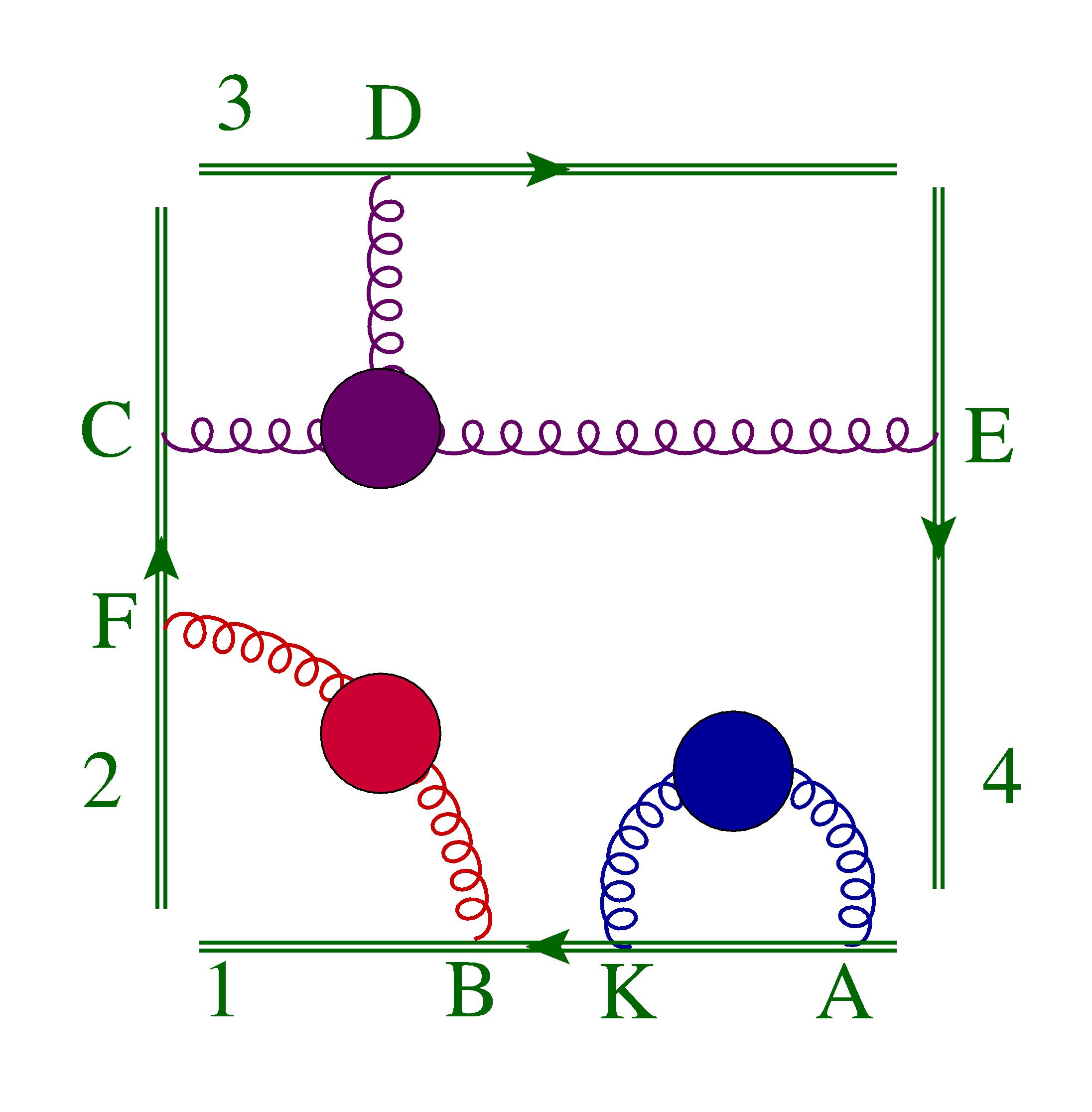} }
	\qquad
	\subfloat[][]{\includegraphics[height=4cm,width=4cm]{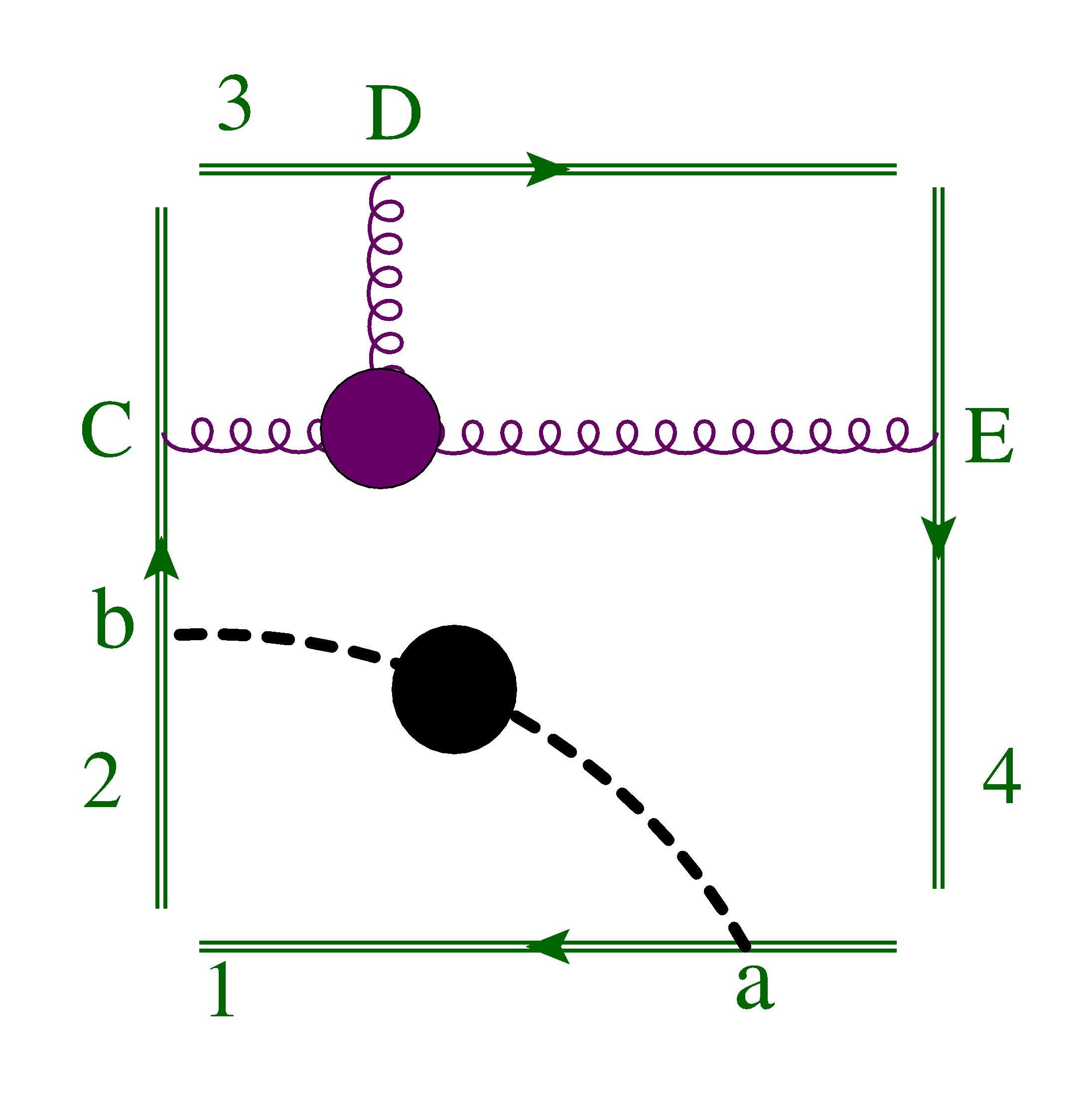} }
	\caption{Cweb $ W^{(2,1)}_{4,\text{II}}(1,1,2,3) $ and the corresponding Fused-Web}
	\label{fig:six-one-web4-8-av}
\end{figure}

This Cweb is made out of one three-point and one two-point gluon correlator along with a boomerang. It and has six diagrams, out of which we only show in fig.~(\ref{fig:six-one-web4-8-av}\txb{a}). %
The shuffle and the $ s $-factors of the Normal ordered diagrams are shown in the following table. 

\begin{table}[H]
	\begin{minipage}[c]{0.5\textwidth}
		\begin{table}[H]
			\begin{center}
				\begin{tabular}{|c|c|c|}
					\hline 
					\textbf{Diagrams}  & \textbf{Sequences}  & \textbf{s-factors}  \\ 
					\hline
					$C_{1}$  & $\lbrace\lbrace ABK\rbrace \lbrace CF \rbrace \rbrace$  & 0 \\ \hline
					$C_{2}$  & $\lbrace\lbrace ABK\rbrace \lbrace FC \rbrace \rbrace$  & 0 \\ \hline
					$C_{3}$  & $\lbrace\lbrace AKB\rbrace \lbrace FC \rbrace \rbrace$  & 1 \\ \hline
				\end{tabular}
			\end{center}
		\end{table}
	\end{minipage}
	\begin{minipage}[c]{0.5\textwidth}
		\begin{table}[H]
			\begin{center}
				\begin{tabular}{|c|c|c|}
					\hline 
					\textbf{Diagrams}  & \textbf{Sequences}  & \textbf{s-factors}  \\ 
					\hline
					$C_{4}$  & $\lbrace\lbrace BAK\rbrace \lbrace FC \rbrace \rbrace$  & 1 \\ \hline
					$C_{5}$  & $\lbrace\lbrace BAK\rbrace \lbrace CF \rbrace \rbrace$  & 2 \\ \hline
					$C_{6}$  & $\lbrace\lbrace AKB\rbrace \lbrace CF \rbrace \rbrace$  & 2 \\ \hline
				\end{tabular}
			\end{center}
		\end{table}
	\end{minipage}
	\caption{Normal ordered diagrams of Cweb $ W^{(2,1)}_{4,\text{II}}(1,1,2,3) $}
	\label{tab:six-one-web4-8-av}
\end{table}

Note that the $ s $-factor for the reducible diagrams for this Cweb is $ \{1,1,2,2\} $. Thus, uniqueness theorem and the properties of $ D $ fixes,
\begin{align}
D\;=\;R(1_2,2_2)\,,\qquad\qquad r(D)=r(R(1_2,2_2))\;=1\,.
\end{align}
To obtain the diagonal blocks of $ A $, we use the Fused-Web formalism~\cite{Agarwal:2022wyk}. This Cweb does not have any completely entangled diagrams thus its mixing matrix is free from identity. The partially entangled diagrams for this Cweb are possible when the red two-point gluon correlator is placed inside the two attachments of the boomerang correlator, which are diagrams $ C_1 $ and $ C_2 $ of  table~\ref{tab:six-one-web4-8-av}. A diagram of the \reducedWeb correspond to this type of entanglement is shown in fig.~(\ref{fig:six-one-web4-8-av}\txb{b}). The Fused-Web for this diagram has $ S=\{1_2\} $ with mixing matrix $ R(1_2) $. Thus, the matrix $ A $ and its rank are given by, 
\begin{align}
A\;=\;R(1_2)\,, \qquad r(A)=1\,.
\end{align}
The above statements are summarized in table~\ref{tab:six-one-web4-8-av-Ent}. 

\noindent The diagonal blocks of the mixing matrix for this Cweb is then given by, 
\begin{align}
R=\left(\begin{array}{ccccccc}
R\,(1_2) & & & \cdots & & \\
\vdots &  &  & R\,(1_2,2_2) & & \\
\end{array}\right )\,
\end{align}
The rank and thus the number of exponentiated colour factor for this Cweb is given by,
\begin{align}
r(R)\,=\,r\left (R(1_2)\right )+r\left (R(1_2,2_2)\right )=2
\end{align}
\begin{table}[H]
	\begin{center}
		\begin{tabular}{|c|c|c|c|c|c|}
			\hline
			Entanglement & Diagrams of  & \reducedWeb & Diagrams in  & $ s $-factors   & $ R $ \\ 
			& Cweb &  & \reducedWeb &    &  \\
			\hline
			First Partial Entangled  & $ C_{1} $, $ C_{2} $ & \ref{fig:six-one-web4-8-av}\textcolor{blue}{b} & $ \{b,\,C\}$ & 1 & $ R(1_2) $ \\ 
			&  & & $ \{C,\,b\} $ & 1 & \\ \hline
		\end{tabular}	
	\end{center}
	\caption{\reducedWebs and their mixing matrices for Cweb $ W^{(2,1)}_{4,\text{II}}(1,1,2,3) $}
	\label{tab:six-one-web4-8-av-Ent}
\end{table}

\vspace{0.5cm}
\textbf{3.}\, $ \textbf{W}^{(2,1)}_{4,\text{I}}(1,1,2,3) $\\

\begin{figure}

	\centering
	\subfloat[][]{\includegraphics[height=4cm,width=4cm]{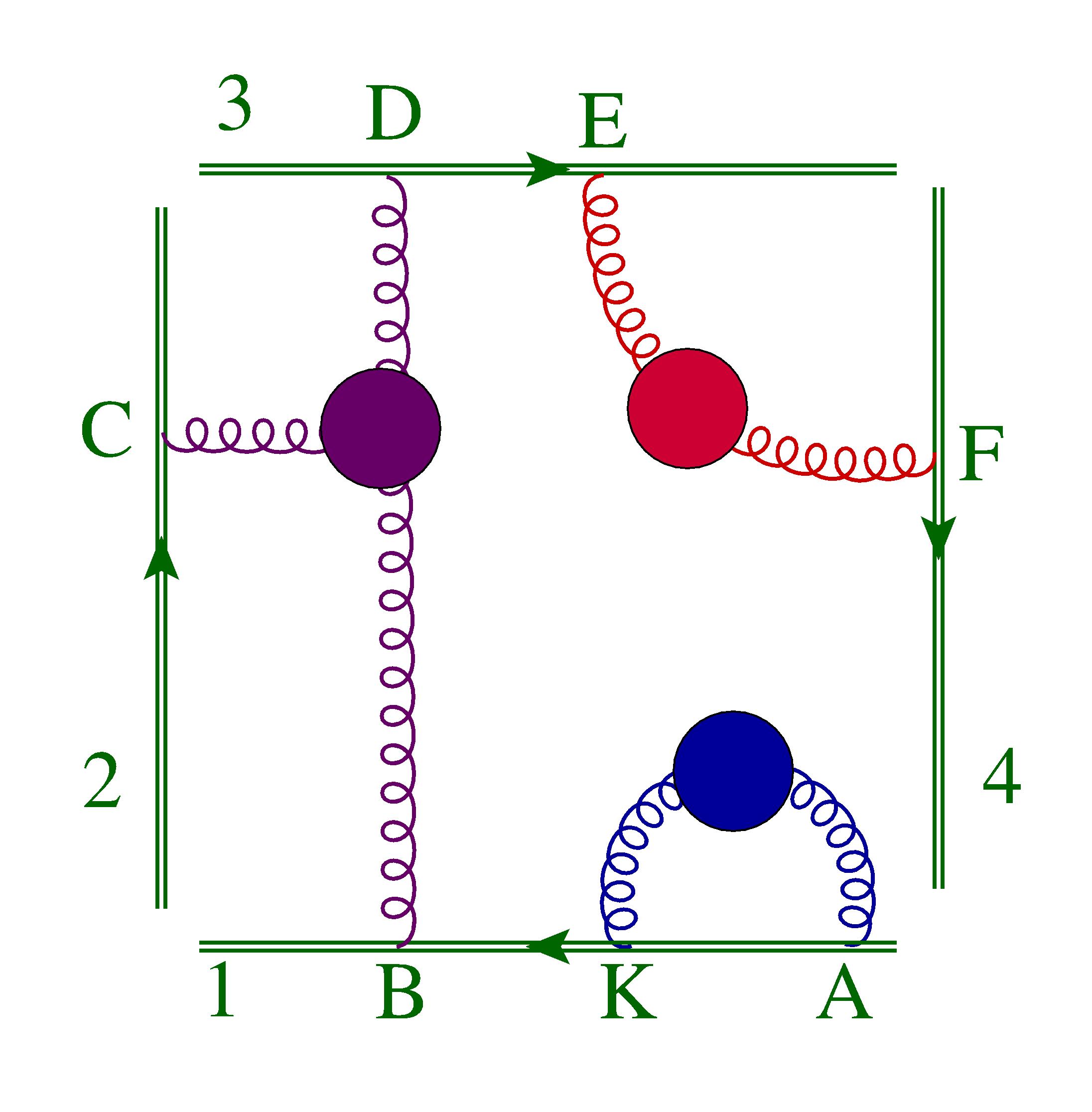} }
	\qquad
	\subfloat[][]{\includegraphics[height=4cm,width=4cm]{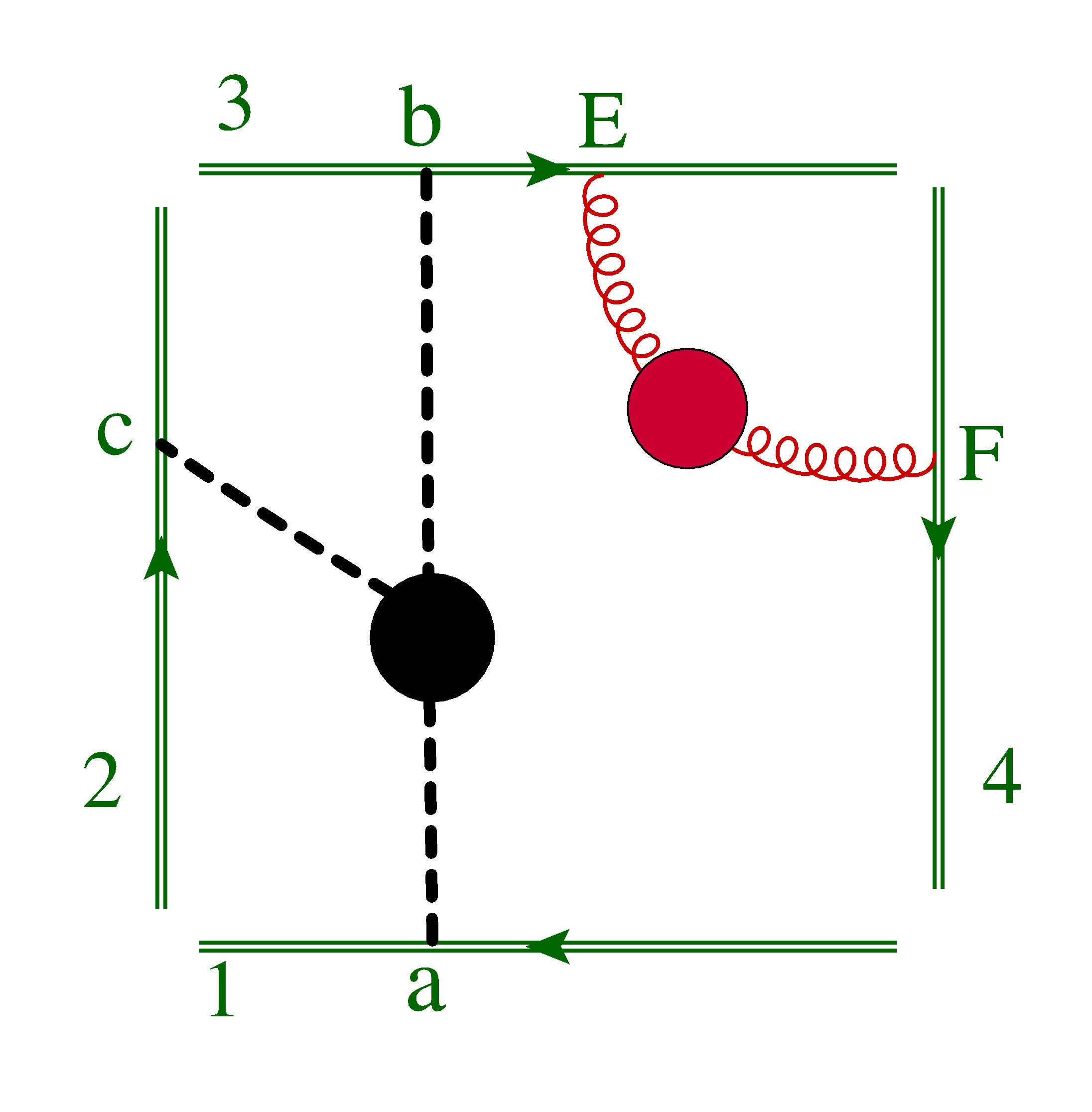} }
	\caption{Cweb $ W^{(2,1)}_{4,\text{I}}(1,1,2,3) $ and its Fused-Web}
	\label{fig:six-one-web4-4-av}
\end{figure}

This Cweb has six diagrams and one of them is shown in fig.~(\ref{fig:six-one-web4-4-av}\txb{a}). The normal ordered diagrams and their corresponding $ s $-factors are shown in the following table. 
\begin{table}[H]
	\begin{minipage}[c]{0.5\textwidth}
		\begin{table}[H]
			\begin{center}
				\begin{tabular}{|c|c|c|}
					\hline 
					\textbf{Diagrams}  & \textbf{Sequences}  & \textbf{s-factors}  \\ 
					\hline
					$C_{1}$  & $\lbrace\lbrace ABK\rbrace, \lbrace DE\rbrace\rbrace$  & 0 \\ \hline
					$C_{2}$  & $\lbrace\lbrace ABK\rbrace, \lbrace ED\rbrace\rbrace$  & 0 \\ \hline
					$C_{3}$  & $\lbrace\lbrace AKB\rbrace, \lbrace DE\rbrace\rbrace$  & 1 \\ \hline
				\end{tabular}
			\end{center}
		\end{table}
	\end{minipage}
	\hspace{1cm}
	\begin{minipage}[c]{0.5\textwidth}
		\begin{table}[H]
			\begin{center}
				\begin{tabular}{|c|c|c|}
					\hline 
					\textbf{Diagrams}  & \textbf{Sequences}  & \textbf{s-factors}  \\ 
					\hline
					$C_{4}$  & $\lbrace\lbrace BAK\rbrace, \lbrace DE\rbrace\rbrace$  & 1 \\ \hline
					$C_{5}$  & $\lbrace\lbrace BAK\rbrace, \lbrace ED\rbrace\rbrace$  & 2 \\ \hline
					$C_{6}$  & $\lbrace\lbrace AKB\rbrace, \lbrace ED\rbrace\rbrace$  & 2 \\ \hline
				\end{tabular}
			\end{center}
		\end{table}
	\end{minipage}
	\caption{Normal ordered diagrams of Cweb $ W^{(2,1)}_{4,\text{I}}(1,1,2,3) $}
	\label{tab:six-one-web4-4-av}
\end{table}
This Cweb has two irreducible diagrams and four reducible diagrams, further the reducible diagrams have $ s $-factors $ \{1_2,2_2\} $. Thus,
using the Uniqueness theorem and the fact that the block $ D $ is itself a mixing matrix, we get $ D $ and rank of $ D $ as,  
\begin{align}
D\;=\;R(1_2,2_2)\,,\qquad\qquad r(D)=r(R(1_2,2_2))\;=1\;
\end{align}
This Cweb does not contain any partially entangled diagrams thus the mixing matrix does not have any identity. The only way that the diagrams of this Cweb are entangled is when the correlators on line $ 1 $ are entangled. Applying the Fused-Web formalism~\cite{Agarwal:2022wyk}, we get the Fused diagram, shown in fig.~(\ref{fig:six-one-web4-4-av}\txb{b}) for this type of entanglement. The shuffle of attachments of this Fused diagram generate a Fused-Web with  $ S=\{1_2\} $ and thus it has the mixing matrix $ R(1_2) $. 

Note that, there is only one type of entanglement is possible for this Cweb, thus the block $ A $ is given by,
\begin{align}
A\,=\,R(1_2)\,.
\end{align}
The above statements are summarized in table~\ref{tab:six-one-web4-4-av-Ent}. The diagonal blocks of the mixing matrix is then given by, 
\begin{align}
R=\left (\begin{array}{cccc}
R(1_2) & &\cdots & \\
\vdots & & R(1_2,2_2)
\end{array}\right )
\end{align}
Thus, the rank of the mixing matrix for this Cweb is given by, 
\begin{align}
r(R)\,=\,r\left (R(1_2)\right )+r\left (R(1_2,2_2)\right )\,=\,2\,.
\end{align}
\begin{table}[H]
	\begin{center}
		\begin{tabular}{|c|c|c|c|c|c|}
			\hline
			Entanglement & Diagrams of  & \reducedWeb & Diagrams in  & $ s $-factors   & $ R $ \\ 
			& Cweb &  & \reducedWeb &    &  \\
			\hline
			First Partial Entangled  & $ C_{1} $, $ C_{2} $ & \ref{fig:six-one-web4-4-av}\textcolor{blue}{b} & $ \{b,\,E\} $ & 1 & $ R(1_2) $ \\ 
			&  & & $ \{E\,b\} $ & 1 & \\ \hline
		\end{tabular}	
	\end{center}
	\caption{\reducedWebs and their mixing matrices for Cweb $ W^{(2,1)}_{4,\text{I}}(1,1,2,3) $}
	\label{tab:six-one-web4-4-av-Ent}
\end{table}

\vspace{0.5cm}
\textbf{4.}\, $ \textbf{W}^{(2,1)}_{4,\text{I}}(1,1,1,4) $\\

\begin{figure}

	\centering
	\subfloat[][]{\includegraphics[height=4cm,width=4cm]{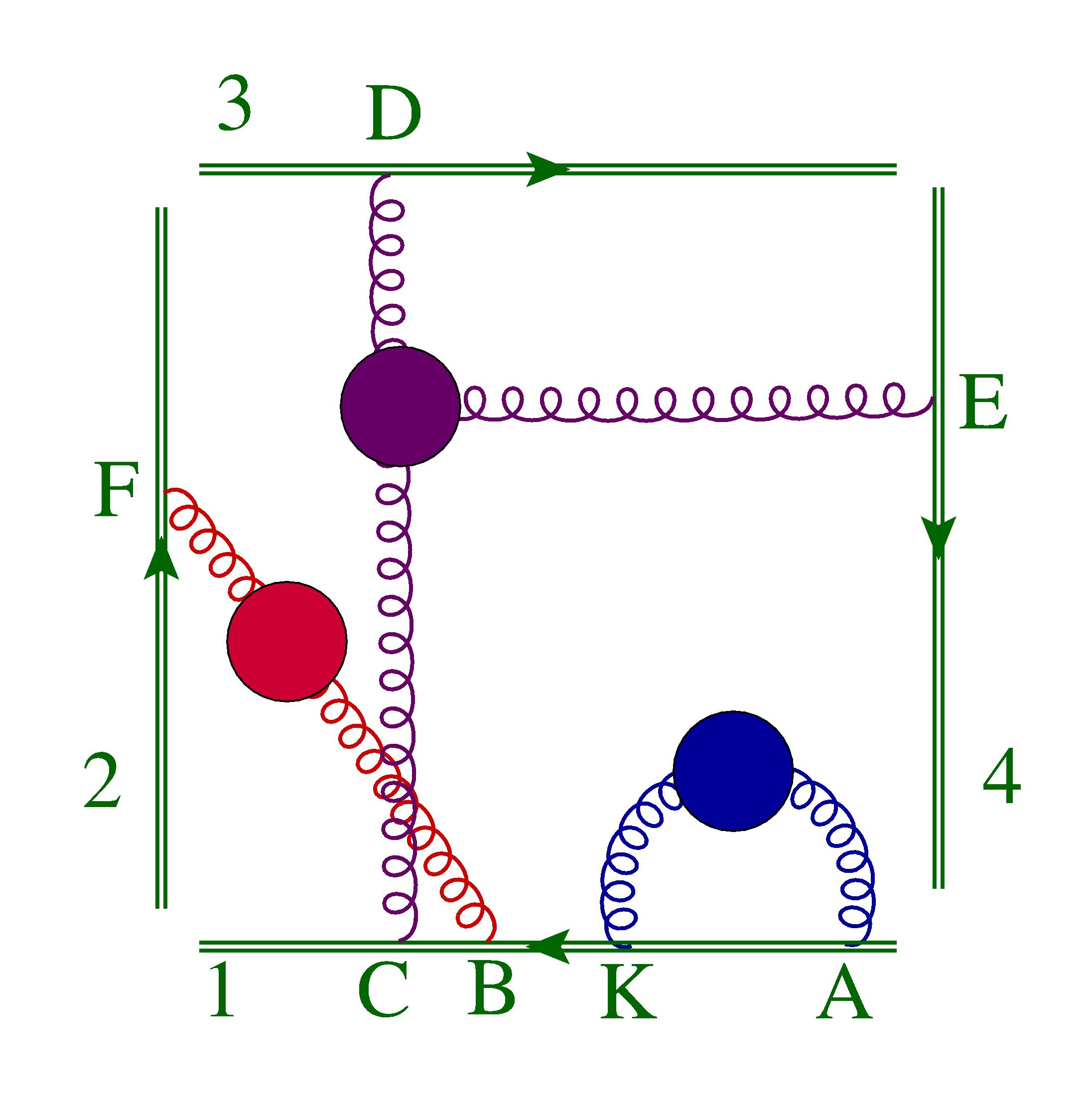} }
	\caption{Cweb $ W^{(2,1)}_{4,\text{I}}(1,1,1,4) $}
	\label{fig:six-one-web4-5-av}
\end{figure}

\vspace{0.2cm}
\noindent A diagram of this Cweb is shown in fig. (\ref{fig:six-one-web4-5-av}). It has twelve diagrams, out of which six are reducible, two are completely entangled, and other four are partially entangled. The Normal ordered diagrams and their $ s $-factors are shown in the following table.

\begin{table}[H]
	\begin{minipage}[c]{0.5\textwidth}
		\begin{table}[H]
			\begin{center}
				\begin{tabular}{|c|c|c|}
					\hline 
					\textbf{Diagrams}  & \textbf{Sequences}  & \textbf{s-factors}  \\ 
					\hline
					$C_{1}$  & $\lbrace\lbrace ABCK\rbrace$  & 0 \\ \hline 
					$C_{2}$  & $\lbrace\lbrace ACBK\rbrace$  & 0 \\ \hline
					$C_{3}$  & $\lbrace\lbrace ABKC\rbrace$  & 0 \\ \hline 
					$C_{4}$  & $\lbrace\lbrace CABK\rbrace$  & 0 \\ \hline
					$C_{5}$  & $\lbrace\lbrace ACKB\rbrace$  & 0 \\ \hline 
					$C_{6}$  & $\lbrace\lbrace BACK\rbrace$  & 0 \\ \hline 
				\end{tabular}
			\end{center}
		\end{table}
	\end{minipage}
	\begin{minipage}[c]{0.5\textwidth}
		\begin{table}[H]
			\begin{center}
				\begin{tabular}{|c|c|c|}
					\hline 
					\textbf{Diagrams}  & \textbf{Sequences}  & \textbf{s-factors}  \\ 
					\hline
					$C_{7}$  & $\lbrace\lbrace AKBC\rbrace$  & 1\\ \hline
					$C_{8}$  & $\lbrace\lbrace AKCB\rbrace$  & 1 \\ \hline 
					$C_{9}$  & $\lbrace\lbrace BAKC\rbrace$  & 1 \\ \hline
					$C_{10}$  & $\lbrace\lbrace BCAK\rbrace$  & 1 \\ \hline
					$C_{11}$  & $\lbrace\lbrace CAKB\rbrace$  & 1 \\ \hline 
					$C_{12}$  & $\lbrace\lbrace CBAK\rbrace$  & 1 \\ \hline   
				\end{tabular}
			\end{center}
		\end{table}
	\end{minipage}
	\caption{Normal ordered diagrams of Cweb $ W^{(2,1)}_{4,\text{I}}(1,1,1,4) $}
	\label{tab:six-one-web4-5-av}
\end{table}
Note that the reducible diagrams of this Cweb has $ S=\{1_6\} $.
The fact that $ D $ follows the properties of mixing matrices and the uniqueness theorem fixes $ D $ and its rank as,
\begin{align}
D\;=\;R(1_{6})\,,\qquad\qquad r(D)=r(R(1_{6}))\;=2\;
\end{align}
\begin{figure}[b]
	\captionsetup[subfloat]{labelformat=empty}
	\centering
	\subfloat[][(a)]{\includegraphics[height=4cm,width=4cm]{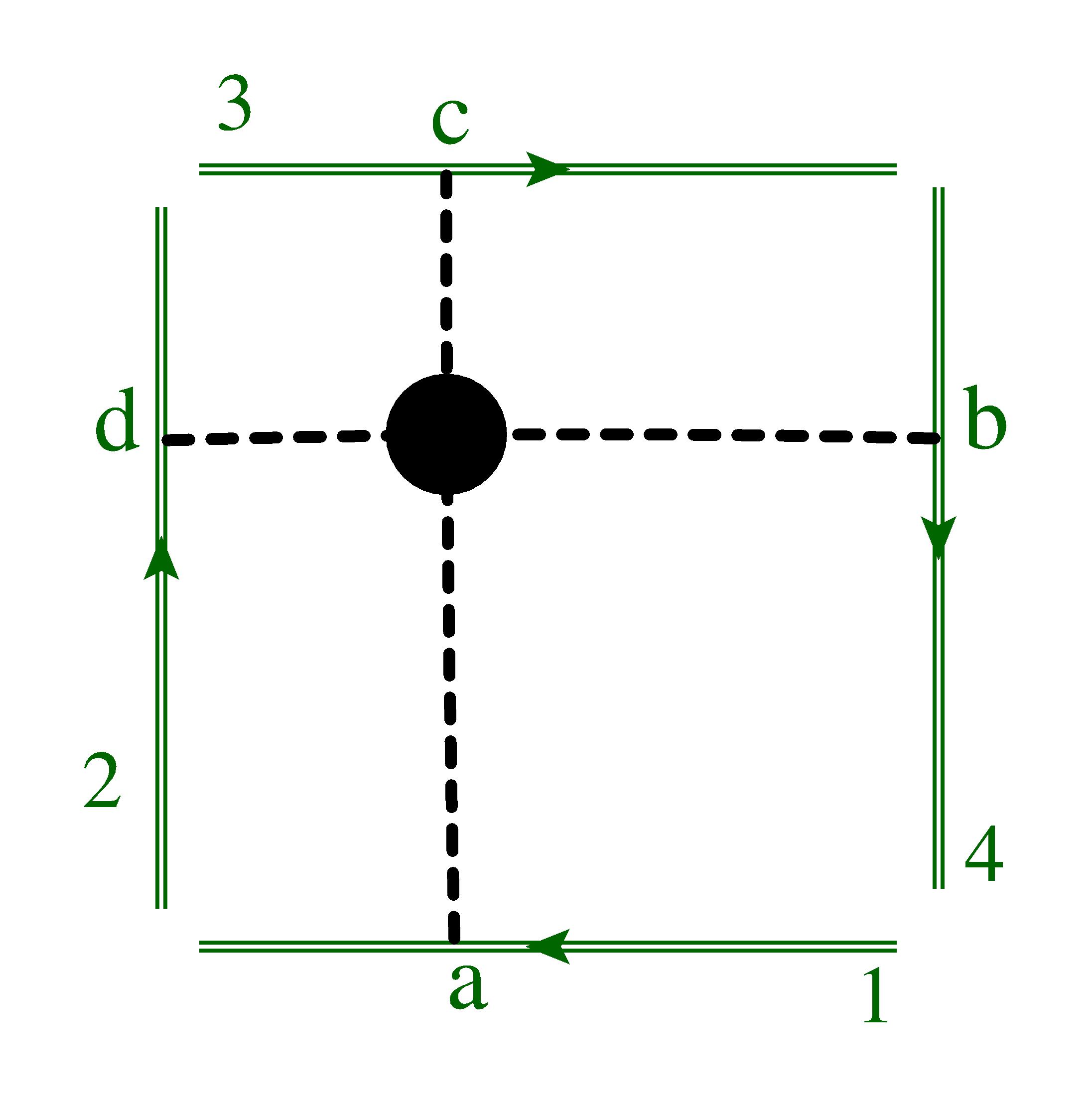} }
	\qquad 
	\subfloat[][(b)]{\includegraphics[height=4cm,width=4cm]{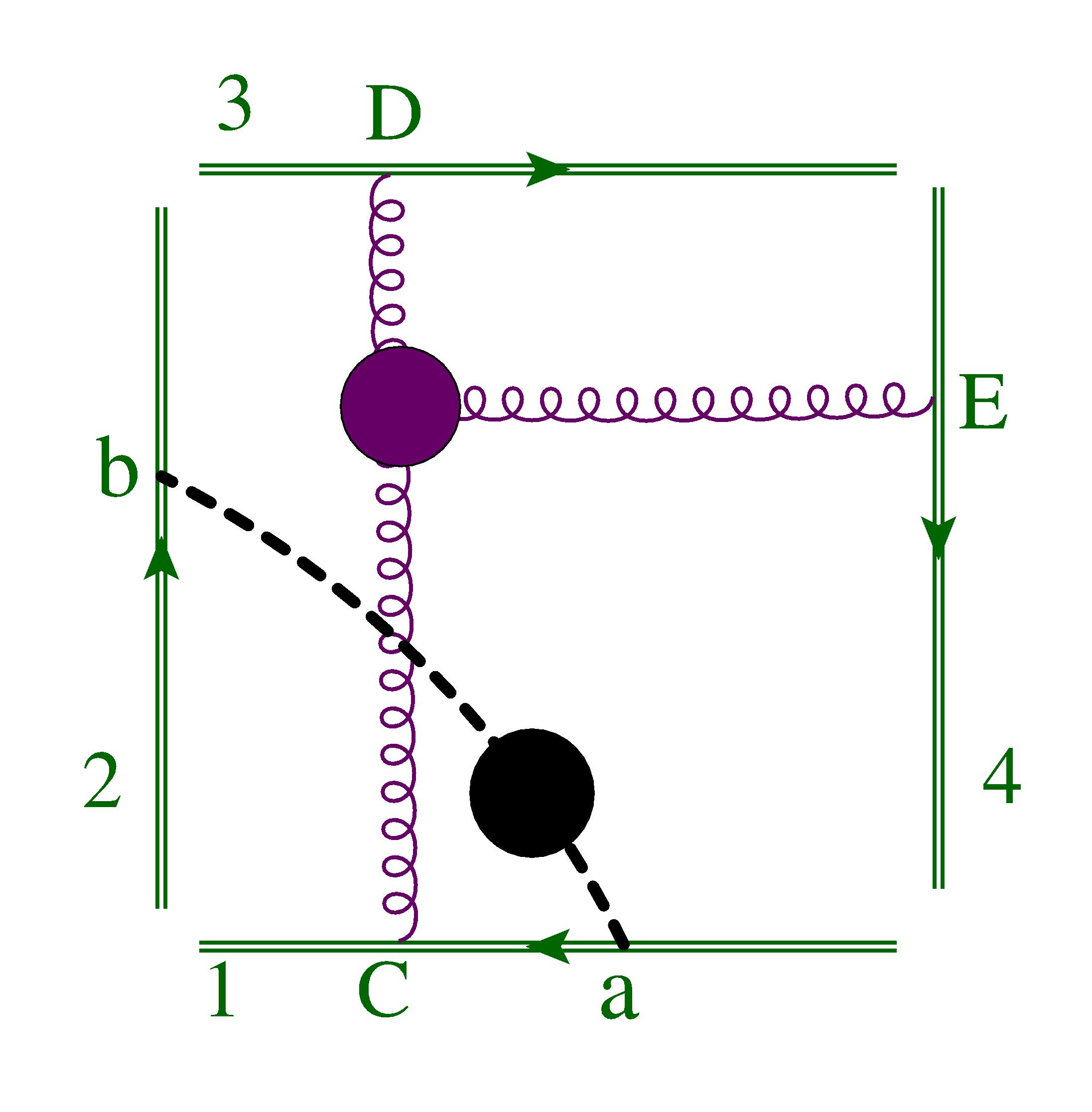} }
	\qquad 
	\subfloat[][(c)]{\includegraphics[height=4cm,width=4cm]{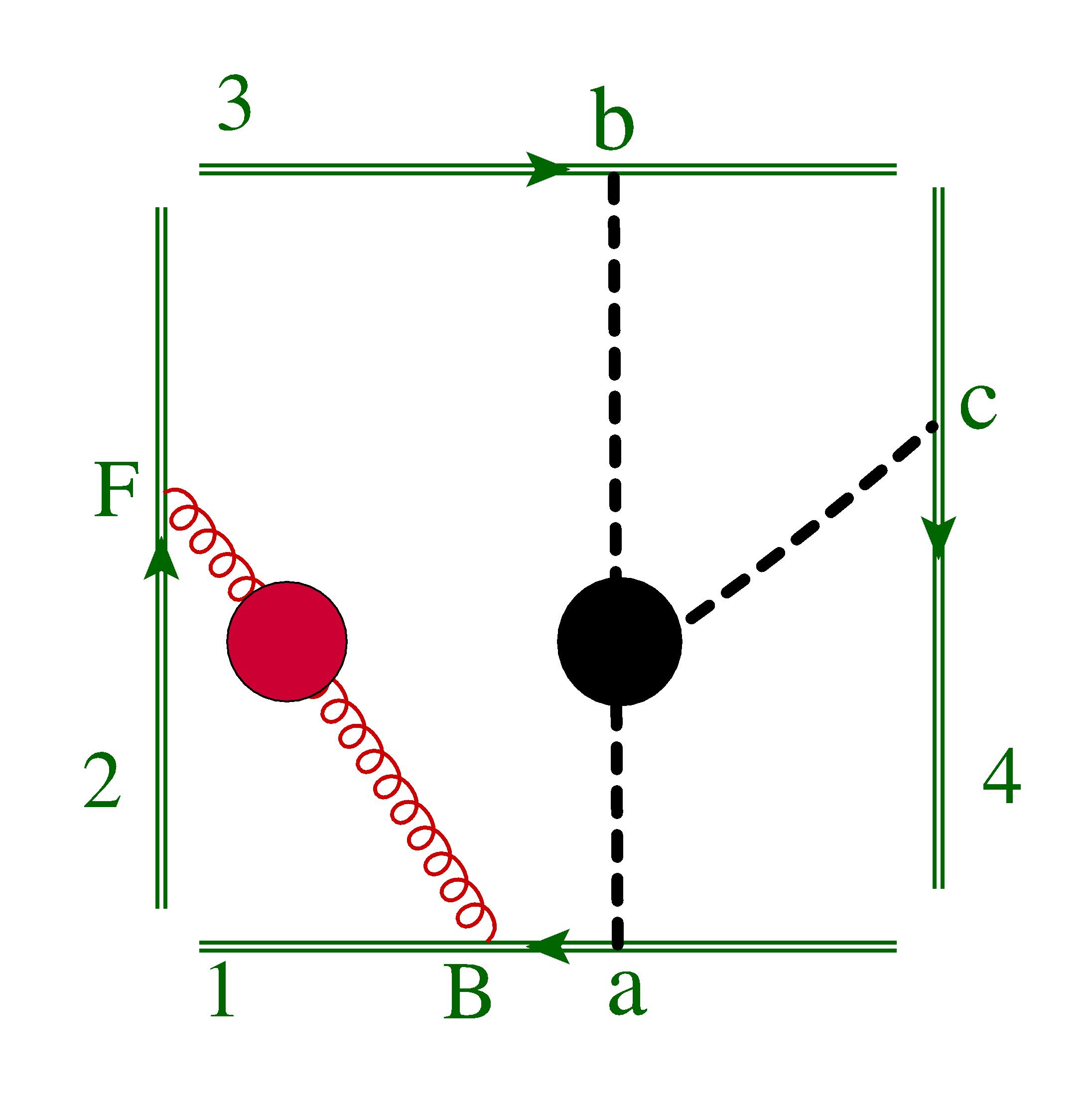} }
	\caption{\reducedWebs for Cweb $ W^{(2,1)}_{4,\text{I}}(1,1,1,4) $ }
	\label{fig:six-one-web4-5-allAVAtar-WEBS}
\end{figure}
The diagrams of this Cweb are completely entangled when all three correlators on line $ 1 $ are entangled. These are $ C_1 $ and $ C_2 $ in table~\ref{tab:six-one-web4-5-av} and corresponds to two identities in the mixing matrix for this Cweb. The Fused diagram is for $ C_1 $ and $ C_2 $ is shown in fig.~(\ref{fig:six-one-web4-5-allAVAtar-WEBS}\txb{a}). 

There are two different ways in which two of the three correlators on Wilson line $ 1 $ are entangled. The Fused diagrams for these two different types of entanglements are shown in fig.~(\ref{fig:six-one-web4-5-allAVAtar-WEBS}\txb{b}) and fig.~(\ref{fig:six-one-web4-5-allAVAtar-WEBS}\txb{c}). Each of these forms a Fused-Web with $ S=\{1_2\} $ and mixing matrix $ R(1_2) $. The table~\ref{tab:six-one-web4-5-av-Ent} summarizes the above statements.   

The order of diagrams in the Cweb given in table \ref{tab:six-one-web4-5-av} is chosen in a way such that diagrams with same kind of entanglement appear together. Then the diagonal blocks of matrix $ A $ becomes,
\begin{align}
A\,=\,\left(\begin{array}{c|cc}
\textbf{I}_{2} & & \cdots\\
\hline 
\textbf{O}_{4\times 2}& & \begin{array}{cc}
R\,(1_2) & \\
& R\,(1_2)
\end{array}	
\end{array}\right)\,,\qquad 
\end{align}
Further, the diagonal blocks of $ R $ is given by, 
\begin{align}
R\,=\,\left(\begin{array}{cccc}
\textbf{I}_{2} & & \cdots &\\
& & \begin{array}{cc}
R\,(1_2) & \\
& R\,(1_2)
\end{array} \\
& & & R(1_6)	
\end{array}\right)\,,\qquad 
\end{align}
The rank of the mixing matrix for this Cweb is given by 
\begin{align}
r(R)\;=\;r(1_2)+2r(R(1_2))+r(R(1_6))\;=\;6\,.
\end{align}

\begin{table}[H]
	\begin{center}
		\begin{tabular}{|c|c|c|c|c|c|}
			\hline
			Entanglement & Diagrams of  & \reducedWeb & Diagrams in  & $ s $-factors   & $ R $ \\ 
			& Cweb &  & \reducedWeb &    &  \\
			\hline
			Complete entangled  & $ C_{1} $, $ C_{2} $  & \ref{fig:six-one-web4-5-allAVAtar-WEBS}\textcolor{blue}{a} & - & 1 & $ I_{2} $ \\ 
			& & & & &\\
			\hline
			First Partial Entangled  & $ C_{3} $, $ C_{4} $ & \ref{fig:six-one-web4-5-allAVAtar-WEBS}\textcolor{blue}{b} & $ \{a,\,C\} $ & 1 & $ R(1_2) $ \\ 
			&  & & $ \{C\,a\} $ & 1 & \\ \hline
			Second Partial Entangled  & $ C_{5} $, $ C_{6} $ & \ref{fig:six-one-web4-5-allAVAtar-WEBS}\textcolor{blue}{c} & $ \{a\,B\} $ & 1 & $ R(1_2) $ \\ 
			&  & & $ \{B\,a\} $ & 1 & \\ \hline
		\end{tabular}	
	\end{center}
	\caption{\reducedWebs and their mixing matrices for Cweb $ W^{(2,1)}_{4,\text{I}}(1,1,1,4) $}
	\label{tab:six-one-web4-5-av-Ent}
\end{table}
\vspace{0.5cm}
\textbf{5.}\,$\textbf{W}\,_{4}^{(4)}(1,1,3,3)$ \\

\begin{figure}[H]

	\centering
	\subfloat[][]{\includegraphics[height=4cm,width=4cm]{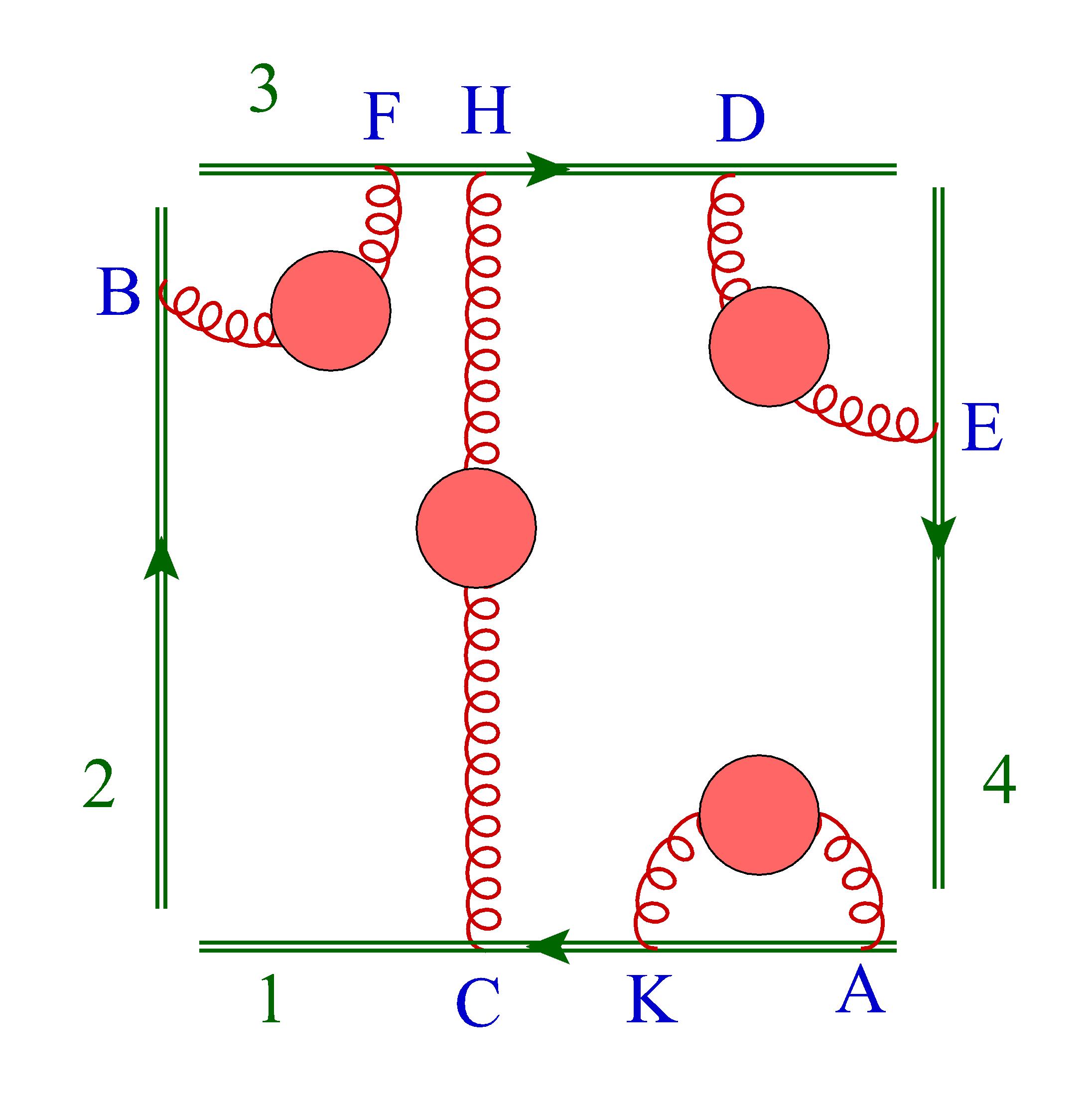} }
	\qquad
	\subfloat[][]{\includegraphics[height=4cm,width=4cm]{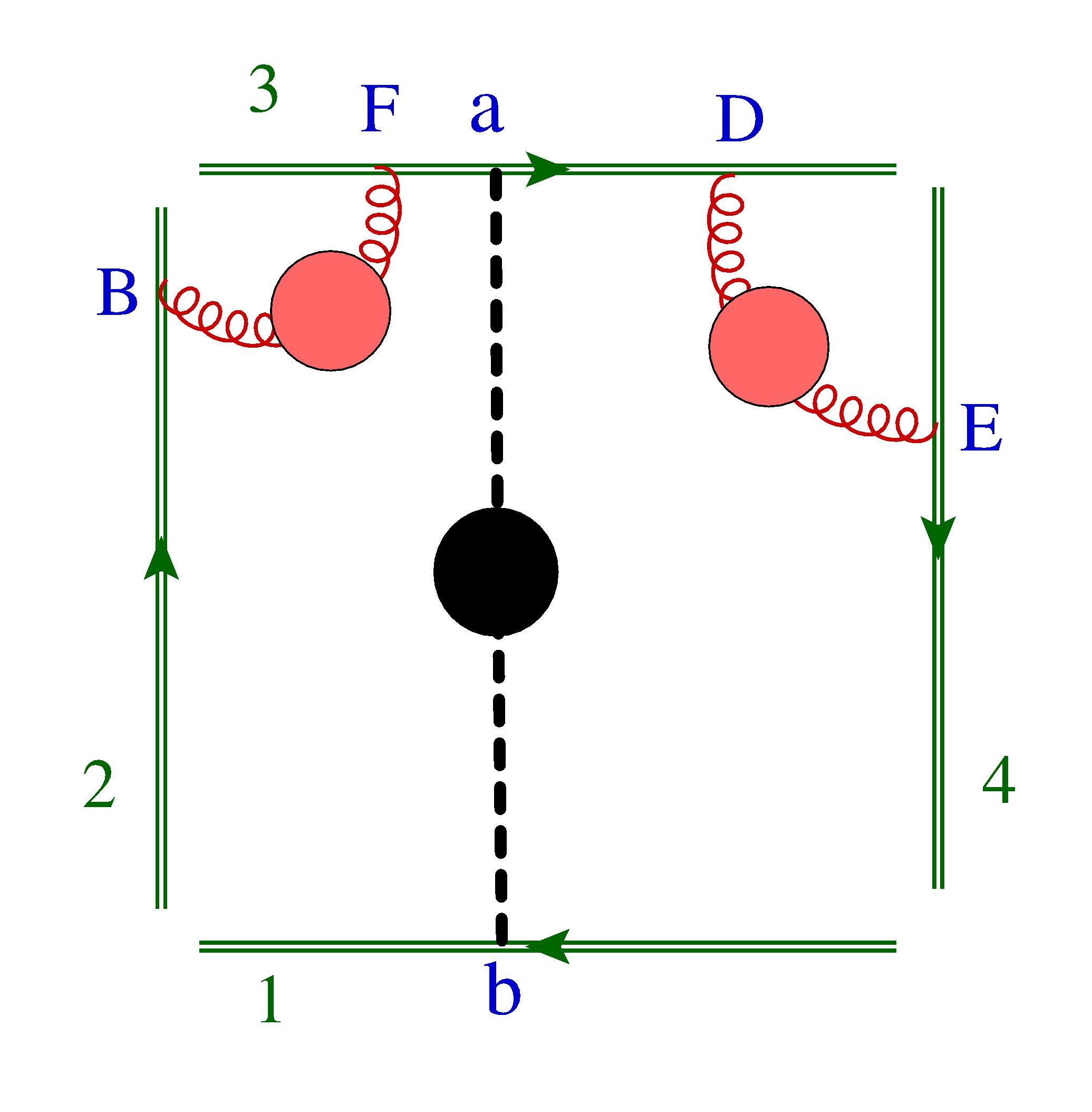} }
	\caption{Cweb $\text{W}\,_{4}^{(4)}(1,1,3,3)$ and its Fused-Web}
	\label{fig:4legsWeb1}
\end{figure}

\vspace{0.2cm}
\noindent This Cweb, shown in fig.~(\ref{fig:4legsWeb1}\txb{a}) has eighteen diagrams, out of which twelve are reducible and remaining six are partially entangled. The Normal ordered diagrams and their $ s $-factors are shown in table \ref{tab:4legsWeb1}.

\begin{table}[H]
	\begin{minipage}[c]{0.5\textwidth}
		\begin{table}[H]
			\begin{center}
				\begin{tabular}{|c|c|c|}
					\hline 
					\textbf{Diagrams}  & \textbf{Sequences}  & \textbf{s-factors}  \\ 
					\hline
					$C_{1}$  & $\lbrace\lbrace ACK\rbrace, \lbrace FHD\rbrace\rbrace$  & 0 \\ \hline
					$C_{2}$  & $\lbrace\lbrace ACK\rbrace, \lbrace FDH \rbrace\rbrace$  & 0 \\ \hline 
					$C_{3}$  & $\lbrace\lbrace ACK\rbrace, \lbrace HFD\rbrace\rbrace$  & 0 \\ \hline
					$C_{4}$  & $\lbrace\lbrace ACK\rbrace, \lbrace HDF\rbrace\rbrace$  & 0 \\ \hline
					$C_{5}$  & $\lbrace\lbrace ACK\rbrace, \lbrace DFH\rbrace\rbrace$  & 0 \\ \hline
					$C_{6}$  & $\lbrace\lbrace ACK\rbrace, \lbrace DHF\rbrace\rbrace$  & 0 \\ \hline 
					$C_{7}$  & $\lbrace\lbrace AKC\rbrace, \lbrace HFD\rbrace\rbrace$  &  1\\ \hline
					$C_{8}$  & $\lbrace\lbrace AKC\rbrace, \lbrace HDF\rbrace\rbrace$  &  1 \\ \hline
					$C_{9}$  & $\lbrace\lbrace CAK\rbrace, \lbrace FDH\rbrace\rbrace$  &  1\\ \hline
				\end{tabular}
			\end{center}
		\end{table}
	\end{minipage}
	\hspace{1cm}
	\begin{minipage}[c]{0.5\textwidth}
		\begin{table}[H]
			\begin{center}
				\begin{tabular}{|c|c|c|}
					\hline 
					\textbf{Diagrams}  & \textbf{Sequences}  & \textbf{s-factors}  \\ 
					\hline
					$C_{10}$  & $\lbrace\lbrace CAK\rbrace, \lbrace DFH\rbrace\rbrace$  & 1 \\ \hline
					$C_{11}$  & $\lbrace\lbrace AKC\rbrace, \lbrace FHD\rbrace\rbrace$  & 2 \\ \hline
					$C_{12}$  & $\lbrace\lbrace AKC\rbrace, \lbrace FDH\rbrace\rbrace$  & 2 \\ \hline
					$C_{13}$  & $\lbrace\lbrace AKC\rbrace, \lbrace DFH\rbrace\rbrace$  & 2 \\ \hline 
					$C_{14}$  & $\lbrace\lbrace AKC\rbrace, \lbrace DHF\rbrace\rbrace$  & 2 \\ \hline 
					$C_{15}$  & $\lbrace\lbrace CAK\rbrace, \lbrace FHD\rbrace\rbrace$  & 2 \\ \hline 
					$C_{16}$  & $\lbrace\lbrace CAK\rbrace, \lbrace HFD\rbrace\rbrace$  & 2 \\ \hline 
					$C_{17}$  & $\lbrace\lbrace CAK\rbrace, \lbrace HDF\rbrace\rbrace$  & 2 \\ \hline 
					$C_{18}$  & $\lbrace\lbrace CAK\rbrace, \lbrace DHF\rbrace\rbrace$  & 2 \\ \hline  			
				\end{tabular}
			\end{center}
		\end{table}
	\end{minipage}
	\caption{Normal ordered diagrams of Cweb $\text{W}\,_{4}^{(4)}(1,1,3,3)$}
	\label{tab:4legsWeb1}
\end{table}

For this case, the $ s $-factors of reducible diagrams forms a column weight vector $ S=\{1_4,2_8\} $ which gives mixing matrix $ R(1_4,2_8) $ using Uniqueness theorem. Thus the block $ D $ of this matrix is,
\begin{align}
D = R(1_4,2_8)\qquad; \qquad r(D)\;=\;r(R(1_4,2_8))\;=\;2\,,
\end{align}
Now we use procedure of Fused-Webs for the determination of diagonal blocks of $ A $. 
\noindent In this Cweb there is no completely entangled diagram. Therefore the identity matrix in $ A $ does not exist.

There is only one kind of entanglement --- involving boomerang and two point correlator on line $ 1 $ --- that appears in partially entangled diagrams. There are six such diagrams, $ C_1, C_2, C_3, C_4 , C_5 $ and $ C_6 $, whose order of attachments are given in table~\ref{tab:4legsWeb1}. The shuffle of attachments of Fused diagram, shown in fig.~(\ref{fig:4legsWeb1}\txb{b}) generates a Fused-Web with column weight vector $ S=\{1_6\} $ and the mixing matrix is $ R(1_6) $, as given in table~\ref{tab:4legsWeb1-Ent}.  
\begin{table}[H]
	\begin{center}
		\begin{tabular}{|c|c|c|c|c|c|}
			\hline
			Entanglement & Diagrams of  & \reducedWeb & Diagrams in  & $ s $-factors   & $ R $ \\ 
			& Cweb &  & \reducedWeb &    &  \\
			\hline 
			Partial Entangled   & $ C_1 $, $ C_2 $,  & \ref{fig:4legsWeb1}\textcolor{blue}{b}& $ \{F,\,a,\,D\} $ & 1 &  \\ 
			& $ C_{3} $, $ C_{4} $ &                                                     & $ \{F,\,D,\,a\} $ & 1 &  \\ 
			& $ C_{5} $, $ C_{6} $ &                                                     & $ \{a,\,F,\,D\} $ & 1 & \\ 
			&  &                                                                         & $ \{a,\,D,\,F\} $ & 1 & $ R(1_6) $ \\ 
			&&                                                                                                    & $ \{D,\,a,\,F\} $ & 1 & \\ 
			&&                                                                                                    & $  \{D,\,F,\,a\}$ & 1 &  \\ \hline
		\end{tabular}	
	\end{center}
	\caption{\reducedWebs and their mixing matrices for Cweb $\text{W}\,_{4}^{(4)}(1,1,3,3)$}
	\label{tab:4legsWeb1-Ent}
\end{table}
\noindent The order of diagrams in the Cweb given in table \ref{tab:4legsWeb1}, is chosen such that all the partially entangled diagrams appear together. Therefore, mixing matrices of Fused-Web is the block $ A $ itself, 
\begin{align}
A\,= R(1_6)
\end{align}
Having obtained the block $ A $, the diagonal blocks of $ R $ are,
\begin{align}
R\,=\,\left( \begin{array}{cc}
R\,(1_6) & \cdots\\
\cdots& R\,(1_4,2_8)
\end{array}\right)\,,\qquad 
\end{align}
the rank of $ R $ is the number of exponentiated colour factors which is given as
\begin{align}
r(R) = r(A)  + r(D)\;=\;2+2  = 4
\end{align}

\vspace{0.5cm}
\textbf{6.}\,$\textbf{W}\,_{4}^{(4)}(1,1,2,4)$ \\

\begin{figure}[H]
	
	\centering
	\subfloat[][]{\includegraphics[height=4cm,width=4cm]{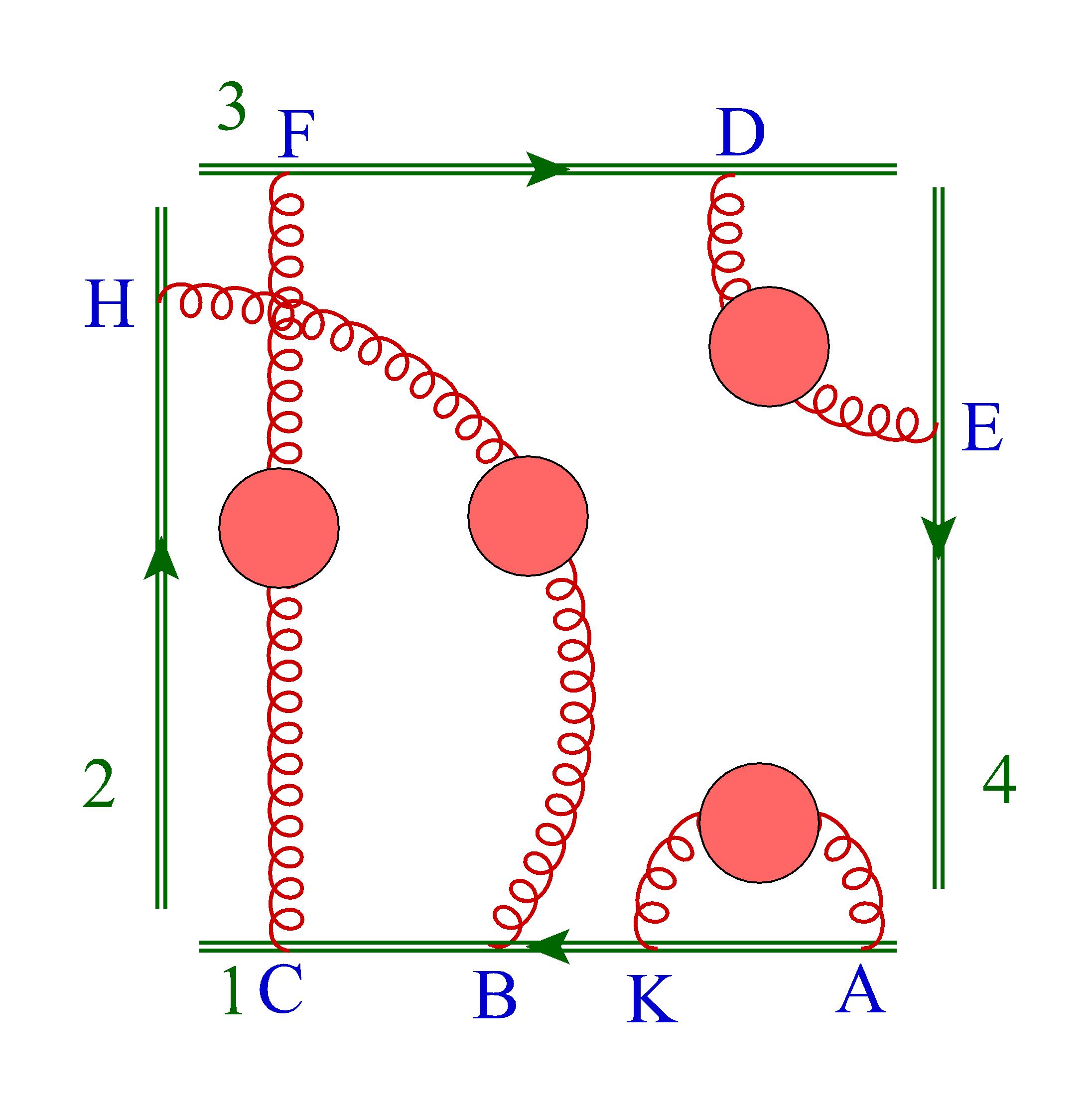} }
	\caption{Cweb $\text{W}\,_{4}^{(4)}(1,1,2,4)$}
	\label{fig:4legsWeb2}
\end{figure}

\vspace{0.2cm}
\noindent This Cweb, shown in fig.~(\ref{fig:4legsWeb2}) has twenty four diagrams, out of which twelve are reducible and remaining twelve are partially entangled. The Normal ordered diagrams and their $ s $-factors are shown in table \ref{tab:4legsWeb2}.

\begin{table}[H]
	\begin{minipage}[c]{0.5\textwidth}
		\begin{table}[H]
			\begin{center}
				\begin{tabular}{|c|c|c|}
					\hline 
					\textbf{Diagrams}  & \textbf{Sequences}  & \textbf{s-factors}  \\ 
					\hline
					$C_{1}$&$\{\{ABCK\},\{ FD\}\}$&0\\\hline
					
					$C_{2}$&$\{\{ABCK\},\{ DF\}\}$&0\\\hline
					
					$C_{3}$&$\{\{ACBK\},\{ FD\}\}$&0\\\hline
					
					$C_{4}$&$\{\{ACBK\},\{ DF\}\}$&0\\\hline
					
					$C_{5}$&$\{\{ABKC\},\{ FD\}\}$&0\\\hline
					
					$C_{6}$&$\{\{ABKC\},\{ DF\}\}$&0\\\hline
					
					$C_{7}$&$\{\{CABK\},\{ FD\}\}$&0\\\hline
					
					$C_{8}$&$\{\{CABK\},\{ DF\}\}$&0\\\hline
					
					$C_{9}$&$\{\{ACKB\},\{ FD\}\}$&0\\\hline
					
					$C_{10}$&$\{\{ACKB\},\{ DF\}\}$&0\\\hline
					
					$C_{11}$&$\{\{BACK\},\{ FD\}\}$&0\\\hline
					
					$C_{12}$&$\{\{BACK\},\{ DF\}\}$&0\\\hline
				\end{tabular}
			\end{center}
		\end{table}
	\end{minipage}
	\hspace{1cm}
	\begin{minipage}[c]{0.5\textwidth}
		\begin{table}[H]
			\begin{center}
				\begin{tabular}{|c|c|c|}
					\hline 
					\textbf{Diagrams}  & \textbf{Sequences}  & \textbf{s-factors}  \\ 
					\hline
					$C_{13}$&$\{\{AKBC\},\{ FD\}\}$&1\\\hline
					
					$C_{14}$&$\{\{BAKC\},\{ FD\}\}$&1\\\hline
					
					$C_{15}$&$\{\{CAKB\},\{ DF\}\}$&1\\\hline
					
					$C_{16}$&$\{\{CBAK\},\{ DF\}\}$&1\\\hline
					
					$C_{17}$&$\{\{AKBC\},\{ DF\}\}$&2\\\hline
					
					$C_{18}$&$\{\{AKCB\},\{ FD\}\}$&2\\\hline
					
					$C_{19}$&$\{\{AKCB\},\{ DF\}\}$&2\\\hline
					
					$C_{20}$&$\{\{BAKC\},\{ DF\}\}$&2\\\hline
					
					$C_{21}$&$\{\{BCAK\},\{ FD\}\}$&2\\\hline
					
					$C_{22}$&$\{\{BCAK\},\{ DF\}\}$&2\\\hline
					
					$C_{23}$&$\{\{CAKB\},\{ FD\}\}$&2\\\hline
					
					$C_{24}$&$\{\{CBAK\},\{ FD\}\}$&2\\\hline	
				\end{tabular}
			\end{center}
		\end{table}
	\end{minipage}
	\caption{Normal ordered diagrams of Cweb $\text{W}\,_{4}^{(4)}(1,1,2,4)$}
	\label{tab:4legsWeb2}
\end{table}

\begin{figure}[H]
	\centering
	\subfloat[][]{\includegraphics[height=4cm,width=4cm]{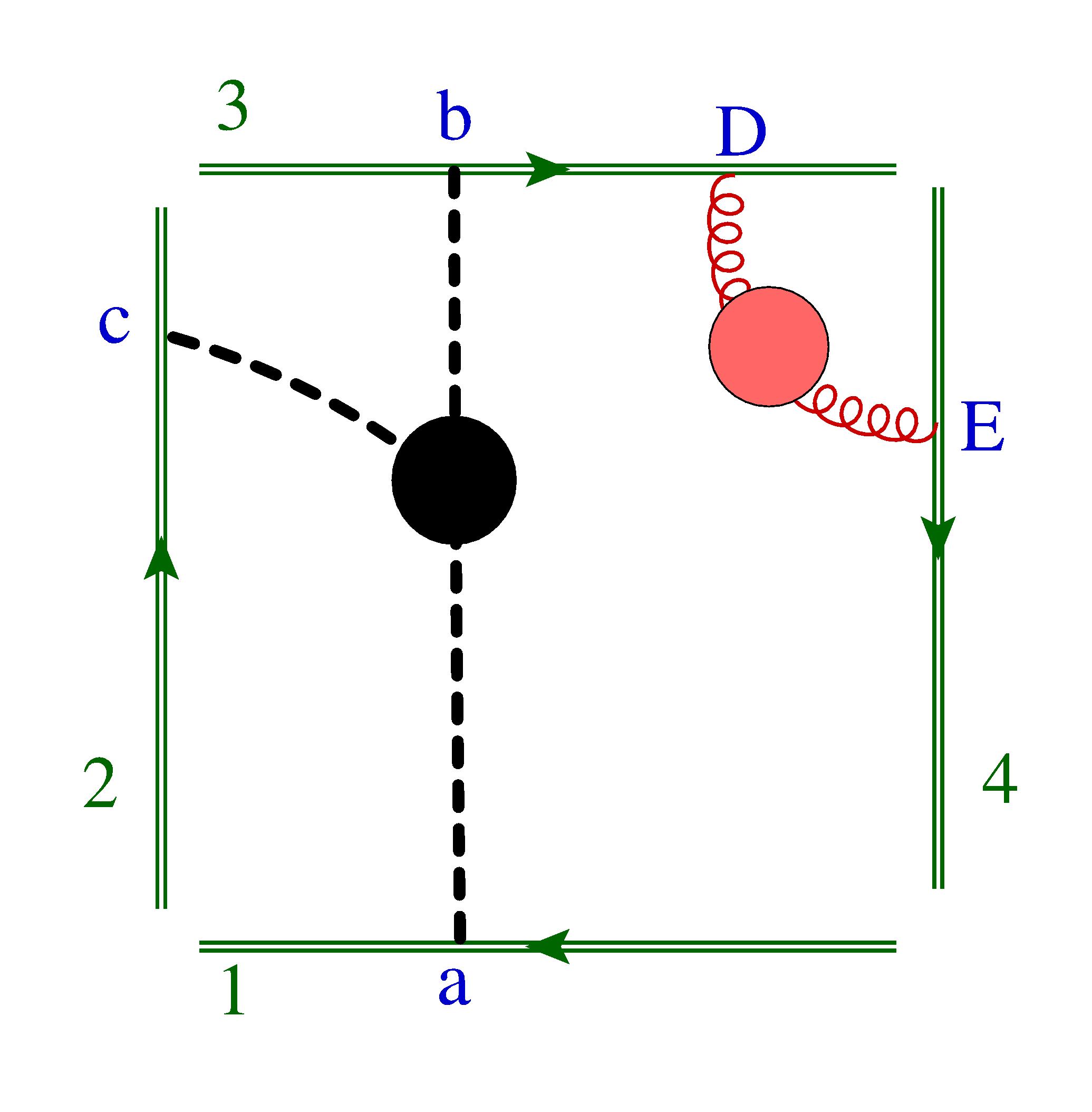} }
	\qquad 
	\subfloat[][]{\includegraphics[height=4cm,width=4cm]{4legsWeb2Av-3} }
	\qquad \qquad\qquad
	\subfloat[][]{\includegraphics[height=4cm,width=4cm]{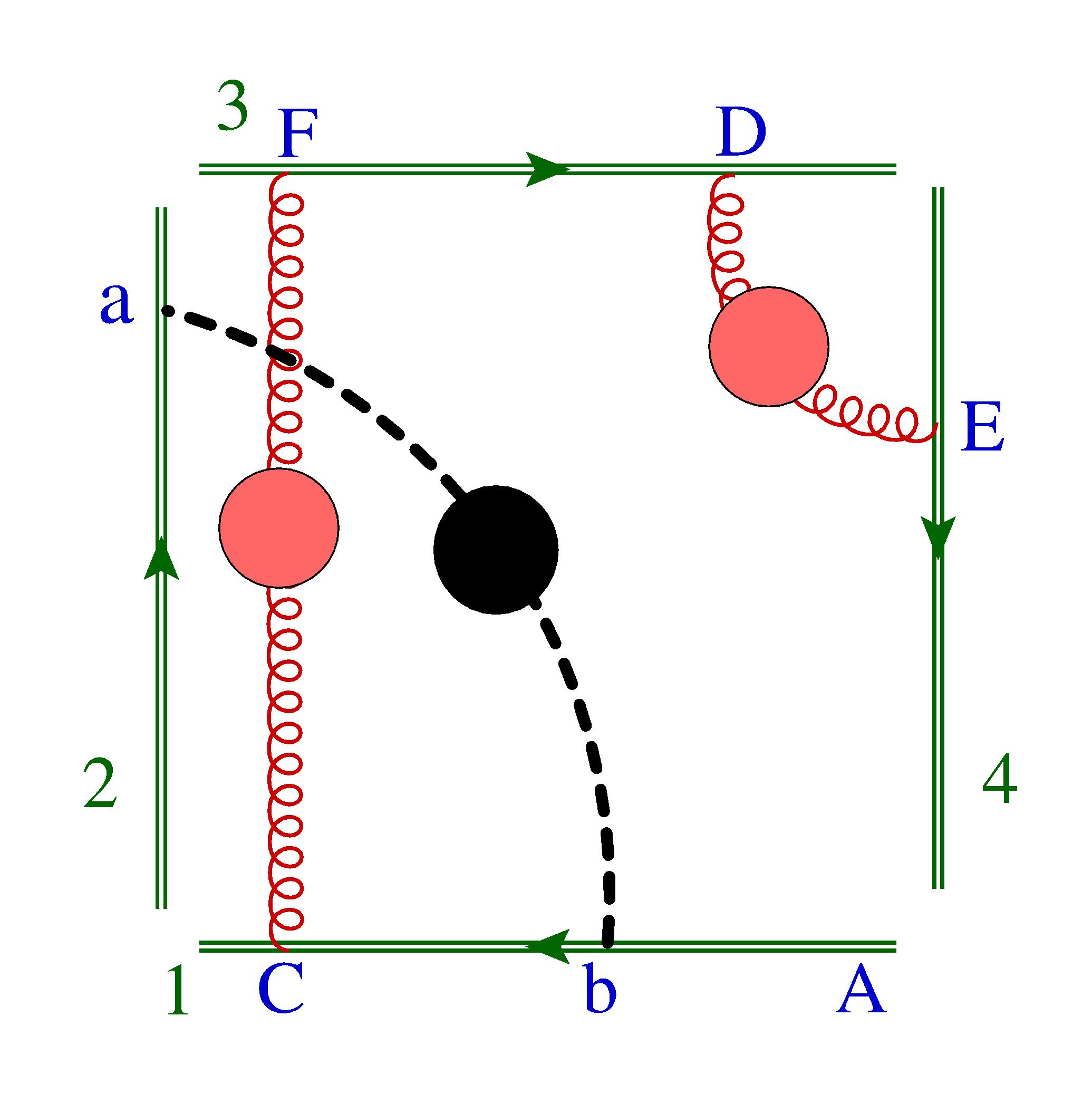} }
	\qquad 
	\subfloat[][]{\includegraphics[height=4cm,width=4cm]{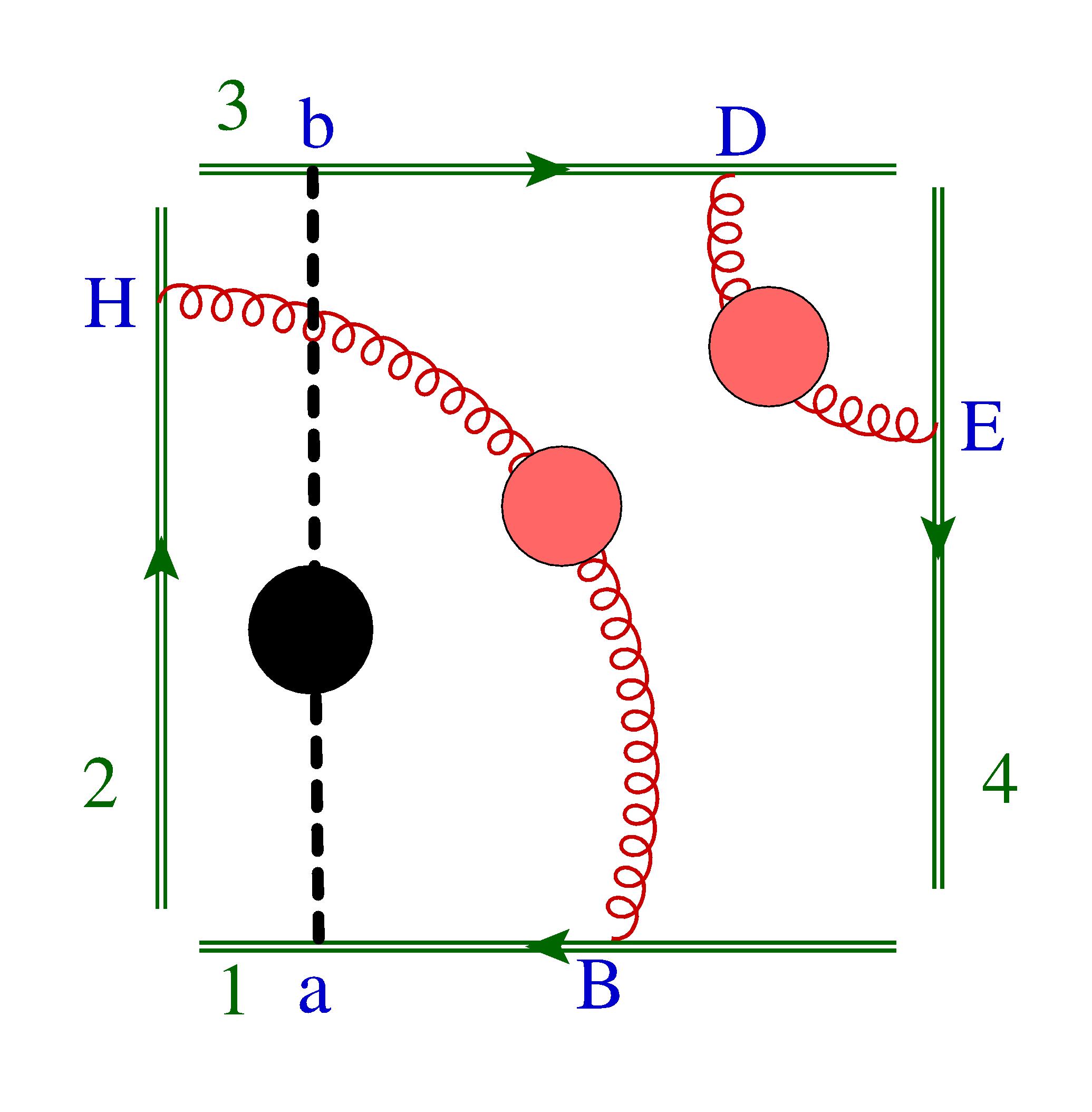} }
	\caption{\reducedWebs for Cweb $\text{W}\,_{4}^{(4)}(1,1,2,4)$}
	\label{fig:4legsWeb2-allAVAtar-WEBS}
\end{figure}

For this case, the $ s $-factors of reducible diagrams forms a column weight vector $ S=\{1_4,2_8\} $. This is a column weight vector of a known Basis web present at four loops, with mixing matrix be $ R(1_4,2_8) $. Thus the block $ D $ of this matrix is,
\begin{align}
D = R(1_4,2_8)\,\qquad r(D)=r(R(1_4,2_8))\,=\,2\,.
\end{align}
We now use procedure of Fused-Webs to determine the diagonal blocks of $ A $. In this Cweb, the identity in $ A $ does not exist as there is no completely entangled diagram. Further, there are two kind of entanglements leading to different partial entanglement. 
\begin{itemize}
	\item First, when one of the two, two-point correlator is entangled with the boomerang on line $ 1 $. These two possibilities of choosing the two point correlator leads to two distinct partial entanglements, appearing in $ C_5,C_6,C_7,C_8 $, and $ C_5,C_6,C_7,C_8 $ respectively. The Fused diagrams associated with $ C_5 $ and $ C_9 $ are shown in fig.~(\ref{fig:4legsWeb2-allAVAtar-WEBS}\txb{c}) and fig.~(\ref{fig:4legsWeb2-allAVAtar-WEBS}\txb{d}). The shuffle of correlators for each of these generate the associated two Fused-Webs. The column weight vector $ S=\{1_2,2_2\} $ and mixing matrix $ R(1_2,2_2) $ are the same for both the Fused-Webs.
	
	\item   The second kind involves entanglement of both the two point correlators along with the boomerang. Two possible order of attachments of two point correlators in between the boomerang define two distinct such partial entanglements appearing in diagrams $ C_1,C_2 $ and $ C_3,C_4 $ respectively. The Fused diagrams corresponding to $ C_1 $ and $ C_3 $ are shown in fig.~(\ref{fig:4legsWeb2-allAVAtar-WEBS}\txb{a}) and fig.~(\ref{fig:4legsWeb2-allAVAtar-WEBS}\txb{b}). The shuffle of correlators for each of these generate the associated two Fused-Webs. Both of these has column weight vector $ S=\{1_2\} $ and mixing matrix is $ R(1_2) $.
\end{itemize}   
Above discussion is mentioned in the table~\ref{tab:4legsWeb2-Ent}. It classifies the irreducible diagrams of the Cweb according to the entangled pieces.
\begin{table}
	\begin{center}
		\begin{tabular}{|c|c|c|c|c|c|}
			\hline
			Entanglement & Diagrams of  & \reducedWeb & Diagrams in  & $ s $-factors   & $ R $ \\ 
			& Cweb &  & \reducedWeb &    &  \\
			\hline 
			First Partial Entangled   & $ C_1 $, $ C_2 $,  & \ref{fig:4legsWeb2-allAVAtar-WEBS}\textcolor{blue}{a} & $ \{b,\,D\} $ & 1 & $ I_{2} $ \\ 
			&                                                                                                      & & $ \{D,\,b\} $& 1&\\
			\hline
			Second Partial Entangled  & $ C_{3} $, $ C_{4} $ & \ref{fig:4legsWeb2-allAVAtar-WEBS}\textcolor{blue}{b} & $ \{b,\,D\} $ & 1 & $ R(1_2) $ \\ 
			&  & & $ \{D,\,b\} $ & 1 & \\ \hline
			Third Partial Entangled  & $ C_{5} $, $ C_{6} $ & \ref{fig:4legsWeb2-allAVAtar-WEBS}\textcolor{blue}{c} & $\{\{b,\,C\},\{ F,\,D\}\}$ & 1 & $ R(1_2,2_2) $ \\ 
			& $ C_{7} $, $ C_{8} $& & $ \{\{b,\,C\},\{ D,\,F\}\}$  & 2 & \\ 
			&  & & $ \{\{C,\,b\},\{ F,\,D\}\}$ & 2 & \\
			&  & & $ \{\{C,\,b\},\{ D,\,F\}\}$ & 1 & \\\hline
			Fourth Partial Entangled  & $ C_{9} $, $ C_{10} $ & \ref{fig:4legsWeb2-allAVAtar-WEBS}\textcolor{blue}{d} & $ \{\{a,\,B\},\{ F,\,D\}\} $ & 2 & $ R(1_2,2_2) $ \\ 
			&  $ C_{11} $, $ C_{12} $& & $ \{\{a,\,B\},\{ D,\,F\}\} $ & 1 & \\
			&  & & $ \{\{B,\,a\},\{ F,\,D\}\} $ & 1 & \\
			&  & & $ \{\{B,\,a\},\{ D,\,F\}\} $ & 2 & \\ \hline
		\end{tabular}	
	\end{center}
	\caption{\reducedWebs and their mixing matrices for Cweb $\text{W}\,_{4}^{(4)}(1,1,2,4)$}
	\label{tab:4legsWeb2-Ent}
\end{table}
\noindent The order of diagrams in the Cweb given in table \ref{tab:4legsWeb2}, is chosen such that diagrams with same kind of entanglement appear together. Therefore, mixing matrices of the \reducedWebs for this Cweb are present as the diagonal blocks of $ A $, given as,
\begin{align}
A\,=\,\left(  \begin{array}{cccc}
R\,(1_2) & &\cdots &\\
\vdots& R\,(1_2)& &\\
&&R(1_2,2_2) &\\
\vdots&&&R(1_2,2_2)
\end{array} \right)\,,\qquad 
\end{align}
The rank of $ A $ is,  
\begin{align}
r(A) &= 2\,\,r(R(1_2)) +  2\,\,r(R(1_2,2_2))=  4\,.\nonumber
\end{align}
The diagonal blocks of $ R $ is then given by
\begin{align}
R\,=\,\left(  \begin{array}{ccccc}
R\,(1_2) & &\cdots &&\cdots\\
\vdots& R\,(1_2)& &&\\
&&R(1_2,2_2) &&\vdots\\
\vdots&&&R(1_2,2_2)&\\
&&&&R(1_4,2_8)
\end{array} \right)\,,\qquad 
\end{align}
\noindent The number of exponentiated colour factors is the rank of the mixing matrix $ R $, which is given as
\begin{align}
r(R) = r(A)  + r(D) \;=\;4 + r(R(1_4,2_8)) = 6\,.
\end{align}

\vspace{0.5cm}
\textbf{7.}\,$\textbf{W}\,_{4}^{(4)}\,(1,2,2,3)$ \\

\begin{figure}[H]
	\centering
	\subfloat[][]{\includegraphics[height=4cm,width=4cm]{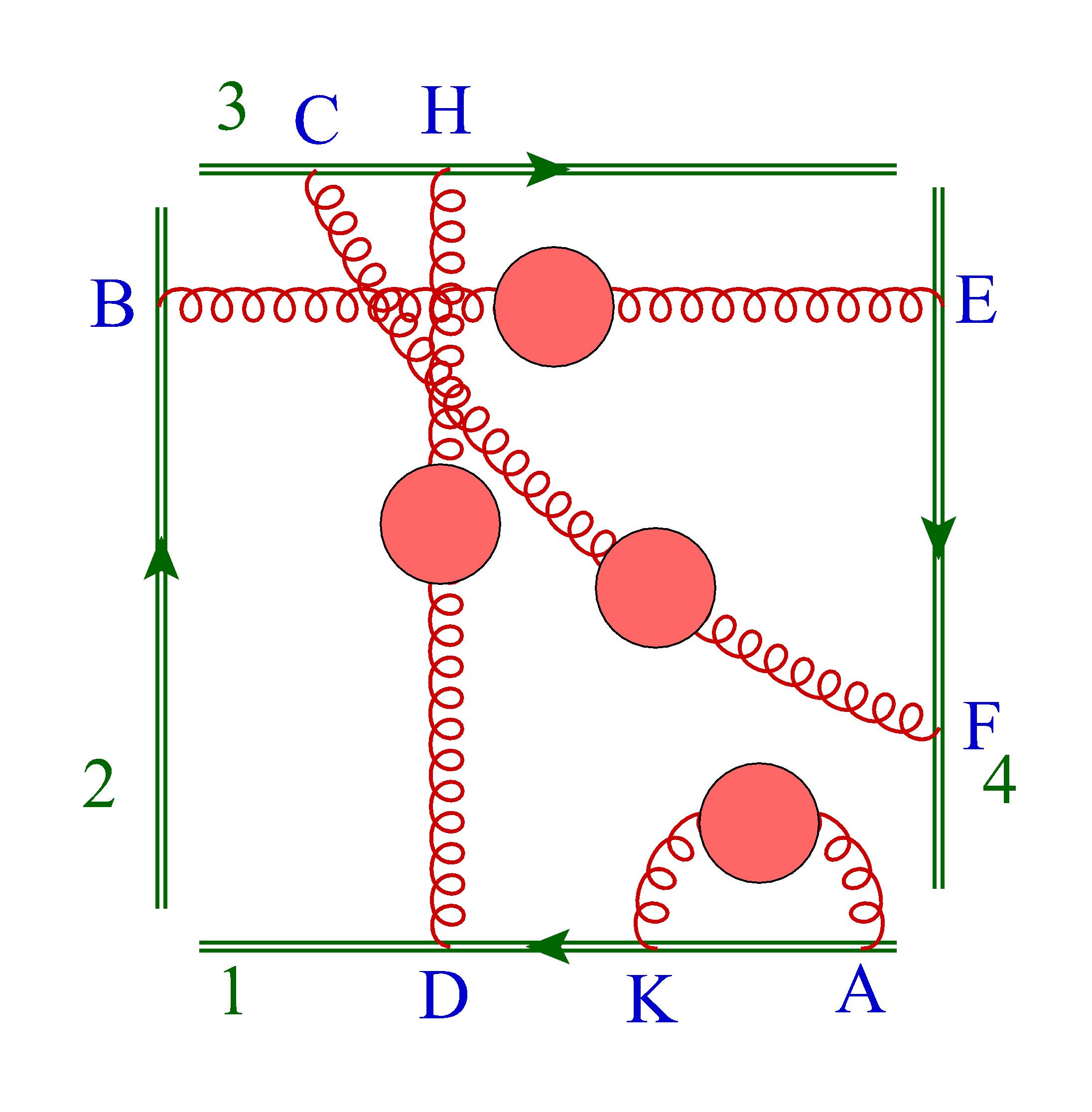} }
	\qquad
	\subfloat[][]{\includegraphics[height=4cm,width=4cm]{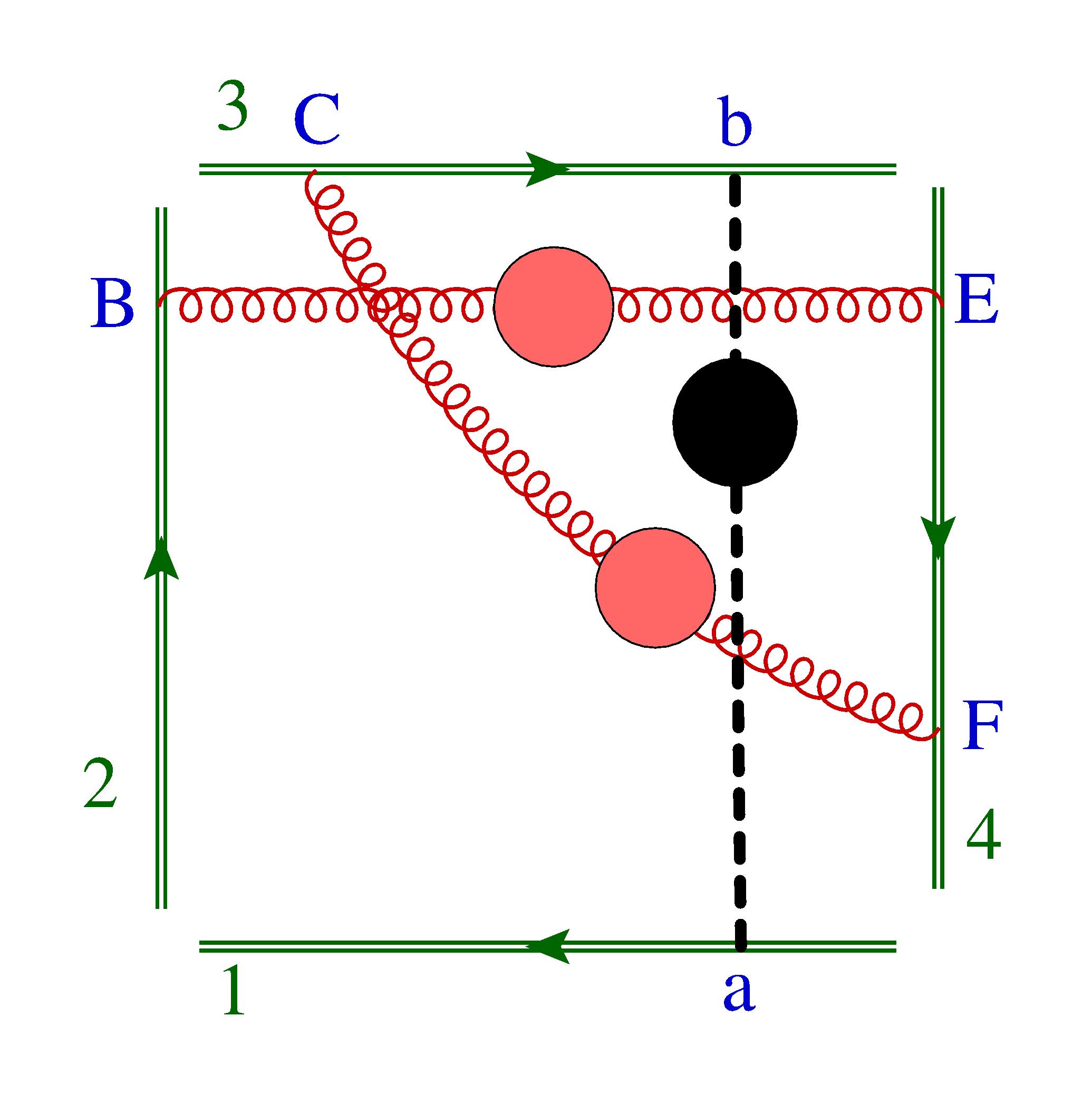} }
	\caption{Cweb $\text{W}\,_{4}^{(4)}\,(1,2,2,3)$ and the associated Fused-Web}
	\label{fig:4legsWeb3}
\end{figure}

\vspace{0.2cm}
\noindent This Cweb, shown in fig. (\ref{fig:4legsWeb3}\txb{a}) has twelve diagrams, out of which eight are reducible and remaining four are partially entangled. The Normal ordered diagrams and their $ s $-factors are shown in table \ref{tab:4legsWeb3}. 
\begin{table}[H]
	\begin{minipage}[c]{0.5\textwidth}
		\begin{table}[H]
			\begin{center}
				\begin{tabular}{|c|c|c|}
					\hline 
					\textbf{Diagrams}  & \textbf{Sequences}  & \textbf{s-factors}  \\ 
					\hline
					$C_{1}$&$\{\{ADK\},\{ CH\},\{EF\}\}$&0\\\hline
					
					$C_{2}$&$\{\{ADK\},\{ CH\},\{FE\}\}$&0\\\hline
					
					$C_{3}$&$\{\{ADK\},\{ HC\},\{EF\}\}$&0\\\hline
					
					$C_{4}$&$\{\{ADK\},\{ HC\},\{FE\}\}$&0\\\hline
					
					$C_{5}$&$\{\{AKD\},\{ HC\},\{FE\}\}$&1\\\hline
					
					$C_{6}$&$\{\{DAK\},\{ CH\},\{EF\}\}$&1\\\hline
				\end{tabular}
			\end{center}
		\end{table}
	\end{minipage}
	\hspace{1cm}
	\begin{minipage}[c]{0.5\textwidth}
		\begin{table}[H]
			\begin{center}
				\begin{tabular}{|c|c|c|}
					\hline 
					\textbf{Diagrams}  & \textbf{Sequences}  & \textbf{s-factors}  \\ 
					\hline
					$C_{7}$&$\{\{AKD\},\{ CH\},\{EF\}\}$&2\\\hline
					
					$C_{8}$&$\{\{DAK\},\{ HC\},\{FE\}\}$&2\\\hline
					
					$C_{9}$&$\{\{AKD\},\{ HC\},\{EF\}\}$&3\\\hline
					
					$C_{10}$&$\{\{DAK\},\{ CH\},\{FE\}\}$&3\\\hline
					
					$C_{11}$&$\{\{AKD\},\{ CH\},\{FE\}\}$&4\\\hline
					
					$C_{12}$&$\{\{DAK\},\{ HC\},\{EF\}\}$&4\\\hline	
				\end{tabular}
			\end{center}
		\end{table}
	\end{minipage}
	\caption{Normal ordered diagrams of Cweb $\text{W}\,_{4}^{(4)}\,(1,2,2,3)$}
	\label{tab:4legsWeb3}
\end{table}

For this case, the $ s $-factors of reducible diagrams forms a column weight vector $ S=\{1_2,2_2,3_2,4_2\} $. This is a column weight vector of a known Basis web present at four loops, with mixing matrix be $ R(1_2,2_2,3_2,4_2) $. Thus the block $ D $ of this matrix is,

\begin{align}
D = R(1_2,2_2,3_2,4_2)\,,\qquad r(D)=r(R(1_2,2_2,3_2,4_2 ))=1
\end{align}
Now we use procedure of Fused-Webs for the determination of diagonal blocks of $ A $. There is no completely entangled diagram in this Cweb which implies that identity in $ A $ does not exist. There is only one kind of entanglement involving boomerang and the two point correlator on line $ 1 $ which appears in all four partially entangled diagrams $ C_1, C_2, C_3, C_4 $. The Fused diagram corresponding to $ C_1 $ is shown in fig.~(\ref{fig:4legsWeb3}\txb{b}). The shuffle of its correlators generates the digrams of Fused-Web with $ S=\{1_2,2_2\} $, which mix through the mixing matrix $ R(1_2,2_2) $. The details of diagrams in Fused-Web, their $ s $-factors and the associated diagrams of Cweb are listed in table~\ref{tab:4legsWeb3-Ent}.  
\begin{table}[H]
	\begin{center}
		\begin{tabular}{|c|c|c|c|c|c|}
			\hline
			Entanglement & Diagrams of  & \reducedWeb & Diagrams in  & $ s $-factors   & $ R $ \\ 
			& Cweb &  & \reducedWeb &    &  \\
			\hline 
			First Partial Entangled  & $ C_{1} $, $ C_{2} $ & \ref{fig:4legsWeb3}\textcolor{blue}{b} & $\{\{ C,\,b\},\{E,\,F\}\}$ & 1 & $ R(1_2,2_2) $ \\ 
			& $ C_{3} $, $ C_{4} $& & $ \{\{ C,\,b\},\{F,\,E\}\}$  & 2 & \\ 
			&  & & $ \{\{ b,\,C\},\{E,\,F\}\}$ & 2 & \\
			&  & & $ \{\{ b,\,C\},\{F,\,E\}\}$ & 1 & \\\hline
		\end{tabular}	
	\end{center}
	\caption{\reducedWebs and their mixing matrices for Cweb $\text{W}\,_{4}^{(4)}\,(1,2,2,3)$}
	\label{tab:4legsWeb3-Ent}
\end{table}
\noindent The order of diagrams in the Cweb given in table \ref{tab:4legsWeb3}, is chosen such that all the partially entangled diagrams appear together. Therefore, mixing matrix of \reducedWeb is the block $ A $, given as, 
\begin{align}
A\,=\,R(1_2,2_2)\nonumber
\end{align}
The diagonal blocks of $ R $ is then given by,
\begin{align}
R \;=\;\left( \begin{array}{cc}
R(1_2,2_2) & \cdots\\
\cdots     & R(1_2,2_2,3_2,4_2)
\end{array}  
\right)
\end{align}
The rank of mixing matrix $ R $ is given as
\begin{align}
r(R) = r(A)  + r(D)\;=\; 1 + 1 = 2.
\end{align}

\textbf{8.}\, $ \textbf{W}^{(4)}_{4}(1,1,1,5) $\\

\begin{figure}[H]
	\captionsetup[subfloat]{labelformat=empty}
	\centering
	\subfloat[][]{\includegraphics[height=4cm,width=4cm]{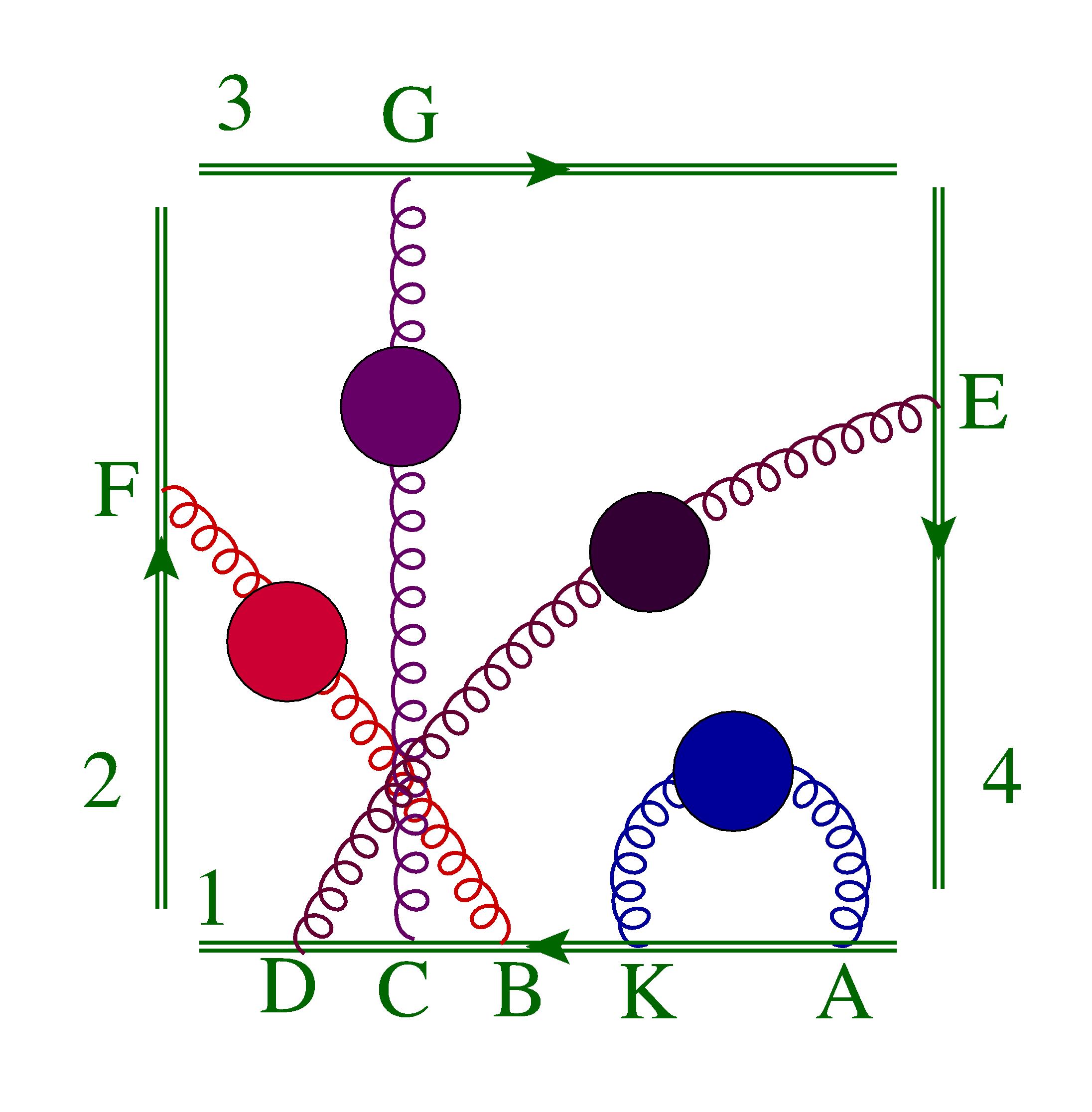} }
	\caption{Cweb $ W^{(4)}_{4}(1,1,1,5) $}
	\label{fig:six-one-web4-7-av}
\end{figure}

\vspace{0.2cm}
\noindent This Cweb, shown in fig. (\ref{fig:six-one-web4-7-av}) has sixty diagrams, out of which twenty four are reducible, six are completely entangled and other thirty are partially entangled. The Normal ordered diagrams and their $ s $-factors are shown in table \ref{tab:six-one-web4-7-av}. 

\begin{table}
	\begin{minipage}[c]{0.5\textwidth}
		\begin{table}[H]
			\begin{center}
				\begin{tabular}{|c|c|c|}
					\hline 
					\textbf{Diagrams}  & \textbf{Sequences}  & \textbf{s-factors}  \\ 
					\hline
					$C_{1}$  & $\lbrace\lbrace ABCDK\rbrace \rbrace$  & 0\\ \hline 
					$C_{2}$  & $\lbrace\lbrace ABCDK\rbrace \rbrace$  & 0 \\ \hline
					$C_{3}$  & $\lbrace\lbrace ACBDK\rbrace \rbrace$  & 0 \\ \hline
					$C_{4}$  & $\lbrace\lbrace ACDBK\rbrace \rbrace$  & 0 \\ \hline
					$C_{5}$  & $\lbrace\lbrace ADBCK\rbrace \rbrace$  & 0 \\ \hline
					$C_{6}$  & $\lbrace\lbrace ADCBK\rbrace \rbrace$  & 0 \\ \hline
					
					$C_{7}$  & $\lbrace\lbrace ABCKD\rbrace \rbrace$  & 0 \\ \hline
					$C_{8}$  & $\lbrace\lbrace DABCK\rbrace \rbrace$  & 0 \\ \hline
					
					$C_{9}$  & $\lbrace\lbrace ABDKC\rbrace \rbrace$  & 0 \\ \hline
					$C_{10}$  & $\lbrace\lbrace CABDK\rbrace \rbrace$  & 0 \\ \hline
					
					$C_{11}$  & $\lbrace\lbrace ACBKD\rbrace \rbrace$ & 0  \\ \hline 
					$C_{12}$  & $\lbrace\lbrace DACBK\rbrace \rbrace$  & 0 \\ \hline
					
					$C_{13}$  & $\lbrace\lbrace ACDKB\rbrace \rbrace$  & 0 \\ \hline
					$C_{14}$  & $\lbrace\lbrace BACDK\rbrace \rbrace$  & 0 \\ \hline
					
					$C_{15}$  & $\lbrace\lbrace ADBKC\rbrace \rbrace$  & 0 \\ \hline
					$C_{16}$  & $\lbrace\lbrace CADBK\rbrace \rbrace$  & 0 \\ \hline
					
					$C_{17}$  & $\lbrace\lbrace ADCKB\rbrace \rbrace$  & 0 \\ \hline
					$C_{18}$  & $\lbrace\lbrace BADCK\rbrace \rbrace$  & 0 \\ \hline
					
					$C_{19}$  & $\lbrace\lbrace BACKD\rbrace \rbrace$  & 0 \\ \hline
					$C_{20}$  & $\lbrace\lbrace ACKBD\rbrace \rbrace$  & 0 \\ \hline
					$C_{21}$  & $\lbrace\lbrace ACKDB\rbrace \rbrace$  & 0 \\ \hline
					$C_{22}$  & $\lbrace\lbrace BDACK\rbrace \rbrace$  & 0 \\ \hline
					$C_{23}$  & $\lbrace\lbrace DACKB\rbrace \rbrace$  & 0 \\ \hline
					$C_{24}$  & $\lbrace\lbrace DBACK\rbrace \rbrace$  & 0 \\ \hline
					
					$C_{25}$  & $\lbrace\lbrace CABKD\rbrace \rbrace$  & 0 \\ \hline	
					$C_{26}$  & $\lbrace\lbrace ABKCD\rbrace \rbrace$  & 0 \\ \hline
					$C_{27}$  & $\lbrace\lbrace ABCDC\rbrace \rbrace$  & 0 \\ \hline
					$C_{28}$  & $\lbrace\lbrace CDABK\rbrace \rbrace$  & 0 \\ \hline
					$C_{29}$  & $\lbrace\lbrace DABKC\rbrace \rbrace$  & 0 \\ \hline
					$C_{30}$  & $\lbrace\lbrace DCABK\rbrace \rbrace$  & 0 \\ \hline

				\end{tabular}
			\end{center}
		\end{table}
	\end{minipage}
	\hspace{1cm}
	\begin{minipage}[c]{0.5\textwidth}
		\begin{table}[H]
			\begin{center}
				\begin{tabular}{|c|c|c|}
					\hline 
					\textbf{Diagrams}  & \textbf{Sequences}  & \textbf{s-factors}  \\ 
					\hline
					$C_{31}$  & $\lbrace\lbrace ADKBC\rbrace \rbrace$  & 0 \\ \hline
					$C_{32}$  & $\lbrace\lbrace ADKCB\rbrace \rbrace$  & 0 \\ \hline
					$C_{33}$  & $\lbrace\lbrace BADKC\rbrace \rbrace$  & 0 \\ \hline
					$C_{34}$  & $\lbrace\lbrace BCADK\rbrace \rbrace$  & 0 \\ \hline
					$C_{35}$  & $\lbrace\lbrace CADKB\rbrace \rbrace$  & 0 \\ \hline
					$C_{36}$  & $\lbrace\lbrace CBADK\rbrace \rbrace$  & 0 \\ \hline
					$C_{37}$  & $\lbrace\lbrace AKBCD\rbrace \rbrace$  & 1 \\ \hline
					$C_{38}$  & $\lbrace\lbrace AKBDC\rbrace \rbrace$  & 1 \\ \hline
					$C_{39}$  & $\lbrace\lbrace AKCBD\rbrace \rbrace$  & 1 \\ \hline
					$C_{40}$  & $\lbrace\lbrace AKCDB\rbrace \rbrace$  & 1 \\ \hline
					$C_{41}$  & $\lbrace\lbrace AKDBC\rbrace \rbrace$  & 1 \\ \hline
					$C_{42}$  & $\lbrace\lbrace AKDCB\rbrace \rbrace$  & 1 \\ \hline
					$C_{43}$  & $\lbrace\lbrace BAKCD\rbrace \rbrace$  & 1 \\ \hline
					$C_{44}$  & $\lbrace\lbrace BAKDC\rbrace \rbrace$  & 1 \\ \hline
					$C_{45}$  & $\lbrace\lbrace BCAKD\rbrace \rbrace$  & 1 \\ \hline
					$C_{46}$  & $\lbrace\lbrace BCDAK\rbrace \rbrace$  & 1 \\ \hline
					$C_{47}$  & $\lbrace\lbrace BDAKC\rbrace \rbrace$  & 1 \\ \hline
					$C_{48}$  & $\lbrace\lbrace BDCAK\rbrace \rbrace$  & 1 \\ \hline
					$C_{49}$  & $\lbrace\lbrace CAKBD\rbrace \rbrace$  & 1 \\ \hline
					$C_{50}$  & $\lbrace\lbrace CAKDB\rbrace \rbrace$  & 1 \\ \hline
					$C_{51}$  & $\lbrace\lbrace CBAKD\rbrace \rbrace$  & 1 \\ \hline
					$C_{52}$  & $\lbrace\lbrace CBDAK\rbrace \rbrace$  & 1 \\ \hline
					$C_{53}$  & $\lbrace\lbrace CDAKB\rbrace \rbrace$  & 1 \\ \hline
					$C_{54}$  & $\lbrace\lbrace CDBAK\rbrace \rbrace$  & 1 \\ \hline
					$C_{55}$  & $\lbrace\lbrace DAKBC\rbrace \rbrace$  & 1 \\ \hline
					$C_{56}$  & $\lbrace\lbrace DAKCB\rbrace \rbrace$  & 1 \\ \hline
					$C_{57}$  & $\lbrace\lbrace DBAKC\rbrace \rbrace$  & 1 \\ \hline
					$C_{58}$  & $\lbrace\lbrace DBCAK\rbrace \rbrace$  & 1 \\ \hline
					$C_{59}$  & $\lbrace\lbrace DCAKB\rbrace \rbrace$  & 1 \\ \hline
					$C_{60}$  & $\lbrace\lbrace DCBAK\rbrace \rbrace$  & 1 \\ \hline
				\end{tabular}
			\end{center}
		\end{table}
	\end{minipage}
	\caption{Normal ordered diagrams of Cweb $ W^{(4)}_{4}(1,1,1,5) $}
	\label{tab:six-one-web4-7-av}
\end{table}
For this case, the $ s $-factors of reducible diagrams forms a column weight vector $ S=\{1_{24}\} $. This is a column weight vector of a known Basis web present at four loops, with mixing matrix be $ R(1_{24}) $. Thus the block $ D $ of this matrix is,
\begin{align}
D\;=\;R(1_{24})\,,\qquad\qquad r(D)=r(R(1_{24}))\;=6\;
\end{align}
\begin{figure}
	\captionsetup[subfloat]{labelformat=empty}
	\centering
	\subfloat[][(a)]{\includegraphics[height=4cm,width=4cm]{RedB4LWany} }
	\qquad 
	\subfloat[][(b)]{\includegraphics[height=4cm,width=4cm]{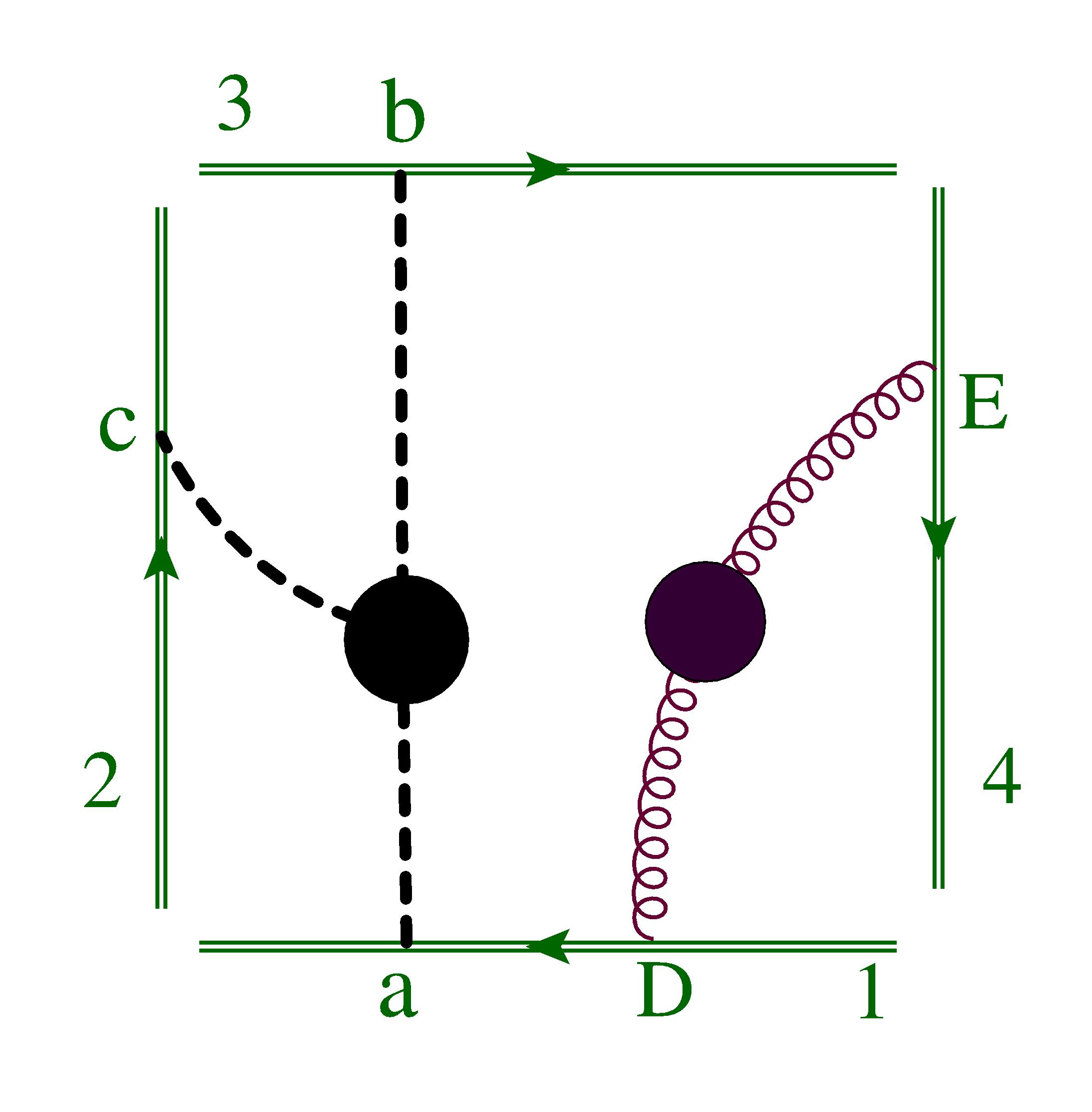} }
	\qquad 
	\subfloat[][(c)]{\includegraphics[height=4cm,width=4cm]{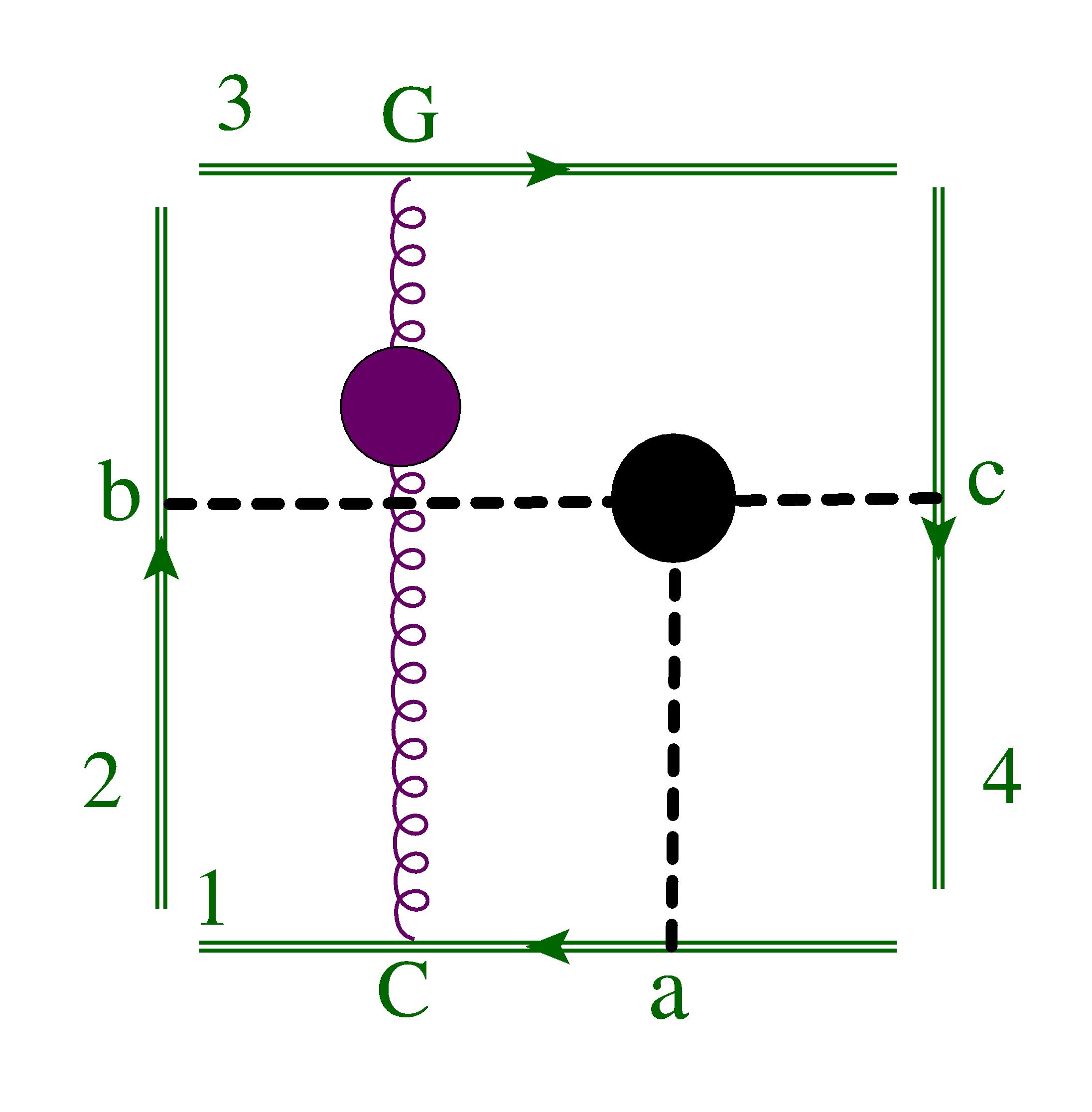} }
	\qquad 
	\subfloat[][(d)]{\includegraphics[height=4cm,width=4cm]{RedB4LW7Ent1-3} }
	\qquad 
	\subfloat[][(e)]{\includegraphics[height=4cm,width=4cm]{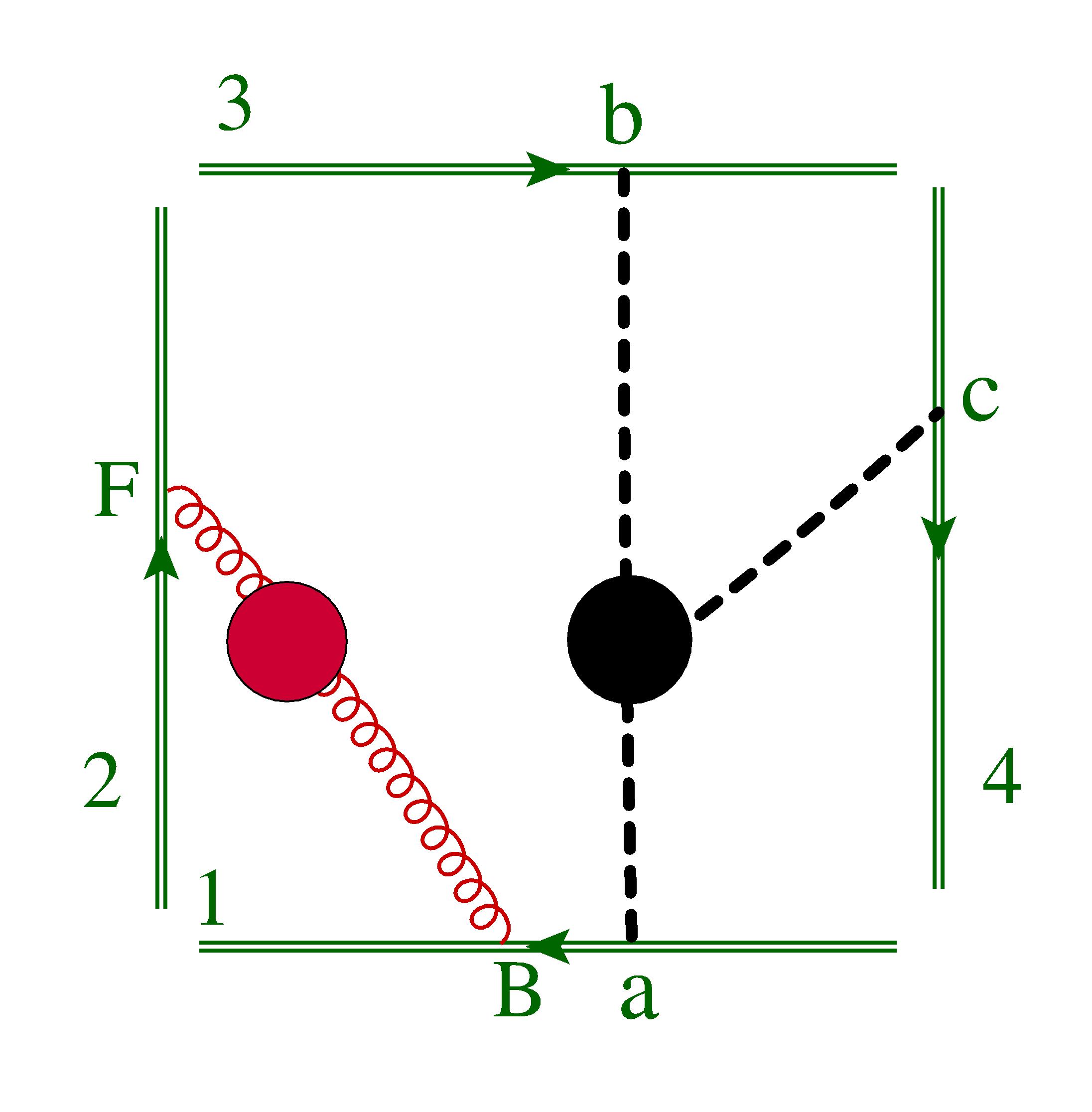} }
	\qquad 
	\subfloat[][(f)]{\includegraphics[height=4cm,width=4cm]{RedB4LW7Ent2-5} }
	\qquad 
	\subfloat[][(g)]{\includegraphics[height=4cm,width=4cm]{RedB4LW7Ent4-6} }
	\qquad 
	\subfloat[][(h)]{\includegraphics[height=4cm,width=4cm]{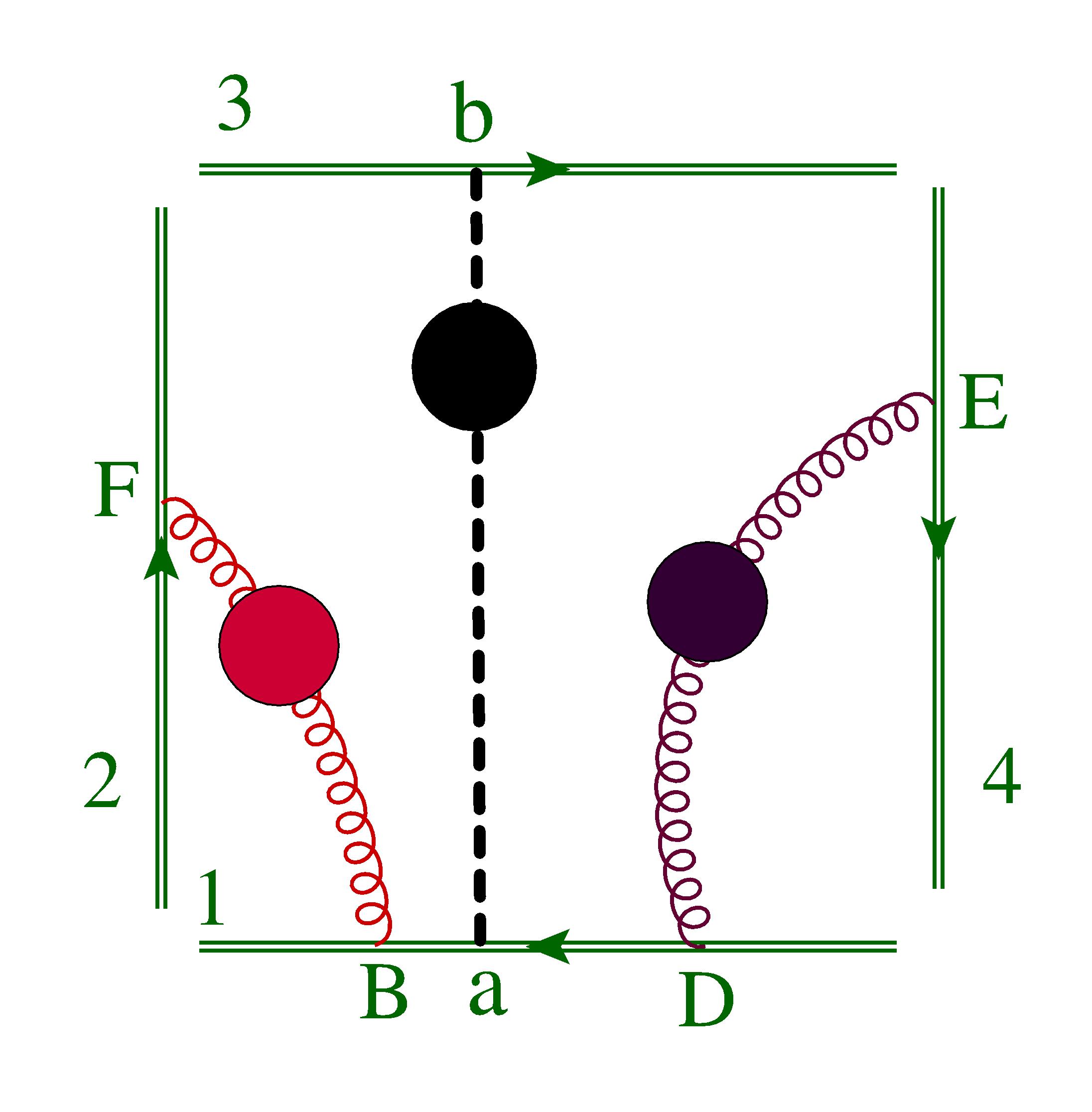} }
	\qquad 
	\subfloat[][(i)]{\includegraphics[height=4cm,width=4cm]{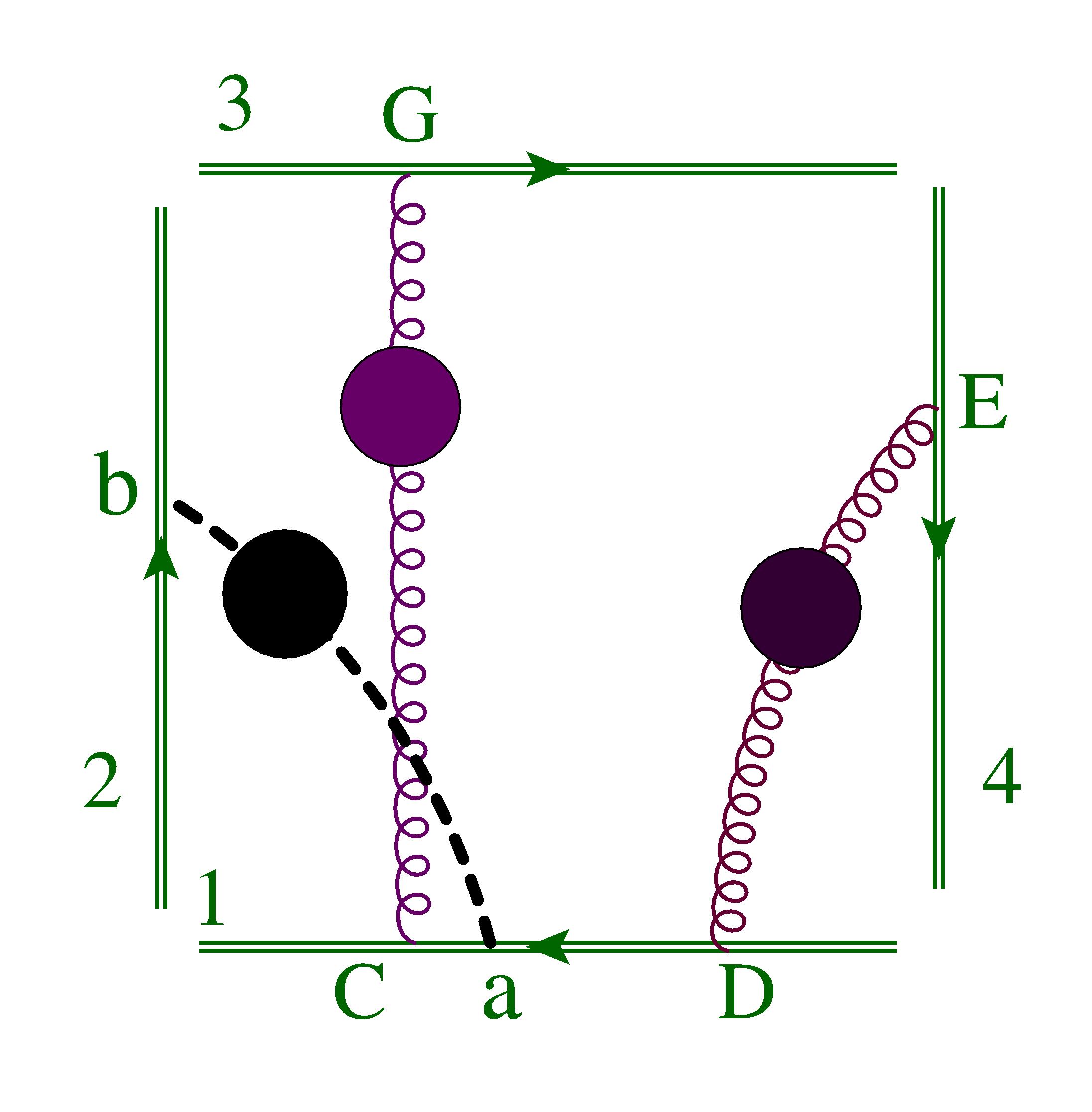} }
	\qquad 
	\subfloat[][(j)]{\includegraphics[height=4cm,width=4cm]{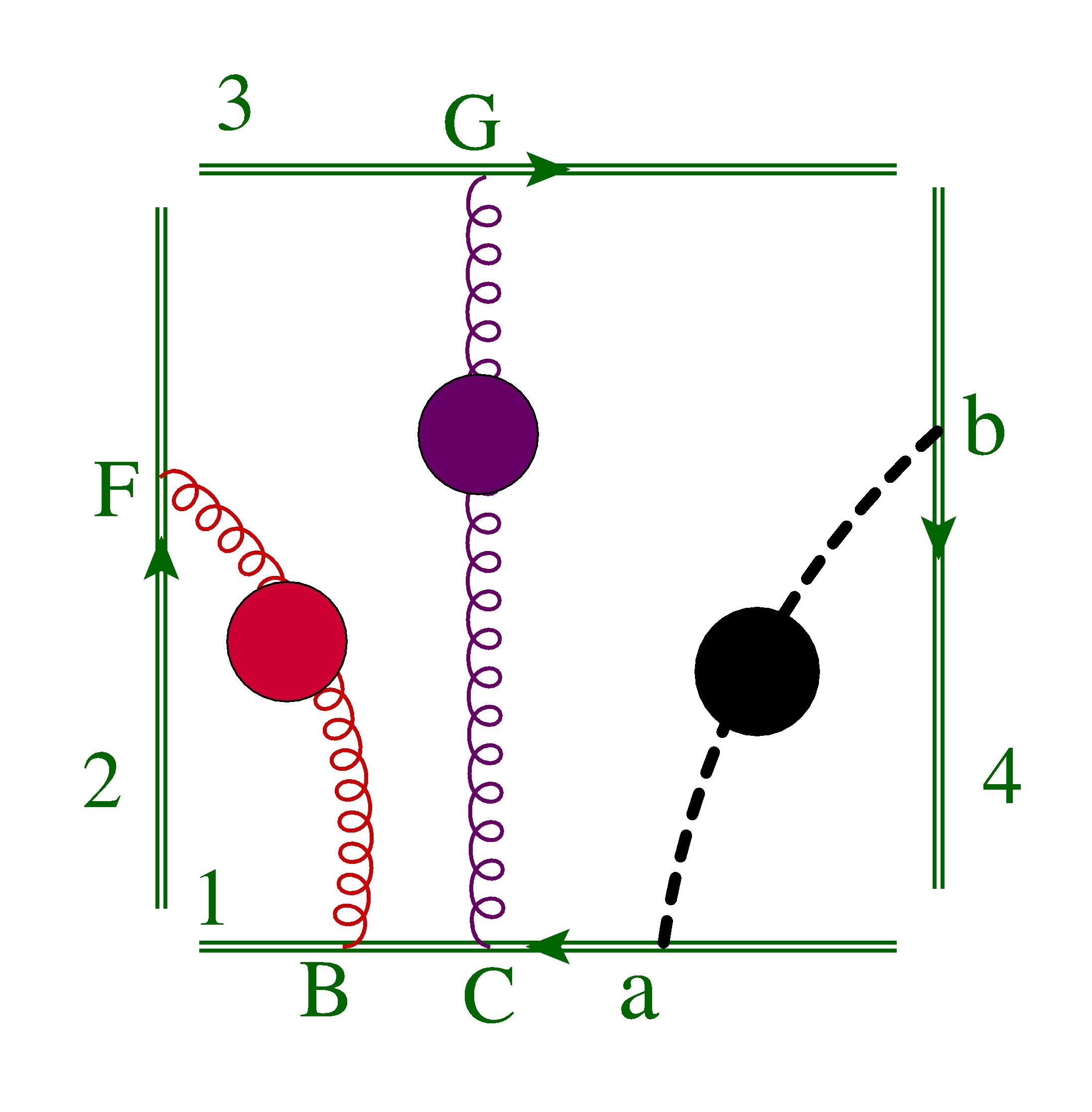} }
	
	\caption{\reducedWebs for Cweb $ W^{(4)}_{4}(1,1,1,5) $ }
	\label{fig:six-one-web4-7-allAVAtar-WEBS}
\end{figure}
We now proceed to determine the diagonal blocks of $ A $ starting with identity matrix. The number of completely entangled digrams is the order of identity matrix. In order to be completely entangled every attachment of two point correlators on line $ 1 $ must be in between the attachments of boomerang. Also, these three attachments can be ordered in $ 3! $ ways, inside boomerang. Thus the order of identity matrix is six which correspond to the first six diagrams of the Cweb with order of attachments given in table~\ref{tab:six-one-web4-4-av}. 
\\
Now we consider partially entangled diagrams. There are two kind of entanglements that lead to such diagrams:
\begin{itemize}
	\item Two of the three two point correlators are entangled with boomerang on line $ 1 $. These two correlators can be chosen in $ \,^3C_2 $ ways. Also, attachments of each pair can be further ordered in two ways inside boomerang. These two are distinct entanglements. In this ways we have following distinct entanglements of this kind
	\begin{align}
	\,^3C_2\times\,2 \nonumber
	\end{align}
	These six are associated with $ \{C_7,C_8\} $, $ \{C_9,C_{10}\} $, $ \{C_{11},C_{12}\} $, $ \{C_{13},C_{14}\} $, $ \{C_{15},C_{16}\} $ and $ \{C_{17},C_{18}\} $. the Fused diagrams for first diagrams of each of these pairs are shown in figs.~(\ref{fig:six-one-web4-5-allAVAtar-WEBS}\txb{b}), (\ref{fig:six-one-web4-5-allAVAtar-WEBS}\txb{c}), (\ref{fig:six-one-web4-5-allAVAtar-WEBS}\txb{d}), (\ref{fig:six-one-web4-5-allAVAtar-WEBS}\txb{e}), (\ref{fig:six-one-web4-5-allAVAtar-WEBS}\txb{f}) and (\ref{fig:six-one-web4-5-allAVAtar-WEBS}\txb{g}) respectively. The shuffle of attachments generate six Fused-Webs, one for every pair. Each of these has $ S=\{1_2\} $, and their mixing matrix is $ R(1_2) $.
	
	\item One of the three two point correlators are entangled with boomerang on line $ 1 $. There are three possibilities for this. Each of which is treated as a distinct entanglement within this kind. The three cases are associated with $ \{C_{19},C_{20},C_{21},C_{22},C_{23},C_{24}\} $, $ \{C_{25},C_{26},C_{27},C_{28},C_{29},C_{30}\} $ and $ \{C_{31},C_{32},C_{33},C_{34},C_{35},C_{36}\} $. For first diagram of each of these sets, the associated Fused diagrams are shown in figs.~(\ref{fig:six-one-web4-5-allAVAtar-WEBS}\txb{h}), (\ref{fig:six-one-web4-5-allAVAtar-WEBS}\txb{i}) and (\ref{fig:six-one-web4-5-allAVAtar-WEBS}\txb{j}). The shuffle of attachments generate three Fused-Webs, one for each of three sets. Every such Fused-Web has $ S=\{1_6\} $, and it has the mixing matrix $ R(1_6) $  
\end{itemize}
Above discussion is given in detail in table~\ref{tab:six-one-web4-7-av-Ent}. It classifies the irreducible diagrams of the Cweb according to the entanglements. This table provides the \reducedWebs with the associated mixing matrices for the Cweb. 
\begin{table}
	\begin{center}
		\begin{tabular}{|c|c|c|c|c|c|}
			\hline
			Entanglement & Diagrams of  & \reducedWeb & Diagrams in  & $ s $-factors   & $ R $ \\ 
			& Cweb &  & \reducedWeb &    &  \\
			\hline
			Complete entangled  & $ C_{1} $, $ C_{2} $,$ C_3 $  & \ref{fig:six-one-web4-7-allAVAtar-WEBS}\textcolor{blue}{a} & - & 1 & $ I_{6} $ \\ 
			& $ C_{4} $, $ C_{5} $,$ C_6 $& & & &\\
			\hline
			First Partial Entangled  & $ C_{7} $, $ C_{8} $ & \ref{fig:six-one-web4-7-allAVAtar-WEBS}\textcolor{blue}{b} & $ \{a,\,D\} $ & 1 & $ R(1_2) $ \\ 
			&  & & $ \{D\,a\} $ & 1 & \\ \hline
			Second Partial Entangled  & $ C_{9} $, $ C_{10} $ & \ref{fig:six-one-web4-7-allAVAtar-WEBS}\textcolor{blue}{c} & $ \{C\,a\} $ & 1 & $ R(1_2) $ \\ 
			&  & & $ \{a\,C\}$ & 1 & \\ \hline
			Third Partial Entangled  & $ C_{11} $, $ C_{12} $ & \ref{fig:six-one-web4-7-allAVAtar-WEBS}\textcolor{blue}{d} & $ \{a,\,D\}$ & 1 & $ R(1_2) $ \\ 
			&  & & $ \{a,\,D\} $ & 1 & \\ \hline
			Fourth Partial Entangled  & $ C_{13} $, $ C_{14} $ & \ref{fig:six-one-web4-7-allAVAtar-WEBS}\textcolor{blue}{e} & $ \{a,\,B\} $ & 1 & $ R(1_2) $ \\ 
			&  & & $ \{B\,a\} $ & 1 & \\ \hline
			Fifth Partial Entangled  & $ C_{15} $, $ C_{16} $ & \ref{fig:six-one-web4-7-allAVAtar-WEBS}\textcolor{blue}{f} & $ \{C\,a\} $ & 1 & $ R(1_2) $ \\ 
			&  & & $ \{a\,C\} $ & 1 & \\ \hline
			Sixth Partial Entangled  & $ C_{17} $, $ C_{18} $ & \ref{fig:six-one-web4-7-allAVAtar-WEBS}\textcolor{blue}{g} & $  \{B\,a\} $ & 1 & $ R(1_2) $ \\ 
			&  & & $ \{a,\,B\} $ & 1 & \\ \hline
			Seventh Partial Entangled  & $ C_{19} $, $ C_{20} $ & \ref{fig:six-one-web4-7-allAVAtar-WEBS}\textcolor{blue}{h} & $ \{B,\,a,\,D\} $ & 1 &  \\ 
			& $ C_{21} $, $ C_{22}$  & & $ \{B,\,D,\,a\} $ & 1 & \\ 
			& $ C_{23} $, $ C_{24}$ & & $ \{a,\,B,\,D\} $ & 1 & $ R(1_6) $\\ 
			&  & & $ \{a,\,D,\,B\} $ & 1 & \\ 
			&  & & $ \{D,\,a,\,B\} $ & 1 & \\ 
			&  & & $ \{D,\,B,\,a\} $ & 1 & \\ 
			\hline
			Eighth Partial Entangled  & $ C_{25} $, $ C_{26} $ & \ref{fig:six-one-web4-7-allAVAtar-WEBS}\textcolor{blue}{i} & $ \{C,\,a,\,D\} $ & 1 &  \\ 
			& $ C_{27} $, $ C_{28}$ & & $ \{C,\,D,\,a\} $ & 1 & \\ 
			& $ C_{29} $, $ C_{30}$ & & $ \{a,\,C,\,D\} $ & 1 & $ R(1_6) $\\ 
			&  & & $ \{a,\,D,\,C\} $ & 1 & \\ 
			&  & & $ \{D,\,a,\,C\} $ & 1 & \\ 
			&  & & $ \{D,\,C,\,a\} $ & 1 & \\ 
			\hline
			Ninth Partial Entangled  & $ C_{31} $, $ C_{32} $ & \ref{fig:six-one-web4-7-allAVAtar-WEBS}\textcolor{blue}{j} & $ \{C,\,a,\,B\} $ & 1 &  \\ 
			& $ C_{33} $, $ C_{34}$ & & $ \{C,\,B,\,a\}$ & 1 & \\ 
			& $ C_{35} $, $ C_{36}$ & & $ \{a,\,C,\,B\} $ & 1 & $ R(1_6) $\\ 
			&  & & $ \{a,\,B,\,C\} $ & 1 & \\ 
			&  & & $ \{B,\,a,\,C\} $ & 1 & \\ 
			&  & & $ \{B,\,C,\,a\} $ & 1 & \\  \hline
		\end{tabular}	
	\end{center}
	\caption{\reducedWebs and their mixing matrices for Cweb $ W^{(4)}_{4}(1,1,1,5) $}
	\label{tab:six-one-web4-7-av-Ent}
\end{table}
\noindent The order of diagrams in the Cweb given in table \ref{tab:six-one-web4-7-av}, is chosen such that diagrams with same kind of entanglement appear together. Therefore, mixing matrices of the \reducedWebs for this Cweb are present as the diagonal blocks of $ A $, which is given as, 
\begin{align}
A\,=\,\left(\begin{array}{c|cc}
\textbf{I}_{6} & & \cdots\\ 
\hline
\textbf{O}_{2\times 4}& & \begin{array}{ccccccccc}
R\,(1_2) & & \cdots &  & & & \cdots &  & \\
& R\,(1_2) &  & \cdots& & & &  &\vdots\\
&  & R\,(1_2) & & \cdots& & &  &\\
\vdots &  &  & R\,(1_2) & & \cdots&  & &\\
&  &  & &  R\,(1_2) & & \cdots &  &\\
&  &  & & & R\,(1_2) &  & \cdots    &   \\
\vdots &  &  & & & &  R\,(1_6)&  &\\
&  &  & & & & &  R\,(1_6) &   \\
&  &  & & & &  & & R\,(1_6) \\
\end{array}	
\end{array}\right)\,.
\end{align}
the rank of $ A $ is,
\begin{align}
r(A)\,&=\,r(\textbf{I}_6)+6\,r(R\,(1_2))+3\, r(R\,(1_6))\,=\,18\,.\nonumber
\end{align}

The diagonal blocks of $ R $ is given as,

\begin{align}
R\,=\,\left(\begin{array}{c|cc}
\textbf{I}_{6} & & \cdots\\ 
\hline
\textbf{O}_{2\times 4}& & \begin{array}{cccccccccc}
R\,(1_2) & & \cdots &  & & & \cdots & & & \\
& R\,(1_2) &  & \cdots& & & &  &&\vdots\\
&  & R\,(1_2) & & \cdots& & &  &&\\
\vdots &  &  & R\,(1_2) & & \cdots& & & &\\
&  &  & &  R\,(1_2) & & \cdots &  &&\\
&  &  & & & R\,(1_2) &  & \cdots    &&   \\
\vdots &  &  & & & &  R\,(1_6)&  &&\\
&  &  & & & & &  R\,(1_6) &  & \\
&  &  & & & &  & & R\,(1_6) &\\
&  &  & & & &  & & & R(1_{24})\\
\end{array}	
\end{array}\right)\,.
\end{align}
\noindent The number of exponentiated colour factors is the rank of the mixing matrix, which is given as
\begin{align}
r(R) &=\; r(A)  + r(R(1_{24})) \;=\; 18  + 6  = 24 
\end{align}


\section{Summary and Outlook}
The logarithm of the Soft function can be expressed as a sum over Cwebs which are sets of Feynman diagrams. This exponentiation in terms of Cwebs allows us to make predictions about the IR structures in the multiparton scattering amplitudes to all orders in the perturbation theory. The diagrams of a Cweb mix via a mixing matrix such that they select only colour factors that correspond to fully connected diagrams.

In this article we have defined boomerang Cwebs as the Cwebs that contain at least one two-point gluon correlator whose both ends are attached to the same Wilson line. The studies of these Cwebs are useful for the calculation of IR structure of scattering amplitudes that involve massive particles~\cite{Gardi:2021gzz}. 

It was shown in~\cite{Agarwal:2022wyk} that the diagonal blocks of the mixing matrices can be constructed using the basis Cwebs. We have found that there is a direct correspondence of basis Cwebs with the tree graphs used in graph theory. 
We have shown that there is a flow of data from the boomerang and non-boomerang Cwebs and vice versa, which helps in predicting the diagonal blocks of their mixing matrices. Using these concepts, we have calculated the diagonal blocks and ranks for all the boomerang Cwebs at four loops that connect four Wilson lines. 
In doing so, we have used Fused-Web formalism to  obtain the general form of four special classes of boomerang Cwebs by determining the types of entanglements and number of diagrams belonging to each of them. 
Note that Cweb and web do not differ in terms of the colour structure, thus colour factors of boomerang Cwebs will exhibit the same structure as boomerang webs; specifically the complete decoupling of  self energy diagrams
	from boomerang webs which was proven in~\cite{Gardi:2021gzz} 
	using combinatorics will hold true for 
	Cwebs as well.
We believe that the general structure of the diagonal blocks of several other classes of Cwebs at all order in the perturbation theory can be obtained following the method described in this paper.  

\section*{Acknowledgements} 

\noindent 
SP would like to thank Physical Research Laboratory (PRL), Department of Space, Govt. of India, for a PDF 
fellowship, AS would like to thank CSIR, Govt. of India, for an SRF fellowship (09/1001(0075)/2020-EMR-I). 
\appendix
\section*{Appendix}
\section{Replica trick}
\label{sec:repl}
One of the powerful techniques in the combinatorial problems in physics, which involves exponentiation is the  replica trick \cite{MezaPariVira}. For Wilson line correlators, the replica trick algorithm was developed in \cite{Gardi:2010rn,Laenen:2008gt}. The same replica trick was adopted in \cite{Agarwal:2020nyc,Agarwal:2021him} for the calculation of the mixing matrices for four-loop Cwebs \cite{Agarwal:2020nyc,Agarwal:2021ais}. Here, we briefly discuss the replica trick algorithm, which was used in the calculation of the mixing matrices for Cwebs at four loops. To start with, we consider the path integral of the Wilson line correlators as, 
\begin{align}
	\mathcal{S}_n(\gamma_i)=\,\int \mathcal{D}A_\mu^a\,\exp(iS(A_\mu ^a)) \prod _{k=1}^n\phi_k(\gamma_k)=\exp[\mathcal{W}_n(\gamma_i)]\,
\end{align}
where $ S(A_\mu ^a) $ is the classical action of the gauge fields. In order to proceed with the replica trick algorithm, one introduces $ N_r $ non-interacting identical copies of each gluon field $ A_\mu $, which means, we replace each $ A_\mu $ by $ A_\mu ^i $, where,  $ i=1,\ldots, N_r $. Now, for each replica, we associate a copy of each Wilson line, thereby, replacing each Wilson line by a product of $ N_r $ Wilson lines. Thus, in the replicated theory, the path integral of the Wilson line correlator can then be written as,
\begin{align}
	{\cal S}_n^{\, {\rm repl.}} \left( \gamma_i \right) \, = \,   \Big[ 
	{\cal S}_n \left( \gamma_i \right) \Big]^{N_r} \, = \, \exp \Big[ N_r \,
	{\cal W}_n (\gamma_i) \Big] \, =  \, {\bf 1} + N_r \, {\cal W}_n (\gamma_i) 
	+ {\cal O} (N_r^2) \, .
	\label{exprepl}
\end{align}
Now, using this equation, one can calculate $ \mathcal{W}_n $ by calculating $ \mathcal{O}(N_r) $ terms of the Wilson line correlator in the replicated theory. The method of replicas involves five steps, which are summarized below. 
\begin{enumerate}
	\item [-] Associate a replica number to each connected gluon correlator in a Cweb. 
	\item[-]  Define a replica ordering operator $ \textbf{R} $, which acts on the colour generators on each Wilson line and order them according to their replica numbers. Thus, if $ \textbf{T}_i $ denotes a colour generator for a correlator belonging to replica number $ i $, then action of $ \textbf{R} $ on $ \textbf{T}_i \textbf{T}_j$ preserves the order for $ i\leq j $, and reverses the order for $ i>j $. Thus, replica ordered colour factor for a diagram in a Cweb will always be a diagram of the same Cweb.
	\item [-] The next step in order to calculate the exponentiated colour factors, one needs to find the hierarchies between the replica numbers present in a Cweb. If a Cweb has $ m $ connected pieces, we call hierarchies $ h(m) $. $ h(m) $ are known as Bell number or Fubini number \cite{IntSeq} in the number theory and combinatorics. The first few Fubini numbers are given by $ h(m)=\{1,1,3,13,75,541\} $ for $ m= 0,1,2,3,4,5$. At four loops, the highest number of correlator in a Cweb is $ m_{\text{max}}=4 $, which corresponds to $ h_{\text{max}}=75 $   
	\item [-]  The next object is to calculate $ M_{N_r}(h) $, which counts the number of appearances of a particular hierarchy in the presence of $ N_r $ replicas.  For a given hierarchy $ h $, which contains $ n_r(h) $ distinct replicas, the multiplicity $ M_{N_r}(h) $ is given by, 
	\begin{align}
		M_{N_r}(h) \, = \, \frac{N_r!}{\big( N_r - n_r(h) \big)! \,\, n_r(h)!}  \,
	\end{align}  
	\item [-] The exponentiated colour factor for a diagram $ d $ is then given by, 
	\begin{align}
		C_{N_r}^{\, {\rm repl.}}  (d) \, = \, \sum_h M_{N_r} (h) \, \textbf{R} \big[ C(d) \big| h 
		\big]  \, ,
		\label{expocolf}
	\end{align}
	where $ \textbf{R} \big[ C(d) \big| h $ is the replica ordered colour factor of diagram $ d $, for hierarchy $ h $. Finally, the exponentiated colour factor for diagram $ d $ is computed by extracting the coefficient of $ \mathcal{O}(N_r) $ terms of the above equation.  	
\end{enumerate}

\newpage

\bibliographystyle{JHEP}
\bibliography{boom}
\end{document}